%% file: RDS_ResubArxiv.tex
\documentclass[aps,reprint,twocolumn,footinbib,longbibliography,superscriptaddress,nobalancelastpage]{revtex4-2}

\input{preamble_arxiv}

\begin{document}

\title{Geometric thermodynamics of reaction-diffusion systems: Thermodynamic trade-off relations and optimal transport for pattern formation}
\date{\today}

\author{Ryuna Nagayama}
\email{ryuna.nagayama@ubi.s.u-tokyo.ac.jp}
\affiliation{Department of Physics, The University of Tokyo, 7-3-1 Hongo, Bunkyo-ku, Tokyo 113-0033, Japan}
\affiliation{Universal Biology Institute, The University of Tokyo, 7-3-1 Hongo, Bunkyo-ku, Tokyo 113-0033, Japan}
\author{Kohei Yoshimura}
\affiliation{Department of Physics, The University of Tokyo, 7-3-1 Hongo, Bunkyo-ku, Tokyo 113-0033, Japan}
\affiliation{Universal Biology Institute, The University of Tokyo, 7-3-1 Hongo, Bunkyo-ku, Tokyo 113-0033, Japan}
\author{Artemy Kolchinsky}
\affiliation{ICREA-Complex Systems Lab, Pompeu Fabra University, 08003 Barcelona, Spain}
\affiliation{Universal Biology Institute, The University of Tokyo, 7-3-1 Hongo, Bunkyo-ku, Tokyo 113-0033, Japan}
\author{Sosuke Ito}
\affiliation{Department of Physics, The University of Tokyo, 7-3-1 Hongo, Bunkyo-ku, Tokyo 113-0033, Japan}
\affiliation{Universal Biology Institute, The University of Tokyo, 7-3-1 Hongo, Bunkyo-ku, Tokyo 113-0033, Japan}

\begin{abstract}
\add{We establish universal relations between pattern formation and dissipation with a geometric approach to nonequilibrium thermodynamics of deterministic reaction-diffusion systems. We first provide a way to systematically decompose the entropy production rate based on the orthogonality of thermodynamic forces, in this way identifying the amount of dissipation caused by each factor. This enables us to extract the excess entropy production rate that genuinely contributes to the time evolution of patterns. We also show that a similar geometric method further decomposes the entropy production rate into detailed contributions, e.g., the dissipation from each point in real or wavenumber space. Second, we relate the excess entropy production rate to the details of the change in patterns through two types of thermodynamic trade-off relations for reaction-diffusion systems: thermodynamic speed limits and thermodynamic uncertainty relations. The former relates dissipation and the speed of pattern formation, and the latter bounds the excess entropy production rate with partial information on patterns, such as specific Fourier components of concentration distributions. In connection with the derivation of the thermodynamic speed limits, we also extend optimal transport theory to reaction-diffusion systems, which enables us to measure the speed of the time evolution. This extension of optimal transport also solves the minimization problem of the dissipation associated with the transition between two patterns, and it constructs energetically efficient protocols for pattern formation. We numerically demonstrate our results using chemical traveling waves in the Fisher--Kolmogorov--Petrovsky--Piskunov equation and changes in symmetry in the Brusselator model. Our results apply to general reaction-diffusion systems and contribute to understanding the relations between pattern formation and unavoidable dissipation.}
\end{abstract}

\maketitle
\section{Introduction}

\add{\subsection{Background and motivation}}

\add{Reaction-diffusion systems (RDSs) have been used to study the formation of various spatiotemporal patterns~\cite{pearson1993complex,belousov1958periodic,huang2003dynamic,kondo2010reaction,murray2003mathematical,cantrell2004spatial,fisher1937wave} since the pioneering work of Turing~\cite{turing1952chemical}. In nature, pattern formation by RDSs achieves a variety of functions. For example, organisms use the reactions and diffusive dynamics of biomolecules for morphogenesis~\cite{kondo2010reaction,murray2003mathematical,wolpert1969positional}. Cells also accelerate biochemical reactions through phase separation, which is interpretable as an RDS~\cite{banani2017biomolecular,castellana2014enzyme}. From an engineering perspective, we can also apply RDSs to biomimetic materials~\cite{yoshida2010self}, computation~\cite{adamatzky2005reaction,gorecki2015chemical,parrilla2020programmable}, and information processing~\cite{chen2016trainable}.}


\add{In achieving such functions, it is generally crucial to minimize costs associated with the functions. One of the fundamental costs is the energy dissipation, measured with the entropy production required to change a given pattern to a desired pattern through reactions and diffusion. In particular, biological systems must reduce energy dissipation to the extent that it does not alter the desired function, since it can access only limited energy resources~\cite{qian2007phosphorylation}. To consider such a minimization of entropy production, we must understand universal relations between the time evolution of patterns and the dissipation with nonequilibrium thermodynamics of RDSs.}

\add{However, most attempts to relate the time evolution of patterns to the dissipation are limited to qualitative observations of some specific systems~\cite{Irvin1988,Hanson1974,Mahara2004,Mahara2010} and are considerably less advanced. This may be due to the historical context: nonequilibrium thermodynamics of RDSs originated in the study of dissipative structures by Prigogine and his collaborators, who focused on steady-state patterns and stability rather than the non-stationary change in patterns~\cite{glansdorff1973thermodynamic, nicolis1977self}. Unfortunately, thermodynamics cannot predict stable steady-state patterns except in some special cases~\cite{glansdorff1973thermodynamic,avanzini2024nonequilibrium}, and is replaced by methods based on dynamical systems~\cite{Kuramoto1984,cross1993pattern,rupe2024principles}. On the other hand, nonequilibrium thermodynamics itself has continued to develop away from dissipative structures. A major development was the establishment of stochastic thermodynamics, which deals with mesoscopic systems~\cite{jarzynski1997nonequilibrium,sekimoto2010stochastic,seifert2012stochastic}. Stochastic thermodynamics has developed some methods to quantitatively connect time evolution with dissipation and has considered minimum dissipation problems for stochastic systems~\cite{Proesmans2016,Gingrich2016,Rotskoff2017,Large2019,remlein2021optimality,ilker2022shortcuts,Blaber2023,Engel2023,aurell2011optimal,aurell2012refined,nakazato2021geometrical,van2021geometrical,dechant2022minimum,van2023thermodynamic,yoshimura2023housekeeping,zhong2024beyond}. More recently, developments from stochastic thermodynamics were imported into thermodynamics of chemical systems, including RDSs~\cite{falasco2018information, rao2016nonequilibrium, ge2016mesoscopic}. This has provided a thermodynamic framework for traveling waves~\cite{avanzini2019thermodynamics} and phase separation~\cite{miangolarra2023nonRDS} and reveals thermodynamic constraints on a particular class of RDSs that conserve mass~\cite{liang2022universal}. Still, the relations between the time evolution of the patterns and the dissipation, which is valid in general RDSs, have not been found.}

\add{In this study, we reveal universal relations between the time evolution of patterns and the dissipation in RDSs by applying the wisdom of stochastic thermodynamics to general deterministic RDSs. Our analysis also extends to systems driven by particle exchange with outside and external mechanical forces~\cite{mielke2011gradient,riaz2006pattern} and nonideal mixtures such as those describing phase separation~\cite{cahn1958free,avanzini2021nonequilibrium,miangolarra2023nonRDS,aslyamov2023nonideal,avanzini2024nonequilibrium}. Our first result is a decomposition of EPR according to different sources, which helps us to understand quantitatively where and how much dissipation occurs during the time evolution. One decomposition that has attracted particular attention is that of excess EPR, which becomes zero in steady state, versus housekeeping EPR, which remains positive even in steady state~\cite{oono1998steady}. Although such a decomposition is not unique~\cite{hatano2001steady,esposito2010three, maes2014nonequilibrium}, we mainly focus on the geometric decomposition because it enables us to extract the part that essentially contributes to time evolution as excess EPR~\cite{dechant2022geometric_E,dechant2022geometric_R}. Our second result is a set of thermodynamic trade-off relations, universal inequalities that connect details of time evolution and dissipation. Although various types of trade-off relations may be derived, we focus in particular on  thermodynamic speed limits (TSLs), that relate speed of time evolution and dissipation~\cite{aurell2011optimal,aurell2012refined,shiraishi2018speed,dechant2019thermodynamic,nakazato2021geometrical,van2021geometrical,dechant2022minimum,yoshimura2023housekeeping,van2023thermodynamic,kolchinsky2022information,van2023topological,miangolarra2023minimal,chennakesavalu2023unified}, and thermodynamic uncertainty relations (TURs), that relate precision and dissipation~\cite{barato2015thermodynamic,horowitz2020thermodynamic}.}

\add{In addition, we extend optimal transport theory, which deals with the transport between probability distributions~\cite{villani2009optimal,santambrogio2015optimal}, to RDSs, where the concentration distribution changes through reactions and diffusion. This extension is directly related to the problem of minimizing dissipation in pattern formation. In particular, we focus on the Wasserstein distance, a fundamental quantity in optimal transport theory~\cite{schiebinger2019optimal,pmlr-v70-arjovsky17a,peyre2019computational} that also plays a central role in modern stochastic thermodynamics. For example, the Wasserstein distance determines the minimum dissipation~\cite{benamou2000computational,aurell2011optimal,aurell2012refined,nakazato2021geometrical,van2021geometrical, dechant2022minimum, yoshimura2023housekeeping, van2023thermodynamic} and has gained attention in various problems, such as optimal control of thermal engines~\cite{fu2021maximal,nakazato2021geometrical, miangolarra2022geometry} and information thermodynamics~\cite{nakazato2021geometrical,fujimoto2021game, van2023thermodynamic,nagase2023thermodynamically}. It also enables us to reinterpret the geometric excess/housekeeping decomposition from the perspective of optimal transport~\cite{dechant2022geometric_E,dechant2022geometric_R,yoshimura2023housekeeping}. Furthermore, measuring the speed of the time evolution with the Wasserstein distance yields various TSLs~\cite{aurell2011optimal,aurell2012refined,dechant2019thermodynamic,nakazato2021geometrical,van2021geometrical,dechant2022minimum, yoshimura2023housekeeping, van2023thermodynamic, kolchinsky2022information, van2023topological,miangolarra2023minimal} for stochastic systems.}

\add{We extend results from stochastic thermodynamics to deterministic RDSs by focusing on the common geometric structure of nonequilibrium thermodynamics, named \textit{geometric thermodynamics}~\cite{ito2023geometric}. This geometric approach makes it possible to extend results for stochastic systems to deterministic systems as follows. The geometric excess/housekeeping decomposition was originally obtained by focusing on a geometry of thermodynamic forces for Langevin systems~\cite{dechant2022geometric_E,dechant2022geometric_R,ito2023geometric}. Because the geometry of thermodynamic forces is common in broad systems, the decomposition has been extended to deterministic CRNs~\cite{yoshimura2023housekeeping} and fluid systems~\cite{yoshimura2023geometric}. The geometric interpretation of thermodynamic forces has also induced TURs for such deterministic systems~\cite{yoshimura2021thermodynamic,yoshimura2023housekeeping,kolchinsky2022information,yoshimura2023geometric}. In addition, generalization of the Wasserstein distance has led to the TSLs for deterministic CRNs by relating the excess EPR and the speed of time evolution~\cite{yoshimura2023housekeeping, van2023topological}.}

\begin{figure*}
    \centering
    \includegraphics[width=\linewidth]{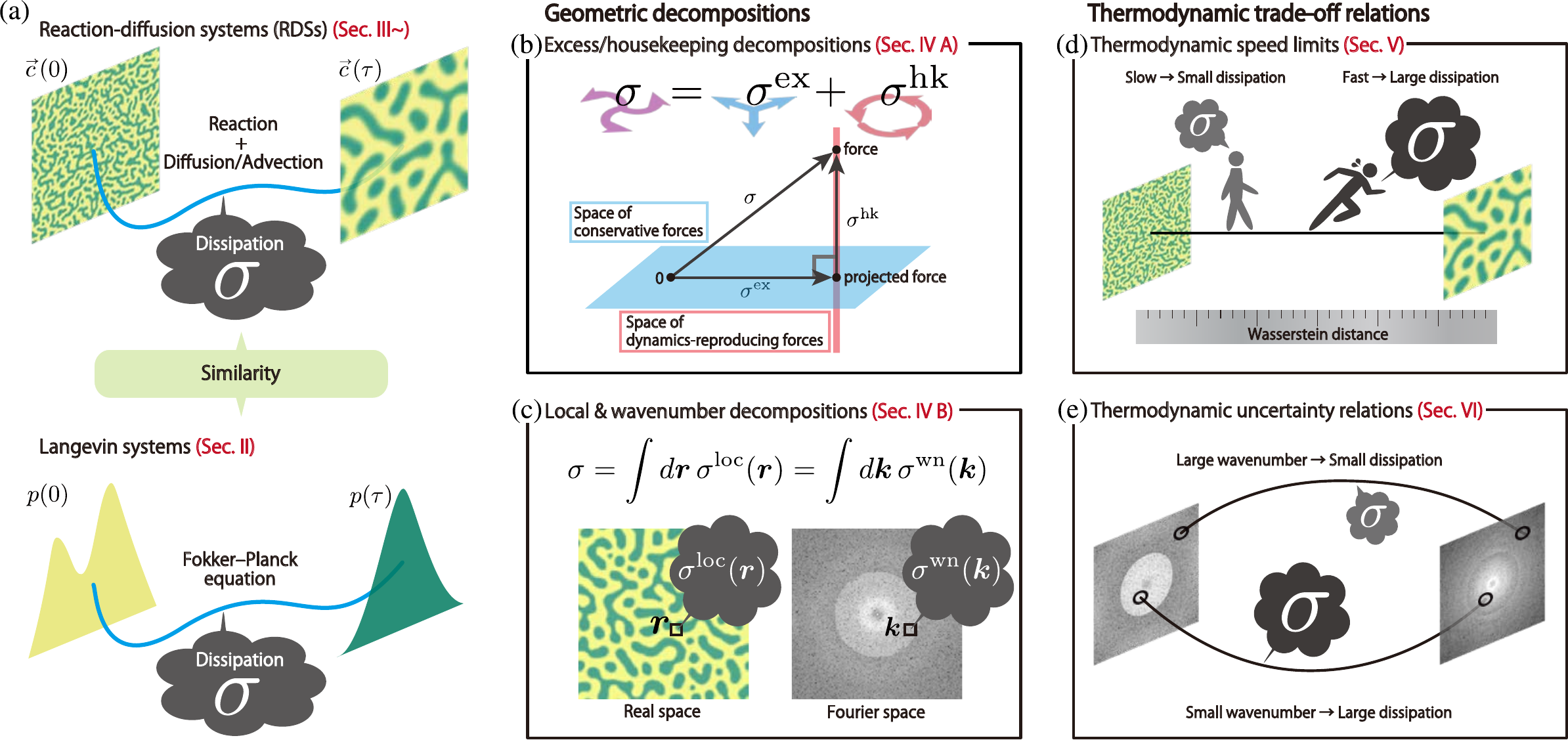}
    \caption{\add{Schematics of the results in this paper. The red letters in the figure indicate the corresponding chapters. (a) The similarity between RDSs and Langevin systems. For Langevin systems, we consider the time evolution from the probability $p(0)$ at time $t=0$ to the probability $p(\tau)$ at time $t=\tau$. For RDSs, we consider the time evolution from the initial concentration $\vec{c}(0)$ to the final concentration $\vec{c}(\tau)$. Dissipation is quantified by the EPR $\sigma$ for the both systems. (b) The excess/housekeeping decomposition, the local decomposition and the wavenumber decomposition. The EPR $\sigma$ is decomposed into the excess EPR $\sigma^{\rm ex}$ and the housekeeping EPR $\sigma^{\rm hk}$, which corresponds to the contributions of the conservative force and the nonconservative force, respectively. This decomposition is introduced by projecting the force onto the space of conservative forces. (c) The EPR $\sigma$ is also given by the integral of the local EPR $\sigma^{\rm loc}(\boldsymbol{r})$ in real space, and the integral of the wavenumber EPR $\sigma^{\rm wn}(\boldsymbol{k})$ in Fourier space. (d) TSLs refer to a trade-off between dissipation and speed. If the speed of the transition between initial and final states is slower, the dissipation can be smaller. The distance between initial and final states is measured by the Wasserstein distance and the minimal amount of dissipation also depends on this distance. (e) TURs refer to the trade-off between dissipation and the change of the spatial pattern. More dissipation is required to change the mode corresponding to smaller wavenumbers.}}
    \label{fig:total_schematic}
\end{figure*}


\add{\subsection{Road map}}

\add{All results in this paper are based on the similarity between Langevin systems and RDSs [Fig.~\ref{fig:total_schematic}(a)]. Before discussing the results for RDSs, we begin this paper by explaining the framework of geometric thermodynamics of Langevin systems and other relevant background, including optimal transport theory, in Sec.~\ref{sec:LS_geomthermo}. Although this part is essentially an aid to understanding the main result for RDSs, it also contains a new result re-imported from RDSs, the wavenumber decomposition of the EPR (Sec.~\ref{sec:local_wn_decomp}).}

\add{In Sec.~\ref{sec:RDS_thermo}, we introduce nonequilibrium thermodynamics of RDSs. This preliminary section also provides vector notation, inner products, and generalizations of differential operators to simplify the description and calculation (Sec.~\ref{sec:vector_fields}). The notation helps us establish an analogy between stochastic systems and deterministic RDSs. In particular, the core of the subsequent results is that the EPR is expressed as the squared norm of the thermodynamic force.}

\add{We provide a way to decompose EPR using orthogonal decompositions of the thermodynamic force in Sec.~\ref{sec:geometric_decomps}. This geometric method enables us to systematically decompose the EPR into various contributions. In particular, extending the geometric excess/housekeeping decomposition~\cite{dechant2022geometric_E,dechant2022geometric_R,ito2023geometric} to RDSs [Fig.~\ref{fig:total_schematic}(b)] is important, since the excess part extracts the dissipation that truly contributes to the time evolution. It is obtained by projecting the thermodynamic force onto the space of conservative forces, which describe the relaxation (Sec.~\ref{sec:ex&hk}). We also show that a similar approach based on orthogonality decomposes EPR into contributions from each point in the real or wavenumber space in Sec.~\ref{sec:spectral_decomp} [Fig.~\ref{fig:total_schematic}(c)]. The latter, named wavenumber decomposition, has been undiscovered even in stochastic thermodynamics. Finally, we numerically demonstrate the usefulness of the decompositions with the Fisher--Kolmogorov--Petrovsky--Piskunov (Fisher--KPP) equation and the Brusselator model, which show the appearance of a chemical traveling wave and change in the symmetry of patterns, respectively (Sec.~\ref{sec:system_for_numex}). The numerical demonstration reveals the difference in how the EPR and the excess EPR reflect the structure of the patterns and its time evolution by combining the excess/housekeeping decomposition with the local and wavenumber decompositions. In subsequent parts of this paper, we mainly study the details of the excess EPR to focus on the relations between the time evolution of patterns and the dissipation.}

\add{In Sec.~\ref{sec:RDS_Wdist&TSL}, we establish the optimal transport theory for RDSs and use it to establish TSLs [Fig.~\ref{fig:total_schematic}(d)] and minimum dissipation formulas for RDSs. In stochastic thermodynamics, two types of Wasserstein distances, the $1$- and $2$-Wasserstein distances, have provided different insights. We extend both of them to RDSs, which enables us to measure the distance between patterns and the speed of change in patterns. The extension of the $2$-Wasserstein distance is an improvement on previous attempts~\cite{liero2013gradient, liero2016optimal} (Sec.~\ref{sec:RDS_2W_dist}), and the extension of the $1$-Wasserstein distance is essentially new (Sec.~\ref{sec:RDS_1W_dist}). The $2$-Wasserstein distance leads to a series of TSLs, because it is closely related to excess EPR. It universally links the dissipation and the speed of the change in patterns. In addition, we obtain another series of TSLs based on the $1$-Wasserstein distance by constructing a new quantity that measures the intensity of diffusion and reaction consistent with thermodynamics. It gives trade-off relations between three pieces of information: the speed of the time evolution, the dissipation, and the intensity of reaction and diffusion (Sec.~\ref{sec:RDS_TSL}). We show that the lower bounds given by the TSLs for the $1$- and $2$-Wasserstein distances are the minimum dissipation achievable under some different conditions (Sec.~\ref{sec:minimumdissipation}). We also provide protocols for achieving the minimum dissipation, confirming that generalized optimal transport theory can be applied to RDSs. Finally, we numerically demonstrate the TSLs and the minimum dissipation with the two systems in Sec.~\ref{sec:numex_TSL}. In particular, we obtain several quantities related to the $1$-Wasserstein distance analytically for the Fisher--KPP equation. The content of Sec.~\ref{sec:RDS_Wdist&TSL} integrates the two different approaches recently developed in stochastic thermodynamics based on optimal transport~\cite{yoshimura2023housekeeping,van2023thermodynamic} in the setting of RDSs.}

\add{We also derive TURs, which bound the excess EPR using partial information about the system, in Sec.~\ref{sec:RDS_TUR}. In particular, we discover a myriad of TURs underlying the TUR for well-mixed CRNs~\cite{yoshimura2021thermodynamic} by considering the Fourier transform of the concentration distribution (Sec.~\ref{sec:Fourier_TUR}). These TURs reveal that more dissipation is required to change the mode corresponding to smaller wavenumbers, i.e., lower spatial frequencies [Fig.~\ref{fig:total_schematic}(e)]. We numerically demonstrate the TURs using the Brusselator model, which shows a notable change in symmetry (Sec.~\ref{sec:numex_TUR}). It confirms that the TURs give lower bounds on the EPR, reflecting the change in the spatial structure of the patterns. In addition, we also compare the TURs with the wavenumber decomposition, which gives lower bounds on the EPR depending on the wavenumber.}

\section{Background: Geometric thermodynamics for Langevin systems}\label{sec:LS_geomthermo}

\add{Before proceeding to RDSs, we briefly introduce geometric thermodynamics for Langevin systems~\cite{ito2023geometric}. Although this section mainly consists of the existing results, it also includes a novel result, the wavenumber decomposition of the EPR in Sec.~\ref{sec:local_wn_decomp}.}

In the following, we consider a Brownian particle in a $d$-dimensional Euclidean space $\mathbb{R}^d$. We assume that the temperature is homogeneous and we set the temperature and Boltzmann's constant to unity for simplicity. The following Langevin equation describes the time evolution of the position of the particle,
\begin{align}\label{Langevin_eq}
    d_t \check{\bm{r}}(t)=-D\bm{\nabla}_{\bm{r}}U(\check{\bm{r}}(t);t)+D\bm{K}^{\rm nc}(\check{\bm{r}}(t);t)+\sqrt{2D} \boldsymbol{\xi}(t),
\end{align}
where $d_t =d/dt$ stands for the time derivative, $\check{\bm{r}}(t)$ indicates the position of the Brownian particle at time $t$, $\bm{\nabla}_{\bm{r}}$ is the differential operator for spatial coordinates $\bm{r}\in\mathbb{R}^d$, $-\bm{\nabla}_{\bm{r}}U(\check{\bm{r}}(t);t)\coloneqq-\left.\bm{\nabla}_{\bm{r}}U(\bm{r};t)\right|_{\bm{r}=\check{\bm{r}}(t)}$ indicates the potential force on the particle, $\bm{K}^{\rm nc}(\check{\bm{r}}(t);t)$ indicates the nonconservative mechanical force on the particle, $D$ is the diffusion constant which is given by the mobility and temperature, and $\boldsymbol{\xi}(t)= (\xi_i(t))_{i=1}^{d}$ is the white Gaussian noise satisfying $\langle \xi_i (t)\rangle=0$ and $\langle \xi_i(t)\xi_j(t')\rangle=\delta_{ij}\delta(t-t')$. In later sections, we will use some of the symbols introduced in this section to represent their counterparts in RDSs.

\subsection{Fokker--Planck equation and entropy production rate}
The probability density $p(\bm{r};t)$ of the Brownian particle at position $\bm{r}$ at time $t$ described by the Langevin equation Eq.~\eqref{Langevin_eq} evolves according to the Fokker--Planck equation, 
\begin{align}\label{Fokker--Planck_eq}
    &\partial_t p(\bm{r};t)=-\bm{\nabla}_{\bm{r}}\cdot\bm{J}(\bm{r};t),\\[5pt]
    &\bm{J}(\bm{r};t)\coloneqq D p(\bm{r};t)\qty[-\bm{\nabla}_{\bm{r}} (U(\bm{r};t) + \ln p(\bm{r};t) )+\bm{K}^{\rm nc} (\bm{r};t) ],
\end{align}
where $\partial_t = \partial/\partial t$ stands for the partial time derivative and $\bm{J}(\bm{r};t)$ is the thermodynamic current. Here, $p(\bm{r};t)$ is the probability density and thus $\int_{\mathbb{R}^d} d \bm{r} p(\bm{r};t)=1$ and $p(\bm{r};t)\geq 0$ hold. In the following, we assume the following boundary conditions: $p(\bm{r};t)$ and its derivatives vanish as $\|\bm{r}\|\to\infty$, where $\|\cdot\|$ is the Euclidean norm.

Defining thermodynamic force $\bm{F}(\bm{r};t)$ as
\begin{align}\label{FP_force}
    \bm{F}(\bm{r};t)\coloneqq -\bm{\nabla}_{\bm{r}} (U(\bm{r};t) + \ln p(\bm{r};t) )+\bm{K}^{\rm nc} (\bm{r};t),
\end{align}
the thermodynamic current $\bm{J}$ and force $\bm{F}$ satisfy a linear relation,
\begin{align}
    \bm{J}(\bm{r};t)=\mathsf{M}(\bm{r};t)\bm{F}(\bm{r};t), 
    \label{FP_linear}
\end{align}
where the $d\times d$ positive-definite matrix $\mathsf{M}_{ij}(\bm{r};t)\coloneqq D p(\bm{r};t)\delta_{ij}$ indicates the mobility tensor. We can rewrite the Fokker--Planck equation in Eq.~\eqref{Fokker--Planck_eq} as $\partial_{t}p=-\bm{\nabla}_{\bm{r}}\cdot\qty(\mathsf{M}\bm{F})$. 

The EPR $\sigma$ for Langevin systems is written as an inner product of $\bm{J}$ and $\bm{F}$ or a squared norm of $\bm{F}$,
\begin{align}
    \sigma&\coloneqq\fip{\bm{J}}{\bm{F}}=\fip{\mathsf{M}\bm{F}}{\bm{F}}=\fip{\bm{F}}{\bm{F}}_{\mathsf{M}}\notag\\
    &=\int_{\mathbb{R}^d}d\bm{r}\,Dp(\bm{r};t)\|\bm{F}(\bm{r};t)\|^2,
    \label{FP_EPR}
\end{align}
where the inner products $\fip{\cdot}{\cdot}$ and $\fip{\cdot}{\cdot}_{\mathsf{M}}$ are defined as $\fip{\bm{J}'}{\bm{F}'}\coloneqq\int_{\mathbb{R}^d} d\bm{r}\bm{J}'(\bm{r})\cdot\bm{F}'(\bm{r})$ and $\fip{\bm{F}'}{\bm{F}''}_{\mathsf{M}}\coloneqq\fip{\mathsf{M}\bm{F}'}{\bm{F}''}$ for all vector-valued functions $\bm{J}'(\bm{r})$, $\bm{F}'(\bm{r})$ and $\bm{F}''(\bm{r})$ that take values at $\mathbb{R}^d$. The mobility tensor $\mathsf{M}$ can be regarded as the metric tensor because $\mathsf{M}$ is positive-definite. We also introduce the EP $\Sigma_{\tau}$ as $\Sigma_{\tau}\coloneqq\int_{0}^{\tau}dt\,\sigma$.

\subsection{Geometric excess/housekeeping decomposition of entropy production rate for Langevin systems}\label{sec:FP_geodecomp}

The Langevin system in Eq.~\eqref{Langevin_eq} is driven by two contributions: one is the conservative contribution, which causes relaxation to the equilibrium state corresponding to $U(\bm{r};t)$, and the other is the nonconservative contribution due to $\bm{K}^{\mathrm{nc}}(\bm{r};t)$, which keeps the system out of equilibrium even in the steady state. The thermodynamic force $\bm{F}$ in Eq.~\eqref{FP_force} contains the two contributions, the conservative force $-\bm{\nabla}_{\bm{r}} (U(\bm{r};t) + \ln p(\bm{r};t) )$, which is the gradient of $-(U(\bm{r};t) + \ln p(\bm{r};t))$, and the nonconservative force $\bm{K}^{\mathrm{nc}}$. Thus, the EPR $\sigma=\fip{\bm{F}}{\bm{F}}_{\mathsf{M}}$ quantifies these two contributions simultaneously.

To quantify the conservative and nonconservative contributions separately, we construct the geometric decomposition of the EPR by utilizing the generalized Pythagorean theorem for the force space with the inner product $\fip{}{}_{\mathsf{M}}$:
\begin{align}\label{FP_pythag}
\fip{\bm{F}}{\bm{F}}_{\mathsf{M}}=\fip{\bm{F}^{\ast}}{\bm{F}^{\ast}}_{\mathsf{M}}+\fip{\bm{F}-\bm{F}^{\ast}}{\bm{F}-\bm{F}^{\ast}}_{\mathsf{M}},
\end{align}
which is valid when we decompose $\bm{F}$ into two orthogonal parts $\bm{F}^{\ast}$ and $\bm{F}-\bm{F}^{\ast}$, satisfying 
\begin{align}\label{FP_ortho}
    \fip{\bm{F}^{\ast}}{\bm{F}-\bm{F}^{\ast}}_{\mathsf{M}}=0.
\end{align}

The force $\bm{F}^{\ast}$ that allows for the Pythagorean theorem~\eqref{FP_pythag} is not unique~\cite{dechant2022geometric_R}. We focus on $\bm{F}^{\ast}$, which enables us to regard $\fip{\bm{F}^{\ast}}{\bm{F}^{\ast}}_{\mathsf{M}}$ and $\fip{\bm{F}-\bm{F}^{\ast}}{\bm{F}-\bm{F}^{\ast}}_{\mathsf{M}}$ as dissipation due to conservative and nonconservative forces, respectively. For this purpose, we assume that $\bm{F}^{\ast}(\bm{r};t)$ is the gradient of a potential as $\bm{F}^{\ast}(\bm{r};t)=\bm{\nabla}_{\bm{r}}\phi^{\ast}(\bm{r};t)$, inspired by the original form of the conservative force $-\bm{\nabla}_{\bm{r}} (U(\bm{r};t) + \ln p(\bm{r};t) )$. Then, we can derive the condition on $\phi^{\ast}$ as the sufficient condition for the orthogonality in Eq.~\eqref{FP_ortho},
\begin{align}\label{FP_EL_eq}
\bm{\nabla}_{\bm{r}}\cdot\qty(\mathsf{M}\bm{\nabla}_{\bm{r}}\phi^{\ast})=\bm{\nabla}_{\bm{r}}\cdot\qty(\mathsf{M}\bm{F}),
\end{align}
which let us determine $\bm{F}^{\ast}$ uniquely (see Appendix~\ref{ap:FP_deriv_EL_eq} for details).

Using the decomposition of $\bm{F}$ into the conservative part $\bm{F}^{\ast}$ and the orthogonal part $\bm{F}-\bm{F}^{\ast}$, we define the excess and housekeeping EPRs as $\sigma^{\mathrm{ex}}\coloneqq\fip{\bm{F}^{\ast}}{\bm{F}^{\ast}}_{\mathsf{M}}$ and $\sigma^{\mathrm{hk}}\coloneqq\fip{\bm{F}-\bm{F}^{\ast}}{\bm{F}-\bm{F}^{\ast}}_{\mathsf{M}}$, respectively. These EPRs are nonnegative because they are represented by the squared norm. Then, the Pythagorean theorem~\eqref{FP_pythag} is a decomposition of EPR into the excess and the housekeeping EPRs,
\begin{align} 
    \sigma = \sigma^{\mathrm{ex}} + \sigma^{\mathrm{hk}}.
    \label{Gdecomposition}
\end{align}
\add{Time integration gives a decomposition of the EP $\Sigma_{\tau}$ into the excess EP $\Sigma_{\tau}^{\mathrm{ex}}$ and the housekeeping EP $\Sigma_{\tau}^{\mathrm{hk}}$, $\Sigma_{\tau} = \Sigma^{\mathrm{ex}}_{\tau} + \Sigma^{\mathrm{hk}}_{\tau}$. Here, the excess EP $\Sigma_{\tau}^{\mathrm{ex}}$ and the housekeeping EP $\Sigma_{\tau}^{\mathrm{hk}}$ are defined as $\Sigma_{\tau}^{\mathrm{ex}}\coloneqq\int_{0}^{\tau}dt\,\sigma^{\mathrm{ex}}$ and $\Sigma_{\tau}^{\mathrm{hk}}\coloneqq\int_{0}^{\tau}dt\,\sigma^{\mathrm{hk}}$.}

\begin{figure}
    \centering
    \includegraphics[width=\linewidth]{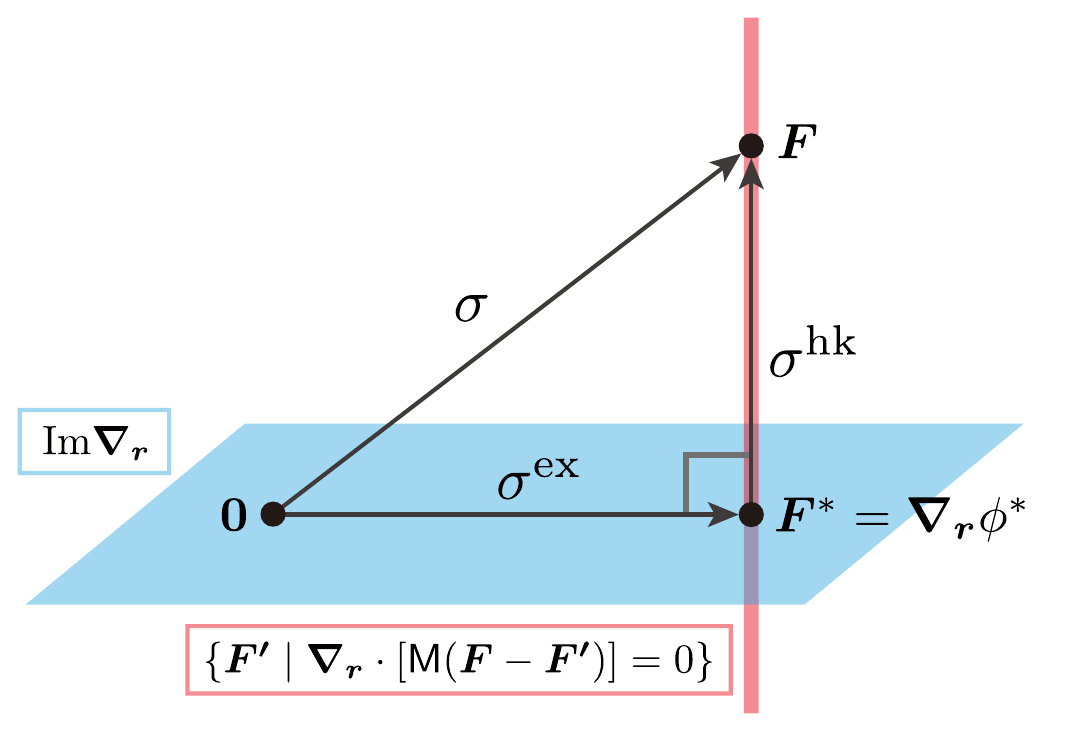}
    \caption{The geometric decomposition of the EPR for Langevin systems. The blue plane indicates $\mathrm{Im}\bm{\nabla}_{\bm{r}}$, and the red line indicates $\{ \bm{F}'|\bm{\nabla}_{\bm{r}} \cdot [\mathsf{M} (\bm{F}-\bm{F}')]= 0\} $. The thermodynamic force, whose squared norm is the EPR as $\sigma=\fip{\bm{F}}{\bm{F}}_{\mathsf{M}}$, is decomposed into the two orthogonal parts: $\bm{F}^{\ast}$ is the projection of $\bm{F}$ onto $\mathrm{Im}\bm{\nabla}_{\bm{r}}$, whose squared norm is the excess EPR as $\sigma^{\mathrm{ex}}=\fip{\bm{F^{\ast}}}{\bm{F^{\ast}}}_{\mathsf{M}}$, and the remaining part  $\bm{F}-\bm{F}^{\ast}$, whose squared norm is the housekeeping EPR as $\sigma^{\mathrm{hk}}=\fip{\bm{F}-\bm{F^{\ast}}}{\bm{F}-\bm{F^{\ast}}}_{\mathsf{M}}$. We can regard $\bm{F}^{\ast}$ as the projection of $\bm{0}$ onto the red line or of $\bm{F}$ onto the blue plane.}
    \label{fig:FP_EPR_decomp}
\end{figure}
Since the geometric decomposition~\eqref{Gdecomposition} is a generalized Pythagorean theorem, we can conceptualize the decomposition geometrically, as summarized in Fig.~\ref{fig:FP_EPR_decomp}. Let us consider the geometric nature of the excess EPR. 
The conservative force $\bm{F}^{\ast}=\bm{\nabla}_{\bm{r}}\phi^{\ast}$, whose squared norm provides the excess EPR, is uniquely given by the minimization problem
\begin{align}\label{FP_another_projected_force}
    \bm{F}^{\ast} = \argmin_{\bm{F}'| \bm{\nabla}_{\bm{r}}\cdot\qty(\mathsf{M}\bm{F}')=\bm{\nabla}_{\bm{r}}\cdot\qty(\mathsf{M}\bm{F})}\fip{\bm{F}'}{\bm{F}'}_{\mathsf{M}}, 
\end{align}
which follows from condition Eq.~\eqref{FP_EL_eq} (see Appendix~\ref{ap:FP_deriv_anotherminimization} for details).
As a result, we can rewrite the excess EPR $\sigma^{\mathrm{ex}}$ as the following variational problem
\begin{align}\label{FP_exEPR}
\sigma^{\mathrm{ex}}=\inf_{\bm{F}'|\bm{\nabla}_{\bm{r}}\cdot\qty(\mathsf{M}\bm{F}')=\bm{\nabla}_{\bm{r}}\cdot\qty(\mathsf{M}\bm{F})}\fip{\bm{F}'}{\bm{F}'}_{\mathsf{M}}.
\end{align}
The minimization problem~\eqref{FP_another_projected_force} means that $\bm{F}^*$ is the closest point to the origin $\bm{0}$ in the affine subspace $\{ \bm{F}'\mid\bm{\nabla}_{\bm{r}} \cdot [\mathsf{M} (\bm{F}-\bm{F}')]= 0\} $. The excess EPR can be seen as the shortest distance between this affine subspace and the origin.

We can also obtain a geometric interpretation of the housekeeping EPR because the set of conservative forces is the image of the gradient operator $\mathrm{Im}\,\bm{\nabla}_{\bm{r}}\coloneqq\qty{\bm{F}'\mid\exists\phi, \;\bm{F}'=\bm{\nabla}_{\bm{r}}\phi}$, and $\bm{F}-\bm{F}'$ for the element $ \bm{F}' \in \{ \bm{F}'\mid\bm{\nabla}_{\bm{r}} \cdot [\mathsf{M} (\bm{F}-\bm{F}')]= 0\} $ is in the orthogonal complement of $\mathrm{Im}\,\bm{\nabla}_{\bm{r}}$ with respect to the inner product $\fip{\cdot}{\cdot}_\mathsf{M}$. 
Consequently, the geometric decomposition~\eqref{Gdecomposition} can also be treated from the viewpoint of the projection onto the subspace $\mathrm{Im}\bm{\nabla}_{\bm{r}}$: the conservative force $\bm{F}^{\ast}$ is given by the minimization problem
\begin{align}\label{FP_projected_force}
    \bm{F}^{\ast} = \argmin_{\bm{F}'\in\mathrm{Im}\bm{\nabla}_{\bm{r}}}\fip{\bm{F}-\bm{F}'}{\bm{F}-\bm{F}'}_{\mathsf{M}}. 
\end{align}
This is derived similarly to Eq.~\eqref{FP_another_projected_force} by using the condition in Eq.~\eqref{FP_EL_eq} (see Appendix~\ref{ap:FP_deriv_anotherminimization} for details). Then, we can rewrite the housekeeping EPR $\sigma^{\mathrm{hk}}$ as the variational problem
\begin{align}\label{FP_hkEPR}
\sigma^{\mathrm{hk}}=\inf_{\phi}\fip{\bm{F}-\bm{\nabla}_{\bm{r}}\phi}{\bm{F}-\bm{\nabla}_{\bm{r}}\phi}_{\mathsf{M}},
\end{align}
which shows that the housekeeping EPR is the squared distance between the actual force and the subspace of the conservative forces. 

In addition to geometric interpretations, the constraint in Eq.~\eqref{FP_exEPR} lets us discuss the physical meaning of the decomposition.
The constraint with the Fokker--Planck equation $\partial_tp(\bm{r};t)=-\bm{\nabla}_{\bm{r}}\cdot(\mathsf{M}\bm{F})$ yields the relation
\begin{align}
\partial_{t}p=-\bm{\nabla}_{\bm{r}}\cdot\qty(\mathsf{M}\bm{F}^{\ast}),
\end{align}
so we can interpret the excess EPR as the minimum dissipation required to reproduce the original dynamics. In contrast, the housekeeping EPR reflects the dissipation caused by the cyclic current that does not affect the dynamics because $\bm{J}^{\mathrm{hk}}\coloneqq\mathsf{M}\qty(\bm{F}-\bm{\nabla}_{\bm{r}}\phi^{\ast})$ satisfies $\bm{\nabla}_{\bm{r}}\cdot\bm{J}^{\mathrm{hk}}=0$. 


If the nonconservative force $\bm{K}^\mathrm{nc}$ in Eq.~\eqref{FP_force} is absent, then the housekeeping EPR always vanishes, and the optimal potential $\phi^{\ast}$ is given by $\phi^{\ast}=-U-\ln p$. Conversely, the excess EPR vanishes when the system is in steady state since the condition in Eq.~\eqref{FP_EL_eq} reduces to $\bm{\nabla}_{\bm{r}}\cdot(\mathsf{M}\bm{F}^{\ast})=0$ in steady state, which $\bm{F}^{\ast}=\bm{0}$ solves.

\subsection{Local decomposition and wavenumber decomposition of entropy production rate for Langevin systems}\label{sec:local_wn_decomp}

The EPR is the volume integral of the positive quantity $\bm{J}(\bm{r};t)\cdot\bm{F}(\bm{r};t)$. From this viewpoint, we can decompose the dissipation at each spatial location. Similarly, it is expected that we can identify the dissipation at each wavenumber in the Fourier space. In this section, we introduce two new geometric decompositions of the EPR. One is a decomposition of the EPR into contributions from each spatial location and the other is a decomposition into contributions from each wavenumber.

\textit{Local decomposition}--- We define the local EPR as
\begin{align}\label{FP_localEPR}
    \sigma^{\mathrm{loc}}(\bm{r})\coloneqq\bm{J}(\bm{r};t)\cdot\bm{F}(\bm{r};t)\geq0,
\end{align}
which satisfies
\begin{align}\label{FP_localdecomp}
    \sigma = \int_{\mathbb{R}^d}d\bm{r}\,
    \sigma^{\mathrm{loc}}(\bm{r}).
\end{align}
The local EPR $\sigma^{\mathrm{loc}}(\bm{r})$ is nonnegative and indicates dissipation at location $\bm{r}$.

We can also decompose the excess and housekeeping EPRs as $\sigma^{\mathrm{ex}} = \int_{\mathbb{R}^d}d\bm{r}\,\sigma^{\mathrm{ex,loc}}(\bm{r})$, and $\sigma^{\mathrm{hk}} = \int_{\mathbb{R}^d}d\bm{r}\,\sigma^{\mathrm{hk,loc}}(\bm{r})$, where the local excess and housekeeping EPRs are defined as
\begin{align}
    \sigma^{\mathrm{ex,loc}}(\bm{r})&\coloneqq\bm{\nabla}_{\bm{r}}\phi^{\ast}(\bm{r};t)\cdot\mathsf{M}(\bm{r};t)\bm{\nabla}_{\bm{r}}\phi^{\ast}(\bm{r};t)\notag\\
    &=\bm{F}^{\ast}(\bm{r};t)\cdot\mathsf{M}(\bm{r};t)\bm{F}^{\ast}(\bm{r};t),\\
    \sigma^{\mathrm{hk,loc}}(\bm{r})&\coloneqq[\bm{F}(\bm{r};t)-\bm{\nabla}_{\bm{r}}\phi^{\ast}(\bm{r};t)]\notag\\
    &\phantom{\coloneqq}\cdot\mathsf{M}(\bm{r};t)[\bm{F}(\bm{r};t)-\bm{\nabla}_{\bm{r}}\phi^{\ast}(\bm{r};t)].
\end{align}
The local excess and housekeeping EPRs are nonnegative because the mobility tensor $\mathsf{M}$ is positive-definite for all $\bm{r}\in V$. Note that $\phi^{\ast}$ is a solution of the partial differential equation in Eq.~\eqref{FP_EL_eq}, which means that we need global information to obtain the local excess and housekeeping EPRs. 

Because these local excess and housekeeping EPRs are not introduced by the geometric decomposition for the local EPR $\sigma^{\mathrm{loc}}$, the geometric excess/housekeeping decomposition can be locally violated as $\sigma^{\mathrm{loc}}(\bm{r})\neq\sigma^{\mathrm{ex,loc}}(\bm{r})+\sigma^{\mathrm{hk,loc}}(\bm{r})$. In other words, there may be a non-zero cross-term $\sigma^{\mathrm{cross}}(\bm{r})\coloneqq\sigma^{\mathrm{loc}}(\bm{r})-\sigma^{\mathrm{ex,loc}}(\bm{r})-\sigma^{\mathrm{hk,loc}}(\bm{r})=2\bm{\nabla}_{\bm{r}}\phi^{\ast}\cdot\mathsf{M}[\bm{F}-\bm{\nabla}_{\bm{r}}\phi^{\ast}]$.
The cross-term may be negative or positive, but it satisfies $\int_{\mathbb{R}^d}d\bm{r}\,\sigma^{\mathrm{cross}}(\bm{r})=0$, thereby guaranteeing that the geometric excess/housekeeping decomposition holds globally as $\sigma=\sigma^{\mathrm{ex}}+\sigma^{\mathrm{hk}}$.

\textit{Wavenumber decomposition}--- Next, we decompose the EPR into nonnegative wavenumber components using Parseval's identity.
Because Parseval's identity can be regarded as a generalization of the Pythagorean theorem, we can regard this decomposition as another kind of geometric decomposition. 
We define the wavenumber EPR as
\begin{align}\label{FP_wnEPR}
\sigma^{\mathrm{wn}}(\bm{k})\coloneqq\frac{1}{(2\pi)^d}\|\hat{\bm{F}}(\bm{k};t)\|^2\geq0,
\end{align}
where we introduced the weighted Fourier transform of a vector field $\bm{F}'$ with a weight $\sqrt{\mathsf{M}_{ii}}=\sqrt{Dp(\bm{r};t)}$ defined as
\begin{align}
    \hat{F}'_{i}(\bm{k};t)\coloneqq\int_{\mathbb{R}^d}d\bm{r}\sqrt{Dp(\bm{r};t)}F'_{i}(\bm{r};t)\mathrm{e}^{-\mathrm{i}\bm{k}\cdot\bm{r}}.
\end{align}
Note that $\hat{\bm{F}}(\bm{k};t)$ is a complex vector and its Euclidean norm is defined as $\|\hat{\bm{F}}\|\coloneqq\sqrt{\sum_{i=1}^{d}\overline{\hat{F}_{i}}\hat{F}_i}$ with the overline indicating complex conjugate.
The wavenumber EPR $\sigma^{\mathrm{wn}}(\bm{k})$ provides a decomposition of EPR as
\begin{align}
    \sigma=\int_{\mathbb{R}^d}d\bm{k}\,\sigma^{\mathrm{wn}}(\bm{k}).
\end{align}
Though this can be understood as a consequence of Parseval's identity, we can show it directly as
\begin{align}\label{FP_drv_wndecomp}
    &\int_{\mathbb{R}^d}d\bm{k}\,\sigma^{\mathrm{wn}}(\bm{k})=\frac{1}{(2\pi)^d}\int_{\mathbb{R}^d}d\bm{k}\|\hat{\bm{F}}(\bm{k};t)\|^2.\notag\\
    &\quad=\int_{\mathbb{R}^d\times\mathbb{R}^d}d\bm{r}d\bm{r}'\left[\left\{\frac{1}{(2\pi)^d}\int_{\mathbb{R}^d}d\bm{k}\mathrm{e}^{\mathrm{i}\bm{k}\cdot(\bm{r}-\bm{r}')}\right\}\right.\notag\\
    &\quad\left.\phantom{\int_{\mathbb{R}^d}d\bm{k}}\times D\sqrt{p(\bm{r};t)p(\bm{r}';t)}\bm{F}(\bm{r};t)\cdot\bm{F}'(\bm{r}';t)\right]\notag\\
    &\quad=\int_{\mathbb{R}^d}d\bm{r}\,Dp(\bm{r};t)\|\bm{F}(\bm{r};t)\|^2=\sigma.
\end{align}
Here, we use the Fourier transform of the delta function $\delta(\bm{r}-\bm{r}')=\int_{\mathbb{R}^d}d\bm{k}\,\mathrm{e}^{\mathrm{i}\bm{k}\cdot(\bm{r}-\bm{r}')}/(2\pi)^d$. 

As we did for local EPR, we can also decompose the excess and housekeeping EPRs into wavenumber contributions as $\sigma^{\mathrm{ex}}=\int_{\mathbb{R}^d}d\bm{k}\,\sigma^{\mathrm{ex,wn}}(\bm{k})$, and $\sigma^{\mathrm{hk}}=\int_{\mathbb{R}^d}d\bm{k}\,\sigma^{\mathrm{hk,wn}}(\bm{k})$ by defining the wavenumber excess and housekeeping EPRs as
\begin{align}
\sigma^{\mathrm{ex,wn}}(\bm{k})&\coloneqq\frac{1}{(2\pi)^d}\|\hat{F}^{\ast}(\bm{k};t)\|^2\\
\sigma^{\mathrm{hk,wn}}(\bm{k})&\coloneqq\frac{1}{(2\pi)^d}\|\hat{F}(\bm{k};t)-\hat{F}^{\ast}(\bm{k};t)\|^2.
\end{align}
We remark that the geometric excess/housekeeping decomposition can also be violated at each wavenumber as $\sigma^{\mathrm{wn}}(\bm{k})\neq\sigma^{\mathrm{ex,wn}}(\bm{k})+\sigma^{\mathrm{hk,wn}}(\bm{k})$, and there is a nonzero cross term $\sigma^{\mathrm{cross,wn}} (\bm{k})\coloneqq \sigma^{\mathrm{wn}}(\bm{k})-\sigma^{\mathrm{ex,wn}}(\bm{k})+\sigma^{\mathrm{hk,wn}}(\bm{k})$ which satisfies $\int_{\mathbb{R}^d}d\bm{k}\,\sigma^{\mathrm{cross,wn}} (\bm{k})=0$.

The wavenumber decomposition is based on the orthonormality of the Fourier basis. Therefore, it may be possible to generalize the geometric decomposition of the EPR using an orthonormal basis other than the Fourier basis, e.g., a wavelet basis~\cite{graps1995introduction,torrence1998practical}. It may also be interesting to consider the spectral decomposition of the EPR~\cite{nardini2017entropy} based on the Harada--Sasa relation~\cite{harada2005equality} in terms of our wavenumber decomposition.

\subsection{Wasserstein distance}\label{sec:W_distance}
The excess EPR obtained in the previous section can be interpreted as a geometric quantity using the Wasserstein geometry developed in optimal transport theory~\cite{villani2009optimal,dechant2022geometric_R}. Here, we briefly review the Wasserstein distance and its dynamical reformulation, which is intrinsically important in thermodynamics. 

The $q$-Wasserstein distance for a positive number $q\geq 1$ between two probability distributions $p_{A}$ and $p_{B}$ is defined as
\begin{align}\label{Lp_W_distance}
    &{W}_{q}(p_{A},p_{B}) \coloneqq \notag
    \\&\left( \inf_{\pi\in\Pi(p_A,p_B)}\int_{\mathbb{R}^d \times \mathbb{R}^d} d\bm{r}d\bm{r'}\|\bm{r}-\bm{r}'\|^q\pi(\bm{r},\bm{r}') \right)^{\frac{1}{q}},
\end{align}
where $\Pi(p_A,p_B)$ is the set of joint probability distributions with marginals $p_A$ and $p_B$:
\begin{align*}
    \Pi(p_A,p_B)\coloneqq&\left\{\pi\middle|\;p_A(\bm{r})=\int_{\mathbb{R}^d}d\bm{r}'\pi(\bm{r},\bm{r}'),\right.\\&\left.p_B(\bm{r}')=\int_{\mathbb{R}^d}d\bm{r}\,\pi(\bm{r},\bm{r}'),\;\pi(\bm{r},\bm{r}')\geq 0\right\}.
\end{align*}
Here we assume that the moments up to the $q$-th order are finite for the two probability distributions $p_{A}$ and $p_{B}$.
We can confirm that ${W}_{q}$ satisfies the axioms of distance. We can also prove 
\begin{align}\label{hierarchical_ineq}
    {W}_{q}(p_A,p_B)\leq {W}_{q'}(p_A,p_B)\;\;\mathrm{for}\;\; q\leq q',
\end{align}
by Hölder’s inequality~\cite{villani2009optimal}.

For every $q$, we can reformulate the $q$-Wasserstein distance as an optimization problem related to the dynamics of a probability distribution subject to a continuity equation. 
In particular, we can obtain the square of the $2$-Wasserstein distance by the minimization problem
\begin{align}\label{Benamou--Brenier}
    {W}_{2}(p_{A},p_{B})^2=\inf_{p,\bm{F}'}\tau\int_{0}^{\tau}dt\int_{\mathbb{R}^d}d\bm{r}\|D \bm{F}'(\bm{r};t)\|^2 p(\bm{r};t)
\end{align}
with the following three constraints
\begin{align}\label{BB_constraints}
    p(\cdot;0)=p_A(\cdot),\;p(\cdot;\tau)=p_B(\cdot),\;\partial_t p=-\bm{\nabla}_{\bm{r}}\cdot\qty(pD\bm{F}').
\end{align}
In other words, we minimize the right-hand side in Eq.~\eqref{Benamou--Brenier} over trajectories of probability distributions that start and end on $p_A$ and $p_B$ and satisfy a continuity equation. 
This reformulation for the case $q=2$ was initially made by Benamou and Brenier, so the equation \eqref{Benamou--Brenier} is called the Benamou--Brenier formula in optimal transport theory~\cite{benamou2000computational}. 
We can also consider an extension of the Benamou--Brenier formula for general $q$~\cite{ambrosio2005gradient, brasco2012survey,evans1999differential}. 
\add{For the special case of $q=1$, we can express the Benamou--Brenier formula as an optimization problem of the current, which is known as the Beckmann problem~\cite{beckmann1952continuous}
\begin{align}\label{Benamou--Brenier_L1}
    {W}_{1}(p_{A},p_{B})=\inf_{\bm{J}'}\int_{0}^{\tau}dt\int_{\mathbb{R}^d}d\bm{r}\|\bm{J}'(\bm{r};t)\|
\end{align}
where we impose the following condition on $\bm{J}'$: there exists a time series of probability distribution $p'$ satisfying
\begin{align}\label{BB_constraints_L1}
    p'(\cdot;0)=p_A(\cdot),\;p'(\cdot;\tau)=p_B(\cdot),\;\partial_t p'=-\bm{\nabla}_{\bm{r}}\cdot\bm{J}'.
\end{align}
}

Both expressions of the $1$-Wasserstein distance in the original definition~\eqref{Lp_W_distance} and the Beckmann problem~\eqref{Benamou--Brenier_L1} are reduced to an expression known as the Kantorovich--Rubinstein duality,
\begin{align}\label{FP_Kantorovich--Rubinstein}
    {W}_{1}(p_{A},p_{B})=\sup_{\phi\in\mathrm{Lip}^1}\qty{\int_{\mathbb{R}^d}d\bm{r}\,\phi\qty(p_B-p_A)},
\end{align}
where the set of $1$-Lipschitz functions is denoted by
\begin{align}\label{1-Lip}
    \mathrm{Lip}^1\coloneqq\qty{\phi\mid \|\bm{\nabla}_{\bm{r}}\phi\|\leq1}.
\end{align}
The derivation of Eq.~\eqref{FP_Kantorovich--Rubinstein} from the original definition of the $1$-Wasserstein distance in Eq.~\eqref{Lp_W_distance} is well-known and based on the method of Lagrange multipliers~\cite{villani2009optimal}. The Beckmann problem~\eqref{Benamou--Brenier_L1} can be directly obtained from Kantorovich--Rubinstein duality~\eqref{FP_Kantorovich--Rubinstein} by again using the method of Lagrange multipliers~\cite{evans1999differential,chen2017matricial}. 

\subsection{Wasserstein geometry and thermodynamic trade-off relations}\label{sec:FP_trade-off}
Considering a trajectory of probability distribution $\left\{p(t)\right\}_{t\in[0,\tau]}$ obeying the Fokker--Planck equation~\eqref{Fokker--Planck_eq}, we can define the length of the trajectory using the $q$-Wasserstein distance as
\begin{align}
    l_{q,\tau}\coloneqq\int_{0}^{\tau}dt\, v_{q}(t) \,,
\end{align}
with $v_{q}(t)$ defined as
\begin{align} \label{definition_speedq}
    v_{q}(t)\coloneqq\lim_{\varDelta t\to0}\frac{{W}_q\qty(p(t),p(t+\varDelta t))}{\varDelta t}.
\end{align}
This quantity $v_{q}(t)$ indicates the speed of the dynamics of $p(t)$ in the space of probability distributions. The form of the mobility tensor $\mathsf{M}$, the Benamou--Brenier formula for the $2$-Wasserstein distance~\eqref{Benamou--Brenier}, and the variational form of the excess EPR~\eqref{FP_exEPR} lead to
\begin{align}\label{FP_speed_and_exEPR}
    \sigma^{\mathrm{ex}}=\frac{v_{2}(t)^2}{D},
\end{align} 
which means the square root of the excess EPR at time $t$ is proportional to the speed of the dynamics of the probability distribution.

The relation between $\sigma^{\mathrm{ex}}$ and $v_{2}(t)$ leads to the hierarchy of TSLs~\cite{nakazato2021geometrical}
\begin{align}\label{FP_L2_TSL}
    \frac{{W}_2(p(0),p(\tau))^2}{D}
    \leq\frac{ l_{2,\tau}^2}{D}\leq\tau\Sigma^{\mathrm{ex}}_{\tau}\leq\tau\Sigma_{\tau}.
\end{align}
The inequality $l_{2,\tau}\geq{W}_2(p(0),p(\tau))$, which comes from the triangle inequality, yields the first inequality in Eq.~\eqref{FP_L2_TSL}. This inequality reflects the fact that ${W}_2(p(0),p(\tau))$ is the geodesic length between $p(0)$ and $p(\tau)$. 
The second inequality is derived from the Cauchy--Schwarz inequality $\left[\int_{0}^{\tau}dt\,v_{2}(t)\right]^2\leq\left[\int_{0}^{\tau}dt\,\right]\left[\int_{0}^{\tau}dt\,v_{2}(t)^2\right]$ and the relation between $v_{2}$ and $\sigma^{\mathrm{ex}}$~\eqref{FP_speed_and_exEPR}. 
The third inequality is a consequence of the nonnegativity of the housekeeping EP $\Sigma^{\mathrm{hk}}_{\tau} \geq 0$ and the decomposition of the EP. 
Overall, the inequalities in Eq.~\eqref{FP_L2_TSL} tell us that transitioning to a more distant distribution in less time requires more dissipation.

The inequality between Wasserstein distances in Eq.~\eqref{hierarchical_ineq} leads to another hierarchy of TSLs as
\begin{align}
    \frac{{W}_1(p(0),p(\tau))^2}{D}
    \leq\frac{ l_{1,\tau}^2}{D}\leq\tau\Sigma^{\mathrm{ex}}_{\tau}\leq\tau\Sigma_{\tau}.
\end{align}
These lower bounds on the EPs are weaker than those in Eq.~\eqref{FP_L2_TSL} because Eq.~\eqref{hierarchical_ineq} shows 
\begin{align} \label{hierachical_TSLs}
\frac{ l_{1,\tau}^2}{D} &\leq \frac{ l_{2,\tau}^2}{D} \leq\tau\Sigma_{\tau}, \nonumber\\
 \frac{{W}_1(p(0),p(\tau))^2}{D} &\leq \frac{{W}_2(p(0),p(\tau))^2}{D} \leq\tau\Sigma_{\tau},
\end{align}
where we used the fact that Eq.~\eqref{hierarchical_ineq} leads to $v_2(t) \geq v_1(t)$ and $l_{2,\tau} \geq l_{1,\tau}$.
The generalization of the TLSs discussed here from Langevin dynamics to Markov jump processes (MJPs) is rather complicated, and we only note that there are multiple ways of generalizing~\cite{yoshimura2023housekeeping,van2023thermodynamic}.

Using the excess EPR, we can obtain a TUR for time-independent observable $\varphi(\bm{r})$ as 
\begin{align}\label{FP_TUR}
    ( d_t \langle\varphi\rangle_{p_t} )^2\leq D\left\langle\|\bm{\nabla}_{\bm{r}}\varphi\|^2\right\rangle_{p_t}\sigma^{\mathrm{ex}},
\end{align}
where the bracket indicates the average over the probability distribution $p(\bm{r};t)$, $\langle \varphi\rangle_{p_t}\coloneqq\int_{\mathbb{R}^d}d\bm{r}\,p(\bm{r};t)\varphi(\bm{r})$~\cite{dechant2022geometric_E}. 
The TUR represents a trade-off relation between the excess EPR, $\sigma^{\mathrm{ex}}$, the speed of the observable, $d_t\langle\varphi\rangle_{p_t}$, and the average squared magnitude of the gradient of the observable, $\left\langle\|\bm{\nabla}_{\bm{r}}\varphi\|^2\right\rangle_{p_t}$. In other words, we need more dissipation to make a flatter observable change faster. 
It is derived from the Cauchy--Schwarz inequality, $\fip{\bm{\nabla}_{\bm{r}}\varphi}{\bm{\nabla}_{\bm{r}}\phi^{\ast}}^2_{\mathsf{M}}\leq\fip{\bm{\nabla}_{\bm{r}}\varphi}{\bm{\nabla}_{\bm{r}}\varphi}_{\mathsf{M}}\fip{\bm{\nabla}_{\bm{r}}\phi^{\ast}}{\bm{\nabla}_{\bm{r}}\phi^{\ast}}_{\mathsf{M}}$, and the fact that $\phi^{\ast}$ reproduces the dynamics as $\partial_t p=-\bm{\nabla}_{\bm{r}}\cdot(\mathsf{M}\bm{\nabla}_{\bm{r}}\phi^{\ast})$. 
Here, the quantities appearing in the Cauchy--Schwarz inequality are given by $\fip{\bm{\nabla}_{\bm{r}}\varphi}{\bm{\nabla}_{\bm{r}}\phi^{\ast}}_{\mathsf{M}}= d_t \langle\varphi\rangle_{p_t}$, $\fip{\bm{\nabla}_{\bm{r}}\varphi}{\bm{\nabla}_{\bm{r}}\varphi}_{\mathsf{M}} = D\left\langle\|\bm{\nabla}_{\bm{r}}\varphi\|^2\right\rangle_{p_t}$ and $\fip{\bm{\nabla}_{\bm{r}}\phi^{\ast}}{\bm{\nabla}_{\bm{r}}\phi^{\ast}}_{\mathsf{M}}= \sigma^{\mathrm{ex}}$, which proves the TUR.

\add{We note our usage of the term TUR. Conventionally, TUR refers to a lower bound on dissipation using the expectation value and the variance of a general current. Although the trade-off in Eq.~\eqref{FP_TUR} appears different from the conventional form of TUR, we refer to it as a TUR. This is because the TUR~\eqref{FP_TUR} represents a specific case of the short-time limit of the conventional TUR for any initial state~\cite{dechant2018current,liu2020thermodynamic,otsubo2020estimating}, as discussed in detail in Ref.~\cite{dechant2022geometric_E}.}

We can also interpret the TUR from the viewpoint of the Wasserstein geometry by rewriting Eq.~\eqref{FP_TUR} as
\begin{align}\label{FP_TUR_CRB}
    v_{\varphi}(t) \coloneqq\frac{|d_t\langle\varphi\rangle_{p_t}|}{\sqrt{\left\langle\|\bm{\nabla}_{\bm{r}}\varphi\|^2\right\rangle_{p_t}}}\leq v_{2}(t),
\end{align}
using the relation between $v_{2}(t)$ and $\sigma^{\mathrm{ex}}$ in Eq.~\eqref{FP_speed_and_exEPR}. Here, we define $v_{\varphi}(t)$ as the speed of the observable $\varphi$ normalized by the spatial fluctuation of $\varphi$. 
Therefore, the TUR means that the normalized speed of an observable is slower than the speed of the probability distribution moving on the manifold of distributions equipped with the Wasserstein metric. 
This is similar to the Cramér--Rao bound~\cite{rao1992information} with parameter $t$, called the information geometric speed limit~\cite{ito2020stochastic,nicholson2020time}, written as
\begin{align}\label{FP_CRB}
    v_{\varphi}^{I}(t)\coloneqq\frac{|d_t\langle\varphi\rangle_{p_t}|}{\sqrt{\mathrm{Var}[\varphi]}}\leq v_{I}(t),
\end{align}
where $v_I(t) \coloneqq\sqrt{\langle(d_t\ln p(t))^2\rangle_{p_t}}$ is the square root of the Fisher information and $\mathrm{Var}[\varphi]$ is the variance defined as $\mathrm{Var}[\varphi]= \langle \varphi^2 \rangle_{p_t} - \langle \varphi \rangle^2_{p_t}$. 
From an information geometric point of view, we can regard $v_{I}(t)$ as the speed of the probability distribution, $p(t)$, on the manifold equipped with the information-geometric (Fisher) metric.

We can derive the inequality $v_1(t)\leq v_2(t)$, which provides the hierarchy of TSLs~\eqref{hierachical_TSLs}, from the TUR by taking the optimal potential of the Kantorovich--Rubinstein duality~\eqref{FP_Kantorovich--Rubinstein} as observable in Eq.~\eqref{FP_TUR_CRB}.

\section{Thermodynamics of reaction-diffusion systems}\label{sec:RDS_thermo}
Hereafter, we will focus on reaction-diffusion systems (RDSs). We will refer to the geometric framework reviewed above and introduce new notions to generalize it. We will need to use many kinds of symbols, which we summarize in Table.~\ref{tab:symbols}. 

This section introduces the two basics of RDSs: dynamics and thermodynamics. 
We begin with a class of RDSs called closed systems in Sec.~\ref{sec:closed_RDS}. 
A closed RDS does not exchange molecules with the outside, as opposed to an open RDS, the other class of RDSs, as explained in Sec.~\ref{sec:open_RDS}. 
Section~\ref{sec:force&EPR} introduces the thermodynamics of RDSs in terms of thermodynamic forces and the EPR.  To simplify the discussion in subsequent sections, we unify quantities associated with reaction and diffusion by introducing appropriate vector fields and operators in Sec.~\ref{sec:vector_fields}. We also introduce the concept of conservative and non-conservative thermodynamic forces for RDSs, which play a central role in the geometric excess/housekeeping decomposition of EPR in Sec.~\ref{sec:conservative_nonconservative_forces}. 
\begin{table}
    \centering
    \begin{tabular}{cccc}
    & Total & Diffusion & Reaction\\
    \hline\hline
    \rowcolor[gray]{.97}[\tabcolsep]
    Value & $\mathbb{R}^{N\times d}\oplus\mathbb{R}^M$ & $\mathbb{R}^{N\times d}$ & $\mathbb{R}^M$ \\[1pt]
    Force & $\mathcal{F}=(\vec{\bm{F}},\bm{f})$ & $\vec{\bm{F}}=\left[\bm{F}_{(\alpha)}\right]_{\alpha=1}^N$ & $\bm{f}=(f_\rho)_{\rho=1}^M$\\[2pt]
    \rowcolor[gray]{.97}[\tabcolsep]
    Current & $\mathcal{J}=(\vec{\bm{J}},\bm{j})$ & $\vec{\bm{J}}=\left[\bm{J}_{(\alpha)}\right]_{\alpha=1}^N$ & $\bm{j}=(j_\rho)_{\rho=1}^M$ \\[2pt]
    \hline
    Mobility & $\mathcal{M}=\dvec{\mathsf{M}}\oplus \mathsf{m}$ & $\dvec{\mathsf{M}}=\left[\mathsf{M}_{(\alpha\beta)}\right]$ & $\mathsf{m}=(\delta_{\rho\rho'}m_{\rho})$ \\[2pt]
    \rowcolor[gray]{.97}[\tabcolsep]
    Gradient & $\Grad\!=\!\bm{\nabla}_{\bm{r}} \oplus \nabla_{\!\mathrm{s}}$ & $\bm{\nabla}_{\bm{r}}$ & $\nabla_{\!\mathsf{s}}=(S_{\rho\alpha})$\\[2pt]
    Potential & \multicolumn{3}{c}{$\vec{\phi}(\bm{r})\in\mathbb{R}^N$} \\[2pt]\hline
    \end{tabular}
    \caption{Summary of essential quantities appearing in the total RDSs, the diffusion part, and the reaction part. Forces and currents take values in the second row at each spatial and temporal point. Operators $\dvec{\mathsf{M}}$, $\mathsf{m}$, and $\mathcal{M}$ map forces to the corresponding currents locally. Gradient operators can generate diffusion and reaction forces from a single potential function as $\vec{\bm{F}}=\bm{\nabla}_{\bm{r}} \vec{\phi}$ and $\bm{f}=\nabla_{\mathsf{s}} \vec{\phi}$.}
    \label{tab:symbols}
\end{table}

\subsection{Closed reaction-diffusion systems}\label{sec:closed_RDS}
We consider an RDS that describes the time evolution of a concentration distribution of $N$ chemical species in the $d$-dimensional area $V\subseteq\mathbb{R}^d$ due to reactions, advection, and diffusion. 
If no particles interact with the outside of the system, the system is called a closed RDS. 
Let $\mathcal{S}$ index the chemical species as $\mathcal{S}\coloneqq\qty{1,2,\cdots,N}$. 
We denote the $\alpha$-th chemical species ($\alpha \in \mathcal{S}$) as $Z_{\alpha}$, and its concentration distribution at location $\bm{r}\in V$ and time $t$ as $c_{\alpha}(\bm{r};t)$. We note that $c_{\alpha}(\bm{r};t)$ is not a probability density and $\int d\bm{r} c_{\alpha}(\bm{r};t)$ is not necessarily equal to $1$  or any other constant.

\add{The time evolution of the concentration of $\alpha$-th chemical species under the assumption of Fick's law are usually given by the reaction-diffusion (RD) equation, 
\begin{align}
    \partial_t c_{\alpha}(\bm{r};t)=D_{\alpha}\bm{\nabla}_{\bm{r}}^2c_{\alpha}(\bm{r};t)+R_{\alpha}(\bm{r};t),
    \label{conventionalrds}
\end{align}
where $D_{\alpha}$ indicates the diffusion constant of $Z_{\alpha}$ and $R_{\alpha}(\bm{r};t)$ represents the effect of reactions.} Note that the reaction term $R_{\alpha}$ can depend on the concentration distribution. 
We can rewrite the first term to $-\bm{\nabla}_{\bm{r}}\cdot\bm{J}_{(\alpha)}^{\mathrm{Fick}} (\bm{r};t)$, where $\bm{J}_{(\alpha)}^{\mathrm{Fick}}(\bm{r};t)\coloneqq-D_{\alpha}\bm{\nabla}_{\bm{r}}c_{\alpha}(\bm{r};t)$ is the current obeying Fick's law.

\add{However,} assuming Fick's law is optional in our discussion. The diffusion current given by Fick's law sometimes fails to describe the dynamics of chemical species, for example, under an electric field, which causes advection, or in a nondilute solution. 
To include a wider range of phenomena, we deal with the following more general RD equation for $\alpha$-th chemical species,
\begin{align}\label{RDS}
    \partial_t c_{\alpha}(\bm{r};t)=-\bm{\nabla}_{\bm{r}}\cdot\bm{J}_{(\alpha)}(\bm{r};t)+R_{\alpha}(\bm{r};t),
\end{align}
where $\bm{J}_{(\alpha)}(\bm{r};t)=[J_{(\alpha)i}(\bm{r};t)]_{i=1,\cdots,d}$ is the general diffusion current for $\alpha$-th species. \add{The diffusion currents can depend on the concentration distribution as in the case of Fick's law. Note that the general RD equation~\eqref{RDS} reproduces the usual one~\eqref{conventionalrds} if the diffusion current obeys Fick's law as $\bm{J}_{(\alpha)}(\bm{r};t) = \bm{J}_{(\alpha)}^{\mathrm{Fick}} (\bm{r};t)$.} 

In this paper, we assume one of the following three boundary conditions on diffusion currents. Note that the boundary conditions on diffusion currents constrain concentration distributions through the dependence of diffusion currents on the concentration distributions. The first one is the no-flux boundary condition, $\bm{J}_{(\alpha)}(\bm{r};t)\cdot\bm{n}(\bm{r})=0$ for all $\alpha\in\mathcal{S}$, $t$, and $\bm{r}\in\partial V$, where $\partial V$ indicates the boundary of $V$, and $\bm{n}(\bm{r})$ is the unit normal vector of the surface at $\bm{r}\in\partial V$. This condition corresponds to considering a chemical reaction system in a container, where the exchange of particles via diffusion with the outside never happens. 
The second one is the periodic boundary condition: 
when the space $V$ has a periodic structure, like a supercube, we may assume that all quantities depending on $\bm{r}$ satisfy the periodic boundary condition. Note that this rule applies not only to the currents but also to the other quantities. 
The third one is the fast decay of the diffusion currents at infinity, which we consider when $V=\mathbb{R}^d$. 
Those conditions can be combined. For example, considering an infinitely long pipe with a square cross section, we may assume the no-flux boundary or the periodic boundary on the sides of the pipe, while supposing the fast decay for the current at infinity.  

We can consider the thermodynamic structure of the RDS by rewriting the reaction term $R_{\alpha}(\bm{r};t)$ with reaction currents based on the details of the reactions, as explained below. 
We consider $M$ reactions indexed by $\mathcal{R}\coloneqq\qty{1,2,\cdots,M}$. We write the $\rho$-th reaction ($\rho \in \mathcal{R}$) as
\begin{align}\label{reaction}
    \ce{$\sum_{\alpha\in\mathcal{S}}\nu_{\alpha\rho}^+Z_{\alpha}$ <=> $\sum_{\alpha\in\mathcal{S}}\nu_{\alpha\rho}^-Z_{\alpha}$},
\end{align}
where $\nu_{\alpha\rho}^{\pm}$ indicates the number of $Z_{\alpha}$ consumed ($+$) or produced ($-$) by the $\rho$-th reaction. 
We assume $\nu_{\alpha\rho}^{\pm}$ is independent of location $\bm{r}$ and time $t$ for all $\alpha$ and $\rho$. 
We write the forward ($+$) and reverse ($-$) fluxes of the $\rho$-th reaction at $(\bm{r};t)$ as $j^{+}_{\rho}(\bm{r};t)$ and $j^{-}_{\rho}(\bm{r};t)$, respectively. 
All the fluxes are always assumed to be positive $j^{\pm}_{ \rho}(\bm{r};t) > 0$. 
They can depend on space and time $(\bm{r};t)$ directly and/or via concentration distributions (e.g., assuming mass action kinetics, a flux is given by $j^{\pm}_{\rho}(\bm{r};t)=\kappa_{\rho}^{\pm}(\bm{r};t)\prod_{\alpha}\qty[c_{\alpha}(\bm{r};t)]^{\nu_{\alpha\rho}^{\pm}}$, where $\kappa_{\rho}^{\pm}(\bm{r};t)$ is the reaction rate constant for the forward/reverse reaction). 
The reaction current of $\rho$-th reaction $j_{\rho}(\bm{r};t)$ is given by $j_{\rho}(\bm{r};t)=j_{\rho}^{+}(\bm{r};t)-j_{\rho}^{-}(\bm{r};t)$. Using these reaction currents, we can rewrite $R_{\alpha}(\bm{r};t)$ as
\begin{align}
    R_{\alpha}(\bm{r};t)=\sum_{\rho\in\mathcal{R}} S_{\alpha\rho} j_{\rho}(\bm{r};t),
\end{align}
where $S_{\alpha\rho}\coloneqq\nu_{\alpha\rho}^{-}-\nu_{\alpha\rho}^{+}$ is the $(\alpha,\rho)$-th element of the stoichiometric matrix, which denotes the net increase of $Z_{\alpha}$ through the $\rho$-th reaction.

\add{Finally, we rewrite the RD equation as a continuity equation, which will be convenient for future calculations. Introducing the vector notation $\vec{c} = ({c}_1, \dots, {c}_N){}^{\top}$, $\vec{\bm{J}} =(\bm{J}_{(1)}, \dots, \bm{J}_{(N)}){}^{\top}$, and $\vec{R} =(R_1, \dots, R_N){}^{\top}$ with transposition ${}^{\top}$, the RD equation~\eqref{RDS} is rewritten in a simpler form $\partial_t \vec{c} =-\bm{\nabla}_{\bm{r}}\cdot \vec{\bm{J}} +\vec{R}$. Here, we make the dependence on $(\bm{r};t)$ implicit to simplify the notation. Rewriting the reaction term as $\vec{R} = \nabla_{\mathsf{s}}^{\top} \bm{j}$ with the vector of the reaction current $\bm{j}\coloneqq(j_{1}, \dots, j_{M}){}^{\top}$ and the matrix $(\nabla_{\mathsf{s}})_{\rho\alpha}\coloneqq S_{\alpha\rho}$, the RD equation reduces to a continuity equation
\begin{align}
    \partial_t \vec{c} =-\bm{\nabla}_{\bm{r}}\cdot \vec{\bm{J}} +\nabla_{\mathsf{s}}^{\top} \bm{j}.
    \label{RDSvec}
\end{align}
Here, the matrix $\nabla_{\mathsf{s}}^{\top}$ is the stoichiometric matrix.
}


\subsection{Open reaction-diffusion systems}\label{sec:open_RDS}
We generally deal with an open RDS, where some of the $N$ species can be exchanged with the outside of the system. 
We classify the species into two categories, internal species, which are not exchanged with the outside of the system, and external species, which are exchanged with the outside of the system. 
Let $N_\mathcal{X}(\leq N)$ denote the number of internal species. 
We index the internal and external species as $\mathcal{X}\coloneqq\qty{1, \dots, N_\mathcal{X}}$ and $\mathcal{Y}\coloneqq\qty{N_\mathcal{X}+1, \dots, N}=\mathcal{S}\setminus\mathcal{X}$, respectively.


Taking into account the decomposition of $\mathcal{S}$ into $\mathcal{X}$ and $\mathcal{Y}$, we introduce the following notation. 
Let $\vec{e} =(e_1, \cdots , e_N){}^{\top}$ be an arbitrary vector consisting of $N$ elements such as $\vec{c}$, $\vec{\bm{J}}$ or $\vec{R}$. 
We define $\vec{e}_{\mathcal{X}}$ and $\vec{e}_{\mathcal{Y}}$ as the vectors of the first $N_{\mathcal{X}}$ elements and the last $N-N_\mathcal{X}$ elements of $\vec{e}$; therefore, they decompose $\vec{e}$ as $\vec{e}= (\vec{e}{}^{\top}_{\mathcal{X}}, \vec{e}{}^{\top}_{\mathcal{Y}}){}^{\top}$.

We also define a subset of $\mathcal{R}$, $\mathcal{R}_{\mathcal{X}}$, as the set of the indexes of reactions that change the concentrations of internal chemical species:
\begin{align}
    \mathcal{R}_{\mathcal{X}}\coloneqq\{\rho\in\mathcal{R}\mid\exists\alpha\in\mathcal{X},\,S_{\alpha\rho}\neq0\}.
\end{align}
Reactions whose index belongs to $\mathcal{R}\setminus\mathcal{R}_{\mathcal{X}}$ only change the concentrations of external species, while reactions whose index belongs to $\mathcal{R}_{\mathcal{X}}$ possibly change the concentrations of external chemical species.

The exchange of external species can be modeled in various ways. 
For example, we can describe the interaction with the outside by fixing the concentrations of the external species on the boundary. 
We can also assume that the concentration distributions of the external species are homogeneous both on the boundary and in the bulk. 
In addition, we can control the concentration distribution of the external species with external currents.

Nonetheless, the time evolution of the internal species is essential for further discussion.
With the notation already introduced in this section, they can be written as 
\begin{align}\label{internal_dynamics}
    \partial_t \vec{c}_{\mathcal{X}}=-\bm{\nabla}_{\bm{r}}\cdot\vec{\bm{J}}_{\mathcal{X}}+\qty(\Grad_{\mathsf{s}}{}^{\top}\bm{j})_{\mathcal{X}}. 
\end{align}
We impose the same boundary conditions on the diffusion current corresponding to the internal species in open systems as in closed systems, e.g., the no-flux boundary condition, the periodic boundary condition, or the fast decay of diffusion currents. In the following results, we only consider the time evolution of internal species, which are described by the continuity-equation-like form~\eqref{internal_dynamics}, so that we do not need to consider how to describe the interaction with the outside of the system. The only exception is a generalization of the $2$-Wasserstein distance, where we have to assume the homogeneity of the external species as discussed in Sec.~\ref{sec:RDS_2W_dist}.

\subsection{Thermodynamic force and entropy production rate}\label{sec:force&EPR}
Here, we introduce the thermodynamic forces and the EPR. 
Corresponding to the diffusion and the reaction currents, two kinds of forces, diffusion and reaction forces, are introduced. 
In the following, we assume that the temperature is homogeneous in $V$, and we choose units so that the product of the temperature $T$ and the gas constant $R_{\rm gas}$ is equal to $1$~\cite{falasco2018information} for simplicity. 

We also assume that the chemical potential can be defined for each species at each location. We let $\mu_{\alpha}(\bm{r};t)$ denote the chemical potential of $\alpha$-th species at $(\bm{r};t)$. \add{The chemical potential can depend on the concentration distribution and its spatial derivatives. For example, the chemical potential in an ideal dilute solution without applied mechanical forces is given by $\mu^{\mathrm{id}}_{\alpha}=\mu^{\circ}_{\alpha}+\ln c_{\alpha}$ with the standard chemical potential $\mu^{\circ}_{\alpha}$, which is independent of $t$ and $\bm{r}$. In the Cahn–Hilliard equation~\cite{cahn1958free}, the chemical potential also depends on the concentration gradient as $\mu_{\alpha}=c_{\alpha}^3-c_{\alpha}-\gamma\bm{\nabla}_{\bm{r}}^2c_{\alpha}$ with a constant $\gamma$.}

\textit{Diffusion force}--- We introduce the diffusion force. We write the diffusion force for the $\alpha$-th species at $(\bm{r};t)$ as $\bm{F}_{(\alpha)}(\bm{r};t)=[F_{(\alpha)i}(\bm{r};t)]_{i=1,\cdots,d}$. The diffusion force is defined by using the chemical potential as
\begin{align}\label{RDS_diffusion_force}
    \bm{F}_{(\alpha)}(\bm{r};t)\coloneqq-\bm{\nabla}_{\bm{r}}\mu_{\alpha}(\bm{r};t)+\bm{K}^{\mathrm{nc}}_{(\alpha)}(\bm{r};t),
\end{align}
where $\bm{K}^{\mathrm{nc}}_{(\alpha)}$ denotes the nonconservative mechanical force on particles of $\alpha$-th chemical species. Note that all mechanical forces on particles of the $\alpha$-th chemical species that can be represented by the gradient of a potential, e.g., the gravity and the Coulomb force, are included in the gradients of the chemical potential $-\bm{\nabla}_{\bm{r}}\mu_{\alpha}(\bm{r};t)$. We introduce the vector notations as $\vec{\bm{F}}\coloneqq(\bm{F}_{(1)}, \dots, \bm{F}_{(N)}){}^{\top}$, $\vec{\mu}\coloneqq(\mu_1,\dots,\mu_{N})^{\top}$, and $\vec{\bm{K}}{}^{\mathrm{nc}}\coloneqq(\bm{K}^{\mathrm{nc}}_{(1)}, \dots, \bm{K}^{\mathrm{nc}}_{(N)}){}^{\top}$, which let us rewrite Eq.~\eqref{RDS_diffusion_force} as
\begin{align}\label{RDS_diffusion_force_vec}
    \vec{\bm{F}}\coloneqq-\bm{\nabla}_{\bm{r}}\vec{\mu}+\vec{\bm{K}}{}^{\mathrm{nc}}.
\end{align}
Here, the first term in the right-hand side, $\bm{\nabla}_{\bm{r}}\vec{\mu}$, indicates $(\bm{\nabla}_{\bm{r}}\mu_1,\dots, \bm{\nabla}_{\bm{r}}\mu_N)^{\top}$.

We assume a linear relation between the diffusion current and the diffusion force as
\begin{align}\label{linear_diff}
\vec{\bm{J}}(\bm{r};t) = \dvec{\mathsf{M}}(\bm{r};t) \vec{\bm{F}}(\bm{r};t),
\end{align}
where $\dvec{\mathsf{M}}(\bm{r};t)=[\mathsf{M}_{(\alpha\beta)}(\bm{r};t)]_{\alpha,\beta\in\mathcal{S}}$ is the mobility tensor, each of whose elements is a $d\times d$ matrix as $\mathsf{M}_{(\alpha\beta)}=[M_{(\alpha\beta)ij}]_{i,j=1,\cdots,d}$. 
It can be rewritten with the elements of current, force, and mobility tensor as $J_{(\alpha)i}\qty(\bm{r};t)=\sum_{\beta\in\mathcal{S}}\sum_{j=1}^{d}M_{(\alpha\beta)ij}(\bm{r};t)F_{(\beta)j}(\bm{r};t)$.
We further assume that $\dvec{\mathsf{M}}(\bm{r};t)$ is symmetric and positive-definite: $M_{(\alpha\beta)ij}=M_{(\beta\alpha)ji}$ holds for all $\alpha,\beta\in\mathcal{S}$ and for all $1\leq i,j\leq d$, and $\vec{\bm{F}}{}'^{\top}\dvec{\mathsf{M}}\vec{\bm{F}}{}'=\sum_{\alpha,\beta\in\mathcal{S}}\sum_{i,j=1}^{d}F'_{(\alpha)i}M_{(\alpha\beta)ij}F'_{(\beta)j}>0$ holds for all $\vec{\bm{F}}{}'\neq\vecrm{\bm{0}}$ ($\vecrm{\bm{0}}$ indicates a diffusion force or current all of whose elements are zero).

The mobility tensor possibly depends on the concentration distribution. 
In the special case where the diffusion current obeys Fick's law $\bm{J}_{(\alpha)}(\bm{r};t) = \bm{J}_{(\alpha)}^{\mathrm{Fick}}(\bm{r};t)$ and the force is given by the chemical potential as $\bm{F}_{(\alpha)}(\bm{r};t)=-\bm{\nabla}_{\bm{r}}\mu^{\mathrm{id}}_{\alpha}(\bm{r};t)$, the mobility tensor becomes
\begin{align}\label{simple_mobility_tensor}
    \mathsf{M}_{(\alpha\beta)}(\bm{r};t) =  D_{\alpha} c_{\alpha} (\bm{r};t) \delta_{\alpha \beta} \mathsf{I},
\end{align}
where $\delta_{\alpha\beta}$ is the Kronecker delta and $\mathsf{I}$ is the $d\times d$ identity matrix. 
In general, $\mathsf{M}_{(\alpha\alpha)}$ will not be proportional to the identity matrix if the mobility of the $\alpha$-th species is not isotropic. 
The off-diagonal entries $\mathsf{M}_{(\alpha\beta)}$ can also be nonzero matrices, which represent inter-species effects on the diffusion currents by the forces~\cite{griepentrog2004unique,liero2013gradient}. 


\textit{Reaction force}--- We next define the reaction force $\bm{f} = (f_{1}, \dots, f_{M}){}^{\top}$. 
The $\rho$-the element $f_\rho(\bm{r};t)$, which gives the reaction force on $\rho$-th reaction at $(\bm{r};t)$, is defined in terms of the chemical potential as
\begin{align}\label{RDS_reaction_force}
    f_{\rho}(\bm{r};t)\coloneqq-\sum_{\alpha\in\mathcal{S}}S_{\alpha\rho}\mu_{\alpha}(\bm{r};t).
\end{align}
Using the vector notations, we can rewrite Eq.~\eqref{RDS_reaction_force} as
\begin{align}\label{RDS_reaction_force_vec}
    \bm{f}\coloneqq-\nabla_{\mathsf{s}}\vec{\mu}.
\end{align}

\add{We also impose the condition of local detailed balance,}
\begin{align}\label{RDS_LDB_condition}
    f_{\rho}(\bm{r};t)= \ln\frac{j_{\rho}^{+}(\bm{r};t)}{j_{ \rho}^{-}(\bm{r};t)},
\end{align}
on the reaction force everywhere for all $\rho\in\mathcal{R}$. It relates the reaction force and the fluxes. In particular, when the fluxes obey the mass action kinetics and the chemical potential is $\mu_{\alpha}^{\mathrm{id}}$, the local detailed balance condition reduces to $\ln(\kappa^{+}_{\rho}/\kappa^{-}_{\rho})=-\nabla_{\mathsf{s}}\vec{\mu}{}^{\circ}$ with the vector $\vec{\mu}{}^{\circ}\coloneqq(\mu^{\circ}_1,\cdots,\mu^{\circ}_N){}^{\top}$. Note that we can utilize the local detailed balance condition in Eq.~\eqref{RDS_LDB_condition} as the definition of the reaction force even for systems where the chemical potential cannot be defined or where some unrecognized chemical species are present.

\add{The local detailed balance condition~\eqref{RDS_LDB_condition} establishes a relationship between the reaction force and current, analogous to the one for diffusion force~\eqref{linear_diff}, as
\begin{align}
    \bm{j}(\bm{r};t)=\mathsf{m}(\bm{r};t)\bm{f}(\bm{r};t),
    \label{linear_react}
\end{align}
with a diagonal matrix $[\mathsf{m}(\bm{r};t)]_{\rho\rho'}\coloneqq\delta_{\rho\rho'}m_{\rho}(\bm{r};t)$. Here, the Kronecker delta is denoted by $\delta_{\rho\rho'}$, and the diagonal elements of $\mathsf{m}(\bm{r};t)$ are defined as
\begin{align}
&m_{\rho}(\bm{r};t)\coloneqq\notag\\&\left\{
\begin{array}{ll}
\dfrac{j^+_{\rho}(\bm{r};t)-j^-_{\rho}(\bm{r};t)}{\ln j^+_{\rho}(\bm{r};t)-\ln j^-_{\rho}(\bm{r};t)} & (j^{+}_{\rho}(\bm{r};t)\neq j^{-}_{\rho}(\bm{r};t)),\\
j^{+}_{\rho}(\bm{r};t) & (j^{+}_{\rho}(\bm{r};t)=j^{-}_{\rho}(\bm{r};t)).
\end{array}
\right.
\label{Def_l}
\end{align}
In the following, we refer to $\mathsf{m}$ as the edgewise Onsager coefficient matrix, inspired by the Onsager coefficient that provides the linear relation between the force and current in a steady state~\cite{onsager1931reciprocal}. We remark that the edgewise Onsager coefficient matrix can depend on the concentration distribution due to the $\vec{c}$-dependence of the fluxes $j^{\pm}_{\rho}$. In this sense, it differs from the mobility tensor, which does not depend on the diffusion current. Despite its dependence on fluxes, $\mathsf{m}$ is a physically fruitful quantity, as shown in previous studies~\cite{van2023thermodynamic,yoshimura2023housekeeping} and in what follows in this paper.}

\add{As the mobility tensor $\dvec{\mathsf{M}}$, the edgewise Onsager coefficient matrix $\mathsf{m}$ is positive definite. This positive definiteness follows from the mathematical fact that the logarithmic mean between two positive numbers $a,b>0$, $(a-b)/\ln(a/b)$, is always positive. This is because the diagonal element of the matrix $m_{\rho}$ is the logarithmic mean between the forward and reverse fluxes $j_{\rho}^{\pm}>0$. Actually, the edgewise Onsager coefficient $m_{\rho}$ can be regarded as a kind of activity, that is, an indicator of the back-and-forth intensity of each reaction measured by the average of the speed of the forward and reverse reactions. For example, activity is conventionally evaluated by (double) the arithmetic mean $j_\rho^+(\bm{r};t)+j_\rho^-(\bm{r};t)$ and the geometric mean $\sqrt{j_\rho^+(\bm{r};t)j_\rho^-(\bm{r};t)}$, which are called the dynamical~\cite{maes2008canonical} and frenetic activity~\cite{maes2017non}, respectively. The edgewise Onsager coefficient is the middle of these conventional activities because the inequality between means, $\sqrt{ab}\leq(a-b)/\ln(a/b)\leq(a+b/2)$, yields
\begin{align}\label{ineq_activity}
    \sqrt{j_\rho^+(\bm{r};t)j_\rho^-(\bm{r};t)}\leq m_{\rho}(\bm{r};t)\leq\frac{j_\rho^+(\bm{r};t)+j_\rho^-(\bm{r};t)}{2}.
\end{align}
}

\textit{Entropy production rate}--- \add{The product between the forces and the currents defines the EPR $\sigma$ as,
\begin{align}
    \sigma &\coloneqq\int_{V} d\bm{r}\left(\sum_{\alpha\in\mathcal{S}}\bm{J}_{(\alpha)}\cdot\bm{F}_{(\alpha)}+\sum_{\rho\in\mathcal{R}}j_{\rho}f_{\rho}\right) \notag \\
    &= \int_{V} d\bm{r}\qty(  \vec{\bm{J}}{}^{\top}  \vec{ \bm{F}} + \bm{j}^{\top} \bm{f}),
    \label{Def_EPR}
\end{align}
whose time integration provides the EP during time duration $[0,\tau]$ as $\Sigma_{\tau}\coloneqq\int_{0}^{\tau}dt\,\sigma$.
The definition of EPR~\eqref{Def_EPR} derives the second law of thermodynamics, since the positivity of $\dvec{\mathsf{M}}$ and the local detailed balance condition~\eqref{RDS_LDB_condition} lead to $\vec{\bm{J}}{}^{\top} \vec{ \bm{F}}= \vec{\bm{F}}{}^{\top} \dvec{\mathsf{M}} \vec{\bm{F}} \geq 0$ and $\bm{j}^{\top} \bm{f}\geq 0$, respectively. The equality of the second law $\sigma=0$ is satisfied if and only if the system is in equilibrium, i.e., $j_{\rho}(\bm{r};t)=0$ and $\bm{J}_{(\alpha)}(\bm{r};t)=\bm{0}$ hold for all $\alpha\in\mathcal{S}$, $\rho\in\mathcal{R}$, and $\bm{r}\in V$. It indicates that the EPR is a measure of the irreversibility of the system.} 

\add{When the temperature is constant, we can regard the EP as dissipated work, which is the difference between the work done on the system and the increased free energy~\cite{falasco2018information,penocchio2019thermodynamic}. Hence, a smaller EP means that the state of the system changes with less work, that is, it is energetically efficient. From this viewpoint, today's nonequilibrium thermodynamics, especially stochastic thermodynamics, usually considers minimum dissipation problems~\cite{Proesmans2016,Gingrich2016,Rotskoff2017,Large2019,remlein2021optimality,ilker2022shortcuts,Blaber2023,Engel2023,aurell2011optimal,aurell2012refined,nakazato2021geometrical,van2021geometrical,dechant2022minimum,van2023thermodynamic,yoshimura2023housekeeping,zhong2024beyond} to control systems optimally. We later see that some results in this paper also directly relate to such minimization problems, which have not been previously discussed for RDSs.}

\add{Since RDSs consist of diffusive dynamics and reactions, the EPR~\eqref{Def_EPR} accounts for the dissipation arising from multiple factors. Thus, it is essential to decompose the EPR into contributions from different factors to understand the thermodynamic properties of a system. One of the simplest decompositions is into the EPR from diffusion $\sigma^{\mathrm{diff}}$ and from reactions $\sigma^{\mathrm{react}}$,
\begin{align}\label{decomp_diff_rct}
    \sigma^{\mathrm{diff}}\coloneqq\int_{V}d\bm{r}\sum_{\alpha\in\mathcal{S}}\bm{J}_{(\alpha)}\cdot\bm{F}_{(\alpha)},\;\sigma^{\mathrm{react}}\coloneqq\int_{V}d\bm{r}\sum_{\rho\in\mathcal{R}}j_{\rho}f_{\rho}.
\end{align}
We provide a unified way to perform decompositions into various factors in Sec.~\ref{sec:geometric_decomps}.}

\subsection{Unifying the diffusion and reaction}\label{sec:vector_fields}


We consider the quantities associated with reaction and diffusion separately in the previous sections. However, treating these quantities together will be useful for further discussion. Here, we introduce two inner products and some operators to handle reactions and diffusion together.

\textit{Forces and currents}--- We introduce the force $\mathcal{F}$ and the current $\mathcal{J}$ by unifying the diffusion force (current) and the reaction force (current) as
\begin{align}\label{Def_current_vec_func}
\mathcal{J} \coloneqq \left(
    \vec{\bm{J}}, \bm{j}\right),\quad
\mathcal{F} \coloneqq \left(
    \vec{\bm{F}},\bm{f}\right).
\end{align}
We refer to $\vec{\bm{F}}{}'$ and $\bm{f}'$ as the diffusion part and the reaction part of $\mathcal{F}'=(\vec{\bm{F}}{}', \bm{f}')$, respectively, and the same is true for the current. We also define the inner product of two vector fields $\mathcal{J}'=(\vec{\bm{J}'}, \bm{j}')$ and $\mathcal{F}'=(\vec{\bm{F}'}, \bm{f}')$ as
\begin{align}\label{Def_innerproduct_jf}
\fip{\mathcal{J}'}{\mathcal{F}'} &\coloneqq\int_{V} d\bm{r}\qty(\sum_{\alpha\in\mathcal{S}}\bm{J}_{\alpha}'\cdot\bm{F}_{\alpha}' +\sum_{\rho\in\mathcal{R}}j_{\rho}' f_{\rho}' ) \nonumber \\
&= \int_{V} d\bm{r}\qty( \vec{\bm{J}'}{}^{\top} \vec{\bm{F}'}+\bm{j}' {}^{\top} \bm{f}'),
\end{align}
which immediately leads to a new expression of the EPR as
\begin{align}
    \sigma=\fip{\mathcal{J}}{\mathcal{F}}.
\end{align}

\textit{Potentials and concentrations}--- We introduce the inner product of two vector fields with $N$ elements, e.g., the chemical potential $\vec{\mu}$, the concentration distribution $\vec{c}$, and its time derivative $\partial_t\vec{c}$, as
\begin{align}
\cip{\vec{\phi}}{\vec{\psi} }\coloneqq\int_{V}d\bm{r}\,\sum_{\alpha\in\mathcal{S}}\phi_{\alpha} \psi_{\alpha}= \int_{V}d\bm{r} \vec{\phi} \:{}^{\top} \vec{\psi},
\end{align} 
with $\vec{\phi}=(\phi_1,\dots,\phi_N)^{\top}$ and $\vec{\psi}=(\psi_1,\dots,\psi_N)^{\top}$. We refer to $\vec{\psi}_{\mathcal{X}}$ and $\vec{\psi}_{\mathcal{Y}}$ as the internal part and the external part of $\vec{\psi}$, e.g., $\partial_{t}\vec{c}_{\mathcal{X}} =(\partial_t c_1,\dots,\partial_t c_{N_{\mathcal{X}}})^{\top}$ is the internal part of $\partial_{t}\vec{c}=(\partial_t c_1,\dots,\partial_t c_N)^{\top}$.

\textit{Generalized gradient and divergence operators}--- 
We define the generalized gradient operator $\Grad$, taking a potential $\vec{\phi} = (\phi_1, \dots, \phi_N){}^{\top}$ to a force, as
\begin{align}\label{Def_Grad}
\Grad \vec{\phi}\coloneqq (\bm{\nabla}_{\bm{r}}\vec{\phi}, \nabla_{\mathsf{s}}\vec{\phi}), 
\end{align}
where $\bm{\nabla}_{\bm{r}}\vec{\phi}$ is defined as $\bm{\nabla}_{\bm{r}}\vec{\phi}\coloneqq(\bm{\nabla}_{\bm{r}} \phi_1, \dots, \bm{\nabla}_{\bm{r}} \phi_N){}^{\top}$. We remark that $\nabla_{\mathsf{s}}\vec{\phi}(\bm{r})$ is an $M$-dimensional vector because $\nabla_{\mathsf{s}}$ is an $M \times N$ matrix and $\vec{\phi}(\bm{r})$ is an $N$-dimensional vector. The generalized gradient operator $\Grad = \bm{\nabla}_{\bm{r}}\oplus \nabla_{\mathsf{s}}$ is regarded as the direct sum between $\bm{\nabla}_{\bm{r}}$ and $\nabla_{\mathsf{s}}$, where $\oplus $ represents the direct sum. The generalized gradient $\Grad$ enables us to unify the relations between the diffusion and reaction forces and the chemical potential in Eqs.~\eqref{RDS_diffusion_force_vec} and ~\eqref{RDS_reaction_force_vec} as
\begin{align}\label{RDS_force_vec}
    \mathcal{F}=-\Grad\vec{\mu}+\mathcal{K}^{\mathrm{nc}},
\end{align}
where we define the nonconservative mechanical force vector $\mathcal{K}^{\mathrm{nc}}$ as $\mathcal{K}^{\mathrm{nc}}\coloneqq(\vec{\bm{K}}{}^{\mathrm{nc}},\bm{0})$.

We also define an operator $\Div$ that maps a current to the time evolution caused by the current as
\begin{align}\label{Def_Div}
    \Div\mathcal{J}' \coloneqq-\bm{\nabla}_{\bm{r}}\cdot \vec{\bm{J}}{}' +\nabla_{\mathsf{s}}^{\top}\bm{j}',
\end{align}
where $\mathcal{J}' =(\vec{\bm{J}'}, \bm{j}')$ is a vector field with the reaction and diffusion parts, and $\Div\mathcal{J}'$ is a vector field with $N$ elements. This definition means $-\Div$ is a generalized divergence operator. The operator $\Div$ simplifies the form of the time evolution~\eqref{internal_dynamics} as
\begin{align}
    \partial_t \vec{c}_{\mathcal{X}}=\qty(\Div\mathcal{J})_{\mathcal{X}},
    \label{rewrite_RDS}
\end{align}
which emphasizes that a continuity equation gives the time evolution of the internal species.

The operator $\Div$ is conjugate to the generalized gradient operator $\Grad$ in terms of the two inner products $\fip{\cdot}{\cdot}$ and $\cip{\cdot}{\cdot}$ as
\begin{align}\label{conjugate}
    \fip{\mathcal{J}'}{\Grad\vec{\phi}}=\cip{\Div\mathcal{J}'}{\vec{\phi}},
\end{align}
where $\vec{\phi}$ is a potential whose external part is the zero vector, and $\mathcal{J}'$ is a current whose diffusion part corresponds to the internal species, $\bm{J}'_{(\alpha)}$ for $\alpha\in\mathcal{X}$, satisfies the boundary conditions on the system. To derive the conjugation relations between $\Grad$ and $\Div$, we do partial integration and use Gauss's theorem to calculate $\cip{\Div\mathcal{J}'}{\vec{\phi}}$ as follows:
\add{\begin{align}
    \cip{\Div\mathcal{J}'}{\vec{\phi}}&=\int_{V}d\bm{r}\sum_{\alpha\in\mathcal{S}}\qty{-\bm{\nabla}_{\bm{r}}\cdot\bm{J}'_{(\alpha)}+\qty(\nabla_{\mathsf{s}}^{\top}\bm{j}')_{\alpha}}\phi_{\alpha}\notag\\
    &=\int_{V}d\bm{r}\left\{\sum_{\alpha\in\mathcal{S}}\qty(\bm{J}'_{(\alpha)}\cdot\bm{\nabla}_{\bm{r}}\phi_{\alpha})+\bm{j}'^{\top}\nabla_{\mathsf{s}}\vec{\phi}\right\}\notag\\
    &\phantom{=\;}-\int_{V}d\bm{r}\sum_{\alpha\in\mathcal{S}}\bm{\nabla}_{\bm{r}}\cdot\qty(\phi_{\alpha}\bm{J}'_{(\alpha)})\notag\\
    &=\fip{\mathcal{J}'}{\Grad\vec{\phi}}-\sum_{\alpha\in\mathcal{S}}\int_{\partial V}d\bm{n}\cdot\qty(\phi_{\alpha}\bm{J}'_{(\alpha)}).
    \label{conj_grad}
\end{align}
Here, the summand in the second term of the last line, $\int_{\partial V}d\bm{n}\cdot(\phi_{\alpha}\bm{J}'_{(\alpha)})$, vanishes for all $\alpha\in\mathcal{S}$ because $\phi_{\alpha}=0$ holds for all $\alpha\in\mathcal{Y}$, and $\bm{J}'_{(\alpha)}$ satisfies the boundary conditions for all $\alpha\in\mathcal{X}$.} Thus, we obtain Eq.~\eqref{conjugate}. We remind the reader that we also impose the periodic boundary condition on every field containing $\vec{\phi}$ if we impose it on the system.


\textit{Onsager operator}--- Unifying the mobility tensor and the edgewise Onsager coefficient matrix, we introduce the Onsager operator $\mathcal{M}\coloneqq\dvec{\mathsf{M}}\oplus\mathsf{m}$ as the direct sum between $\dvec{\mathsf{M}}$ and $\mathsf{m}$, which maps forces to currents as
\begin{align}\label{Def_Ons_Ope}
\mathcal{M}\mathcal{F}' =
\left(\dvec{\mathsf{M}} \vec{\bm{F}}{}', \mathsf{m} \bm{f}'\right)
\end{align}
The Onsager operator $\mathcal{M}$ possibly depends on the concentration distribution in the same way that the mobility tensor and the edgewise Onsager coefficient matrix do. 

The Onsager operator allows us to unify the linear relations between the diffusion and reaction forces and currents in Eqs.~\eqref{linear_diff} and ~\eqref{linear_react} as 
\begin{align}\label{linear_unify}
    \mathcal{J}(\bm{r};t)=\mathcal{M}(\bm{r};t)\mathcal{F}(\bm{r};t),
\end{align}
and the positive-definiteness of $\dvec{\mathsf{M}}$ and $\mathsf{m}$ makes it invertible. The linear relation~\eqref{linear_unify} lets us rewrite the dynamics of the internal species as 
\begin{align}\label{rewrite_RDS_force}
    \partial_{t}\vec{c}_{\mathcal{X}}=\qty(\Div\mathcal{M}\mathcal{F})_{\mathcal{X}}.
\end{align}

Since $\dvec{\mathsf{M}}$ and $\mathsf{m}$ are symmetric, we obtain $\fip{\mathcal{M}\mathcal{F}'}{\mathcal{F}''} =\fip{\mathcal{F}'}{\mathcal{M} \mathcal{F}''}$ for any forces $\mathcal{F}'$ and $\mathcal{F}''$, which means that $\mathcal{M}$ is a self-adjoint operator. This property and the positive definiteness of $\mathcal{M}$ let us define a new inner product $\fip{\cdot}{\cdot}_{\mathcal{M}}$ as $\fip{\mathcal{F}'}{\mathcal{F}''}_{\mathcal{M}}\coloneqq\fip{\mathcal{M}\mathcal{F}'}{\mathcal{F}''} = \fip{\mathcal{F}'}{\mathcal{M} \mathcal{F}''}$.
\add{The positive-definiteness also guarantees that the inner product is nondegenerate, i.e., $\fip{\mathcal{F}'}{\mathcal{F}'}_{\mathcal{M}}>0$ holds for any $\mathcal{F}'\neq (\vec{\boldsymbol{0}}, \boldsymbol{0})$.} This inner product induced by $\mathcal{M}$ rewrites the EPR as the squared norm of the force as
\begin{align}
\sigma=\fip{\mathcal{J}}{\mathcal{F}}=\fip{\mathcal{M}\mathcal{F}}{\mathcal{F}}=\fip{\mathcal{F}}{\mathcal{F}}_{\mathcal{M}}.
\label{EPR_norm}
\end{align}
Now, the second law of thermodynamics is given by the nonnegativity of the norm, $\fip{\mathcal{F}}{\mathcal{F}}_{\mathcal{M}} \geq 0$.

\subsection{Conservative and nonconservative forces}\label{sec:conservative_nonconservative_forces}

RDSs are driven by two types of forces: one is the force solely due to the chemical potential of the internal species, and the other is the force due to the interaction with outside of the system, i.e., the chemical potential of the external species and the nonconservative mechanical force $\vec{\bm{K}}{}^{\mathrm{nc}}$.

From this viewpoint, we can rewrite the force~\eqref{RDS_force_vec} as
\begin{align}\label{RDS_force_rewrite}
    \mathcal{F}=-\Grad\begin{pmatrix}
 \vec{\mu}_{\mathcal{X}}\\\vecrm{0}_{\mathcal{Y}}
\end{pmatrix}-\Grad\begin{pmatrix}
 \vecrm{0}_{\mathcal{X}}\\\vec{\mu}_{\mathcal{Y}}
\end{pmatrix}+\mathcal{K}^{\mathrm{nc}}.
\end{align}
Here, the first term is determined solely by the chemical potential of the internal species. The remaining two terms are the contributions from the chemical potential of the external species and the nonconservative mechanical forces. 

Inspired by the form in Eq.~\eqref{RDS_force_rewrite}, we can decompose the force $\mathcal{F}$ into two parts as
\begin{align}\label{force_con_noncon}
    \mathcal{F}=\Grad\vec{\phi}+\mathcal{F}^{\mathrm{nc}},
\end{align}
where $\vec{\phi}$ in the first term is a potential whose external part is the zero vector, $\vec{\phi}_{\mathcal{Y}}=\vecrm{0}_{\mathcal{Y}}$. Here, the second term $\mathcal{F}^{\mathrm{nc}}$ is the remainder $\mathcal{F}-\Grad\vec{\phi}$. We refer to $\Grad\vec{\phi}$ and $\mathcal{F}^{\mathrm{nc}}$ as the conservative force and the nonconservative force, respectively. We remark that such a decomposition of the force into conservative and nonconservative forces is not unique. The representation with the chemical potential in Eq.~\eqref{RDS_force_rewrite} corresponds to the case where $\vec{\phi}=-(\vec{\mu}{}^{\top}_{\mathcal{X}},\vecrm{0}{}^{\top}_{\mathcal{Y}}){}^{\top}$ in Eq.~\eqref{force_con_noncon}. It indicates that $-\vec{\phi}$ may be easier to interpret thermodynamically than $\vec{\phi}$.


Using the decomposition in Eq.~\eqref{force_con_noncon}, we can rewrite the EPR $\sigma=\fip{\mathcal{J}}{\mathcal{F}}$ as 
\begin{align}\label{EPR_con_noncon}
    \sigma&=\fip{\mathcal{J}}{\Grad\vec{\phi}+\mathcal{F}^{\mathrm{nc}}}\notag\\
    &=\cip{\partial_{t}\vec{c}}{\vec{\phi}}+\fip{\mathcal{J}}{\mathcal{F}^{\mathrm{nc}}},
\end{align}
where we use the conjugation relation~\eqref{conjugate} and the assumption that $\phi_{\alpha}=0$ for all $\alpha\in\mathcal{Y}$. If the system is driven solely by the conservative force $\Grad\vec{\phi}$, the second term in Eq.~\eqref{EPR_con_noncon} vanishes so that the EPR becomes zero at a steady state, i.e., the system is in equilibrium at the steady state. The conservative force $\Grad\vec{\phi}$ drives relaxation to a state corresponding to $\vec{\phi}$ (see also Appendix~\ref{ap:relax_to_pcan}). On the other hand, the second term in Eq.~\eqref{EPR_con_noncon} provided by the nonconservative force $\mathcal{F}^{\mathrm{nc}}$ may not be zero at the steady state. The nonconservative force $\mathcal{F}^{\mathrm{nc}}$ maintains the system out of equilibrium even at the steady state.

\section{Geometric decompositions of entropy production rate for reaction-diffusion systems}\label{sec:geometric_decomps}
One way to understand the thermodynamics of RDSs is to decompose dissipation into contributions from different causes. \add{Such decompositions can be performed by focusing on the geometry of the forces as follows. Since the EPR is given by the squared norm of the force as $\sigma=\fip{\F}{\F}_{\M}$, we can decompose it by using the Pythagorean theorem as
\begin{align}
\fip{\F}{\F}_{\M}=\fip{\F^{\perp}}{\F^{\perp}}_{\M}+\fip{\F-\F^{\perp}}{\F-\F^{\perp}}_{\M},
\label{general_pythag}
\end{align}
with the force $\F^{\perp}$ satisfying the orthogonality $\fip{\F^{\perp}}{\F-\F^{\perp}}_{\M}=0$. For example, the decomposition into the contributions from diffusion and reactions~\eqref{decomp_diff_rct} is rewritten using the Pythagorean theorem as
\begin{align}
    \fip{\F}{\F}_{\M}=\fip{\F^{\mathrm{diff}}}{\F^{\mathrm{diff}}}_{\M}+\fip{\F^{\mathrm{react}}}{\F^{\mathrm{react}}}_{\M},
    \label{geometric RD decomp}
\end{align}
where the forces $\F^{\mathrm{diff}}\coloneqq(\vec{\bm{F}},\bm{0})$ and $\F^{\mathrm{react}}\coloneqq(\vecrm{\bm{0}},\bm{f})$ provide an orthogonal decomposition of the force as $\F=\F^{\mathrm{diff}}+\F^{\mathrm{react}}$ with $\fip{\F^{\mathrm{diff}}}{\F^{\mathrm{react}}}_{\M}=0$.}


\add{In this section, we introduce more complicated decompositions, using the same geometric method. In Sec.~\ref{sec:ex&hk}, we derive the geometric excess/housekeeping decomposition of the EPR by constructing an orthogonal decomposition of the force into the conservative and nonconservative parts. Notably, the excess EPR plays a central role in relating time evolution and dissipation, since it extracts the part of dissipation rate that contributes to the change in the pattern. We also develop the local decomposition and the wavenumber decomposition, which enable us to identify the dissipation at each point in real space and Fourier space, in Sec.~\ref{sec:spectral_decomp}. We demonstrate our geometric decompositions using two models, which are simple but present typical behaviors of RD systems, in Sec.~\ref{sec:system_for_numex}.} We remark that we can further decompose the decompositions obtained in this section into contributions from reaction and diffusion in the same way as in Eq.~\eqref{decomp_diff_rct}.

\subsection{Excess and housekeeping entropy production rate for reaction-diffusion systems}\label{sec:ex&hk}

\add{
The EPR $\sigma=\fip{\mathcal{F}}{\mathcal{F}}_{\mathcal{M}}$ includes contributions from both conservative and nonconservative sources, 
as shown in Eq.~\eqref{force_con_noncon}. To quantify these two contributions separately, we construct the geometric excess/housekeeping decomposition of EPR for RDSs by using the Pythagorean theorem,
\begin{align}
\fip{\mathcal{F}}{\mathcal{F}}_{\mathcal{M}}=\fip{\mathcal{F}^{\ast}}{\mathcal{F}^{\ast}}_{\mathcal{M}}+\fip{\mathcal{F}-\mathcal{F}^{\ast}}{\mathcal{F}-\mathcal{F}^{\ast}}_{\mathcal{M}},
\label{pythag}
\end{align}
where $\F^{\ast}$ is a conservative force satisfying the orthogonality $\fip{\F^{\ast}}{\F-\F^{\ast}}$. In contrast to the simple case~\eqref{geometric RD decomp}, it is difficult to find $\F^{\ast}$ directly. Instead, we obtain it by projecting the force $\F$ onto the conservative force space, which is defined as
\begin{align}
    \mathrm{Im}_{\mathcal{X}}\Grad\coloneqq\{\Grad\vec{\phi}\mid\forall\alpha\in\mathcal{Y},\,\phi_{\alpha}=0\}.
    \label{conservative_force_space}
\end{align}
Here, we use that conservative forces can be written as the generalized gradient of a potential whose external part is the zero vector, as discussed in Sec.~\ref{sec:conservative_nonconservative_forces}. As we will confirm shortly afterward, the conservation force $\F^{\ast}$ obtained by the projection satisfies the orthogonality and allows us to use the Pythagorean theorem~\eqref{pythag} to decompose EPR into contributions from conservative and nonconservative forces.
}

\textit{Projection of the force onto the conservative force space}--- 
\add{
The projection of the force onto the conservative force space is defined by a variational problem}
\begin{align}
\mathcal{F}^{\ast}\coloneqq \argmin_{\mathcal{F}'\in\mathrm{Im}_{\mathcal{X}}\Grad}\fip{\mathcal{F}-\mathcal{F}'}{\mathcal{F}-\mathcal{F}'}_{\mathcal{M}},
\label{Def_opt_force}
\end{align}
where we impose the same boundary condition on the diffusion part of $\mathcal{M}\mathcal{F}'$ as we do on $\mathcal{J}=\mathcal{M}\mathcal{F}$. For example, if we consider a system with the no-flux boundary condition on the diffusion currents of the internal species, then we also impose the same condition on $(\dvec{\mathsf{M}}\vec{\bm{F}}{}')_{\mathcal{X}}$ in the minimization problem. By definition, $\mathcal{F}^{\ast}$ can be given as $\mathcal{F}^{\ast}=\Grad\vec{\phi}{}^{\ast}$ with
\begin{align}
\vec{\phi}{}^{\ast} \in \argmin_{\vec{\phi}\mid\vec{\phi}_{\mathcal{Y}}=\vecrm{0}_{\mathcal{Y}}}\fip{\mathcal{F}-\Grad\vec{\phi}}{\mathcal{F}-\Grad\vec{\phi}}_{\mathcal{M}}.
\label{Def_optimalpot}
\end{align}
Here, we also impose the same boundary condition on the diffusion part of $\mathcal{M}\Grad\vec{\phi}$ as we do on $\mathcal{J}=\mathcal{M}\mathcal{F}$ in the minimization problem. The minimization problem in Eq.~\eqref{Def_optimalpot} reduces to solving the partial differential equation, \add{which is obtained as the Euler--Lagrange equation,
\begin{align}
\qty(\Div\mathcal{M}\mathcal{F})_{\mathcal{X}}=\qty(\Div\mathcal{M}\Grad\vec{\phi}{}^{\ast})_{\mathcal{X}},
\label{EL_eq}
\end{align}
with the same boundary condition on $(\dvec{\mathsf{M}}\bm{\nabla}_{\bm{r}}\vec{\phi})_{\mathcal{X}}$ as we imposed on the original dynamics of internal species. Note that we can uniquely determine $\mathcal{F}^{\ast}$ since the inner product $\fip{\cdot}{\cdot}_{\M}$ is nondegenerate (see Appendix~\ref{ap:projection_conservative} for details).
}

A notable property of the projected conservative force $\mathcal{F}^{\ast}=\Grad \vec{\phi}{}^*$ is that it preserves the time evolution of the original dynamics of internal species, since Eq.~\eqref{EL_eq} provides the same time evolution of $\vec{c}_{\mathcal{X}}$:
\begin{align}
   \partial_t \vec{c}_{\mathcal{X}}= \qty(\Div\mathcal{M} \mathcal{F})_{\mathcal{X}}=\qty(\Div\mathcal{M} \Grad \vec{\phi}{}^*)_{\mathcal{X}}.
   \label{sametimeevolutionRDS}
\end{align}
From this dynamics-conservation viewpoint, we may come up with another representation of $\mathcal{F}^{\ast}$ as 
\begin{align}\label{another_form_PF}
\mathcal{F}^{\ast}=\argmin_{\mathcal{F}'|(\Div\mathcal{M}\mathcal{F}')_{\mathcal{X}}=(\Div\mathcal{M}\mathcal{F})_{\mathcal{X}}}\fip{\mathcal{F}'}{\mathcal{F}'}_{\mathcal{M}},
\end{align}
with the same boundary condition as Eq.~\eqref{Def_opt_force}.
We can actually check that the minimization problem in Eq.~\eqref{another_form_PF} leads to the same Euler--Lagrange equation as Eq.~\eqref{EL_eq}. Note that the potential $\vec{\phi}$, whose external part is the zero vector, is introduced as the Lagrange multiplier for the constraint $(\Div\mathcal{M}\mathcal{F}')_{\mathcal{X}}=(\Div\mathcal{M}\mathcal{F})_{\mathcal{X}}$ in Eq.~\eqref{another_form_PF} (see Appendix~\ref{ap:minimum_dissipation}).

\add{As already mentioned, the projected conservative force $\mathcal{F}^{\ast}$ is orthogonal to $\mathcal{F}-\mathcal{F}^{\ast}$ with respect to the inner product $\fip{\cdot}{\cdot}_{\mathcal{M}}$ as $\fip{\mathcal{F}-\mathcal{F}^{\ast}}{\mathcal{F}^{\ast}}_{\mathcal{M}}=
\fip{\mathcal{M}\qty(\mathcal{F}-\mathcal{F}^{\ast})}{\Grad\vec{\phi}{}^{\ast}}=\cip{\Div\mathcal{M}\qty(\mathcal{F}-\mathcal{F}^{\ast})}{\vec{\phi}{}^{\ast}}=0$. Here, we used the boundary condition on $(\mathcal{M}\nabla\vec{\phi}{}^{\ast})_\mathcal{X}$, the condition that $\phi^{\ast}_{\alpha}=0$ for all $\alpha\in\mathcal{Y}$, and Eq.~\eqref{conjugate} in the second transformation. 
We also used the condition $\phi^{\ast}_{\alpha}=0$ for all $\alpha\in\mathcal{Y}$ and the Euler--Lagrange equation~\eqref{EL_eq} in the third transformation. 
This orthogonality finally leads to the Pythagorean theorem~\eqref{pythag}, which enables us to define a decomposition of the EPR.} 

\begin{figure}
    \centering
    \includegraphics[width=\linewidth]{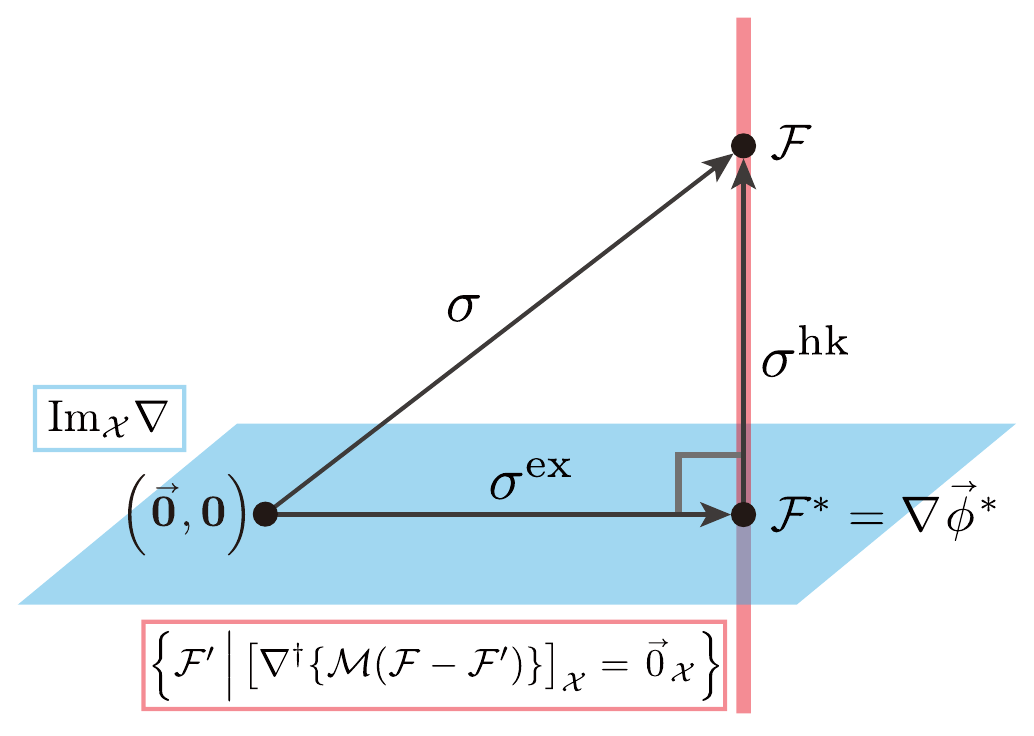}
    \caption{The geometric decomposition of the EPR for RDSs. Here, the blue plane indicates $\mathrm{Im}_{\mathcal{X}}\Grad$, and the red line indicates $\{ \mathcal{F}'|[\Div\{\mathcal{M} (\mathcal{F}-\mathcal{F}')\}]_{\mathcal{X}}= \vecrm{0}_{\mathcal{X}}\} $. The thermodynamic force, whose squared norm is the EPR  $\sigma=\fip{\mathcal{F}}{\mathcal{F}}_{\mathcal{M}}$, is decomposed into the two orthogonal parts: the projection of $\mathcal{F}$ onto $\mathrm{Im}_{\mathcal{X}}\Grad$, $\mathcal{F}^{\ast}$, whose squared norm is the excess EPR  $\sigma^{\mathrm{ex}}=\fip{\mathcal{F}^{\ast}}{\mathcal{F}^{\ast}}_{\mathcal{M}}$, and the remaining part, $\mathcal{F}-\mathcal{F}^{\ast}$, whose squared norm is the housekeeping EPR  $\sigma^{\mathrm{hk}}=\fip{\mathcal{F}-\mathcal{F}^{\ast}}{\mathcal{F}-\mathcal{F}^{\ast}}_{\mathcal{M}}$. This description is parallel to the case of the Langevin systems shown in Fig.~\ref{fig:FP_EPR_decomp}.}
    \label{fig:RDS_EPR_decomp}
\end{figure}
\textit{Excess/housekeeping decomposition of EPR}--- 
\add{Defining the excess and housekeeping EPRs (see also Fig.\ref{fig:RDS_EPR_decomp}) as
\begin{align}
\sigma^{\mathrm{ex}}\coloneqq\fip{\mathcal{F}^{\ast}}{\mathcal{F}^{\ast}}_{\mathcal{M}},\;\sigma^{\mathrm{hk}}\coloneqq\fip{\mathcal{F}-\mathcal{F}^{\ast}}{\mathcal{F}-\mathcal{F}^{\ast}}_{\mathcal{M}},
\label{Def_ex_hk}
\end{align}
the Pythagorean theorem~\eqref{pythag} leads to the geometric excess/housekeeping decomposition of the EPR
\begin{align}
    \sigma=\sigma^{\mathrm{ex}}+\sigma^{\mathrm{hk}}.
    \label{EPR_decomposition}
\end{align}
Here, the nonnegativity of the EPRs is ensured, since each term is the squared norm of the corresponding force. The time integration of the decomposition~\eqref{EPR_decomposition} immediately yields the geometric excess/housekeeping decomposition of the EP as
\begin{align}
    \Sigma_{\tau}=\Sigma^{\mathrm{ex}}_{\tau}+\Sigma^{\mathrm{hk}}_{\tau}.
\end{align}
The physical meaning of the excess and housekeeping EPRs is explained on the basis of the properties of $\F^{\ast}$ in the following.
}

\add{We focus on the excess EPR, which plays a central role in the following sections. The minimization problems, Eq.~\eqref{Def_opt_force} and Eq.~\eqref{another_form_PF}, imply that the excess EPR $\sigma^{\mathrm{ex}}$ is the minimum dissipation incurred by any conservative force that induces the original dynamics of the internal species. Indeed, the property of $\F^{\ast}$~\eqref{another_form_PF} gives the excess EPR by the following optimization problem, 
\begin{align}
\sigma^{\mathrm{ex}}=\inf_{\mathcal{F}'|(\Div\mathcal{M}\mathcal{F}')_{\mathcal{X}}=(\Div\mathcal{M}\mathcal{F})_{\mathcal{X}}}\fip{\mathcal{F}'}{\mathcal{F}'}_{\mathcal{M}},
\label{another_form_exEPR} 
\end{align}
which means that excess EPR extracts the unavoidable dissipation due to the time evolution of the pattern. The variational expression~\eqref{another_form_exEPR} also shows that the excess EPR vanishes at the steady state when the system is closed, since the zero function $\mathcal{F}^{\ast}= (\vec{\boldsymbol{0}}, \boldsymbol{0})$ then satisfies the constraints. 
If the system is open, the excess EPR will be zero as long as the concentrations of the internal species are stationary, even if those of the external species may not. We also remark that we can rewrite $\sigma^{\mathrm{ex}}$ as the time derivative of a quantity, defined in terms of the conservative force, that drives relaxation~\cite{yoshimura2023housekeeping} (see also Appendix~\ref{ap:gradientflow}).}

\add{We emphasize that the excess EPR defined here is different from Prigogine's one for RDSs~\cite{glansdorff1973thermodynamic}. The latter is defined only near the steady state and introduced to predict the stability of the steady state by its sign. On the other hand, the former is always nonnegative and can be defined far from the steady state or even in systems without steady states. In essence, it identifies the minimum dissipation required for time evolution. The common feature of both is that they become zero in the steady state.}

\add{We also reveal the physical meaning of the remaining contribution, namely, the housekeeping EPR
\begin{align}
\sigma^{\mathrm{hk}} &=\inf_{\mathcal{F}'\in\mathrm{Im}_{\mathcal{X}}\Grad}\fip{\mathcal{F}-\mathcal{F}'}{\mathcal{F}-\mathcal{F}'}_{\mathcal{M}}\notag\\
&= \inf_{\vec{\phi}|\vec{\phi}_{\mathcal{Y}}=\vecrm{0}_{\mathcal{Y}}}\fip{\mathcal{F}-\Grad\vec{\phi}}{\mathcal{F}-\Grad\vec{\phi}}_{\mathcal{M}}.
\label{another_form_hkEPR}
\end{align}
Although this paper mainly focuses on the excess EPR, the remaining housekeeping EPR also has an important physical meaning: it is the dissipation due to the current that maintains the pattern without changing it. To show this, we rewrite the Euler--Lagrange equation~\eqref{EL_eq} with the current corresponding to the projected conservative force $\mathcal{J}^{\ast} = \mathcal{M} \mathcal{F}^{\ast}$ as $\qty(\Div \mathcal{J})_{\mathcal{X}}=\qty(\Div \mathcal{J}^{\ast})_{\mathcal{X}}$. This indicates that $\mathcal{J}-\mathcal{J}^{\ast}$ is a \textit{cyclic current}, which does not affect the dynamics of the internal species as
\begin{align}
   \partial_t \vec{c}_{\mathcal{X}}= \qty(\Div \mathcal{J}^{\ast})_{\mathcal{X}} + \qty(\Div \qty[\mathcal{J}-\mathcal{J}^{\ast}])_{\mathcal{X}} = \qty(\Div \mathcal{J}^{\ast})_{\mathcal{X}}.
\end{align}
It is worth noting that while the cyclic currents are cyclic in terms of internal species, physically they are driven by external factors, such as external species or nonconservative mechanical forces. 
This is also the case with homogeneous CRNs without detailed balance, where external species are often made implicit. 
If no external species exist, the system will be detailed balanced, and no cyclic motion will be observed.} 

\add{We also remark that the cyclic current $\mathcal{J}-\mathcal{J}^{\ast}$ can affect the concentrations of the external species. Observe that $\partial_t \vec{c}_{\mathcal{Y}}=\qty(\Div \mathcal{J}^{\ast})_{\mathcal{Y}} + \qty(\Div \qty[\mathcal{J}-\mathcal{J}^{\ast}])_{\mathcal{Y}}$, where the second term in the right-hand side does not vanish generally. In other words, the housekeeping EPR consists of the cyclic contribution plus the contribution from the diffusion of external species. 
Suppose that the mobility tensor has no direct interaction terms between internal and external species, $\mathsf{M}_{(\alpha,\beta)}=\mathsf{O}$ if $(\alpha,\beta)\in\mathcal{X}\times\mathcal{Y}$ or $\mathcal{Y}\times\mathcal{X}$, where $\mathsf{O}$ is the zero matrix. Then, the interpretation is clearly depicted by the decomposition $\sigma^{\mathrm{hk}}=\sigma_{\mathcal{X}}^{\mathrm{cyc}}+\sigma_{\mathcal{Y}}^{\mathrm{diff}}$ with two nonnegative contributions
\begin{align}
    \sigma_\mathcal{X}^\mathrm{cyc}
    &\coloneqq\int_Vd\bm{r}\left[(\bm{f}-\nabla_\mathsf{s}\vec{\phi}{}^\ast)^\top\mathsf{m}(\bm{f}-\nabla_\mathsf{s}\vec{\phi}{}^\ast)
    \phantom{\sum_{\alpha,\beta\in\mathcal{X}}}
    \right.\notag\\
    &\left.+\sum_{\alpha,\beta\in\mathcal{X}}
    (\bm{F}_{(\alpha)}-\bm{\nabla}_{\bm{r}}\phi_\alpha^\ast)^\top
    \mathsf{M}_{(\alpha\beta)}
    (\bm{F}_{(\beta)}-\bm{\nabla}_{\bm{r}}\phi_\beta^\ast)\right]\geq0\\
    \sigma_\mathcal{Y}^\mathrm{diff}
    &\coloneqq\int_Vd\bm{r}\sum_{\alpha\in\mathcal{Y}}\bm{J}_{(\alpha)}\cdot\bm{F}_{(\alpha)}\geq0.
\end{align}
The first term $\sigma_\mathcal{X}^\mathrm{cyc}$ reflects the cyclic motion of the internal species, arising from the cyclic current $\mathcal{J}-\mathcal{J}^\ast$. 
The remainder reflects the dissipation stemming from  diffusion of the external species. 
This term is particular to RDSs, as the housekeeping EPR in a homogeneous CRN can be written only with cyclic contributions~\cite{yoshimura2023housekeeping}. 
The decomposition is straightforwardly proved using the definition of $\sigma^\mathrm{hk}$ and the fact that $\bm{F}_{(\alpha)}-\boldsymbol{\nabla}_{\boldsymbol{r}} \phi^*_\alpha =\bm{F}_{(\alpha)} $ for any $\alpha \in \mathcal{Y}$ because $\phi_\alpha^\ast=0$ for any $\alpha\in\mathcal{Y}$ with the assumption $\mathsf{M}_{(\alpha,\beta)}=\mathsf{O}$ if $(\alpha,\beta)\in\mathcal{X}\times\mathcal{Y}$ or $(\alpha,\beta)\in\mathcal{Y}\times\mathcal{X}$.}

\subsection{Local decomposition and wavenumber decomposition of entropy production rate}\label{sec:spectral_decomp}
In this section, we discuss the local and wavenumber decomposition of EPR for RDSs, analogous to the decomposition for Langevin systems from Sec.~\ref{sec:local_wn_decomp}. The local and wavenumber EPRs enable us to quantify the dissipation at each point in the real and wavenumber spaces, respectively. Thus, we can detect local dissipation due to pattern formation via the local and wavenumber EPRs for RDSs. Further below, we use local EPR decomposition to quantify the spatial dissipation in numerical examples of pattern formation. 

\textit{Local decomposition}---We define the local EPR at the position $\bm{r}$ as 
\begin{align}
 \sigma^{\mathrm{loc}} (\bm{r})\coloneqq \vec{\bm{J}}{}^{\top}(\bm{r};t)\vec{\bm{F}}(\bm{r};t) + \bm{j}^{\top}(\bm{r};t) \bm{f}(\bm{r};t) .
\end{align}
The nonnegativity $\sigma^{\mathrm{loc}} (\bm{r}) \geq 0$ holds locally as it is locally that $\dvec{\mathsf{M}}$ is assumed to be positive-definite and $\bm{j}$ and $\bm{f}$ have the same sign. 
The volume integral gives the EPR,
\begin{align}\label{RDS_localdecomp}
     \sigma=\int_{V}d\bm{r}\,\sigma^{\mathrm{loc}} (\bm{r}).
\end{align}
This equality~\eqref{RDS_localdecomp} indicates a decomposition of the EPR into the local EPRs, the dissipation at each location.
 
We also decompose the excess and housekeeping EPRs into the contributions from each location as
\begin{align}
\sigma^{\mathrm{ex},\mathrm{loc}}(\bm{r})&\coloneqq\qty(\bm{\nabla}_{\bm{r}}\vec{\phi}{}^{\ast}){}^{\top}\dvec{\mathsf{M}}\bm{\nabla}_{\bm{r}}\vec{\phi}{}^{\ast}+\qty(\nabla_{\mathsf{s}}\vec{\phi}{}^{\ast}){}^{\top}\mathsf{m}\nabla_{\mathsf{s}}\vec{\phi}{}^{\ast},\label{partial_ex}\\
\sigma^{\mathrm{hk},\mathrm{loc}}(\bm{r})&\coloneqq\qty(\vec{\bm{F}}-\bm{\nabla}_{\bm{r}}\vec{\phi}{}^{\ast}){}^{\top}\dvec{\mathsf{M}}\qty(\vec{\bm{F}}-\bm{\nabla}_{\bm{r}}\vec{\phi}{}^{\ast})\notag\\
&\phantom{\coloneqq}+\qty(\bm{f}-\nabla_{\mathsf{s}}\vec{\phi}{}^{\ast}){}^{\top}\mathsf{m}\qty(\bm{f}-\nabla_{\mathsf{s}}\vec{\phi}{}^{\ast})\label{partial_hk},
\end{align}
which satisfy $\int_{V}d\bm{r}\,\sigma^{\mathrm{ex},\mathrm{loc}}(\bm{r})=\sigma^{\mathrm{ex}}$ and $\int_{V}d\bm{r}\,\sigma^{\mathrm{hk},\mathrm{loc}}(\bm{r})=\sigma^{\mathrm{hk}}$.
The local excess and housekeeping EPRs are nonnegative because $\dvec{\mathsf{M}}$ and $\mathsf{m}$ are positive-definite. Thus, we can interpret $\sigma^{\mathrm{ex,loc}}(\bm{r})$ as the EPR due to the projected conservative force $\mathcal{F}^{\ast}$ at the location $\bm{r}$. We can also regard $\sigma^{\mathrm{hk,loc}}(\bm{r})$ as the EPR due to the cyclic current at the location $\bm{r}$.
Note that $\sigma^{\mathrm{ex},\mathrm{loc}}(\bm{r})=0$ and $\sigma^{\mathrm{hk},\mathrm{loc}}(\bm{r})=\sigma^{\mathrm{loc}}(\bm{r})$ hold for all $\bm{r}\in V$ if and only if $\Grad\vec{\phi}{}^*=(\vec{\boldsymbol{0}}, \boldsymbol{0})$. The time evolution of the internal species needs to be stationary to achieve this condition because $\partial_t \vec{c}_{\mathcal{X}} = \qty(\Div\mathcal{M} \Grad \vec{\phi}{}^*)_{\mathcal{X}}= \vecrm{0}_{\mathcal{X}}$ holds when $\Grad\vec{\phi}{}^*=(\vec{\boldsymbol{0}}, \boldsymbol{0})$. 

We remark that the definitions of the local excess and housekeeping EPRs include $\vec{\phi}{}^{\ast}$, which is defined in terms of global information about the RDS. For this reason, the local excess and housekeeping EPRs cannot be defined solely from local information. Reflecting this nonlocality, the local excess and housekeeping EPRs do not sum up to the local EPR $\sigma^{\mathrm{loc}}(\bm{r})  \neq \sigma^{\mathrm{ex},\mathrm{loc}}(\bm{r}) + \sigma^{\mathrm{hk},\mathrm{loc}}(\bm{r})$. 
The non-zero cross-term can be obtained as $\sigma^{\mathrm{cross}}(\bm{r})\coloneqq  \sigma^{\mathrm{loc}}(\bm{r}) - \sigma^{\mathrm{ex},\mathrm{loc}}(\bm{r})  - \sigma^{\mathrm{hk},\mathrm{loc}}(\bm{r})= 2(\bm{\nabla}_{\bm{r}}\vec{\phi}{}^{\ast}){}^{\top}\dvec{\mathsf{M}}(\vec{\bm{F}}-\bm{\nabla}_{\bm{r}}\vec{\phi}{}^{\ast})+2(\nabla_{\mathsf{s}}\vec{\phi}{}^{\ast}){}^{\top}\mathsf{m}(\bm{f}-\nabla_{\mathsf{s}}\vec{\phi}{}^{\ast})$, which may be positive or negative in sign. It also satisfies $\int_{V}d\bm{r}\,\sigma^{\mathrm{cross}}(\bm{r})=0$, which maintains the geometric decomposition globally as $\sigma = \sigma^{\mathrm{ex}}+\sigma^{\mathrm{hk}}$.

\textit{Wavenumber decomposition}---
\add{We also provide the wavenumber decomposition of the EPR using Parseval's identity. We define the weighted Fourier transform of forces $\hat{\vec{\bm{F}}}{}'$ and $\hat{\bm{f}}'$ as
\begin{align}
    \displaystyle\hat{F}'_{(\alpha)i}(\bm{k};t)&\coloneqq\displaystyle\int_{V}d\bm{r}\,[\dvec{\mathsf{M}}{}^{\frac{1}{2}}\vec{\bm{F}}{}']_{(\alpha)i}\mathrm{e}^{-\mathrm{i}\bm{k}\cdot\bm{r}},\\
    \displaystyle\hat{f}'_{\rho}(\bm{k};t)&\coloneqq\displaystyle\int_{V}d\bm{r}\,[\mathsf{m}^{\frac{1}{2}}\bm{f}']_{\rho}\mathrm{e}^{-\mathrm{i}\bm{k}\cdot\bm{r}}
    \label{weighted_FT}. 
\end{align}
Here, we use the square root of the mobility tensor $\dvec{\mathsf{M}}{}^{\frac{1}{2}}$ and the edgewise Onsager matrix $\mathsf{m}^{\frac{1}{2}}$, which satisfies $\sum_{\gamma\in\mathcal{S}}\sum_{k=1}^{d}[\dvec{\mathsf{M}}{}^{\frac{1}{2}}]_{(\alpha\gamma) ik}[\dvec{\mathsf{M}}{}^{\frac{1}{2}}]_{(\gamma\beta) kj}=M_{(\alpha\beta) ij}$ and $[\mathsf{m}^{\frac{1}{2}}]_{\rho\rho'}=\sqrt{m_{\rho}}\delta_{\rho\rho'}$. The existence of these operators is guaranteed by the positive-definiteness of $\dvec{\mathsf{M}}$ and $\mathsf{m}$. Note that the elements of $\dvec{\mathsf{M}}{}^{\frac{1}{2}}$ satisfy $[\dvec{\mathsf{M}}{}^{\frac{1}{2}}]_{(\alpha\beta) ij}(\bm{r};t)=\sqrt{D_{\alpha}c_{\alpha}(\bm{r};t)}\delta_{\alpha\beta}\delta_{ij}$ if the mobility tensor has the simple form in Eq.~\eqref{simple_mobility_tensor}.}

\add{Since periodic boundary conditions discretize the wavenumber, the details of the wavenumber decomposition differ slightly between cases where periodic boundary conditions are imposed and cases where they are not. If we do not impose periodic boundary conditions on the system, we define the wavenumber EPR as
\begin{align}
    \sigma^{\mathrm{wn}}(\bm{k})&\coloneqq\frac{1}{(2\pi)^d}\left[\hat{\vec{\bm{F}}}{}^{\dag}(\bm{k};t)\hat{\vec{\bm{F}}}(\bm{k};t)+\hat{\bm{f}}{}^{\dag}(\bm{k};t)\hat{\bm{f}}(\bm{k};t)\right]\notag\\
    &=\frac{1}{(2\pi)^d}\left[\sum_{\alpha\in\mathcal{S}}\sum_{i=1}^{d}\overline{\hat{F}_{(\alpha)i}}\hat{F}_{(\alpha)i}+\sum_{\rho}\overline{\hat{f}_{\rho}}\hat{f}_{\rho}\right]\geq0,
    \label{RDS_wavenumberEPR}
\end{align}
where the superscript $\dag$ indicates conjugate transpose. We can obtain the decomposition
\begin{align}
    \sigma=\int_{\mathbb{R}^d}d\bm{k}\,\sigma^{\mathrm{wn}}(\bm{k}),
    \label{RDS_wavenumberdecomp}
\end{align}
by Parseval's identity. It follows from the Fourier transform of the delta function, as was the case for Langevin systems in Eq.~\eqref{FP_drv_wndecomp} (see also Appendix~\ref{ap:wavenumber_derivation} for the derivation). On the other hand, if we consider a system with periodic boundaries, we define the wavenumber EPR as}
\begin{align}
    \sigma^{\mathrm{wn}}(\bm{k})&\coloneqq\frac{1}{|V|}\left[\hat{\vec{\bm{F}}}{}^{\dag}(\bm{k};t)\hat{\vec{\bm{F}}}(\bm{k};t)+\hat{\bm{f}}{}^{\dag}(\bm{k};t)\hat{\bm{f}}(\bm{k};t)\right]\notag\\
    &=\frac{1}{|V|}\left[\sum_{\alpha\in\mathcal{S}}\sum_{i=1}^{d}\overline{\hat{F}_{(\alpha)i}}\hat{F}_{(\alpha)i}+\sum_{\rho}\overline{\hat{f}_{\rho}}\hat{f}_{\rho}\right]\geq0,
    \label{RDS_wavenumberEPR_PBC}
\end{align}
Here, we let $|V|$ denote the volume of the space $V$. We consider discrete wavenumbers because of the periodic boundary conditions. In this case, we can also obtain the decomposition
\begin{align}
    \sigma=\sum_{\bm{k}}\sigma^{\mathrm{wn}}(\bm{k}),
    \label{RDS_wavenumberdecomp_PBC}
\end{align}
using the Fourier series expansion of the delta function, $\delta(\bm{r})=\sum_{\bm{k}}\mathrm{e}^{\mathrm{i}\bm{k}\cdot\bm{r}}/|V|$ (see also Appendix~\ref{ap:wavenumber_derivation}).

We can also define the wavenumber decomposition of the excess and housekeeping EPRs, $\sigma^{\mathrm{ex,wn}}$ and $\sigma^{\mathrm{hk,wn}}$, using $(\hat{\vec{\bm{F}}}{}^{\ast},\hat{\bm{f}}{}^{\ast})$ and $(\hat{\vec{\bm{F}}}-\hat{\vec{\bm{F}}}{}^{\ast},\hat{\bm{f}}-\hat{\bm{f}}{}^{\ast})$ instead of $(\hat{\vec{\bm{F}}},\hat{\bm{f}})$, respectively. We can interpret $\sigma^{\mathrm{ex,wn}}(\bm{k})$ as the EPR due to the projected conservative force $\mathcal{F}^{\ast}$ at the wavenumber $\bm{k}$. We can also regard $\sigma^{\mathrm{hk,wn}}(\bm{k})$ as the EPR due to the cyclic current at the wavenumber $\bm{k}$. Note that the geometric excess/housekeeping decomposition can be violated at each wavenumber, $\sigma^{\mathrm{wn}}(\bm{k})\neq\sigma^{\mathrm{ex,wn}}(\bm{k})+\sigma^{\mathrm{hk,wn}}(\bm{k})$, as in the case for Langevin systems.

\add{The local and wavenumber decompositions are based on the orthonormality of the basis. Therefore, it is also possible to decompose the EPR using an orthonormal basis other than the Fourier basis. For example, a wavelet basis~\cite{graps1995introduction,torrence1998practical} may allow us to quantify the dissipation corresponding to particular wave packets.}

\subsection{Numerical examples: geometric decompositions}\label{sec:system_for_numex} 
\add{Here we show numerical examples of the geometric decompositions for open RDSs. We discuss the Fisher--KPP equation and the Brusselator model in one dimension, $V_1=[-0.5,0.5]$. When discussing numerical results here and in the following sections, we use $r$, $k$, $\partial_{r}$, $J_{(\alpha)}$ and $F_{(\alpha)}$ instead of $\bm{r}$, $\bm{k}$, $\bm{\nabla}_{\bm{r}}$, $\bm{J}_{(\alpha)}$ and $\bm{F}_{(\alpha)}$ respectively, because we consider only one dimensional systems. We use the same models with the same parameters for other numerical results in Sec.~\ref{sec:numex_TSL} and Sec.~\ref{sec:numex_TUR}.}

\textit{Fisher--KPP equation}.---
The Fisher--KPP equation consists of an internal species $Z_1$, an external species $Z_2$, and an autocatalytic reaction
\begin{align}\label{reaction_FKPP}
    \ce{$Z_1+Z_2$ <=>C[$\kappa_1^+$][$\kappa_1^-$] $2Z_1$}.
\end{align}
Now, the index sets are $\mathcal{S}=\{1,2\}$, $\mathcal{X}=\{1\}$, $\mathcal{Y}=\{2\}$, $\mathcal{R}=\{1\}$, and $\mathcal{R}_{\mathcal{X}}=\{1\}=\mathcal{R}$, and the vectors $\vec{c}=(c_1,c_2)^{\top}$, $\vec{c}_{\mathcal{X}}=(c_1)$, and $\vec{c}_{\mathcal{Y}}=(c_2)$. The stoichiometric matrix is
\begin{align}
    \nabla_{\mathsf{s}}^{\top}=\mqty(1\\-1).
\end{align}
In this system, we assume that the concentration of the external species is kept homogeneous by the interaction with the outside: $c_2(r;t)=1$ holds for all $r\in V_1$. We let the mobility tensor take the simple form in Eq.~\eqref{simple_mobility_tensor}, and assume Fick's law for the diffusion currents, $J_{(\alpha)}=-D_{\alpha}\partial_{r}c_{\alpha}$. Here, $J_{(2)}=0$ because $c_2 (r;t)$ is homogeneous and $\partial_{r}c_{2}=0$. We also assume mass action kinetics for the reaction fluxes: $j^{+}_1(r;t)=\kappa_1^{+}c_1(r;t)c_2(r;t)=\kappa_1^{+}c_1(r;t)$, $j^{-}_1(r;t)=\kappa_1^{-}c_1(r;t)^2$. Then, we can write the dynamics as
\begin{align}
    \partial_tc_1=D_1\partial_r^2c_1+\kappa_1^{+}c_1-\kappa_1^{-}c_1^2.
\end{align}
We impose the no-flux boundary condition $\left. \partial_r c_1(r;t) \right|_{r=\pm{0.5}}=0$. We also use the parameters $D_1=10^{-4}$ and $(\kappa_1^{+}, \kappa_1^{-})=(1,1)$.

In the Fisher--KPP equation, we can explicitly write down the condition to determine the potential $\vec{\phi}{}^{\ast}$~\eqref{EL_eq} as
\begin{align}\label{EL_eq_FKPP}
    &D_1\partial_r^2c_1+\kappa_1^{+}c_1-\kappa_1^{-}c_1^2\notag\\
    &=-\partial_r\qty(D_1c_1\partial_r\phi^{\ast}_1)+\frac{\kappa_1^{+}c_1-\kappa_1^{-}c_1^2}{\ln{(\kappa_1^{+}c_1)}-\ln{(\kappa_1^{-}c_1^2)}}\phi_1^{\ast},
\end{align}
which $\phi^{\ast}_1=\ln (\kappa_1^+/(\kappa_1^-c_1))$ solves. 

Therefore, the excess EPR is 
\begin{align*}
    \sigma^{\mathrm{ex}}&=\int_{-0.5}^{0.5}dr\left[D_1c_1(\partial_r\phi^{\ast}_1)^2+\frac{\qty(\kappa_1^{+}c_1-\kappa_1^{-}c_1^2)\qty(\phi_1^{\ast})^2}{\ln{(\kappa_1^{+}c_1)}-\ln{(\kappa_1^{-}c_1^2)}}\right]\\&=\int_{-0.5}^{0.5}dr\left[D_1c_1(\partial_r\ln c_1)^2+\qty(\kappa_1^{+}c_1-\kappa_1^{-}c_1^2)\ln \frac{\kappa_1^+}{\kappa_1^-c_1}\right]
\end{align*}
This is the same as the total EPR $\sigma$ because $F_{(1)}=\partial_r\ln c_1$ and $f_1=\ln (\kappa_1^+/(\kappa_1^-c_1))$.

\begin{figure}
    \centering
    \includegraphics[width=\linewidth]{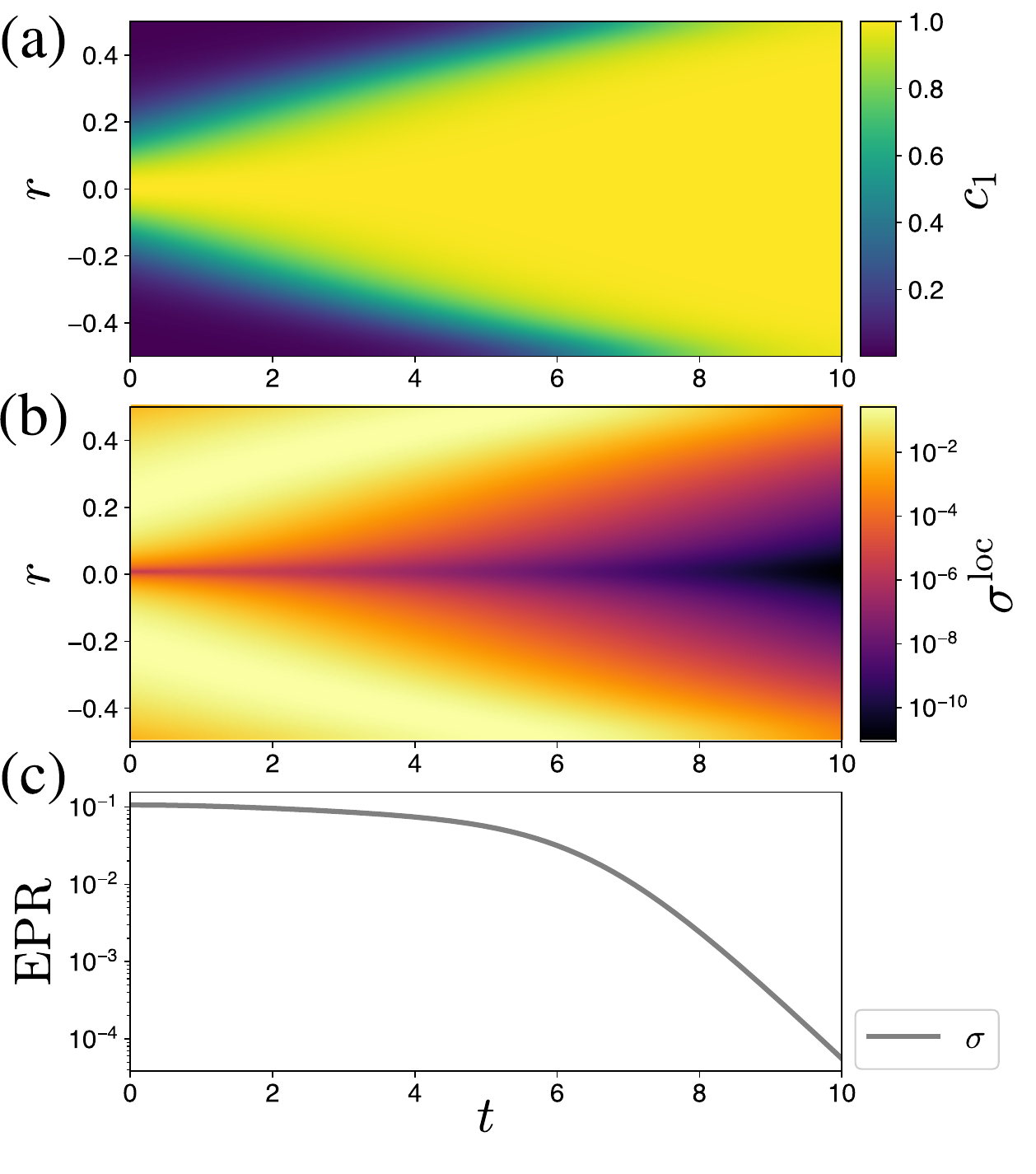}
    \caption{(a) The time series of $c_1$ in the Fisher--KPP equation. The area with high $c_1$ (yellow) expands as time passes. (b) The local EPR  $\sigma^{\mathrm{loc}}$ in the Fisher--KPP equation. The local EPR is colored on a logarithmic scale.  Comparing (a) with (b), we can see that dissipation occurs at the boundary between high- and low-concentration areas. (c) The EPR $\sigma$ in the Fisher--KPP equation. Reflecting the relaxation to the equilibrium state, the EPR is monotonically decreasing in time.}
    \label{fig:EPR_KF}
\end{figure}

In Fig.~\ref{fig:EPR_KF}, we numerically show the time series of the local EPR $\sigma^{\mathrm{loc}}$ and the concentration distribution of the internal species $c_1$. As is well known, in the Fisher--KPP equation, the area of high concentration of $c_1$ spreads over time~\cite{murray2002mathematical}: unlike normal diffusion, the total concentration is not conserved. The reaction does not occur inside the high-concentration area because $c_1=1$ is the equilibrium concentration of the reaction in Eq.~\eqref{reaction_FKPP} with given parameters. Diffusion also does not occur inside the high-concentration area because the concentration gradient disappears there. Thus, the local EPR is larger at the boundary between the high- and low-concentration areas, and no dissipation occurs inside the high-concentration area.

\textit{Brusselator model}.--- The Brusselator model consists of two internal species, $Z_1$ and $Z_2$, an external species $Z_3$, and three reactions,
\begin{align}
    \ce{$Z_3$ <=>C[$\kappa_1^+$][$\kappa_1^-$]  $Z_1$},\quad\ce{$Z_1$ <=>C[$\kappa_2^+$][$\kappa_2^-$]  $Z_2$},\quad\ce{$2Z_1+Z_2$ <=>C[$\kappa_3^+$][$\kappa_3^-$]  $3Z_1$},
\end{align}
where we label the reactions $\rho=1$, $2$, $3$ from left to right. The index sets are $\mathcal{S}=\{1,2,3\}$, $\mathcal{X}=\{1,2\}$, $\mathcal{Y}=\{3\}$, $\mathcal{R}=\{1, 2, 3\}$, and $\mathcal{R}_{\mathcal{X}}=\{1,2,3\}=\mathcal{R}$, and the vectors $\vec{c}=(c_1,c_2,c_3){}^{\top}$, $\vec{c}_{\mathcal{X}}=(c_1,c_2){}^{\top}$, and $\vec{c}_{\mathcal{Y}}=(c_3)$. The stoichiometric matrix is
\begin{align}
    \nabla_{\mathsf{s}}^{\top}=\mqty(1&-1&1\\0&1&-1\\-1&0&0).
\end{align}
The concentration of the external species is again assumed to be homogeneous due to the interaction with the outside: $c_3(r;t)=1$ for all $r\in V_1$. We let the mobility tensor have the simple form in Eq.~\eqref{simple_mobility_tensor} and assume Fick's law for the diffusion currents, $J_{(\alpha)}=-D_{\alpha}\partial_{r}c_{\alpha}$. Here, $J_{(3)}=0$ because $c_3$ is homogeneous and $\partial_r c_3=0$. We also assume the mass action kinetics for the reaction fluxes: $j^{+}_1(r;t)=\kappa_1^{+}c_3(r;t)=\kappa_1^{+}$, $j^{-}_1(r;t)=\kappa_1^{-}c_1(r;t)$, $j^{+}_2(r;t)=\kappa_2^{+}c_1(r;t)$, $j^{-}_2(r;t)=\kappa_2^{-}c_2(r;t)$, $j^{+}_3(r;t)=\kappa_3^{+}c_1(r;t)^2c_2(r;t)$, and $j^{-}_1(r;t)=\kappa_3^{-}c_1(r;t)^3$. Then, the dynamics are given by
\begin{numcases}{}
\partial_tc_1 =D_1\partial_r^2c_1+\kappa_1^{+}-\kappa_1^{-}c_1\nonumber\\ \phantom{\partial_tc_1 =}-\kappa_{2}^{+}c_1+\kappa_{2}^{-}c_2+\kappa_3^{+}c_1^2c_2-\kappa_3^{-}c_1^3\label{dynamics_c1}\\
\partial_tc_2 =D_2\partial_r^2c_2+\kappa_{2}^{+}c_1-\kappa_{2}^{-}c_2-\kappa_3^{+}c_1^2c_2+\kappa_3^{-}c_1^3.\label{dynamics_c2}
\end{numcases}
In the following numerical examples, we use the parameters as follows: $D_1=1.6\times10^{-4}$, $D_2=10^{-3}$, and $(\kappa_1^{+},\kappa_1^{-},\kappa_2^{+},\kappa_2^{-},\kappa_3^{+},\kappa_3^{-})=(1,1,10,0.1,1,1)$. \add{We also impose the periodic boundary conditions, which let each wavenumber be determined by an integer $n$ as
\begin{align}
    k(n)=\frac{2n\pi}{|V_1|}=2n\pi,
    \label{wavenumber-n}
\end{align}
where $|V_1|=1$ is the system size. In the following, we abbreviate the wavenumber EPRs $\sigma^{\mathrm{wn}}(\bm{k})$ and $\sigma^{\mathrm{ex,wn}}(\bm{k})$ as $\sigma^{\mathrm{wn}}(n)$ and $\sigma^{\mathrm{ex,wn}}(n)$.}

\begin{figure*}
    \centering
    \includegraphics[width=\linewidth]{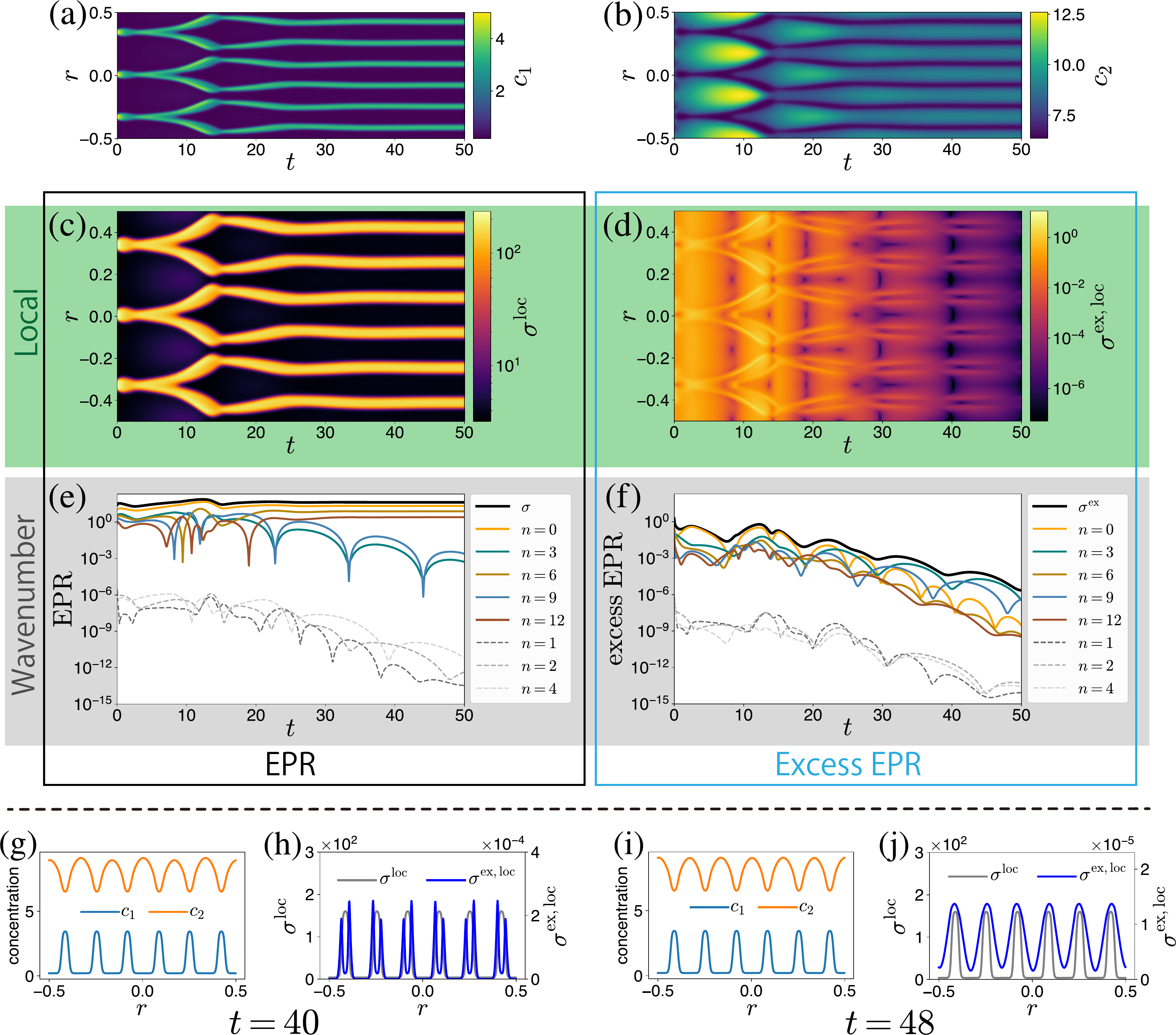}
    \caption{\add{The comparison of the EPR and the excess EPR in the Brusselator model. (a) The time series of $c_1$. (b) The time series of $c_2$. The symmetry of the pattern changes from $3$-fold with time evolution. It reaches $6$-fold symmetry at around $t=25$. (c) The local EPR $\sigma^{\mathrm{loc}}$ colored on a logarithmic scale. In contrast to the Fisher--KPP equation [Fig~\ref{fig:EPR_KF}(b)], $\sigma^{\mathrm{loc}}$ is large on the peaks of the pattern (areas where $c_1$ is high), not on the edges of the pattern. (d) The local excess EPR $\sigma^{\mathrm{ex,loc}}$ colored on a logarithmic scale. It tends to be larger on the edge of the pattern. (e) The EPR and the wavenumber EPRs for some modes. The EPR $\sigma$ (black line) does not vanish to maintain the pattern. Because of the $3$-fold symmetry of the pattern, the wavenumber EPR $\sigma^{\mathrm{wn}}(n)$ is small if $n\neq 3n'$ for any integer $n'$. Further, while the wavenumber EPRs of the form $\sigma^{\mathrm{wn}}(6n')$ do not decay, those with $n\neq 6n'$ decay even if $n$ is a multiple of three. It reflects that the symmetry of the pattern becomes $6$-fold as the system approaches the steady state. (f) The excess EPR and the wavenumber excess EPRs for some modes. The excess EPR $\sigma^{\mathrm{ex}}$ (black line) is considerably smaller than the EPR and decreases as the system approaches the steady state. The $3$-fold symmetry of the pattern makes the wavenumber excess EPR to be small for $n$ that are not multiples of three, as in the case of the wavenumber EPRs. In contrast to the EPR, there is a qualitative order $\sigma^{\mathrm{ex,wn}}(6n')\leq \sigma^{\mathrm{ex,wn}}(3n')$. This is caused by two facts: (1) the excess EPR reflects the change in pattern, and (2) the Fourier components corresponding to wavenumbers that are multiples of three and not six decay as the system approaches the steady state, while those corresponding to multiples of six barely change. (g) Concentrations $c_1(r;t)$ (cyan line) and $c_2(r;t)$ (orange line) as a function of spatial location $r$ at $t=40$. (h) Values of $\sigma^{\mathrm{loc}}(r)$ (gray line) and $\sigma^{\mathrm{ex,loc}}(r)$ (blue line) as a function of spatial location $r$ at $t=40$. The local EPR becomes large on the peaks of the pattern, while the local excess EPR has strong peaks at the edge of the pattern. (i) Values of $\vec{c}_{\mathcal{X}}(t)$ as a function of spatial location $r$ at $t=48$. (j) The values $\sigma^{\mathrm{loc}}(r)$ (gray line) and $\sigma^{\mathrm{ex,loc}}(r)$ (blue line) as a function of spatial location $r$ at $t=48$. Both of the local EPRs become large on the peaks of the pattern.}}
    \label{fig:EPR_BL}
\end{figure*}

\add{We demonstrate how the EPR and the excess EPR change quantitatively with the time evolution of the concentration distribution. Note that we numerically compute the excess EPR since the condition that determines the potential $\phi^{\ast}$~\eqref{EL_eq} for the Brusselator model is difficult to solve analytically, unlike the case of the Fisher--KPP equation. The time series of the concentration distribution and the EPRs are shown in Fig.~\ref{fig:EPR_BL}.}

We can see that the excess EPR \add{[black line in Fig.~\ref{fig:EPR_BL}(f)]} decreases (non-monotonically) as the system approaches the steady state. In addition, the excess EPR $\sigma^{\mathrm{ex}}$ is much smaller than the EPR $\sigma$ \add{[black line in Fig.~\ref{fig:EPR_BL}(e)]} as the majority of the dissipation is the housekeeping EPR, which is caused by cyclic currents that do not affect the time evolution. \add{In other words, for these parameters values, more dissipation is spent maintaining the pattern than changing it.}

\add{The local EPRs show the difference between the EPR and the excess EPR in more detail.} In contrast to the previous model, the local EPR $\sigma^{\mathrm{loc}}$ \add{[black line in Fig.~\ref{fig:EPR_BL}(e)]} is larger on the pattern \add{[areas where $c_1$ is high in Fig.~\ref{fig:EPR_BL}(a)]}, rather than the edges of the pattern. \add{This tendency reflects that the violation of the detailed balance requires dissipation to maintain the pattern, even when the change in the concentration is stationary.} On the other hand, the local excess EPR $\sigma^{\mathrm{ex,loc}}$ is large at the edges of the pattern unless the system is close to the steady state. \add{This reflects the following fact: when the system is far from the steady state, the time evolution tends to be faster at the edges of the pattern, where the gradients of concentrations are larger. We also show the detail of these tendencies in Fig.~\ref{fig:EPR_BL}(g-j): at $t=40$, the local excess EPR is large on the edge of the pattern, while the local EPR is large on the pattern, as shown in Fig.~\ref{fig:EPR_BL}(h). On the other hand, both local EPRs take a large value on the pattern once the system approaches steady state ($t=48$), as shown in Fig.~\ref{fig:EPR_BL}(j).}

\add{The wavenumber decomposition shows the difference between the EPR and the excess EPR more explicitly. This is because the symmetry of the pattern is clear in this numerical example. Here, we focus on the change in the symmetry of the pattern from $3$-fold to $6$-fold. Near steady state, the wavenumber EPR $\sigma^{\mathrm{wn}}(n)$ is larger at $n$ that are multiples of six than at other multiples of three [Fig.~\ref{fig:EPR_BL}(e)]. This reflects that the symmetry of the pattern is $6$-fold at the steady state. On the other hand, near steady state, the wavenumber excess EPR is larger at $n$ that are multiples of three but not multiples of six [Fig.~\ref{fig:EPR_BL}(f)]. This reflects the faster decay of the wavenumber components of the pattern corresponding to $n$ that are multiples of three rather than multiples of six.}

\section{Optimal transport and Thermodynamic speed limits for reaction-diffusion systems}\label{sec:RDS_Wdist&TSL}

We can understand dissipation in RDSs from the perspective of thermodynamic trade-off relations, which quantify the minimum dissipation required to achieve an objective. In particular, we focus on the thermodynamic speed limits (TSLs), which are trade-off relations between the speed of the dynamics and dissipation. They are geometric relations since they typically use some measure of ``distance'' between the initial and final patterns to quantify the speed. 
We measure the distance between two patterns of an RDS with the Wasserstein distance, similar to Langevin systems and MJPs~\cite{aurell2011optimal,aurell2012refined,dechant2019thermodynamic,nakazato2021geometrical,van2021geometrical,dechant2022minimum, yoshimura2023housekeeping, van2023thermodynamic, kolchinsky2022information, van2023topological,miangolarra2023minimal}.

An RDS is a composite of chemical reactions and diffusive dynamics. Since some kinds of Wasserstein distance have been studied for both types of dynamics, we can generalize the $1$- and $2$-Wasserstein distances to RDSs to derive TSLs. 
Sections~\ref{sec:RDS_1W_dist} and \ref{sec:RDS_2W_dist} are dedicated to generalization of the Wasserstein distances, while some differences between these distances are discussed in Sec.~\ref{sec:properties_of_Wdist}. 
We derive TSLs with the $1$- and $2$-Wasserstein distances for RDSs using a connection between the $2$-Wasserstein distance and the excess EPR in Sect.~\ref{sec:RDS_TSL}. \add{We revisit the TSLs in terms of minimum dissipation in Sec.~\ref{sec:minimumdissipation}.}
The final section~\ref{sec:numex_TSL} provides numerical demonstrations of the TSLs.

Note that we need to specify boundary conditions to define the Wasserstein distance variationally (otherwise, it will not be well defined). 
We adopt here the boundary conditions discussed in Sec.~\ref{sec:open_RDS} for quantities that are considered as currents, e.g., a quantity obtained by acting the mobility tensor on a force. 

\subsection{1-Wasserstein distance for reaction-diffusion systems}\label{sec:RDS_1W_dist}

\add{Fixing the boundary conditions and stoichiometry, we define the $1$-Wasserstein distance by generalizing the Beckmann problem~\eqref{Benamou--Brenier_L1} as 
\begin{align}\label{Def_L1}
{W}_{1,\mathcal{X}}\qty(\vec{c}^{A},\vec{c}^{B})\coloneqq\inf_{\mathcal{J}'}\qty{\int^{\tau}_0dt|\mathcal{J}'|_{\rm RD}},
\end{align}
with the norm of current $|\cdot|_{\RD}$ defined as
\begin{align}
|\mathcal{J}'|_{\rm RD}\coloneqq\int_{V}dr\left[\sum_{\alpha\in\mathcal{S}}\|\bm{J}'_{(\alpha)}\|+\sum_{\rho\in\mathcal{R}}|j'_{\rho}|\right].
\end{align}
Here, we impose the following condition on $\J'$: there exists a time series of concentration distributions $\vec{c}'$ satisfying
\begin{align}\label{constraints_L1}
 \partial_{t}\vec{c}'_\mathcal{X}=\qty(\Div\mathcal{J}')_{\mathcal{X}},\;\vec{c}'_{\mathcal{X}}(0)=\vec{c}^{A}_{\mathcal{X}},\vec{c}'_{\mathcal{X}}(\tau)=\vec{c}^{B}_{\mathcal{X}}. 
\end{align}
The condition means that the minimization is performed over all time series that obey the continuity equation with current $\mathcal{J}'$ connecting $\vec{c}^{A}$ and $\vec{c}^{B}$ only with respect to the internal species.} Note that the concentrations of the external species are irrelevant in the formula. As a result, the $1$-Wasserstein distance can be zero even if $\vec{c}^{A} \neq \vec{c}^{B}$ as long as $\vec{c}^{A}_{\mathcal{X}}=\vec{c}^{B}_{\mathcal{X}}$. 
Therefore, it should be regarded as a distance between concentration distributions of internal species rather than whole concentration profiles.

We can reduce the optimization for time in Eq.~\eqref{Def_L1} as
\begin{align}\label{another_Def_L1}
    {W}_{1,\mathcal{X}}\qty(\vec{c}^{A},\vec{c}^{B})=\inf_{\mathcal{U}}|\mathcal{U}|_{\rm RD}.
\end{align}
with the constraint that $\U$ satisfies
\begin{align}\label{another_condition_L1}
    \vec{c}^{B}_{\mathcal{X}}-\vec{c}^{A}_{\mathcal{X}}=\qty(\Div\mathcal{U})_{\mathcal{X}},
\end{align}
and the boundary condition on the diffusion part of $\mathcal{U}$ for the internal species. Using this reduced optimization problem, we can compute ${W}_{1,\mathcal{X}}$ numerically with less computational complexity. We provide the derivation of Eq.~\eqref{another_Def_L1} \add{and its geometric interpretation} in Appendix~\ref{ap:reduction_L1}.

\add{In addition, we can also generalize the Kantorovich--Rubinstein duality~\eqref{FP_Kantorovich--Rubinstein} to RDSs by considering the dual problem of the minimization problem in Eq.~\eqref{another_Def_L1}  (see the details in Appendix~\ref{ap:KR_dual}) as
\begin{align}\label{RDS_Kantorovich--Rubinstein}
{W}_{1,\mathcal{X}}\qty(\vec{c}^A,\vec{c}^B)=\sup_{\vec{\phi}\in\mathrm{Lip}_{\mathcal{X}}^1}\cip{\vec{\phi}}{\vec{c}^B-\vec{c}^A}.
\end{align}
Here, the set $\mathrm{Lip}^{1}_{\mathcal{X}}$ appearing in the conditions of optimization is a generalization of the set of $1$-Lipschitz functions $\mathrm{Lip}^{1}$ in Eq.~\eqref{1-Lip}, and defined as
\begin{align}\label{RDS_1_Lip}
    \mathrm{Lip}^{1}_{\mathcal{X}}\coloneqq&\left\{\vec{\phi}\;\middle|\;\forall\alpha\in\mathcal{X},\;\|\bm{\nabla}_{\bm{r}}\phi_{\alpha}\|\leq1; \right.\notag\\
    &\phantom{\Big\{\vec{\phi}\mid\;\;}\kern-.5pt\forall\alpha\in\mathcal{Y},\;\phi_{\alpha}=0;\notag\\
    &\phantom{\Big\{\vec{\phi}\mid\;}\kern-.5pt\left.\forall\rho\in\mathcal{R}_{\mathcal{X}},\;\left|(\nabla_{\mathsf{s}}\vec{\phi})_{\rho}\right|\leq1\right\}.
\end{align}
This representation supports the justification for generalizing the $1$-Wasserstein distance as the form in Eq.~\eqref{Def_L1}. It was also used to obtain an analytical form of the $1$-Wasserstein distance in the numerical example that appeared in Sec.~\ref{sec:numex_TSL}.
}

\subsection{2-Wasserstein distance for reaction-diffusion systems}\label{sec:RDS_2W_dist}

We define the $2$-Wasserstein distance between concentration distributions of the internal species \add{while fixing the boundary conditions, the stoichiometry, the $\vec{c}$-dependence of the Onsager operator, and the concentration distribution of the external species} as 
\begin{align}\label{Def_L2}
{W}_{2,\mathcal{X}}\qty(\vec{c}^{A},\vec{c}^{B}\middle|\vec{b}_{\mathcal{Y}})^2\coloneqq\inf_{\vec{c}',\mathcal{F}'
}\qty{\tau\int^{\tau}_0dt\fip{\mathcal{F}'}{\mathcal{F}'}_{\mathcal{M}_{\vec{c}'}}},
\end{align}
where we impose the conditions
\begin{align}\label{constraints_L2}
\partial_{t}\vec{c}'_{\mathcal{X}}=\qty(\Div\mathcal{M}_{\vec{c}'}\mathcal{F}')_{\mathcal{X}},
\end{align}\begin{align}\label{ext_constraints_L2}
    \vec{c}'_{\mathcal{X}}(0)=\vec{c}^{A}_{\mathcal{X}},\;\vec{c}'_{\mathcal{X}}(\tau)=\vec{c}^{B}_{\mathcal{X}},
\end{align}
and
\begin{align}\label{ext_constraints_L2_2}
\vec{c}'_{\mathcal{Y}}(t)=\vec{c}^A_{\mathcal{Y}}=\vec{c}^{B}_{\mathcal{Y}}=\vec{b}_{\mathcal{Y}},
\end{align}
on $\vec{c}'$ and $\mathcal{F}'$. In particular, the conditions in Eq.~\eqref{ext_constraints_L2_2} correspond to the concentration of the external species being fixed. Here, we write $\mathcal{M}_{\vec{c}'}$ instead of $\mathcal{M}$ because we want to emphasize its dependence on concentration $\vec{c}'$. 
In addition, we assume that $\dvec{\mathsf{M}}$ and $\mathsf{m}$ depend on time only through concentration and not explicitly, so that ${W}_{2,\mathcal{X}}$ is invariant to changes of the parameter $\tau$. \add{Note that the $2$-Wasserstein distance ${W}_{2,\mathcal{X}}(\vec{c}^{A},\vec{c}^{B}|\vec{b}_{\mathcal{Y}})$ is not a distance between concentration distributions of all species, but only those of the internal species, as in the case of the $1$-Wasserstein distance. This is because the condition~\eqref{ext_constraints_L2_2} fixes the concentration distribution of the external species to $\vec{b}_{\mathcal{Y}}$.}

This definition generalizes the Benamou--Brenier formula of the $2$-Wasserstein distance for Langevin systems in Eq.~\eqref{Benamou--Brenier}, MJPs~\cite{maas2011gradient}, and CRNs~\cite{yoshimura2023housekeeping}. \add{This $2$-Wasserstein distance also extends the dissipation distance~\cite{liero2013gradient,liero2016optimal}, defined in the context of the gradient flow structure of detailed-balanced RDSs, to general open systems.}

We can also rewrite Eq.~\eqref{Def_L2} in the form of an optimization problem not for the force $\mathcal{F}'$ but for the potential $\vec{\phi}$ as 
\begin{align}\label{another_Def_L2}
&{W}_{2,\mathcal{X}}\qty(\vec{c}^{A},\vec{c}^{B}\middle|\vec{b}_{\mathcal{Y}})^2\coloneqq \inf_{\vec{\phi},\vec{c}'
}\qty{\tau\int^{\tau}_0dt\fip{\Grad\vec{\phi}}{\Grad\vec{\phi}}_{\mathcal{M}_{\vec{c}'}}},
\end{align}
with the conditions
\begin{align}\label{another_condition_L2}
\partial_{t}\vec{c}'_{\mathcal{X}}=\qty(\Div\mathcal{M}_{\vec{c}'}\Grad\vec{\phi})_{\mathcal{X}},\,\vec{\phi}_{\mathcal{Y}}=\vecrm{0}_{\mathcal{Y}},
\end{align}
and the same conditions as in Eqs.~\eqref{ext_constraints_L2} and~\eqref{ext_constraints_L2_2}  (see also Appendix~\ref{ap:optimizer_L2} and \ref{ap:reformulate_L2}). \add{This representation originates from the following two facts: the definition of the $2$-Wasserstein distance corresponds to the minimization problem of EP, and such the minimization is achieved by a conservative force in short-time limit~\eqref{another_form_exEPR}.}

\subsection{Features of $1$- and $2$-Wasserstein distances}\label{sec:properties_of_Wdist}

The $1$- and $2$-Wasserstein distances are not distances between concentration distributions of all the species but between those of the internal species, as mentioned in previous sections. 
In fact, we can prove that the axioms of distance hold for ${W}_{1,\mathcal{X}}$ and ${W}_{2,\mathcal{X}}$ as distances between concentration distributions of the internal species (see also Appendix~\ref{ap:aximos_distance}).

In addition, these distances cannot be defined between arbitrary concentration distributions. The constraints on dynamics in Eq.~\eqref{constraints_L1} or Eq.~\eqref{constraints_L2} let us define the $1$- and $2$-Wasserstein distances between the concentration distributions $\vec{c}^{A}$ and $\vec{c}^{B}$ satisfying
\begin{align}
    \vec{c}^{B}-\vec{c}^{A}\in\mathrm{Im}_{\mathcal{X}}\Div,
\end{align}
where the set $\mathrm{Im}_{\mathcal{X}}\Div$ is defined as
\begin{align}\label{Def_ImX_Div}
    \mathrm{Im}_{\mathcal{X}}\Div\coloneqq\qty{\vec{c}'\;\middle|\;\exists\mathcal{J}'\;, \vec{c}'_{\mathcal{X}}=\qty(\Div\mathcal{J}')_{\mathcal{X}}}.
\end{align}
Here, we impose the boundary condition that was imposed on the RDS on the diffusion part of $\mathcal{J}'$ in Eq.~\eqref{Def_ImX_Div}. In other words, we can define ${W}_{1,\mathcal{X}}(\vec{c}^{A},\vec{c}^B)$ and ${W}_{2,\mathcal{X}}(\vec{c}^{A},\vec{c}^B|\vec{b}_{\mathcal{Y}})$ only for $\vec{c}^B\in\vec{c}^{A}+\mathrm{Im}_{\mathcal{X}}\Div\coloneqq\{\vec{c}'\mid\vec{c}'-\vec{c}^{A}\in\mathrm{Im}_{\mathcal{X}}\Div\}$, which is a generalization of the concentration space of a CRN restricted by stoichiometry. We call this affine space the \textit{stoichiometric manifold} following Ref.~\cite{kobayashi2022hessian} (it is also called the stoichiometric compatibility class~\cite{feinberg2019foundations}). Thus, the form of the operator $\Div$ and the boundary conditions determine whether the Wasserstein distances between a given pair of concentration distributions are well-defined. Note that the Wasserstein distances between concentration distributions belonging to the same time series obtained by the time evolution according to the RDSs are always well-defined.

The $1$- and $2$-Wasserstein distances require different information to compute. We need two distributions of the internal species $\vec{c}^A_{\mathcal{X}}$ and $\vec{c}^B_{\mathcal{X}}$, the operator $\Grad$ (or $\Div$) and the boundary conditions to obtain the $1$-Wasserstein distance. In contrast, we need the time-independent concentration distribution of the external species $\vec{b}_{\mathcal{Y}}$ and the form of $\mathcal{M}$ as a functional of the concentration distributions, in addition to them, to obtain the $2$-Wasserstein distance. We can regard $\Grad$ and the boundary conditions as having information on the topology of the CRN and the topology of the space where diffusion occurs, and $\mathcal{M}$ as having information on the kinetic aspect of diffusion and reactions. Therefore, the topology determines the $1$-Wasserstein distance, and the topology and the kinetic properties determine the $2$-Wasserstein distance.

Due to the difference in the information required to define the $1$- and $2$-Wasserstein distances, there is no simple inequality between ${W}_{1,\mathcal{X}}$ and ${W}_{2,\mathcal{X}}$, as seen in Eq.~\eqref{hierarchical_ineq}. \add{To compare the $1$- and $2$-Wasserstein distances, we define the following functional of the Onsager operator, which represents the intensity of reactions and diffusive dynamics,
\begin{align}\label{Def_A}
    \mnorm{\M}\coloneqq\int_Vd\bm{r}\qty[\diffmax(\bm{r};t)+\sum_{\rho\in\mathcal{R}_{\mathcal{X}}}m_{\rho}(\bm{r};t)].
\end{align}
Here the intensity of diffusive dynamics of the internal species $\diffmax(\bm{r};t)$ is defined by evaluating the magnitude of the mobility tensor as
\begin{align}\label{max_mobility}
    \diffmax(\bm{r};t)\coloneqq\max_{\vec{\bm{F}}{}'}\vec{\bm{F}}{}'{}^{\top}\dvec{\mathsf{M}}(\bm{r};t)\vec{\bm{F}}{}',
\end{align}
under the following conditions:
\begin{align}
    \forall\alpha\in\mathcal{X},\|\bm{F}'_{(\alpha)}\|=1,\;\forall\alpha\in\mathcal{Y},\bm{F}'_{(\alpha)}=\bm{0}.
\end{align}
These conditions allow for the equal incorporation of contributions from each internal chemical species when measuring the intensity of diffusion. When we consider the simple mobility tensor~\eqref{simple_mobility_tensor}, our definition of $\diffmax(\bm{r};t)$ reduces to
\begin{align}
    \diffmax(\bm{r};t)=\sum_{\alpha\in\mathcal{X}}D_{\alpha}c_{\alpha}(\bm{r};t).
\end{align}
This form confirms that $\diffmax(\bm{r};t)$ represents the intensity of diffusion, since it becomes large at $(\bm{r};t)$ where the concentration of species with a large diffusion coefficient is high. We also regard $m_{\rho}$ as the intensity of the $\rho$-th reaction since it measures the average reaction rate of the forward and reverse fluxes with the logarithmic mean.}

\add{We consider the following situation: (i) the $\vec{c}$-dependence of the Onsager operator is fixed, and (ii) the $t$-dependence of the Onsager operator is based only on the $t$-dependence of $\vec{c}$. We represent its dependence as $\mathcal{M}_{\vec{c}}$. The intensity of the Onsager operator enables us to obtain an inequality between the $1$- and $2$-Wasserstein distances as
\begin{align}\label{compare_L1_L2}
\frac{{W}_{1,\mathcal{X}}\qty(\vec{c}^A,\vec{c}^B)^2}{\left\langle\mnorm{\M_{\vec{c}^{\star}}}\right\rangle_{\tau}}\leq{W}_{2,\mathcal{X}}\qty(\vec{c}^A,\vec{c}^B\middle|\vec{b}_{\mathcal{Y}})^2
\end{align}
under the same topology, namely, the same boundary conditions and the same stoichiometry (see Appendix~\ref{ap:deriv_compare_L1_L2} for the proof). Here, the concentration distribution $\vec{c}^{\star}(t)$ is the optimizer for the $2$-Wasserstein distance~\eqref{Def_L2}, which satisfies $\vec{c}^{\star}_{\mathcal{X}}(0) = \vec{c}^A_{\mathcal{X}}$, $\vec{c}^{\star}_{\mathcal{X}}(\tau) = \vec{c}^B_{\mathcal{X}}$, and $\vec{c}^{\star}_{\mathcal{Y}}(t) = \vec{b}_{\mathcal{Y}}$. We also let $\langle \cdots \rangle_{\tau}$ denote the time average as $\langle\cdots \rangle_{\tau} = (1/\tau) \int_0^{\tau} dt \cdots$.
}

\subsection{Thermodynamic speed limits with Wasserstein distances}\label{sec:RDS_TSL}

We can find a relation between the excess EPR and the $2$-Wasserstein distance for RDSs as in the case of Langevin systems, MJPs, and CRNs~\cite{dechant2022geometric_E,dechant2022geometric_R,yoshimura2023housekeeping}. The relation leads to TSLs based on the $2$-Wasserstein distance. Moreover, the inequality between the $1$- and $2$-Wasserstein distances enables us to obtain TSLs based on the $1$-Wasserstein distance.

\begin{figure}
    \centering
    \includegraphics[width=\linewidth]{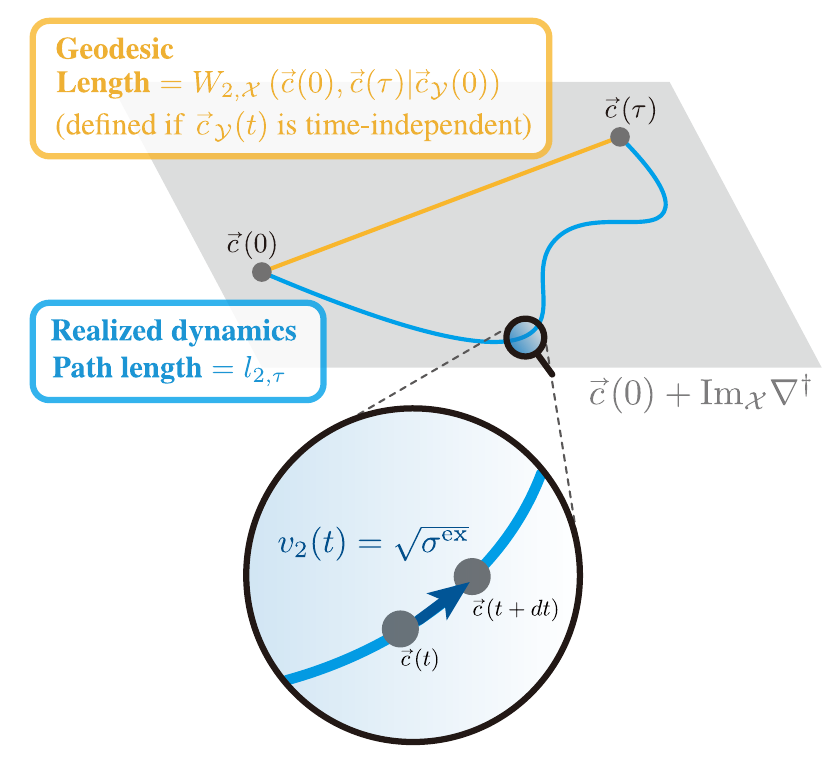}
    \caption{Schematic illustration of the relation between the $2$-Wasserstein distance and the excess EPR. The trajectory of the realized dynamics during the period $[0, \tau]$ (light-blue curve) is a curve with length $l_{2,\tau}$ on the stoichiometric manifold $\vec{c}(0)+\mathrm{Im}_{\mathcal{X}}\Div$ (gray space). We can define the geodesic between the initial and final concentration distributions, $\vec{c}(0)$ and $\vec{c}(\tau)$ (yellow line) if the concentrations of the external species are constant in time as $\vec{c}_{\mathcal{Y}}(t)=\vec{c}_{\mathcal{Y}}(0)$. The geodesic also lies on the stoichiometric manifold and has length ${W}_{2,\mathcal{X}}\qty(\vec{c}(0),\vec{c}(\tau)|\vec{c}_{\mathcal{Y}}(0))$. The speed of the concentration distribution moving on the stoichiometric manifold $v_{2}(t)$ equals the square root of the excess EPR, $\sqrt{\sigma^{\mathrm{ex}}}$.}
    \label{fig:RDS_TSL_scheme}
\end{figure}
We focus on the $2$-Wasserstein distance. We define the path length between the initial and final concentration distributions of internal species induced by the $2$-Wasserstein distance as
\begin{align}\label{RDS_L2_line_length}
    l_{2,\tau}\coloneqq\int_{0}^{\tau}dt\, v_{2}(t),
\end{align}
with the speed of the dynamics of $\vec{c}(t)$ on the set $\vec{c}(0)+\mathrm{Im}_{\mathcal{X}}\Div$ (see also Fig.~\ref{fig:RDS_TSL_scheme}),  
\begin{align}\label{RDS_L2_line_element}
v_{2}(t)\coloneqq\lim_{\varDelta t\to0}\frac{{W}_{2,\mathcal{X}}\qty(\vec{c}(t),\vec{c}(t+\varDelta t)\middle|\vec{c}_{\mathcal{Y}}(t))}{\varDelta t}.
\end{align}
Unlike the case of the $2$-Wasserstein distance, we can define $v_2(t)$ even if $\M$ explicitly depends on time by fixing $\M$ as the value at time $t$.

The speed of the dynamics squared equals the excess EPR,
\begin{align}\label{speed_and_exEPR}
    \sigma^{\mathrm{ex}}=v_{2}(t)^2.
\end{align}
We can prove this relation as follows. Taking $\varDelta t\ll 1$, the definition of the $2$-Wasserstein distance in Eq.~\eqref{Def_L2} and the constraints in Eqs.~\eqref{constraints_L2},~\eqref{ext_constraints_L2} and~\eqref{ext_constraints_L2_2} lead to
\begin{align}\label{infinitesimal_time_L2}
    {W}_{2,\mathcal{X}}&\qty(\vec{c}(t),\vec{c}(t+\varDelta t)\middle|\vec{c}_{\mathcal{Y}}(t))^2=\notag\\&\varDelta t^2\inf_{\mathcal{F}'}\fip{\mathcal{F}'}{\mathcal{F}'}_{\mathcal{M}_{\vec{c}(t)}}+o\qty(\varDelta t^2),
\end{align}
with the constraint
\begin{align}\label{infinitesimal_time_L2_constrains}
    \vec{c}_{\mathcal{X}}(t+\varDelta t)-\vec{c}_{\mathcal{X}}(t)=\varDelta t\qty(\Div\mathcal{M}_{\vec{c}(t)}\mathcal{F}')_{\mathcal{X}}+o\qty(\varDelta t).
\end{align}
We can obtain
\begin{align}\label{limit_W_dist}
    &\lim_{\varDelta t\to0}\frac{{W}_{2,\mathcal{X}}\qty(\vec{c}(t),\vec{c}(t+\varDelta t)\middle|\vec{c}_{\mathcal{Y}}(t))^2}{\varDelta t^2}\notag
    \\&=\inf_{\mathcal{F}'|\partial_t\vec{c}_{\mathcal{X}}(t)=\qty(\Div\mathcal{M}_{\vec{c}(t)}\mathcal{F}')_{\mathcal{X}}}\fip{\mathcal{F}'}{\mathcal{F}'}_{\mathcal{M}_{\vec{c}(t)}}
\end{align}
by taking limit $\varDelta t\to0$ after dividing both sides of Eq.~\eqref{infinitesimal_time_L2} by $\varDelta t^2$ and both sides of Eq.~\eqref{infinitesimal_time_L2_constrains} by $\varDelta t$. Then we derive the relation between the speed of the dynamics and the excess EPR~\eqref{speed_and_exEPR} by comparing the right-hand side of Eq.~\eqref{limit_W_dist} and a form of the excess EPR~\eqref{another_form_exEPR} because $\partial_t\vec{c}_{\mathcal{X}}(t)=\qty(\Div\mathcal{M}_{\vec{c}(t)}\mathcal{F})_{\mathcal{X}}$ holds for the force $\mathcal{F}$ in the original dynamics.

This relation between the speed of the dynamics and the excess EPR leads to the TSL,
\begin{align}\label{TSL_L2}
l_{2,\tau}^2\leq\tau\Sigma_{\tau}^{\mathrm{ex}}\leq\tau\Sigma_{\tau}.
\end{align}
The first inequality is derived from the Cauchy--Schwarz inequality $\qty[\int_{0}^{\tau}dt\,v_{2}(t)]^2\leq\qty[\int_{0}^{\tau}dt]\qty[\int_{0}^{\tau}dt\,v_{2}(t)^2]$ and the property of the excess EPR in Eq.~\eqref{speed_and_exEPR}. \add{We remark that the equality of the first inequality holds if and only if $v_2(t)$ (or equivalently $\sigma^{\mathrm{ex}}$) is independent of time, which follows from the conditions for the equality of the Cauchy--Schwarz inequality to hold.} 

\add{We also obtain a lower bound determined from more limited information, only the initial and final distributions, 
\begin{align}\label{TSL_L2_extconst}
{W}_{2,\mathcal{X}}(\vec{c}(0),\vec{c}(\tau)|\vec{c}_{\mathcal{Y}}(0))^2\leq l_{2,\tau}^2\leq\tau\Sigma_{\tau}^{\mathrm{ex}}\leq\tau\Sigma_{\tau},
\end{align}
if we consider the system where the concentrations of all the external species and the concentration dependence of $\mathcal{M}$ are independent of time.} Here, the first inequality is a consequence of the triangle inequality for the $2$-Wasserstein distance, which is proved in Appendix~\ref{ap:aximos_distance}. The TSL for $l_{2,\tau}$ is tighter than the one for  ${W}_{2,\mathcal{X}}(\vec{c}(0),\vec{c}(\tau))$ reflecting the fact that the path $\{\vec{c}_{\mathcal{X}}\}_{t\in[0,\tau]}$ is generally not the geodesic, whose length is ${W}_{2,\mathcal{X}}(\vec{c}(0),\vec{c}(\tau))$.

The TSL for RDSs~\eqref{TSL_L2} implies a trade-off between the dissipation due to pattern formation and the change speed of the pattern because the TSL in Eq.~\eqref{TSL_L2} is rewritten as $(l_{2,\tau}/\tau)^2 \leq (\Sigma_{\tau}^{\mathrm{ex}}/\tau)$, where $l_{2,\tau}/\tau$ means the change speed from the initial pattern at $t=0$ to the final pattern at $t=\tau$, and $\Sigma_{\tau}^{\mathrm{ex}}/\tau=\langle\sigma^{\mathrm{ex}}\rangle_{\tau}$ means the time average of dissipation due to the time evolution of the pattern. This trade-off relation means that the slower the speed of pattern formation, the smaller the dissipation can be.

The $1$-Wasserstein distance provides a different series of TSLs. Similar to Eq.~\eqref{RDS_L2_line_length}, we define the path length between the initial and final distributions with the $1$-Wasserstein distance as
\begin{align}\label{RDS_L1_line_length}
    l_{1,\tau}\coloneqq\int_{0}^{\tau}dt\,v_{1}(t),
\end{align}
where the integrand $v_1(t)$ indicates the speed of dynamics measured with ${W}_{1,\mathcal{X}}$,
\begin{align}\label{RDS_L1_line_element}
    v_{1}(t)\coloneqq\lim_{\varDelta t\to0}\frac{{W}_{1,\mathcal{X}}(\vec{c}(t),\vec{c}(t+\varDelta t))}{\varDelta t}.
\end{align}

\add{The inequality between the $1$- and $2$-Wasserstein distances~\eqref{compare_L1_L2} and this speed $v_1$ provide a lower bound of the excess EPR,
\begin{align}\label{exEPR_L1_line_element}
    \sigma^{\mathrm{ex}}=v_{2}(t)^2\geq\frac{v_1(t)^2}{\mnorm{\M }}.
\end{align}
This inequality leads to the TSLs based on the $1$-Wasserstein distance (see the derivations in Appendix~\ref{ap:deriv_L1_TSL}),
\begin{align}\label{TSL_L1}
\frac{{W}_{1,\mathcal{X}}\qty(\vec{c}(0),\vec{c}(\tau))^2}{\left\langle\mnorm{\M }\right\rangle_{\tau}}\leq\frac{l_{1,\tau}^2}{\left\langle\mnorm{\M}\right\rangle_{\tau}}\leq\tau\Sigma_{\tau}^{\mathrm{ex}}\leq\tau\Sigma_{\tau},
\end{align}
which generalize TSLs obtained for MJPs~\cite{van2023thermodynamic} to RDSs and tighten them by using the excess EP and the line length $l_{1,\tau}$. As in the case of MJPs, equality between the leftmost and rightmost sides is achievable as $\tau\Sigma_{\tau}={W}_{1,\mathcal{X}}\qty(\vec{c}(0),\vec{c}(\tau))^2/\left\langle\mnorm{\M}\right\rangle_{\tau}$. We discuss this equality from the viewpoint of minimum dissipation in the next section.}

\add{In contrast to the TSLs for the $2$-Wasserstein distance~\eqref{TSL_L2_extconst}, we can always consider the leftmost term in Eq.~\eqref{TSL_L1}. This is because we can define the $1$-Wasserstein distance even when $\vec{c}_{\mathcal{Y}}$ and the $\vec{c}$-dependence of $\M$ change in time. We also remark that the TSL for $l_{1,\tau}$ is tighter than the one for ${W}_{1,\mathcal{X}}(\vec{c}(0),\vec{c}(\tau))$. This reflects the fact that the path $\{\vec{c}_{\mathcal{X}}\}_{t\in[0,\tau]}$ is generally not the geodesic, whose length is ${W}_{1,\mathcal{X}}(\vec{c}(0),\vec{c}(\tau))$, as in the case of the $2$-Wasserstein distance~\eqref{TSL_L2_extconst}.}

\add{The TSL for the $1$-Wasserstein distance provides a more detailed physical insight than the one with the $2$-Wasserstein distance because the $1$-Wasserstein distance lets us separately treat the kinetic parameters and the speed of the time-evolution, which are merged in the case of the $2$-Wasserstein distance. To obtain this physical insight, we rewrite the TSL $l_{1,\tau}^2/\langle\mnorm{\M}\rangle_{\tau}\leq\tau\Sigma^{\mathrm{ex}}_{\tau}$ in Eq.~\eqref{TSL_L1} as $[(l_{1,\tau}/\tau)^2/\langle\mnorm{\M}\rangle_{\tau}]\leq(\Sigma^{\mathrm{ex}}_{\tau}/\tau)$, where $l_{1,\tau}/\tau$ is the change speed of the pattern measured with the $1$-Wasserstein distance, and $\langle\mnorm{\M}\rangle_{\tau}$ indicates the time average of intensity of reactions and diffusive dynamics. This rewriting allows us to regard the TSLs as trade-off relations between the dissipation due to pattern formation, speed of pattern change, and intensity of reactions and diffusion. Simply put, smaller dissipation is possible when the speed of pattern change is slower, or when the intensity of reactions and diffusion is higher.}

\add{We also remark that it is not obvious which lower bound of $\tau\Sigma^{\mathrm{ex}}_{\tau}$ is tighter: ${W}_{2,\mathcal{X}}(\vec{c}(0),\vec{c}(\tau)|\vec{c}_{\mathcal{Y}}(0))^2$ in Eq.~\eqref{TSL_L2_extconst} or $W_{1,\mathcal{X}}(\vec{c}(0),\vec{c}(\tau))^2/\langle\mnorm{\M}\rangle_{\tau}$ in Eq.~\eqref{TSL_L1}, even though there is the hierarchy in short-time limit~\eqref{exEPR_L1_line_element}. This is because the denominator in the left-hand side of the inequality between ${W}_{1,\mathcal{X}}(\vec{c}(0),\vec{c}(\tau))$ and ${W}_{2,\mathcal{X}}(\vec{c}(0),\vec{c}(\tau))$~\eqref{compare_L1_L2} is different from the denominator on the left-hand side of the TSL for ${W}_{1,\mathcal{X}}(\vec{c}(0),\vec{c}(\tau))$~\eqref{TSL_L1}: the former refers to the optimal time series of concentration distributions for the $2$-Wasserstein distance, $\vec{c}^{\star}$, while the latter refers to the original time series, $\vec{c}$. Similarly, it is not clear which lower bound is tighter: $l_{2,\tau}^2$ in Eq.~\eqref{TSL_L2_extconst} or $l_{1,\tau}^2/\langle\mnorm{\M}\rangle_{\tau}$ in Eq.~\eqref{TSL_L1} (see also Appendix~\ref{ap:deriv_L1_TSL}).}

\add{\subsection{Minimum dissipation and optimal transport}\label{sec:minimumdissipation}}

\add{We consider the minimum dissipation required to evolve the concentration distribution of the internal species from the initial distribution $\vec{c}_{\mathcal{X}}(0)$ to the final distribution $\vec{c}_{\mathcal{X}}(\tau)$ over time $\tau$. Here we regard the dissipation as the functional of the force $\F'$ and the Onsager operator $\M'$ as
\begin{align}
\Sigma_{\tau}\left[\M',\F'\right]=\int_{0}^{\tau}dt\fip{\F'}{\F'}_{\M'},
\end{align}
based on the form of the EPR~\eqref{EPR_norm}.}

\add{Actually, the minimum dissipation can be made to vanish by letting the intensity of reactions and diffusive dynamics $\mnorm{\M}$ be infinite. Therefore, minimum dissipation must be considered under some constraints on the Onsager operator $\M$. Here, we show that the $1$- and $2$-Wasserstein distances provide the minimum dissipation under different physical constraints on the Onsager operator $\M$. 
}

\add{
We can relate the $2$-Wasserstein distance to the minimum dissipation as follows. We consider the following minimization problem
\begin{align}
    \inf_{\vec{c}',\F'}\Sigma_{\tau}\left[\M_{\vec{c}'},\F'\right]
\end{align}
under four conditions: \\
(i) the internal part of the concentration distribution $\vec{c}'_{\mathcal{X}}$ satisfies 
\begin{align}
    \vec{c}'_{\mathcal{X}}(0)=\vec{c}_{\mathcal{X}}(0),\;\vec{c}'_{\mathcal{X}}(\tau)=\vec{c}_{\mathcal{X}}(\tau),
\end{align}
(ii) the time evolution of $\vec{c}'_{\mathcal{X}}$ is given by
\begin{align}
    \partial_t\vec{c}'_{\mathcal{X}}=\left(\Div\M_{\vec{c}'}\F'\right),
\end{align}
(iii) the external part of the concentration distribution $\vec{c}'_{\mathcal{Y}}$ is independent of time, i.e., it satisfies
\begin{align}
    \vec{c}'_{\mathcal{Y}}(t)=\vec{c}_{\mathcal{Y}}(0),
\end{align}
for all $t\in[0,\tau]$, and \\
(iv) the Onsager operator $\M_{\vec{c}'}$ depends on time only through $\vec{c}'$. This minimization is related to the $2$-Wasserstein distance as follows,
\begin{align}
    \inf_{\vec{c}',\F'}\Sigma_{\tau}\left[\M_{\vec{c}'},\F'\right]=\frac{{W}_{2,\mathcal{X}}\qty(\vec{c}(0),\vec{c}(\tau)\middle|\vec{c}_{\mathcal{Y}}(0))^2}{\tau}.
    \label{minimumEP_2Wasserstein}
\end{align}
It is directly derived from the definition of the $2$-Wasserstein distance~\eqref{Def_L2}.}

\add{
The optimal force of the minimization problem~\eqref{minimumEP_2Wasserstein} is given by the one of the minimization problem in the definition of the $2$-Wasserstein distance~\eqref{Def_L2} and it is conservative. The conservativeness is verified by the fact that the $2$-Wasserstein distance corresponds to the minimization problem over conservative forces~\eqref{another_Def_L2}. When the system is driven by the optimal force, the concentration distribution moves on the geodesic determined by the $2$-Wasserstein distance with constant speed,
\begin{align}
    v_2=\frac{W_{2,\mathcal{X}}(\vec{c}(0),\vec{c}(\tau)|\vec{c}_{\mathcal{Y}}(0))}{\tau}.
    \label{constantspeed_2W}
\end{align}
This is verified by considering the optimizer of the $2$-Wasserstein distance (see also Appendix~\ref{ap:reformulate_L2}).
}

\add{The minimum dissipation~\eqref{minimumEP_2Wasserstein} ensures that the equalities of the TSLs for the $2$-Wasserstein distance in Eq~\eqref{TSL_L2_extconst} are achievable. It lets us regard the TSL for the $2$-Wasserstein distance as an achievable lower bound of dissipation. The properties of the optimizer of the minimum dissipation~\eqref{minimumEP_2Wasserstein} yield the following consequences when the system achieves the equality of the TSL for the $2$-Wasserstein distance: The conservativeness of the optimal force makes the housekeeping EPR disappear, and the constant speed~\eqref{constantspeed_2W} makes the EPR independent of time.}

\add{We can also relate the $1$-Wasserstein distance to the minimum dissipation when the dependence of the Onsager operator on the concentration distribution is controllable. We consider the following minimization problem
\begin{align}
\inf_{\M',\F'}\Sigma_{\tau}\left[\M',\F'\right]
\end{align}
under two conditions: \\
(i) the current $\M'\F'$ changes the concentration distribution from the initial to the final distribution,
\begin{align}
    \vec{c}_{\mathcal{X}}(\tau)-\vec{c}_{\mathcal{X}}(0)=\left(\Div\int_{0}^{\tau}dt\,\M'\F'\right)_{\mathcal{X}},
    \label{constraint:time_evolution}
\end{align}
(ii) the intensity of reactions and diffusive dynamics has an upper bound $\mmax$,
\begin{align}
    \langle\mnorm{\M'}\rangle_{\tau}\leq\mmax.
    \label{constraint:intensity_bound}
\end{align}
This minimization is related to the $1$-Wasserstein distance as follows (see Appendix~\ref{ap:minimum dissipation 1-Wasserstein derivation} for the proof),
\begin{align}
\inf_{\M',\F'}\Sigma_{\tau}\left[\M',\F'\right] = \frac{W_{1,\mathcal{X}}\qty(\vec{c}(0),\vec{c}(\tau))^2}{\tau\mmax}.
    \label{minimumEP_1Wasserstein}
\end{align}
}


\add{As an optimizer of the minimization problem in Eq.~\eqref{minimumEP_1Wasserstein}, we can take an Onsager operator $\M^{\diamond}=\dvec{\mathsf{M}}{}^{\diamond}\oplus\mathsf{m}^{\diamond}$ and a force $\F^{\diamond} =(\vec{\bm{F}}^{\diamond}, \bm{f}^{\diamond})$ that are independent of time and satisfy $|\tau\M^{\diamond}\F^{\diamond}|_{\RD}=W_{1,\mathcal{X}}(\vec{c}(0),\vec{c}(\tau))$ (see also Appendix~\ref{ap:minimum dissipation 1-Wasserstein derivation}). Thus, the concentration distribution moves on the geodesic determined by the $1$-Wasserstein distance with the constant speed,
\begin{align}
    v_1=\frac{W_{1,\mathcal{X}}(\vec{c}(0),\vec{c}(\tau))}{\tau},
\end{align}
when the system is driven by the optimizer $(\M^{\diamond},\F^{\diamond})$. We remark that the time independence of the optimizer explicitly provides the geodesic as follows: let $\vec{c}^{\diamond}$ denote the concentration distribution whose internal part evolves according to $\partial_{t}\vec{c}^{\diamond}_{\mathcal{X}}=(\Div\M^{\diamond}\F^{\diamond})_{\mathcal{X}}$, then the internal part of $\vec{c}^{\diamond}$ satisfies
\begin{align}
    \vec{c}^{\diamond}_{\mathcal{X}}(t)=\left(1-\frac{t}{\tau}\right)\vec{c}_{\mathcal{X}}(0)+\frac{t}{\tau}\vec{c}_{\mathcal{X}}(\tau),
    \label{geodesic_1Wasserstein}
\end{align}
since $\M^{\diamond}\F^{\diamond}$ does not depend on time.
}

\add{We can also reinterpret the optimizer $\left(\M^{\diamond},\F^{\diamond}\right)$ from an operational viewpoint. We recombine the reaction part of the optimizer $\mathsf{m}^{\diamond}$ and $\bm{f}^{\diamond}$ into the forward and reverse fluxes as below. Let us define the forward and reverse fluxes $j^{\pm\diamond}_{\rho}$ as
\begin{align}
    j^{+\diamond}_{\rho}\coloneqq\frac{\mathrm{e}^{f^{\diamond}_{\rho}}}{\mathrm{e}^{f^{\diamond}_{\rho}}-1}m^{\diamond}_{\rho}f^{\diamond}_{\rho},\;\;j^{-\diamond}_{\rho}\coloneqq\frac{1}{\mathrm{e}^{f^{\diamond}_{\rho}}-1}m^{\diamond}_{\rho}f^{\diamond}_{\rho}.
    \label{fluxes_1Wasserstein}
\end{align}
Then, it satisfies the following relations: $f^{\diamond}_{\rho}=\ln (j^{+\diamond}_{\rho}/j^{-\diamond}_{\rho})$ and $m^{\diamond}_{\rho}=(j^{+\diamond}_{\rho}-j^{-\diamond}_{\rho})/\ln(j^{+\diamond}_{\rho}/j^{-\diamond}_{\rho})$. In particular, if the reactions obey mass action kinetics, we can obtain the reaction rate constants that realize the fluxes $j^{\pm\diamond}_{\rho}$. Similarly, if the mobility tensor has the simple form~\eqref{simple_mobility_tensor} and the system is an ideal dilute solution, we can also obtain the diffusion coefficients and the mechanical forces that realize $\dvec{\mathsf{M}}{}^{\diamond}$ and $\vec{\bm{F}}{}^{\diamond}$ (see also Appendix~\ref{ap:minimum dissipation 1-Wasserstein operation}).}

\add{The minimum dissipation~\eqref{minimumEP_1Wasserstein} also ensures that the equalities of the TSLs for the 1-Wasserstein distance in Eq.~\eqref{TSL_L1} are achievable by setting $\mmax=\langle\mnorm{\M}\rangle_{\tau}$. It provides the TSL for the $1$-Wasserstein distance with physical meaning, as an achievable lower bound on dissipation. When the system achieves the equality of the TSL for the $1$-Wasserstein distance, the EPR is independent of time, since the optimizer of the minimum dissipation with the $1$-Wasserstein distance~\eqref{minimumEP_1Wasserstein} does not depend on time. We also remark that the EP equals the excess EP under the optimal protocol since the leftmost and rightmost sides in the TSLs~\eqref{TSL_L1} coincide. It implies that we can choose a conservative force as an optimizer of the minimum dissipation problem~\eqref{minimumEP_1Wasserstein}, which is verified with the Kantorovich--Rubinstein duality (see also Appendix~\ref{ap:minimum dissipation 1-Wasserstein conservativeness}).}


\add{\subsection{Numerical examples: optimal transport and thermodynamic speed limits}\label{sec:numex_TSL}}

\begin{figure*}
    \centering
    \includegraphics[width=\linewidth]{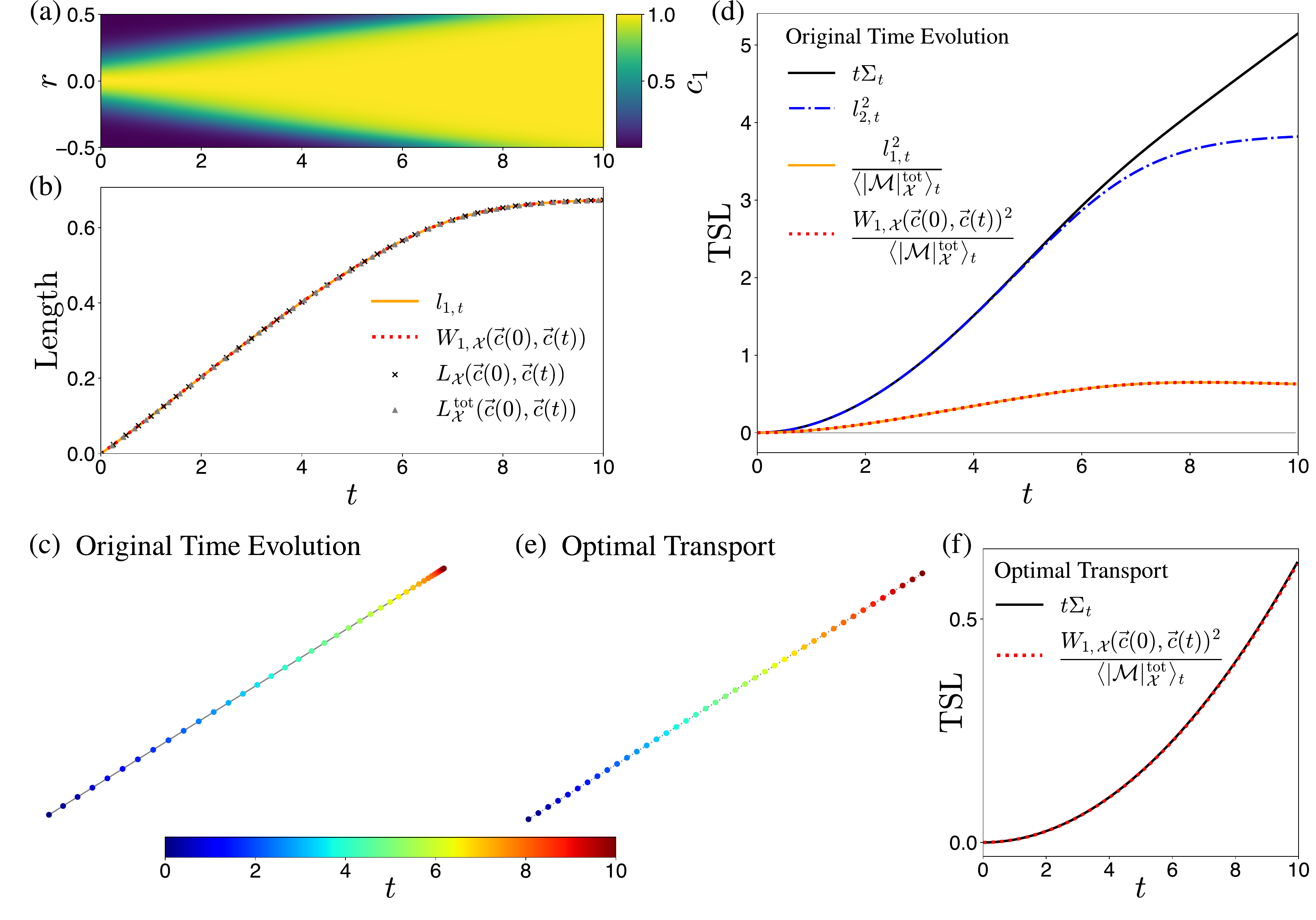}
    \caption{\add{The TSLs and optimal transport in the Fisher--KPP equation. (a) The time series of $c_1$. The figure is the same as Fig.~\ref{fig:EPR_KF}(a). (b) Various lengths between $\vec{c}(0)$ and $\vec{c}(t)$: $l_{1,t}$, ${W}_{1,\mathcal{X}}(\vec{c}(0),\vec{c}(t))$, $L_{\mathcal{X}}(\vec{c}(0),\vec{c}(t))$, and $L_{\mathcal{X}}^{\mathrm{tot}}(\vec{c}(0),\vec{c}(t))$. They have the same value for all time in the time series. (c) The original time series of the concentration distribution embedded in a two-dimensional plane by the multidimensional scaling, which preserves the $1$-Wasserstein distance between distributions as much as possible. Each point represents the concentration distribution at the time specified by the color. In this case, the time series is placed strictly on a straight line since $l_{1,t}=W_{1,\mathcal{X}}(\vec{c}(0),\vec{c}(t))$ holds for all $t$. (d) The TSLs. All the TSLs are confirmed to hold, and the one based on the path length with the $2$-Wasserstein distance $l_{2,t}$ is relatively tight. The gray line indicates zero. (e) The time series generated by the optimal protocol that achieves the minimum dissipation given by the $1$-Wasserstein distance~\eqref{minimumEP_1Wasserstein}. We use $\tau=10$ and $\mmax=\langle\mnorm{\M}\rangle_{\tau}$, which is obtained by the original time evolution. It is embedded in the same two-dimensional plane as (c). The optimal protocol moves the pattern along the shortest path from the initial distribution $\vec{c}(0)$ to the destination $\vec{c}(\tau)$ with a constant speed. (f) The achievement of the equality of the TSL for the $1$-Wasserstein distance by the same optimal protocol as the one used in (e). The black line and the red dotted line are $t\Sigma_t$ and the TSL for the $1$-Wasserstein distance under the optimal protocol, respectively.}}
    \label{fig:TSL_numerics_KF}
\end{figure*}

\begin{figure*}
    \centering
    \includegraphics[width=\linewidth]{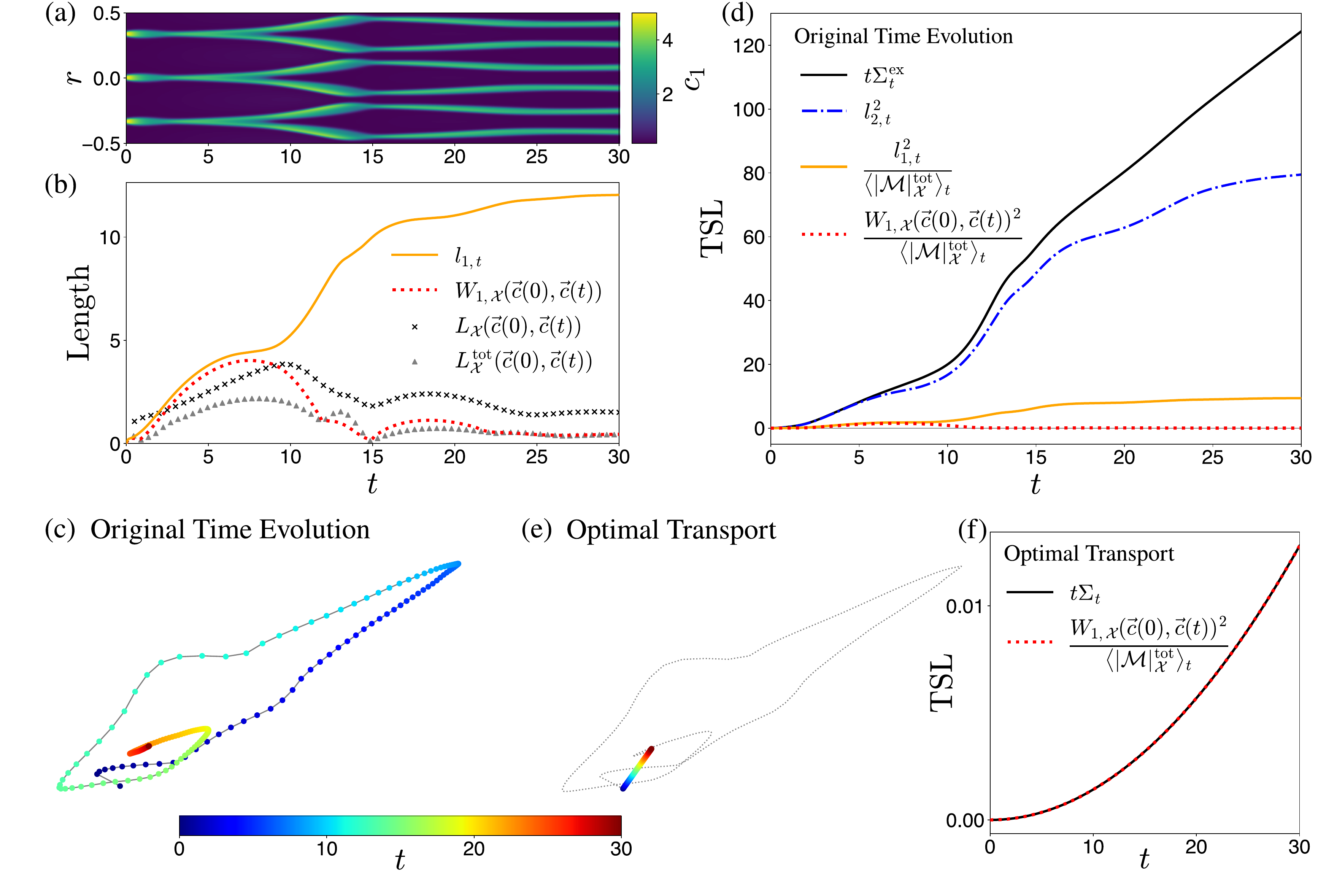}
    \caption{\add{The TSLs and optimal transport in the Brusselator model. (a) The time series of $c_1$. The figure shows the time interval $[0,30]$ of Fig.~\ref{fig:EPR_BL}(a). (b) Various lengths between $\vec{c}(0)$ and $\vec{c}(t)$: $l_{1,t}$, ${W}_{1,\mathcal{X}}(\vec{c}(0),\vec{c}(t))$, $L_{\mathcal{X}}(\vec{c}(0),\vec{c}(t))$, and $L_{\mathcal{X}}^{\mathrm{tot}}(\vec{c}(0),\vec{c}(t))$. We can see that the $1$-Wasserstein distance satisfies the triangle inequality $l_{1,t} \geq {W}_{1,\mathcal{X}}(\vec{c}(0),\vec{c}(t))$ and has no obvious relations with the other distances. (c) The original time series of the concentration distribution embedded in a two-dimensional plane by the multidimensional scaling, which preserves the $1$-Wasserstein distance between distributions as much as possible. Each point represents the concentration distribution at the time specified by the color. The concentration first moves away from and then moves back toward the initial distribution. (d) The TSLs. All the TSLs are confirmed to hold, and the one based on the path length with the $2$-Wasserstein distance $l_{2,t}$ is relatively tight. The gray line indicates zero. (e) The time series generated by the optimal protocol that achieves the minimum dissipation given by the $1$-Wasserstein distance~\eqref{minimumEP_1Wasserstein}. We use $\tau=30$ and $\mmax=\langle\mnorm{\M}\rangle_{\tau}$, which is obtained by the original time evolution. It is embedded in the same two-dimensional plane as (c). The optimal protocol moves the pattern along the shortest path from the initial distribution $\vec{c}(0)$ to the destination $\vec{c}(\tau)$ with a constant speed. (f) The achievement of the equality of the TSL for the $1$-Wasserstein distance by the same optimal protocol as the one used in (e). The black line and the red dotted line are $t\Sigma_t$ and the TSL for the $1$-Wasserstein distance under the optimal protocol, respectively.}}
    \label{fig:TSL_numerics_BL}
\end{figure*}

We show numerical results for the TSLs [Eqs.~\eqref{TSL_L1} and \eqref{TSL_L2}] by using the same time series as in Sec.~\ref{sec:system_for_numex}. \add{We show the results for the Fisher--KPP equation in Fig.~\ref{fig:TSL_numerics_KF}, and the ones for the Brusselator in Fig.~\ref{fig:TSL_numerics_BL}.} We compute the $1$-Wasserstein distance by the primal-dual algorithm~\cite{chambolle2011first,li2018parallel}. \add{We utilize the multidimensional scaling~\cite{borg2007modern} to embed the orbit of the time evolution on the stoichiometric manifold into a two-dimensional plane. This method keeps the pairwise $1$-Wasserstein distances between the concentration distributions at any two times as close as possible (for detail, see Appendix~\ref{ap:embed}).} To compare with the behavior of the $1$-Wasserstein distance, we use the $L^1$ distance between concentration distributions of the internal species,
\begin{align}
    {L}_{\mathcal{X}}(\vec{c}(0),\vec{c}(\tau))\coloneqq\sum_{\alpha\in\mathcal{X}}\int_{V}d\bm{r}\left|c_{\alpha}(\bm{r};\tau)-c_{\alpha}(\bm{r};0)\right|,
\end{align}
and the $L^1$ distance between total concentrations of the internal species,
\begin{align}
    {L}_{\mathcal{X}}^{\mathrm{tot}}(\vec{c}(0),\vec{c}(\tau))\coloneqq\sum_{\alpha\in\mathcal{X}}\left|\int_{V}d\bm{r}\,c_{\alpha}(\bm{r};\tau)-\int_{V}d\bm{r}\,c_{\alpha}(\bm{r};0)\right|.
\end{align}
Note that ${L}_{\mathcal{X}}$ accounts for not only changes in total concentrations but also changes in the shape of the pattern, which is not taken into account in ${L}_{\mathcal{X}}^{\mathrm{tot}}$. The triangle inequality implies ${L}_{\mathcal{X}}(\vec{c}(0),\vec{c}(\tau))\geq{L}_{\mathcal{X}}^{\mathrm{tot}}(\vec{c}(0),\vec{c}(\tau))$. If the system is well-mixed, these distances become equivalent.


\textit{Fisher--KPP equation}.---In Fig.~\ref{fig:TSL_numerics_KF}(b), we show the four lengths between $\vec{c}(0)$ and $\vec{c}(t)$ of the Fisher--KPP equation, $l_{1,t}$, ${W}_{1,\mathcal{X}}(\vec{c}(0),\vec{c}(t))$, $L_{\mathcal{X}}(\vec{c}(0),\vec{c}(t))$, and $L_{\mathcal{X}}^{\mathrm{tot}}(\vec{c}(0),\vec{c}(t))$. We can see that the lengths have the same value for all $t$. This equivalence of lengths is due to the following two conditions: (i) the system consists of only one internal species $Z_1$ and only one reaction, and (ii) the concentration is monotonically increasing for all locations and time, i.e., $\partial_{t}c_1\geq0$ holds for all $\bm{r}$ and $t$ (see Appendix~\ref{ap:FisherKPP_equivalence} for the proof). \add{From the viewpoint of the path on the stoichiometric manifold, the equivalence $W_{1,\mathcal{X}}(\vec{c}(0),\vec{c}(t))=l_{1,t}$ implies that the pattern evolves on the geodesic as shown in Fig.~\ref{fig:TSL_numerics_KF}(c). We also remark that the speed of the time evolution is not constant, i.e., it slows down over time.}

Reflecting the monotonic increase of the area with a high concentration of $c_1$ shown in Fig.~\ref{fig:TSL_numerics_KF}(a), the lengths between $\vec{c}(0)$ and $\vec{c}(t)$ increase monotonically in time. In particular, the lengths increase approximately in proportion to $t$ when the concentration of $Z_1$ is not saturated (roughly $0\leq t\leq 5$). We remark that the lengths are strictly proportional to $t$ if the concentration distribution is a traveling wave solution with one wavefront, which is a simpler case than the numerical example (see also Appendix~\ref{ap:FisherKPP_chemical_wave}).

In Fig.~\ref{fig:TSL_numerics_KF}(d), we demonstrate the TSLs using the Fisher--KPP equation. We use the EP $\Sigma_t$ instead of the excess EP $\Sigma_t^{\mathrm{ex}}$ because $\Sigma_t=\Sigma_t^{\mathrm{ex}}$ holds as explained in Sec.~\ref{sec:system_for_numex}. The squared length $l_{2,t}^2$ bounds $t\Sigma_t$ especially tight when the concentration of $Z_1$ is not saturated (roughly $0\leq t\leq 5$). This is because the EPR for the traveling wave solution of the Fisher--KPP equation is independent of time~\cite{avanzini2019thermodynamics} so that $v_2(t)$ satisfies the condition for the equality of the TSL. However, in the time region where the change in concentration distribution is small, all of the TSLs become looser because $t\Sigma_t$ keeps increasing in proportion to time while the lengths get saturated as the system approaches the steady state.

\add{We can construct an optimal protocol that reduces the thermodynamic cost of changing the pattern from $\vec{c}_{\mathcal{X}}(0)$ to $\vec{c}_{\mathcal{X}}(\tau)$ to the minimum value provided by the TSL for the $1$-Wasserstein distance as discussed in Sec.~\ref{sec:minimumdissipation}. It is obtained by solving the minimum dissipation problem~\eqref{minimumEP_1Wasserstein} with $M_0=\langle\mnorm{\M}\rangle_{\tau}$. In the following, let us fix $\tau=10$. From a geometric point of view, this protocol moves the pattern from $\vec{c}(0)$ to $\vec{c}(\tau)$ along the geodesic~\eqref{geodesic_1Wasserstein} as in the original time evolution; the difference with the original time evolution is that the speed on the stoichiometric manifold is constant as shown in Fig.~\ref{fig:TSL_numerics_KF}(e).}

\add{In Fig.~\ref{fig:TSL_numerics_KF}(f), we show that $t\Sigma_t$ and the TSL for the $1$-Wasserstein distance coincide under the optimal protocol. This is verified as follows. Since the optimizer of the minimization problem in Eq.~\eqref{minimumEP_1Wasserstein} is independent of time, we obtain
\begin{align}
    t\Sigma_t=\left(\frac{t}{\tau}\right)^2\tau\Sigma_{\tau}.
    \label{cost_underOT}
\end{align}
The time-independence of the optimizer also lets $\langle\mnorm{\M}\rangle_{t}=\langle\mnorm{\M}\rangle_{\tau}$ hold for all $t$. We also have $W_{1,\mathcal{X}}(\vec{c}(0),\vec{c}(t))=(t/\tau)W_{1,\mathcal{X}}(\vec{c}(0),\vec{c}(\tau))$ since the concentration distribution moves on the geodesic with a constant speed. Thus, the lower bound provided by the TSL for the $1$-Wasserstein distance satisfies
\begin{align}
    \frac{W_{1,\mathcal{X}}(\vec{c}(0),\vec{c}(t))^2}{\langle\mnorm{\M}\rangle_{t}}=\left(\frac{t}{\tau}\right)^2\frac{W_{1,\mathcal{X}}(\vec{c}(0),\vec{c}(\tau))^2}{\langle\mnorm{\M}\rangle_{\tau}}.
    \label{bound_underOT}
\end{align}
Combining Eq.~\eqref{cost_underOT} and Eq.~\eqref{bound_underOT} concludes that $t\Sigma_t=W_{1,\mathcal{X}}(\vec{c}(0),\vec{c}(t))^2/\langle\mnorm{\M}\rangle_{t}$ holds for all $t$ since the optimal protocol achieves the minimum dissipation~\eqref{minimumEP_1Wasserstein} as $\tau\Sigma_{\tau}=W_{1,\mathcal{X}}(\vec{c}(0),\vec{c}(\tau))^2/\langle\mnorm{\M}\rangle_{\tau}$.}

\textit{Brusselator model}.--- The time series used in the following is the same as the one used in Sec.~\ref{sec:system_for_numex}. We will focus on the time interval $t\in[0,30]$ as shown in Fig.~\ref{fig:TSL_numerics_BL}(a), where the concentrations change significantly.

In Fig.~\ref{fig:TSL_numerics_BL}(b), we show the four lengths between $\vec{c}(0)$ and $\vec{c}(t)$ of the Brusselator. Unlike the Fisher--KPP equation, they behave differently, with only $l_{1,t}$ increasing monotonically. There is no definite order either between ${W}_{1,\mathcal{X}}(\vec{c}(0),\vec{c}(t))$ and ${L}_{\mathcal{X}}(\vec{c}(0),\vec{c}(t))$ or between ${W}_{1,\mathcal{X}}(\vec{c}(0),\vec{c}(t))$ and ${L}_{\mathcal{X}}^{\mathrm{tot}}(\vec{c}(0),\vec{c}(t))$. In particular, the $1$-Wasserstein distance decreases to almost zero at time $t=15$ after increasing. This is because the total concentrations at $t=0$ and $t=15$ are very close, which is evident from $L_{\mathcal{X}}^{\mathrm{tot}}(\vec{c}(0),\vec{c}(15))\simeq0$. From the viewpoint of the path on the stoichiometric manifold, the behavior of the $1$-Wasserstein distance implies that the pattern goes back near the initial state after once moving away from the initial state. \add{The time series of the concentration distribution embedded in a two-dimensional plane [Fig.~\ref{fig:TSL_numerics_BL}(c)] verifies this behavior.} The path of the pattern is not a geodesic of the $1$-Wasserstein distance as it follows from the fact that $l_{1,t}$ and ${W}_{1,\mathcal{X}}(\vec{c}(0),\vec{c}(t))$ are different.

In Fig.~\ref{fig:TSL_numerics_BL}(d), we demonstrate the TSLs using the Brusselator. As in the case of the Fisher--KPP equation, the TSL for $l_{2,t}$ is tighter than the TSLs for $l_{1,t}$ and ${W}_{1,\mathcal{X}}(\vec{c}(0),\vec{c}(t))$. We also remark that the TSL for $l_{1,t}$ is tighter than the one with ${W}_{1,\mathcal{X}}(\vec{c}(0),\vec{c}(t))$. This is because the path of the time series $\{\vec{c}(t)\}_{t\in[0,30]}$ is not a geodesic as mentioned above. 
As in the case of the Fisher--KPP equation, we can also see that all of the TSLs become looser when the system approaches the steady state. It is caused by the fact that the increase proportional to $t$ in $t\Sigma_t^{\mathrm{ex}}$ persists while the pattern stop changing.

\add{As in the case of the Fisher--KPP equation, we can construct an optimal protocol that reduces the thermodynamic cost to the minimum value provided by the TSL for the $1$-Wasserstein distance. In the following, we take $\tau=30$ as the final time in Eq.~\eqref{minimumEP_1Wasserstein}. Since the TSL for the $1$-Wasserstein distance is weak, the optimal protocol can significantly decrease the dissipation required to change the pattern. This reduction of dissipation is due to the following circumstances: as already discussed, the time series of the concentration distribution takes a detour in the present system [Fig.~\ref{fig:TSL_numerics_BL}(c)]. In contrast, the optimal protocol moves the pattern along the shortest path to the destination, which is close to the initial distribution, as shown in Fig.~\ref{fig:TSL_numerics_BL}(e). Thus, the optimal protocol significantly reduces the total change in the pattern, which leads to low dissipation. In Fig.~\ref{fig:TSL_numerics_BL}(f), we also show the TSL for the $1$-Wasserstein distance under the optimal protocol. It behaves the same way as in the case of the Fisher--KPP equation: equality always holds, and $t\Sigma_{t}$ and the bound are proportional to $t^2$.}

\section{Thermodynamic Uncertainty relations for reaction-diffusion systems}\label{sec:RDS_TUR}

\add{We now provide another thermodynamic trade-off relation, the thermodynamic uncertainty relation (TUR). Here, we refer to the lower bound of the instantaneous (excess) EPR by partial information about the pattern and system fluctuations as TUR following previous studies~\cite{yoshimura2021thermodynamic,kolchinsky2022information,yoshimura2023housekeeping}. This is in contrast to the TSLs, which bound the (excess) EP arising in finite time using information about changes of the whole pattern. After providing the TUR for general observables in Sec.~\ref{sec:general_TUR}, we obtain TUR for the Fourier coefficients of the concentration distribution as a special case of the general TUR in Sec.~\ref{sec:Fourier_TUR}. The TUR for the Fourier coefficients generalizes the TUR for CRNs found in the previous study~\cite{yoshimura2021thermodynamic} by reflecting the spatial structure of the pattern.}

\subsection{General thermodynamic uncertainty relation for reaction-diffusion systems}\label{sec:general_TUR}
\add{Here, we derive the TUR for a general observable, which can take complex values. The term \textit{observable} is used to refer to complex-valued, time-independent vector functions whose external parts are zero; i.e., an observable belongs to the set $\qty{\vec{\varphi}:\mathbb{R}^d\to\mathbb{C}^{N}\mid\forall\alpha\in\mathcal{Y},\;\varphi_{\alpha}=0}$. We impose periodic boundary condition on $\vec{\varphi}$ if we consider systems with periodic boundaries. Since we extend the scope of treatment to complex-valued functions, we also extend the inner products $\fip{\cdot}{\cdot}$ and $\cip{\cdot}{\cdot}$ to complex-valued functions by taking conjugate transpose, denoted by $\dag$, as
\begin{align}
     \fip{\mathcal{J}'}{\mathcal{F}'}\coloneqq\int_{V}d\bm{r}\qty(\vec{\bm{J}}{}^{'\dag}\vec{\bm{F}'}+\bm{j}'^{\dag}\bm{f}'),
\end{align}
and
\begin{align}
    \cip{\vec{\phi}}{\vec{\psi}}\coloneqq\int_{V}d\bm{r}\,\vec{\phi}{}^{\dag}\vec{\psi}.
\end{align}
}


\add{Under these preparations, we obtain the TUR for RDSs as a generalization of Eq.~\eqref{FP_TUR},
\begin{align}
    \left|d_t\!\cip{\vec{c}}{\vec{\varphi}}\right|^2\leq \mathfrak{D}_{\vec{\varphi}}\sigma^{\mathrm{ex}},
    \label{general_TUR_RDS}
\end{align}
where the indicator of the fluctuation of $\vec{\varphi}$ is defined as
\begin{align}
\mathfrak{D}_{\vec{\varphi}}\coloneqq\int_{V}d\bm{r}\qty[\qty(\bm{\nabla}_{\bm{r}}\vec{\varphi})^{\dag}\dvec{\mathsf{M}}\bm{\nabla}_{\bm{r}}\vec{\varphi}+\vec{\varphi}^{\dag}\scd\vec{\varphi}].
\label{Def_fluct}
\end{align}
Here, the first term on the right-hand side of Eq.~\eqref{Def_fluct} originates from the spatial inhomogeneity of $\vec{\varphi}$ and the mobility of the system. The matrix in the second term $\scd$ is the scaled diffusion coefficient matrix defined as
\begin{align}
    \scd_{\alpha\beta}(\bm{r};t)&\coloneqq\sum_{\rho\in\mathcal{R}}\frac{j_{\rho}^{+}(\bm{r};t)+j_{\rho}^{-}(\bm{r};t)}{2}S_{\alpha\rho}S_{\beta\rho}\\
    &=\sum_{\rho\in\mathcal{R}}a_{\rho}(\bm{r};t)S_{\alpha\rho}S_{\beta\rho}.
    \label{Def_SDC}
\end{align}
This matrix appears in the diffusion coefficient matrix of the chemical Langevin equation, which is a microscopic description of chemical reactions. In the chemical Langevin equation, the diffusion coefficient matrix is inversely proportional to the system size, and $\scd$ is its proportional coefficient. This implies that the scaled diffusion coefficient reflects the intrinsic fluctuations of chemical reactions ~\cite{yoshimura2021thermodynamic}. Thus, the second term in $\mathfrak{D}_{\vec{\varphi}}$~\eqref{Def_fluct} corresponds to the fluctuation of the reactions.
}

The TUR~\eqref{general_TUR_RDS} is derived as follows. We can obtain an inequality between the scaled diffusion coefficient and the edgewise Onsager coefficient as
\begin{align}\label{ineq_SDC}
\scd_{\alpha\beta}\geq\sum_{\rho\in\mathcal{R}}m_{\rho}S_{\alpha\rho}S_{\beta\rho}=\sum_{\rho,\rho'\in\mathcal{R}}[\mathsf{m}]_{\rho\rho'}S_{\alpha\rho}S_{\beta\rho'},
\end{align}
by using the inequality in Eq.~\eqref{ineq_activity}. This yields 
\begin{align}\label{ineq_fluct}
\mathfrak{D}_{\vec{\varphi}}\geq\fip{\Grad\vec{\varphi}}{\Grad\vec{\varphi}}_{\mathcal{M}}.
\end{align}
Then, we can derive the TUR as
\begin{align*}
\left|d_t\!\cip{\vec{c}}{\vec{\varphi}}\right|^2&=
\left|\cip{\Div\mathcal{M}\mathcal{F}}{\vec{\varphi}}\right|^2=
\left|\cip{\Div\mathcal{M}\mathcal{F}^{\ast}}{\vec{\varphi}}\right|^2\\&=
\left|\fip{\mathcal{M}\mathcal{F}^{\ast}}{\Grad\vec{\varphi}}\right|^2=
\left|\fip{\mathcal{F}^{\ast}}{\Grad\vec{\varphi}}_{\mathcal{M}}\right|^2\\
&\leq\fip{\Grad\vec{\varphi}}{\Grad\vec{\varphi}}_{\mathcal{M}}\fip{\mathcal{F}^{\ast}}{\mathcal{F}^{\ast}}_{\mathcal{M}}\\&\leq \mathfrak{D}_{\vec{\varphi}}\sigma^{\mathrm{ex}},
\end{align*}
where we use Eq.~\eqref{rewrite_RDS} and Eq.~\eqref{EL_eq} in the first line, the boundary condition for $\mathcal{F}^{\ast}$ between the first and second line, the Cauchy--Schwarz inequality for the inner product $\fip{\cdot}{\cdot}_{\mathcal{M}}$ between the second and third line, and Eq.~\eqref{ineq_fluct} in the last transformation.

We can also interpret the TUR from the viewpoint of the Wasserstein geometry, as we did for the TUR for Langevin systems in Eq.~\eqref{FP_TUR_CRB}. By rewriting Eq.~\eqref{general_TUR_RDS}, we obtain
\begin{align}\label{RDS_TUR_CRB}
    v_{\vec{\varphi}}\coloneqq\frac{|d_t\!\cip{\vec{c}}{\vec{\varphi}}|}{\sqrt{\mathfrak{D}_{\vec{\varphi}}}}\leq v_{2},
\end{align}
where we used $\sigma^{\rm ex} = v_{2}^2$.
This inequality means that the speed of the observable change $v_{\vec{\varphi}}$, normalized by the indicator of fluctuation $\mathfrak{D}_{\vec{\varphi}}$, can always be upper bounded by the speed of the concentration distribution $v_{2}$ measured with the $2$-Wasserstein distance.

\subsection{Thermodynamic uncertainty relations for Fourier component of concentration distribution}\label{sec:Fourier_TUR}
Here, we prove a TUR for the Fourier transform of a concentration distribution by properly choosing the observable in the general TUR~\eqref{general_TUR_RDS}. In the following part, we only consider internal species, $\alpha\in\mathcal{X}$. We define the Fourier transform of the concentration distribution as
\begin{align}
\tilde{c}_{\alpha}(\bm{k};t)\coloneqq\int_{V}d\bm{r}\,c_{\alpha}(\bm{r};t)\mathrm{e}^{-\mathrm{i}\bm{k}\cdot\bm{r}},
\label{Fourier_concentration}
\end{align}
where we regard $c_{\alpha}(\bm{r};t)$ as zero outside of $V$. If we consider the system with the periodic boundary condition, we obtain
$c_{\alpha}(\bm{r};t)=\sum_{\bm{k}}\tilde{c}_{\alpha}(\bm{k};t)\mathrm{e}^{\mathrm{i}\bm{k}\cdot\bm{r}}/|V|$. In this case, the vector $\bm{k}$ takes discrete values depending on the details of $V$.

\begin{figure*}
    \centering
    \includegraphics[width=\linewidth]{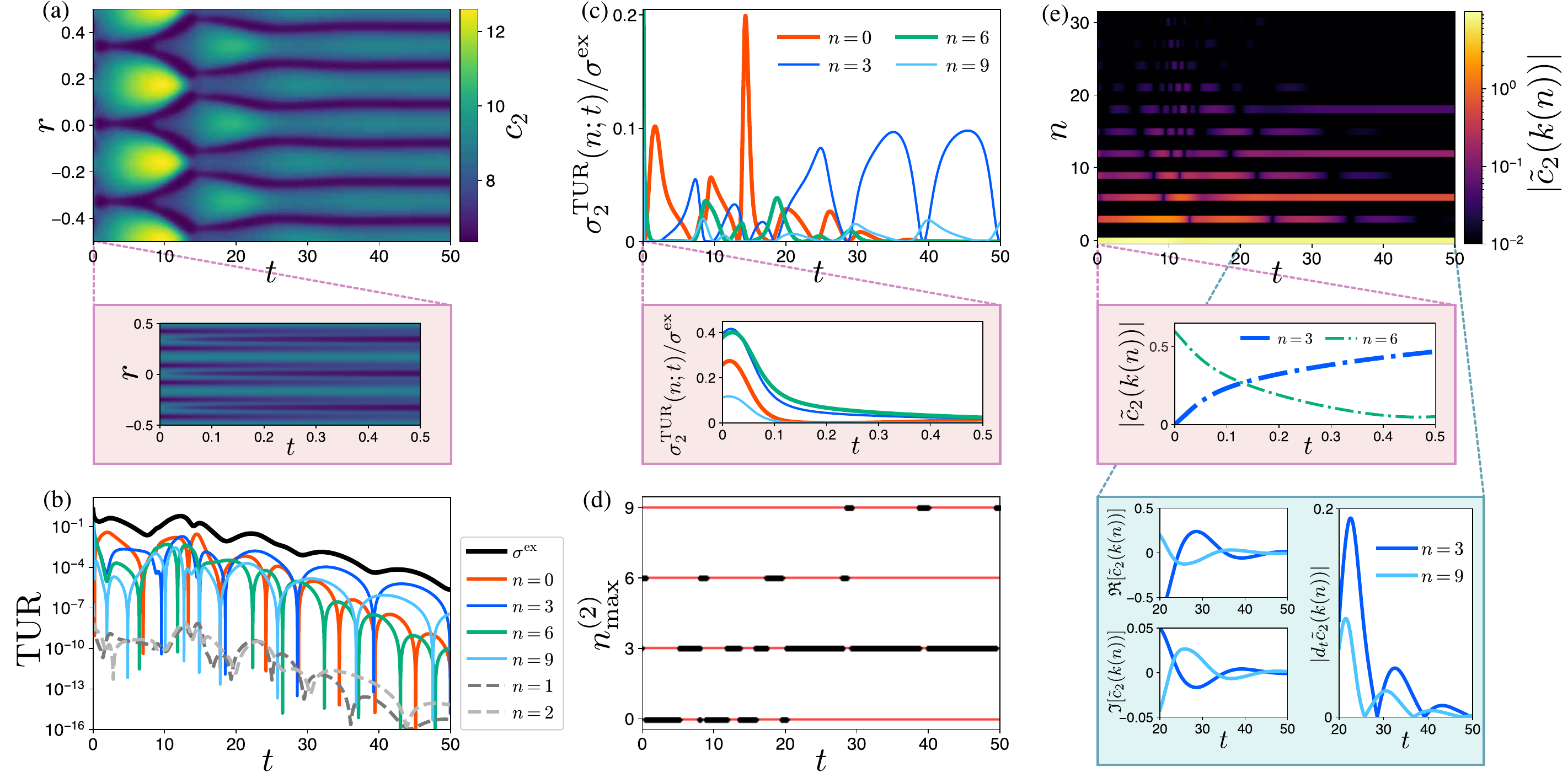}
    \caption{The TUR~\eqref{Fourier_TUR_simple} in the Brusselator. (a) The time series of $c_{2}$ for comparison. In the early stage of the time evolution ($t\in[0,0.5]$), the symmetry of the pattern instantly changes from $6$-fold to $3$-fold, as shown in the pink panel. After that, the pattern goes to the steady state with $6$-fold symmetry. (b) Semilog plot of the excess EPR (black line) and the lower bounds $\sigma^{\mathrm{TUR}}_2(n;t)$ for various $n$. The lower bounds corresponding to multiples of $3$ ($n=0,3,6,9$) are tighter than the bounds corresponding to $n=1,\,2$. (c) The ratio of $\sigma_2^{\mathrm{TUR}}(n;t)$ to the excess EPR for $n=0,3,6,9$. In the early stage ($t\in[0,0.5]$), $\sigma^{\mathrm{TUR}}_{2}(3;t)$ or $\sigma^{\mathrm{TUR}}_{2}(6;t)$ provides the tightest bound as shown in the pink panel. (d) $n^{(2)}_{\mathrm{max}}(t)$ (black dots). Reflecting the symmetry of the pattern, $n^{(2)}_{\mathrm{max}}(t)$ are multiples of three for all time $t$ (red lines). Near the stationary pattern, $n^{(2)}_{\mathrm{max}}(t)=3$ for almost all $t$. (e) The time series of $|\tilde{c}_{2}(k(n))|$. We omit $|\tilde{c}_{2}(k(n))|$ for $n\geq 32$ because they are sufficiently small. The $3$-fold symmetry of the pattern lets $|\tilde{c}_2(k(n))|$ have large values when $n=3n'$ for any integer $n'$. Since the symmetry of the pattern transitions to $6$-fold as approaching the steady state, $|\tilde{c}_2(k(n))|$ decays if $n$ is a multiple of three but not a multiple of six at large $t$. In contrast, the magnitude $|\tilde{c}_2(k(6))|$ decays while $|\tilde{c}_2(k(3))|$ increases in the early stage of the pattern formation (pink panel) reflecting the instant transition from the pattern with $6$-fold symmetry to the pattern with $3$-fold symmetry. The light-blue panel shows the asynchronous oscillations of the Fourier components corresponding to $n=3,\,9$ in $t\in[20,50]$. Here, we let $\mathfrak{R}[z]$ and $\mathfrak{I}[z]$ denote the real part and imaginary part of the complex number $z$, respectively. The oscillations of the two modes are not synchronized. Due to the asynchronous nature of the oscillations, the times when $\left|d_t\tilde{c}_2(k(3);t)\right|$ vanishes differ from the times when $\left|d_t\tilde{c}_2(k(9);t)\right|$ vanishes.}
    \label{fig:TUR_numerics_2}
\end{figure*}

Fixing $\alpha\in\mathcal{X}$ and letting $\vec{\varphi}$ satisfy $\qty(\vec{\varphi}(\bm{r}))_{\beta}=\delta_{\alpha\beta}e^{-\mathrm{i}\bm{k}\cdot\bm{r}}$ in Eq.~\eqref{general_TUR_RDS} yields a TUR for the Fourier transform,
\begin{align}\label{Fourier_TUR}
    \frac{\left|d_t\tilde{c}_{\alpha}(\bm{k};t)\right|^2}{\bm{k}\cdot\mathsf{M}_{(\alpha\alpha)}^{\mathrm{tot}}(t)\bm{k}+\scd^{\mathrm{tot}}_{\alpha\alpha}(t)}\leq\sigma^{\mathrm{ex}},
\end{align}
where $\mathsf{M}_{(\alpha\alpha)}^{\mathrm{tot}} (t) \coloneqq\int_{V}d\bm{r}\,\mathsf{M}_{(\alpha\alpha)}(\bm{r};t)$ and $\scd^{\mathrm{tot}}_{\alpha\alpha}(t)\coloneqq\int_{V}d\bm{r}\,\scd_{\alpha\alpha}(\bm{r};t)$. This TUR generalizes the one for well-mixed CRNs studied in the previous work~\cite{yoshimura2021thermodynamic} because it reduces to the same form as the TUR for CRNs,
\begin{align}\label{TUR_CRNform}
    \frac{\left|d_tc^{\mathrm{tot}}_{\alpha}(t)\right|^2}{\scd^{\mathrm{tot}}_{\alpha\alpha}(t)}\leq\sigma^{\mathrm{ex}},
\end{align}
if $\bm{k}$ is $\bm{0}$. Here, we let $c^{\mathrm{tot}}_{\alpha}(t)$ denote the total concentration of $\alpha$-th species at time $t$ as $c^{\mathrm{tot}}_{\alpha}(t)\coloneqq\int_{V}d\bm{r}c_{\alpha}(\bm{r};t)$.

If the mobility tensor has the simple form in Eq.~\eqref{simple_mobility_tensor}, we obtain a simpler form of the TUR as
\begin{align}\label{Fourier_TUR_simple}
    \frac{\left|d_t\tilde{c}_{\alpha}(\bm{k};t)\right|^2}{\|\bm{k}\|^2D_{\alpha}c^{\mathrm{tot}}_{\alpha}(t)+\scd^{\mathrm{tot}}_{\alpha\alpha}(t)}\leq\sigma^{\mathrm{ex}}.
\end{align}
The first term of the denominator in Eq.~\eqref{Fourier_TUR_simple} monotonically increases for $\|\bm{k}\|$, while the second term is independent of $\|\bm{k}\|$. \add{Thus, the TUR~\eqref{Fourier_TUR_simple} indicates that we can realize large $|d_t\tilde{c}(\bm{k};t)|$ with small dissipation if $\|\bm{k}\|$ is large.} We can also detect which mode and which species is dominant in the time evolution by considering the tightness of the TUR~\eqref{Fourier_TUR_simple} depending on wavenumber $\bm{k}$.

In the context of pattern formation, the trade-off relation in Eq.~\eqref{Fourier_TUR_simple} means that spatially global pattern formation, which is given by the mode change $|d_t\tilde{c}_{\alpha}(\bm{k};t)|$ corresponding to smaller wavenumber $\bm{k}$, is more dissipative rather than spatially local pattern formation, which is given by the mode change $|d_t\tilde{c}_{\alpha}(\bm{k}';t)|$ corresponding to larger wavenumber $\bm{k}'$ ($\| \bm{k}'\| > \| \bm{k}\|$). Thus, the TUR~\eqref{Fourier_TUR_simple} quantifies a required dissipation to form spatial patterns according to its spatial structure. 


We can also obtain a TSL-like thermodynamic trade-off relation
\begin{align}
    \frac{\left|\tilde{c}_{\alpha}(\bm{k};\tau)-\tilde{c}_{\alpha}(\bm{k};0)\right|^2}{\bm{k}\cdot\langle\mathsf{M}_{(\alpha\alpha)}^{\mathrm{tot}}\rangle_{\tau}\bm{k}+\langle\scd^{\mathrm{tot}}_{\alpha\alpha}\rangle_{\tau}}\leq\tau\Sigma^{\mathrm{ex}}_{\tau},
    \label{TSL_from_TUR}
\end{align}  
which is derived by integrating the TUR in Eq.~\eqref{Fourier_TUR} over time and using the triangle inequality for time integration and the Cauchy--Schwarz inequality (see also Appendix~\ref{ap:derivation_trade-off} for details).
Note that the initial time $0$ can be arbitrarily set.

\subsection{Numerical examples: thermodynamic uncertainty relations}\label{sec:numex_TUR}
We demonstrate the TUR for the Fourier transform of a concentration distribution~\eqref{Fourier_TUR_simple}, using the same time series of the Brusselator as in Sec.~\ref{sec:system_for_numex}, shown in Fig.~\ref{fig:EPR_BL}. Because of the periodic boundary condition imposed on the system, we can obtain the Fourier transform of the concentration distribution as
\begin{align}
    \tilde{c}_{\alpha}(k(n);t)=\int_{-0.5}^{0.5}dr\,c_{\alpha}(r;t)\mathrm{e}^{-\mathrm{i}k(n)r}
\end{align}
with discrete wavenumbers $k(n)$~\eqref{wavenumber-n}. To discuss the property of the TUR, we introduce the mode $n^{(\alpha)}_{\mathrm{max}}(t)$ that provides the tightest lower bound on dissipation at time $t$ in the sense of the TUR as
\begin{align}
    n^{(\alpha)}_{\mathrm{max}}(t)\coloneqq\argmax_{n}\sigma^{\mathrm{TUR}}(n;t),
\end{align}
with
\begin{align}
    \sigma^{\mathrm{TUR}}_{\alpha}(n;t)\coloneqq\frac{\left|d_t\tilde{c}_{\alpha}(k(n);t)\right|^2}{k(n)^2D_{\alpha}c^{\mathrm{tot}}_{\alpha}(t)+\scd^{\mathrm{tot}}_{\alpha\alpha}(t)}.
\end{align}

Here we focus on the chemical species $Z_2$, as indexed by $\alpha=2$ (see Appendix \ref{ap:numex_anotherTUR} for results for $Z_1$, corresponding to $\alpha=1$) and discuss the relations between the pattern dynamics and the lower bound of dissipation.

In the following, we show that the lower bound in the TUR reflects the symmetry of the pattern [Fig.~\ref{fig:TUR_numerics_2}]. In this numerical example, the pattern always has a $3$-fold symmetry. Thus, the pattern changes occur only in modes corresponding to multiples of three. This leads to large $\sigma_2^{\mathrm{TUR}}(n;t)$ for $n$ that are multiples of three, as shown in (b). After the pattern with $6$-fold symmetry is formed ($t>20$), $|\tilde{c}_2(k(n))|$ decays if $n$ is a multiple of three but not a multiple of six [(e)]. In particular, the decay of the mode $n=3$ is significant. It leads to $n^{(2)}_{\mathrm{max}}(t)=3$ for almost all $t>20$ [(c, d)]. 

Rapid changes in the pattern have a significant impact on the TUR. In the early stages of the time evolution, the symmetry of the pattern instantly changes from $6$-fold to $3$-fold, as shown in the pink panel in (a). This change is accompanied by a rapid decay of $|\tilde{c}_2(k(6))|$ and a rapid increase of $|\tilde{c}_2(k(3))|$ [the pink panel in (e)]. It makes $\sigma^{\mathrm{TUR}}_2(n;t)$ be tighter for $n=3$ and $n=6$ than other modes, as shown in the pink panel in (c).

The lower bound in the TUR also reflects oscillations in the Fourier components of the pattern. The speed of change of the Fourier component $|d_t\tilde{c}_2(k(n);t)|$ intermittently goes to zero due to the oscillatory behavior of $\tilde{c}_2(k(n);t)$. This causes vanishing of the lower bounds $\sigma_2^{\mathrm{TUR}}(n;t)$, shown as sharp peaks in the semilog plot [(b)]. In particular, the asynchronous oscillations of $\tilde{c}_2(k(n);t)$ shown in the light-blue panel of (e) induce the temporary deviations from $n^{(2)}_{\mathrm{max}}(t)=3$ in $t\geq 20$, shown in (d). For example, the light-blue panel of (e) shows that $|d_t\tilde{c}_2(k(3);t)|$ vanishes at $t\simeq40$, while $|d_t\tilde{c}_2(k(9);t)|$ does not. It lets $n^{(2)}_{\mathrm{max}}(t)$ deviate to nine from three at $t\simeq39$ in (d).

\add{These observations imply that the new TUR determined by the change of each Fourier component~\eqref{Fourier_TUR} may provide a tighter bound of EPR than the TUR of the well-mixed CRN~\cite{yoshimura2021thermodynamic} if we consider chemical systems with pattern changes.} Since the denominator of $\sigma_2^{\mathrm{TUR}}(n;t)$ is smaller for smaller $n$, the case $n=0$ (corresponding simply to the TUR of the well-mixed CRN as Eq.~\eqref{TUR_CRNform}) gives the tightest bound if $|d_t\tilde{c}_{2}(k(n);t)|$ is independent of $n$. In actual pattern formation, however, the magnitude of $d_t\tilde{c}_{2}(k(n);t)$ is biased for each $n$, reflecting the spatiotemporal structure of the pattern. It allows $\sigma_2^{\mathrm{TUR}}(n;t)$ be tighter when $n$ is not zero.

\begin{figure}
    \centering
    \includegraphics[width=\linewidth]{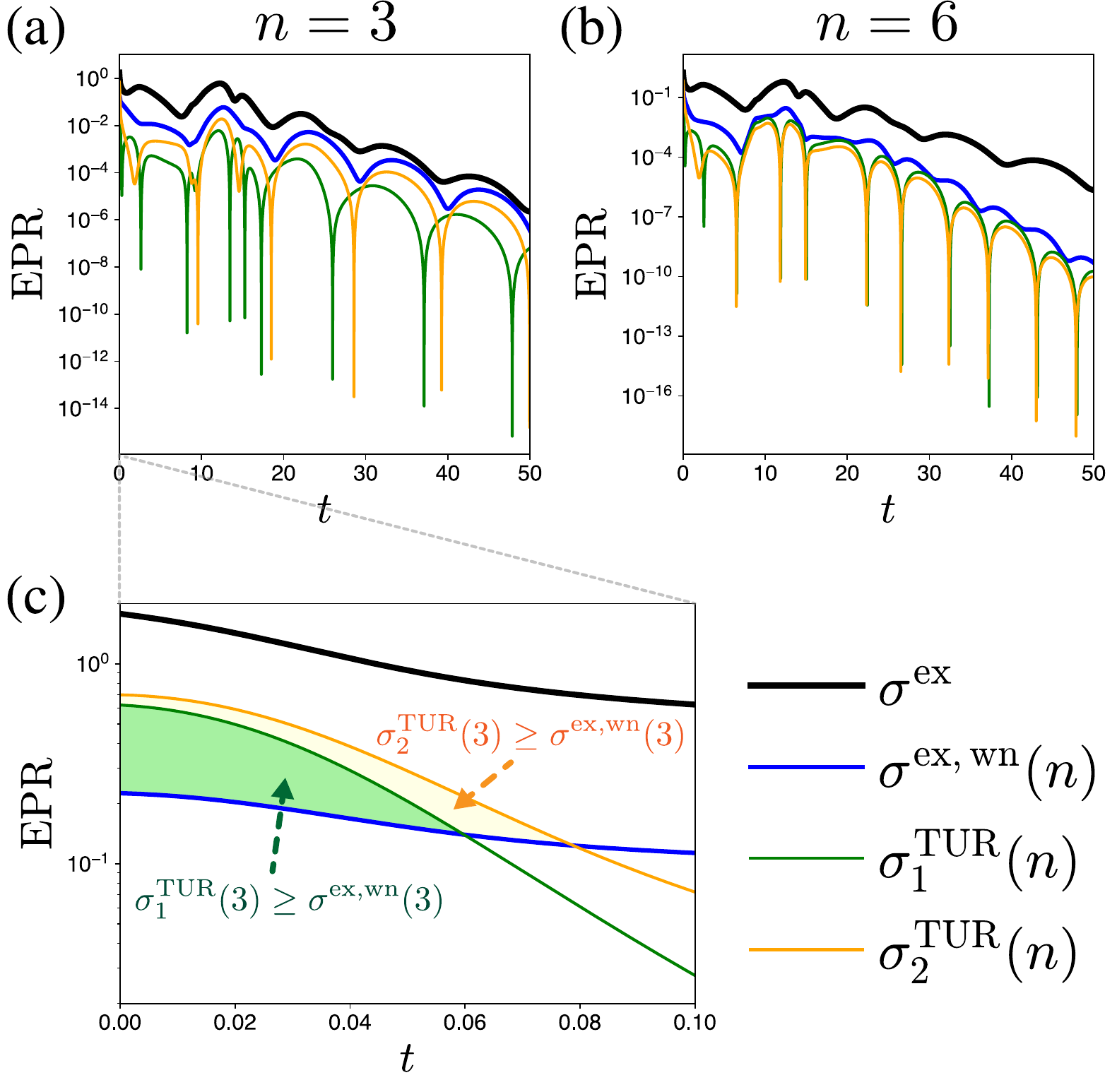}
    \caption{\add{The comparison of the TUR~\eqref{Fourier_TUR} and the wavenumber decomposition of the excess EPR. (a) The behaviour of the excess EPR and its lower bounds ($n=3$). The three bounds $\sigma^{\mathrm{ex,wn}}(3)$ (blue line), $\sigma^{\mathrm{TUR}}_{1}(3;t)$ (green line), and $\sigma^{\mathrm{TUR}}_{2}(3;t)$ (orange line) are smaller than the excess EPR (black line) for all $t$. In addition, the lower bounds provided by the TUR are close to the wavenumber excess EPR rather than the excess EPR. (b) The behaviour of the excess EPR and its lower bounds ($n=6$). We can observe a similar trend as in the case of $n=3$. (c) The behaviour of the excess EPR and its lower bounds near the initial time ($n=3$). The areas filled in light green and light yellow indicate $\sigma^{\mathrm{TUR}}_{1}(3;t)\geq\sigma^{\mathrm{ex,wn}}(3)$ and $\sigma^{\mathrm{TUR}}_{2}(3;t)\geq\sigma^{\mathrm{ex,wn}}(3)$, respectively.}}
    \label{fig:compare_TUR_wndecomp}
\end{figure}

\add{We also compare the lower bounds provided by the TUR and the wavenumber decomposition in Fig.~\ref{fig:compare_TUR_wndecomp}. Since the wavenumber excess EPR is nonnegative at each mode, we obtain
\begin{align}
    \sigma^{\mathrm{ex}}=\sum_{n'}\sigma^{\mathrm{ex,wn}}(n')\geq\sigma^{\mathrm{ex,wn}}(n).
\end{align}
Thus, the value of the wavenumber excess EPR provides a lower bound of the excess EPR. The bound determined by wave number as well as the TUR. We can see that the bounds provided by the TUR are closer to the wavenumber excess EPR on the corresponding mode than the excess EPR itself, except for the points where $|d_t\tilde{c}_{\alpha}(n)|$ vanishes. It implies that the TUR can estimate the wavenumber excess EPR rather than the excess EPR. Note that, however, the lower bounds provided by the TUR possibly become larger than the wavenumber excess EPR at the corresponding mode. For instance, we show that both $\sigma^{\mathrm{TUR}}_{1}(3;t)$ and $\sigma^{\mathrm{TUR}}_{2}(3;t)$ are larger than $\sigma^{\mathrm{ex,wn}}(3)$ at small $t$, where the pattern changes rapidly, in Fig.~\ref{fig:compare_TUR_wndecomp} (c).
}

\section{Discussion}

\add{In this paper, we established universal relations between the time evolution of patterns and dissipation by extending the framework of geometric thermodynamics~\cite{ito2023geometric, dechant2022geometric_E, dechant2022geometric_R, yoshimura2023housekeeping, kolchinsky2022information} to RDSs. Our results enable the treatment of deterministic pattern dynamics, which previous geometric thermodynamics cannot address.}

\add{In particular, we constructed the geometric excess/housekeeping decomposition of EPR and showed that the excess EPR can be understood as the minimum dissipation rate required to reproduce the original time evolution. We also proposed geometric decomposition of EPR, excess EPR, and housekeeping EPR into nonnegative contributions from different spatial locations and Fourier modes (wave numbers). Our decompositions were illustrated using detailed numerical examples.
}

\add{In addition, the excess EPR was related to the details of the pattern and its time evolution by the newly derived TSLs and TURs for RDSs. TSLs are trade-off relations between the speed of the time evolution and the dissipation. These inequalities characterize the dissipation in a finite duration using information about the whole pattern. In contrast, TURs bound the instantaneous excess EPR using partial information about the pattern, such as the Fourier components of the concentration distribution. Our trade-off relations are applicable to deterministic systems with large degrees of freedom, in this way differing from the existing attempts that treat pattern formation in the presence of fluctuations within stochastic thermodynamics~\cite{niggemann2020field,rana2020precision,song2021cost,nguyen2021organization}. }

\add{Our results advance nonequilibrium thermodynamics of RDSs, suggesting engineering applications such as quantifying energy efficiency and optimizing control. In this way, we  go beyond traditional approach, which aim to predict stationary patterns and analyze their stability. This goal cannot be achieved in general cases, which may give the impression that thermodynamics is useless for understanding pattern formation. However, once RDSs are considered objects of control, thermodynamics becomes an essential tool because it quantifies the required dissipation to achieve a desired change in patterns. We have taken the first step in developing this tool, by developing inequalities and extending optimal transport theory so as to relate dissipation with the change in patterns. We believe this new direction will become increasingly important as RDSs undergo exploration by experimentalists and engineers.}

\add{Our results also provide new insights into stochastic thermodynamics. For instance, we have introduced the wavenumber decomposition for Langevin systems. In addition, our results indicate that  optimal transport theory, which in stochastic thermodynamics has been studied separately for systems with discrete and continuous variables, can be applied to general Markov processes with both kinds of variables. To be precise, the master equation for such a general system consists of diffusion terms for the continuous variables and Markov jump terms for the discrete variables. It is equivalent to an RD equation since the MJP is formally the same as an unimolecular CRN. Therefore, all of our results, derived here for RD equations, are also applicable to general master equations and bring new fundamental tools, e.g., decompositions of EPR and thermodynamic trade-off relations, for biological processes modeled by the Markov processes with discrete and continuous variables, such as  F1-ATPase~\cite{zimmermann2012efficiencies}.}

\add{One of the main contributions of this paper is to extend optimal transport theory by generalizing the $1$- and $2$-Wasserstein distances to RDSs. Optimal transport solves the problem of minimizing the dissipation necessary to transition between two given patterns, and it provides an operational way to quantify the speed of evolution, as required for the TSLs. Previously, the $2$-Wasserstein distance was extended to deterministic chemical systems by using its consistency with gradient structures~\cite{liero2013gradient,liero2016optimal,yoshimura2023housekeeping}, but the generalizations does not treat open systems explicitly. The generalization of the $1$-Wasserstein distance to deterministic chemical systems have not received as much attention even though the usefulness of the $1$-Wasserstein distance in stochastic thermodynamics has been revealed~\cite{dechant2022minimum,van2023thermodynamic}. We have extended the $1$- and $2$-Wasserstein distances to deterministic open RDSs and showed that they relate to the EP associated with pattern formation. The Wasserstein distances have different advantages: the $1$-Wasserstein distance allows us to measure the distance between patterns even if the $2$-Wasserstein distance cannot be defined, and it enables us to separately treat the distance and the kinetic information~\eqref{Def_A}. On the other hand, the $2$-Wasserstein distance gives the excess EPR a geometric interpretation and provides tight TSLs.}

\add{Another development relating to the $1$-Wasserstein distance is the intensity of reaction and diffusive dynamics, which represents the kinetic information and links the distance to thermodynamics. The intensity of reactions measures the average rate of forward and reverse reactions with the logarithmic mean. It corresponds to the dynamical state mobility defined in MJPs~\cite{van2023thermodynamic}. The intensity of diffusion is defined by a variational problem, consistent with the diffusion coefficient in Langevin systems. This definition enables us to quantify the intensity of diffusion even for system with spatial anisotropy and interactions between species.}

\add{We remark that our geometric decompositions were derived by expressing dissipation as the squared norm of the thermodynamic forces. In fact, it is possible to express dissipation in terms of other functionals, leading to other kinds of geometric decompositions. In particular, geometric decompositions may be derived by using the fact that information geometry and Hessian geometry generalize orthogonality and the Pythagorean theorem~\cite{amari2000methods,amari2016information,shima2007geometry}. Previously, this was used to derive different extensions of the excess/housekeeping decomposition by Maes and Neto\v{c}n\'{y} from Langevin systems to MJPs and CRNs. One method involves fixing another mean of the fluxes instead of the logarithmic mean $m_{\rho}$ and constructing a nonlinear relation between the force and current based on Hessian geometry~\cite{kobayashi2022hessian}. Another method uses information geometry to study one-way fluxes, instead of currents~\cite{kolchinsky2022information}. We may also generalize these methods to RDSs. However, if we adopt the first alternative, we can no longer regard the excess EPR as the minimum EPR required to realize the original instantaneous change in patterns. It is not in keeping with our original goal of clarifying the relations between the time evolution of patterns and the unavoidable dissipation. The second alternative avoids this problem, but does not directly link the excess EPR to the Wasserstein distance. Thus, we have taken the method based on the quadratic form of thermodynamic forces~\cite{yoshimura2023housekeeping}.}

\add{Finally, we introduce some prospects for future research. First, the wavenumber decomposition can be extended to bases other than the Fourier basis (e.g., the wavelet basis).  Second, while our results are valid for general RDSs, they may be specialized in interesting ways to individual systems; for instance, the TUR for a general observable~\eqref{general_TUR_RDS} will yield interesting bounds on EPR by considering specific observables depending on the nature of the system. In addition, it may be possible to explore other expressions of minimum dissipation  by appropriately tailoring the allowed constraints. This may allow quantifying minimum dissipation in specific systems, such as active phase separation~\cite{weber2019physics,kirschbaum2021controlling,bauermann2022energy,zwicker2022intertwined} and systems used for computation~\cite{adamatzky2005reaction,gorecki2015chemical,parrilla2020programmable,chen2016trainable}, thus highlighting system-specific thermodynamics bounds on pattern formation. Third, it may be possible to utilize the lower bounds of EPR to estimate the dissipation of actual pattern formations as studied in the growing field of thermodynamic inference~\cite{otsubo2020estimating,manikandan2020inferring}. Finally, it would also be meaningful to extend our results to RDSs on general spaces, such as curved surfaces~\cite{plaza2004effect,nishide2022pattern} or graphs~\cite{othmer1971instability,nakao2010turing}. This may be done by replacing the differential operator $\bm{\nabla}_{\bm{r}}$ with an appropriate counterpart for curved surfaces or graphs. This generalization will reveal trade-offs and help in an energetic understanding of pattern formations in living systems, e.g., chemical waves on the cell membrane~\cite{tan2020topological} and epidemics in metapopulation models~\cite{colizza2007reaction}.}

\begin{acknowledgments}
The authors thank Naruo Ohga for fruitful discussions on thermodynamic bounds. K.~Y., A.~K. and S.~I. thank Andreas Dechant for fruitful discussions on the geometric decomposition. R.~N. thanks Yasushi Okada for fruitful discussions on thermodynamics of RDSs. S.~I. thanks Masafumi Oizumi and Daiki Sekizawa for fruitful discussions on the geometric decomposition. \add{This research was supported by JSR Fellowship, the University of Tokyo.}
K.~Y.\ is supported by Grant-in-Aid for JSPS Fellows Grant No.~22J2161. 
A.~K. received funding from the European Union’s Horizon 2020 research and innovation programme under the Marie Sklodowska-Curie Grant Agreement No.~101068029 and support from Grant 62417 from the John Templeton Foundation. The opinions expressed in this publication are those of the authors and do not necessarily reflect the views of the John Templeton Foundation.
S.~I. is supported by JSPS KAKENHI Grants No. 19H05796,
No.~21H01560, No.~22H01141, and No.~23H00467, \add{No.~24H00834}, JST ERATO Grant No.~JPMJER2302, and UTEC-UTokyo
FSI Research Grant Program. 
\end{acknowledgments}

\appendix
\section{Details of the projected force in Langevin systems}
In this section, we provide the details of the conservative force $\bm{F}^{\ast}$. We derive the condition in Eq.~\eqref{FP_EL_eq} from the orthogonality~\eqref{FP_ortho} and the uniqueness of $\bm{F}^{\ast}$ using the condition in Appendix~\ref{ap:FP_deriv_EL_eq}. We also derive the minimization problem in Eq.~\eqref{FP_another_projected_force} and Eq.~\eqref{FP_projected_force} in Appendix~\ref{ap:FP_deriv_anotherminimization}. In Appendix~\ref{ap:approach_by_variational_calculus}, we provide the way to obtain Eq.~\eqref{FP_EL_eq} from the minimization problems in Eq.~\eqref{FP_another_projected_force} and Eq.~\eqref{FP_projected_force}.

\subsection{Sufficiency of Eq.~\eqref{FP_EL_eq} for orthogonality, and the uniqueness of its solution}\label{ap:FP_deriv_EL_eq}

Letting $\bm{F}^{\ast}$ be the gradient of a potential as $\bm{F}^{\ast}=\bm{\nabla}_{\bm{r}}\phi^{\ast}$, we can rewrite the inner product between $\bm{F}^{\ast}$ and $\bm{F}-\bm{F}^{\ast}$ as
\begin{align}\label{FP_check_ortho}
    \fip{\bm{F}^{\ast}}{\bm{F}-\bm{F}^{\ast}}_{\mathsf{M}}&=\fip{\bm{\nabla}_{\bm{r}}\phi^{\ast}}{\bm{F}-\bm{\nabla}_{\bm{r}}\phi^{\ast}}_{\mathsf{M}}\notag\\
    &=\int_{\mathbb{R}^d}d\bm{r}\,\bm{\nabla}_{\bm{r}}\phi^{\ast}\cdot\mathsf{M}(\bm{F}-\bm{\nabla}_{\bm{r}}\phi^{\ast})\notag\\&=-\int_{\mathbb{R}^d}d\bm{r}\,\phi^{\ast}\bm{\nabla}_{\bm{r}}\cdot[\mathsf{M}(\bm{F}-\bm{\nabla}_{\bm{r}}\phi^{\ast})],
\end{align}
where we used the boundary condition for $p(\bm{r};t)$ to ignore the surface term in the third line. Then, the orthogonality in Eq.~\eqref{FP_ortho} reduces to 
\begin{align}
    \int_{\mathbb{R}^d}d\bm{r}\,\phi^{\ast}\bm{\nabla}_{\bm{r}}\cdot[\mathsf{M}(\bm{F}-\bm{\nabla}_{\bm{r}}\phi^{\ast})]=0,
\end{align}
which shows that Eq.~\eqref{FP_EL_eq} is a sufficient condition of the orthogonality. 

Let $\phi^1$ and $\phi^2$ be solutions of Eq.~\eqref{FP_EL_eq}, and $\bm{F}^1$ and $\bm{F}^2$ the corresponding gradient forces. The norm of the difference between these forces is zero because
\begin{align}
    &\fip{\bm{F}^1-\bm{F}^2}{\bm{F}^1-\bm{F}^2}_{\mathsf{M}}\notag\\&=\int_{\mathbb{R}^d}d\bm{r}\,\bm{\nabla}_{\bm{r}}(\phi^1-\phi^2)\cdot\mathsf{M}\bm{\nabla}_{\bm{r}}(\phi^1-\phi^2)\notag\\
    &=-\int_{\mathbb{R}^d}d\bm{r}\,(\phi^1-\phi^2)\bm{\nabla}_{\bm{r}}\cdot[\mathsf{M}\{\bm{\nabla}_{\bm{r}}(\phi^1-\phi^2)\}]\notag\\
    &=\int_{\mathbb{R}^d}d\bm{r}\,(\phi^1-\phi^2)\bm{\nabla}_{\bm{r}}\cdot[\mathsf{M}(\bm{F}-\bm{F})]=0
\end{align}
where we used the boundary condition for $p(\bm{r};t)$ to ignore the surface term in the third line, and Eq.~\eqref{FP_EL_eq} in the fourth line. 
Therefore, the two forces are identical as the norm is nondegenerate. 
Note that the potential satisfying Eq.~\eqref{FP_EL_eq} is not unique because $\phi^{\ast}+C$, where $C$ is a constant, is a solution of Eq.~\eqref{FP_EL_eq} when $\phi^{\ast}$ is a solution.

\subsection{The derivation of minimization problems [Eqs.~\eqref{FP_another_projected_force} and \eqref{FP_projected_force}]}\label{ap:FP_deriv_anotherminimization}

The condition in Eq.~\eqref{FP_EL_eq} also leads to the orthogonality
\begin{align}\label{FP_ortho_anotherprojection}
    \fip{\bm{F}^{\ast}}{\bm{F}'-\bm{F}^{\ast}}_{\mathsf{M}}=0,
\end{align}
where $\bm{F}'$ satisfies the condition in Eq.~\eqref{FP_another_projected_force}.
We can check it by calculating in a similar way as in Eq.~\eqref{FP_check_ortho}. Using the orthogonality in Eq.~\eqref{FP_ortho_anotherprojection}, we obtain the inequality
\begin{align}\label{FP_ineq_anotherprojection}
    \fip{\bm{F}'}{\bm{F}'}_{\mathsf{M}}&=\fip{\bm{F}^{\ast}}{\bm{F}^{\ast}}_{\mathsf{M}}+\fip{\bm{F}'-\bm{F}^{\ast}}{\bm{F}'-\bm{F}^{\ast}}_{\mathsf{M}}\notag\\
    &\geq\fip{\bm{F}^{\ast}}{\bm{F}^{\ast}}_{\mathsf{M}}
\end{align}
for all $\bm{F}'$ satisfying the condition in Eq.~\eqref{FP_another_projected_force}. It immediately leads to the minimization problem in Eq.~\eqref{FP_another_projected_force} because the equality in Eq.~\eqref{FP_ineq_anotherprojection} is achieved if and only if $\bm{F}'=\bm{F}^{\ast}$.

Similarly, the condition in Eq.~\eqref{FP_EL_eq} also leads to another orthogonality
\begin{align}\label{FP_ortho_projection}
    \fip{\bm{F}-\bm{F}^{\ast}}{\bm{F}^{\ast}-\bm{F}'}_{\mathsf{M}}=0,
\end{align}
for all $\bm{F}'\in\mathrm{Im}\bm{\nabla}_{\bm{r}}$. It yields the inequality
\begin{align}\label{FP_ineq_projection}
    &\fip{\bm{F}-\bm{F}'}{\bm{F}-\bm{F}'}_{\mathsf{M}}\notag\\
    &=\fip{\bm{F}-\bm{F}^{\ast}}{\bm{F}-\bm{F}^{\ast}}_{\mathsf{M}}+\fip{\bm{F}^{\ast}-\bm{F}'}{\bm{F}^{\ast}-\bm{F}'}_{\mathsf{M}}\notag\\
    &\geq\fip{\bm{F}-\bm{F}^{\ast}}{\bm{F}-\bm{F}^{\ast}}_{\mathsf{M}},
\end{align}
which leads to the minimization problem in Eq.~\eqref{FP_projected_force} because the equality in Eq.~\eqref{FP_ineq_projection} is achieved if and only if $\bm{F}'=\bm{F}^{\ast}$.

\subsection{The derivation of the condition Eq.~\eqref{FP_EL_eq} as the Euler--Lagrange equation}\label{ap:approach_by_variational_calculus}
We remark that the conditions Eq.~\eqref{FP_EL_eq} and $\bm{F}^{\ast} = \bm{\nabla}_{\bm{r}} \phi^{\ast}$ are conversely obtained from two variational problems Eqs.~\eqref{FP_exEPR} and \eqref{FP_hkEPR}. By considering the action functionals
\begin{align}
\mathcal{I}_{\rm hk} [\phi]&\coloneqq \frac{1}{2} \fip{\bm{F}-\bm{\nabla}_{\bm{r}}\phi}{\bm{F}-\bm{\nabla}_{\bm{r}}\phi}_{\mathsf{M}} = \int d\bm{r}\,I_{\rm hk},\\
\mathcal{I}_{\rm ex} [\bm{F}', \phi] &\coloneqq\frac{1}{2} \fip{\bm{F}'}{\bm{F}'}_{\mathsf{M}} + \int d\bm{r}\,\phi \{ \bm{\nabla}_{\bm{r}}\cdot\qty[\mathsf{M}\qty(\bm{F}-\bm{F}')]\}\nonumber \\
&= \int d\bm{r}\,I_{\rm ex} 
\label{functionalex}
\end{align}
with
\begin{align}
 I_{\rm hk} &\coloneqq\frac{1}{2}(\bm{F}-\bm{\nabla}_{\bm{r}}\phi) \cdot \mathsf{M} (\bm{F}-\bm{\nabla}_{\bm{r}}\phi), \\
 I_{\rm ex} &\coloneqq \frac{1}{2} \bm{F}' \cdot \mathsf{M} \bm{F}' - \bm{\nabla}_{\bm{r}} \phi \cdot \mathsf{M} (\bm{F}- \bm{F}'),
\end{align}
the two variational problems, Eqs.~\eqref{FP_exEPR} and \eqref{FP_hkEPR}, are solved by the Euler--Lagrange equations 
\begin{align}
&\left. \frac{\delta \mathcal{I}_{\rm hk} [\phi]}{\delta \phi} \right|_{\phi = \phi^{\ast}}  =\left. \left[ \frac{\partial I_{\rm hk}}{\partial  \phi } -  \bm{\nabla}_{\bm{r}} \cdot \frac{\partial I_{\rm hk}}{\partial ( \bm{\nabla}_{\bm{r}} \phi) } \right]\right|_{\phi = \phi^{\ast}} = 0,  \label{firstELeq} \\
&\left. \frac{ \delta \mathcal{I}_{\rm ex}  [\bm{F}', \phi]}{\delta \bm{F}' }\right|_{\bm{F}'= \bm{F}^{\ast}, \phi = \phi^{\ast}} = \left. \frac{\partial {I}_{\rm ex} }{ \partial \bm{F}'} \right|_{\bm{F}'= \bm{F}^{\ast}, \phi = \phi^{\ast}}= \bf{0}. \label{secondELeq}
\end{align}
Here, $\phi$ in Eq.~\eqref{functionalex} is the Lagrange multiplier which gives the condition in Eq.~\eqref{FP_EL_eq} from the variation $\left. \delta \mathcal{I}_{\rm ex} [\bm{F}', \phi^{\ast}]/\delta \phi^* \right|_{\bm{F}'= \bm{F}^{\ast}, \phi = \phi^{\ast}} =0$.
The first Euler--Lagrange equation~\eqref{firstELeq} directly provides the condition in Eq.~\eqref{FP_EL_eq}. On the other hand, the second Euler--Lagrange equation~\eqref{secondELeq} provides $\bm{F}^{\ast} = \bm{\nabla}_{\bm{r}} \phi^{\ast}$. Substituting it into the constraint $\bm{\nabla}_{\bm{r}}\cdot\qty[\mathsf{M}\qty(\bm{F}-\bm{F}')] =\bm{0}$, we obtain Eq.~\eqref{FP_EL_eq}.

\section{Details of the projected force in reaction-diffusion systems}\label{ap:optimal_condition}
Here, we derive the Euler--Lagrange equation in Eq.~\eqref{EL_eq} from the optimization problem in Eq.~\eqref{Def_optimalpot} or Eq.~\eqref{another_form_PF}. We also derive the uniqueness of $\mathcal{F}^{\ast}$. The derivation reduces to the case of Langevin systems in Appendix~\ref{ap:approach_by_variational_calculus} by considering a particular situation: There is only one species and no chemical reactions. 

\subsection{The Euler--Lagrange equation for the projection and the uniqueness of the projected conservative force}\label{ap:projection_conservative}

First, we derive Eq.~\eqref{EL_eq} from Eq.~\eqref{Def_optimalpot}. We define a functional to minimize as $\mathcal{I}_{\mathrm{hk}}[\vec{\phi}]\coloneqq\fip{\mathcal{F}-\Grad\vec{\phi}}{\mathcal{F}-\Grad\vec{\phi}}_{\mathcal{M}}/2=\int_{V}d\bm{r}I_{\mathrm{hk}}$ with
\begin{align}
    I_{\mathrm{hk}}\coloneqq\notag& \frac{1}{2}\qty[\vec{\bm{F}}-\bm{\nabla}_{\bm{r}}\vec{\phi}]^{\top}\dvec{\mathsf{M}}\qty[\vec{\bm{F}}-\bm{\nabla}_{\bm{r}}\vec{\phi}]\\
    &+\frac{1}{2}\qty[\bm{f}-\nabla_{\mathsf{s}}\vec{\phi}]^{\top}\mathsf{m}\qty[\bm{f}-\nabla_{\mathsf{s}}\vec{\phi}].
\end{align}
The functional derivative of $\mathcal{I}_{\mathrm{hk}}$ leads to the condition to be satisfied by the optimal potential $\vec{\phi}{}^{\ast}$ as
\begin{align}\label{Functional_derivative_Iex}
    &\left.\frac{\delta\mathcal{I}_{\mathrm{hk}}}{\delta\phi_{\alpha}}\right|_{\vec{\phi}=\vec{\phi}{}^{\ast}}=
  \left.\qty(\pdv{I_{\mathrm{hk}}}{\phi_{\alpha}}-\bm{\nabla}_{\bm{r}}\cdot\pdv{I_{\mathrm{hk}}}{\qty[\bm{\nabla}_{\bm{r}}\phi_{\alpha}]})\right|_{\vec{\phi}=\vec{\phi}{}^{\ast}}\notag\\
    &=\qty[-\nabla_{\mathsf{s}}^{\top}\mathsf{m}\qty(\bm{f}-\nabla_{\mathsf{s}}\vec{\phi}{}^{\ast})+\bm{\nabla}_{\bm{r}}\cdot\qty{\dvec{\mathsf{M}}\qty(\vec{\bm{F}}-\bm{\nabla}_{\bm{r}}\vec{\phi}{}^{\ast})}]_{\alpha}\notag\\
    &=\qty[-\Div\mathcal{M}\qty(\mathcal{F}-\Grad\vec{\phi}{}^{\ast})]_{\alpha}=0,
\end{align}
for all $\alpha\in\mathcal{X}$, while $\phi^{\ast}_{\alpha}=0$ holds for all $\alpha\in\mathcal{Y}$. This condition~\eqref{Functional_derivative_Iex} is nothing more than the condition indicated in Eq.~\eqref{EL_eq}. 

Second, we show that the projected conservative force $\mathcal{F}^{\ast}$ is unique, while $\vec{\phi}{}^{\ast}$ is not. If we consider a constant vector $\vec{C}$ satisfying $\vec{C}_{\mathcal{Y}}=\vecrm{0}_{\mathcal{Y}}$, whose gradient $\Grad\vec{C}$ is a zero vector, $\vec{\phi}{}^{\ast} + \vec{C}$ can also be a solution of the minimization problem in Eq.~\eqref{Def_optimalpot} and thus $\vec{\phi}{}^{\ast}$ is not unique. Supposing Eq.~\eqref{EL_eq} has two solutions, $\vec{\phi}{}^{1}$ and $\vec{\phi}{}^{2}$, which satisfy $\vec{\phi}{}^{1}_{\mathcal{Y}}=\vec{\phi}{}^{2}_{\mathcal{Y}}=\vecrm{0}_{\mathcal{Y}}$, under the boundary condition, we obtain
\begin{align*}\label{uniq}
    \fip{\mathcal{F}^1-\mathcal{F}^2}{\mathcal{F}^1-\mathcal{F}^2}_{\mathcal{M}}&=\fip{\mathcal{M}\Grad (\vec{\phi}{}^1-\vec{\phi}{}^2)}{\Grad(\vec{\phi}{}^1-\vec{\phi}{}^2)}\\
    &=\cip{\Div\mathcal{M} \Grad (\vec{\phi}{}^1-\vec{\phi}{}^2)}{\vec{\phi}{}^1-\vec{\phi}{}^2}\\
    &=\cip{\Div\mathcal{M}\qty(\mathcal{F}-\mathcal{F})}{\vec{\phi}{}^1-\vec{\phi}{}^2} \nonumber \\
    &=0,
    \stepcounter{equation}\tag{\theequation}
\end{align*}
where we define $\mathcal{F}^{i}\coloneqq\Grad\vec{\phi}{}^{i}$ for $i=1,2$. The second transformation in Eq.~\eqref{uniq} is allowed because we are imposing the appropriate boundary condition, and the third line follows from the Euler--Lagrange equation~\eqref{EL_eq}. 
We also use the condition, $\phi^1_{\alpha}-\phi^2_{\alpha}=0$ for all $\alpha\in\mathcal{Y}$, in these transformations. 
Therefore, the nondegenerateness of the inner product $\fip{\cdot}{\cdot}_{\M}$ concludes $\mathcal{F}^1=\mathcal{F}^2$, that is, the uniqueness of the projected potential force.

\subsection{The Euler--Lagrange equation for the minimum dissipation}\label{ap:minimum_dissipation}
We derive the condition~\eqref{EL_eq} from the minimization problem in Eq.~\eqref{another_form_PF}. To solve this constraint minimization problem, we execute the method of Lagrange multiplier with the multiplier $\vec{\phi}$, whose external part is the zero vector as $\vec{\phi}_{\mathcal{Y}}=\vecrm{0}_{\mathcal{Y}}$. Then, the functional to optimize is
\begin{align}
    \mathcal{I}_{\mathrm{ex}}[\mathcal{F}',\vec{\phi}]&\coloneqq\frac{1}{2}\fip{\mathcal{F}'}{\mathcal{F}'}_{\mathcal{M}}+\cip{\vec{\phi}}{\Div\mathcal{M}\qty(\mathcal{F}-\mathcal{F}')}\notag\\
    &=\frac{1}{2}\fip{\mathcal{F}'}{\mathcal{F}'}_{\mathcal{M}}+\fip{\Grad\vec{\phi}}{\mathcal{F}-\mathcal{F}'}_{\mathcal{M}}.
\end{align}
We obtain the conditions to be satisfied by the optimizer $\mathcal{F}^{\ast}$ and $\vec{\phi}{}^{\ast}$ by taking the functional derivative of $\mathcal{I}_{\mathrm{ex}}[\mathcal{F}',\vec{\phi}]$ as
\begin{align}\label{force&pot1}
    &\left.\frac{\delta\mathcal{I}_{\mathrm{ex}}[\mathcal{F}',\vec{\phi}]}{\delta\bm{F}'_{(\alpha)}}\right|_{\vec{\phi}=\vec{\phi}{}^{\ast},\mathcal{F}'=\mathcal{F}^{\ast}} \nonumber \\
    &=\sum_{\beta\in\mathcal{S}}\mathsf{M}_{(\alpha\beta)}\qty(\bm{F}^{\ast}_{(\beta)}-\bm{\nabla}_{\bm{r}}\phi^{\ast}_{\beta})= \bm{0},
\end{align}
for all $\alpha\in\mathcal{S}$ and
\begin{align}\label{force&pot2}
    \left.\frac{\delta\mathcal{I}_{\mathrm{ex}}[\mathcal{F}',\vec{\phi}]}{\delta f'_{\rho}} \right|_{\vec{\phi}=\vec{\phi}{}^{\ast},\mathcal{F}'=\mathcal{F}^{\ast}}=m_{\rho}\qty{f^{\ast}_{\rho}-\qty(\nabla_{\mathsf{s}}\vec{\phi}{}^{\ast})_{\rho}}=0,
\end{align}
for all $\rho\in\mathcal{R}$. The positive definiteness of $\dvec{\mathsf{M}}$ and $\mathsf{m}$ enables us to obtain $\vec{\bm{F}}^{\ast}=\bm{\nabla}_{\bm{r}}\vec{\phi}{}^{\ast}$ and $\bm{f}^{\ast}=\nabla_{\mathsf{s}}\vec{\phi}{}^{\ast}$ from Eq.~\eqref{force&pot1} and Eq.~\eqref{force&pot2}. Unifying these results yields
\begin{align}\label{ex_opt_cond}
    \mathcal{F}^{\ast}=\Grad\vec{\phi}{}^{\ast}.
\end{align}
We obtain the condition in Eq.~\eqref{EL_eq} by substituting Eq.~\eqref{ex_opt_cond} into the constraint $\qty(\Div\mathcal{M}\mathcal{F})_{\mathcal{X}}=\qty(\Div\mathcal{M}\mathcal{F}^{\ast})_{\mathcal{X}}$.

\section{Gradient flow and relaxation in reaction-diffusion systems}\label{ap:gradientflow}

Gradient flow is the flow that causes a specific function (functional) to become smaller. The gradient flow structure of RDSs~\cite{mielke2011gradient,liero2013gradient} is deeply related to the thermodynamics of RDSs, e.g., the relaxation to the equilibrium of the closed ideal dilute solution is described by the gradient flow to the equilibrium concentration distribution (Appendix~\ref{ap:relax_to_eq}). We can also verify that the conservative force $\Grad\vec{\phi}$ provides a gradient flow, which describes the relaxation to a state determined by the potential $\vec{\phi}$ (Appendix~\ref{ap:relax_to_pcan}). Moreover, considering the potential $\vec{\phi}{}^{\ast}$, which provides the excess EPR, the time evolution of the concentration distribution of the internal species can be represented only by a gradient flow. The gradient flow structure allows us to rewrite the excess EPR in a similar form to the EPR of the ideal dilute solutions, which relax to equilibrium (Appendix~\ref{ap:gradientflow_excess}).

\subsection{Relaxation to equilibrium and gradient flow of the ideal solutions}\label{ap:relax_to_eq}

We consider relaxation to the equilibrium of a closed ideal dilute solution without mechanical forces, where the chemical potential is written as $\mu_{\alpha}^{\mathrm{id}}=\mu^{\circ}_{\alpha}+\ln c_{\alpha}$. 
Since the system is closed and there is no mechanical force applied to the system, there exists an equilibrium concentration distribution $\vec{c}^{\mathrm{eq}}$. The equilibrium concentration distribution provides zero force as $-\Grad\vec{\mu}^{\mathrm{id,eq}}=(\vecrm{\bm{0}},\bm{0})$, where $\vec{\mu}^{\mathrm{id,eq}}$ is the chemical potential of the equilibrium state $\mu_{\alpha}^{\mathrm{id,eq}}=\mu^{\circ}_{\alpha}+\ln c^{\mathrm{eq}}_{\alpha}$.

In this situation, the force $\mathcal{F}$ is written as
\begin{align}
    \mathcal{F}=-\Grad\left(\vec{\mu}^{\mathrm{id}}-\vec{\mu}^{\mathrm{id,eq}}\right).
\end{align}
Focusing on the concrete form of $\vec{\mu}^{\mathrm{id}}$, we can rewrite the difference between the chemical potentials $\vec{\mu}^{\mathrm{id}}-\vec{\mu}^{\mathrm{id,eq}}$ as
\begin{align}
    \mu^{\mathrm{id}}_{\alpha}-\mu^{\mathrm{id,eq}}_{\alpha}=\ln\frac{c_{\alpha}}{c_{\alpha}^{\mathrm{eq}}}=\frac{\delta}{\delta c_{\alpha}}D_{\mathrm{KL}}(\vec{c}\|\vec{c}^{\mathrm{eq}}),
\end{align}
where $D_{\mathrm{KL}}(\cdot\|\cdot)$ is the generalized Kullback–Leibler
(KL) divergence,
\begin{align}
    D_{\mathrm{KL}}(\vec{c}\|\vec{c}')\coloneqq\int_{V}d\bm{r}\sum_{\alpha\in\mathcal{S}}\left(c_{\alpha}\ln\frac{c_{\alpha}}{c'_{\alpha}}-c_{\alpha}+c'_{\alpha}\right).
\end{align}
Introducing the vector $\delta/\delta\vec{c}=(\delta/\delta c_{1},\dots, \delta/\delta c_{N})^{\top}$, we can rewrite the force as
\begin{align}
    \mathcal{F}=-\Grad\frac{\delta}{\delta \vec{c}}D_{\mathrm{KL}}(\vec{c}\|\vec{c}^{\mathrm{eq}}),
\end{align}
which rewrites the RD equation as the gradient flow toward the equilibrium~\cite{mielke2011gradient},
\begin{align}\label{gradientflow_to_equilibrium}
    \partial_{t}\vec{c}=-\Div\mathcal{M}\Grad\frac{\delta}{\delta \vec{c}}D_{\mathrm{KL}}(\vec{c}\|\vec{c}^{\mathrm{eq}}).
\end{align}

We can also rewrite the EPR as
\begin{align}\label{EPR_KL_div}
    \sigma&=-\fip{\mathcal{J}}{\Grad\left(\vec{\mu}^{\mathrm{id}}-\vec{\mu}^{\mathrm{id,eq}}\right)}\notag\\
    &=-\cip{\partial_t\vec{c}}{\vec{\mu}^{\mathrm{id}}-\vec{\mu}^{\mathrm{id,eq}}}\notag\\
    &=-\int_{V}d\bm{r}\sum_{\alpha\in\mathcal{S}}(\partial_{t}c_{\alpha})\ln\frac{c_{\alpha}}{c_{\alpha}^{\mathrm{eq}}}\notag\\
    &=-\partial_{t}D_{\mathrm{KL}}(\vec{c}\|\vec{c}^{\mathrm{eq}}),
\end{align}
where we can regard the KL divergence $D_{\mathrm{KL}}(\vec{c}\|\vec{c}^{\mathrm{eq}})$ as the Gibbs free energy difference between the state $\vec{c}$ and the equilibrium state $\vec{c}^{\mathrm{eq}}$.

\subsection{Relaxation due to the conservative force}\label{ap:relax_to_pcan}
In Sec.~\ref{sec:conservative_nonconservative_forces}, we decompose the force at time $t$ as 
\begin{align}
    \mathcal{F}=\Grad\vec{\phi}(t)+\mathcal{F}^{\mathrm{nc}}.
\end{align}
Note that $\mathcal{F}$ and $\mathcal{F}^{\mathrm{nc}}$ also depend on time, and the external part of the potential $\vec{\phi}(t)$ is the zero vector. 

Using the potential $\vec{\phi}(t)$, we can introduce a pseudo-canonical distribution corresponding to $\vec{\phi}$ ~\cite{yoshimura2023housekeeping} as 
\begin{align}\label{Def_pseudocanonical}
    c^{\mathrm{pcan}}_{\alpha}(\bm{r};t)\coloneqq
    c_{\alpha}(\bm{r};t)\dfrac{\mathrm{e}^{\phi_{\alpha}(\bm{r};t)}}{\Xi_{\alpha}(t)}
\end{align}
Here, the parameter $\Xi_{\alpha}(t)$ can be freely chosen to satisfy the following conditions
\begin{align}\label{conservationlaw_Xi}
    \quad\forall\rho,\quad\sum_{\alpha\in\mathcal{X}}S_{\alpha\rho}\ln \Xi_{\alpha}(t)=0
\end{align}
for $\alpha\in\mathcal{X}$, and $\Xi_\alpha(t)=1$ for $\alpha\in\mathcal{Y}$. 
This can be interpreted as meaning that each $\ln\Xi(s)=(\ln\Xi_\alpha(s))$ at time $s$ is a conservation law. That is, at any time $t$, we have
\begin{align}\label{verify_conservationlaw_Xi}
    &\partial_{t}\int_{V}d\bm{r}\sum_{\alpha\in\mathcal{X}}c_{\alpha}(\bm{r};t)\ln \Xi_{\alpha}(s)\notag\\
    &=\int_{V}d\bm{r}\sum_{\alpha\in\mathcal{X}}\left\{-\bm{\nabla}_{\bm{r}}\cdot\bm{J}_{(\alpha)}+(\nabla_{\mathsf{s}}\bm{j})_{\alpha}\right\}\ln \Xi_{\alpha}(s)\notag\\
    &=\int_{V}d\bm{r}\sum_{\rho\in\mathcal{R}}j_{\rho}\left(\sum_{\alpha\in\mathcal{X}}S_{\alpha\rho}\ln\Xi_{\alpha}(s)\right)=0,
\end{align}
hence, $\int_V d\bm{r}\sum_{\alpha}c_\alpha(\bm{r};t)\ln\Xi_\alpha(s)$ does not depend on $t$. 
Here, we ignore the surface term by using the boundary condition on the current and use $\bm{\nabla}_{\bm{r}}\ln\Xi_{\alpha}(s)=0$ in the second line. 
We can use the parameter $\Xi_{\alpha}(t)$ to let conserved quantities, quantities kept constant over time evolution, have the same value in $\vec{c}$ and $\vec{c}^{\mathrm{pcan}}\coloneqq(c_1^{\mathrm{pcan}},\dots c_N^{\mathrm{pcan}})^{\top}$. We may also choose $\Xi_{\alpha}(t)=1$ simply for all $\alpha\in\mathcal{X}$. When we consider closed ideal solutions, the pseudo-canonical distribution $\vec{c}^{\mathrm{pcan}}$ reduces to $\vec{c}^{\mathrm{eq}}$ using $\vec{\phi}=-(\vec{\mu}^{\mathrm{id}}-\vec{\mu}^{\mathrm{id,eq}})$ and $\Xi_{\alpha}(t)=1$.

Using the pseudo-canonical distribution, we obtain
\begin{align}\label{variation_KL_pcan}
    \frac{\delta}{\delta c_{\alpha}}D_{\mathrm{KL}}(\vec{c}\|\vec{c}^{\mathrm{pcan}})&=\ln\dfrac{c_{\alpha}}{c_{\alpha}^{\mathrm{pcan}}}\notag\\
    &=-\phi_{\alpha}+\ln \Xi_{\alpha},
\end{align}
where we consider $\vec{c}$ and $\vec{c}^{\mathrm{pcan}}$ to be independent, i.e., we ignore $c_{\alpha}$ appearing in the definition of $c_{\alpha}^{\mathrm{pcan}}$ in Eq.~\eqref{Def_pseudocanonical} in the variational calculations. Note that $-\phi_{\alpha}+\ln\Xi_{\alpha}$ is zero for all $\alpha\in\mathcal{Y}$ because $\phi_{\alpha}=0$ and $\Xi_{\alpha}=1$ hold for all $\alpha\in\mathcal{Y}$. Then, we can rewrite the RD equation for the internal species as
\begin{align}\label{RD_relax_form}
    \partial_{t}\vec{c}_{\mathcal{X}}&=\left(\Div\mathcal{M}\Grad\vec{\phi}+\Div\mathcal{M}\mathcal{F}^{\mathrm{nc}}\right)_{\mathcal{X}}\notag\\&=\left(-\Div\mathcal{M}\Grad\frac{\delta}{\delta\vec{c}}D_{\mathrm{KL}}(\vec{c}\|\vec{c}^{\mathrm{pcan}})+\Div\mathcal{M}\mathcal{F}^{\mathrm{nc}}\right)_{\mathcal{X}},
\end{align}
where we use Eq.~\eqref{variation_KL_pcan}, Eq.~\eqref{conservationlaw_Xi}, and $\bm{r}$-independence of $\Xi_{\alpha}(t)$. The rewritten form of the RD equation allows us to regard the conservative force $\Grad\vec{\phi}$ as driving the relaxation to the pseudo-canonical distribution corresponding to $\vec{\phi}$, since the first term in the second line in Eq.~\eqref{RD_relax_form} is written as the gradient flow. This gradient flow is similar to the one for the relaxation to the equilibrium in Eq.~\eqref{gradientflow_to_equilibrium}.

\subsection{The excess entropy production rate and gradient flow}\label{ap:gradientflow_excess}

To obtain the geometric excess/housekeeping EPR, we decompose the force as 
\begin{align}
    \mathcal{F}=\Grad\vec{\phi}{}^{\ast}+(\mathcal{F}-\mathcal{F}^{\ast}),
\end{align}
where $\mathcal{F}^{\ast}=\Grad\vec{\phi}{}^{\ast}$. In the following, $\vec{\phi}{}^{\ast}(t)$ indicates the potential for the excess EPR at time $t$. Using the pseudo-canonical distribution corresponding to $\vec{\phi}{}^{\ast}(t)$, we can rewrite the RD equation for the internal species as the gradient flow,
\begin{align}\label{RD_only_relax_form}
    \partial_{t}\vec{c}_{\mathcal{X}}&=\left(-\Div\mathcal{M}\Grad\frac{\delta}{\delta\vec{c}}D_{\mathrm{KL}}(\vec{c}\|\vec{c}^{\mathrm{pcan}})\right)_{\mathcal{X}},
\end{align}
where we use $[\Div\mathcal{M}(\mathcal{F}-\mathcal{F}^{\ast})]_{\mathcal{X}}=\vecrm{0}_{\mathcal{X}}$. The gradient flow~\eqref{RD_only_relax_form} means that the time evolution of the concentration distribution of the internal species can be written solely in terms of the relaxation to the pseudo-canonical distribution corresponding to $\vec{\phi}{}^{\ast}$ determined independently at each time.

Corresponding to Eq.~\eqref{RD_only_relax_form} that the time evolution of the concentration distribution of the internal species can be described only by the relaxation to different states at different times, the excess EPR has a similar informational geometric form to Eq.~\eqref{EPR_KL_div},
\begin{align}\label{exEPR_KL}
    \sigma^{\mathrm{ex}}=\left.-\partial_{t}D_{\mathrm{KL}}(\vec{c}(t)\|\vec{c}^{\mathrm{pcan}}(s))\right|_{s=t},
\end{align}
where $\vec{c}^{\mathrm{pcan}}$ indicates the pseudo-canonical distribution corresponding to $\vec{\phi}{}^{\ast}$. We can verify the representation of the excess EPR in Eq.~\eqref{exEPR_KL} as
\begin{align}
    &\left.-\partial_{t}D_{\mathrm{KL}}(\vec{c}(t)\|\vec{c}^{\mathrm{pcan}}(s))\right|_{s=t}\notag\\
    &=\int_{V}d\bm{r}\sum_{\alpha\in\mathcal{X}}(\partial_tc_{\alpha})(\phi^{\ast}_{\alpha}-\ln\Xi_{\alpha})\notag\\
    &=\int_{V}d\bm{r}\sum_{\alpha\in\mathcal{X}}(\partial_tc_{\alpha})\phi^{\ast}_{\alpha}\notag\\
    &=\cip{\partial_{t}\vec{c}}{\vec{\phi}{}^{\ast}}=\cip{\Div\mathcal{M}\Grad\vec{\phi}{}^{\ast}}{\vec{\phi}{}^{\ast}}\notag\\
    &=\fip{\mathcal{M}\Grad\vec{\phi}{}^{\ast}}{\Grad\vec{\phi}{}^{\ast}}=\fip{\Grad\vec{\phi}{}^{\ast}}{\Grad\vec{\phi}{}^{\ast}}_{\mathcal{M}}\notag\\
    &=\sigma^{\mathrm{ex}}.
\end{align}
In the calculation, we use Eq.~\eqref{verify_conservationlaw_Xi} in the second line, the assumption $\phi^{\ast}_{\alpha}=0$ for all $\alpha\in\mathcal{Y}$ in the third line, and the condition on $\vec{\phi}{}^{\ast}$~\eqref{EL_eq} in the fourth line.

\section{The derivation of the wavenumber decomposition of the entropy production rate in reaction-diffusion systems}\label{ap:wavenumber_derivation}

Here, we provide the derivation of the wavenumber decomposition of EPR in Eq.~\eqref{RDS_wavenumberdecomp}. For simplicity, we abbreviate the weighted Fourier transform in Eq.~\eqref{weighted_FT} as $\hat{\vec{\bm{F}}}=\int_{V}d\bm{r}\,\dvec{\mathsf{M}}{}^{\frac{1}{2}}\vec{\bm{F}}\mathrm{e}^{-\mathrm{i}\bm{k}\cdot\bm{r}}$ and $\hat{\bm{f}}=\int_{V}d\bm{r}\,\mathsf{m}^{\frac{1}{2}}\bm{f}\mathrm{e}^{-\mathrm{i}\bm{k}\cdot\bm{r}}$. If we do not impose the periodic boundary condition on the system, the Fourier transform of the delta function leads to
\begin{align}
    &\int_{\mathbb{R}^d}d\bm{k}\,\sigma^{\mathrm{wn}}(\bm{k})=\frac{1}{(2\pi)^d}\int_{\mathbb{R}^d}d\bm{k}\left[\hat{\vec{\bm{F}}}{}^{\dag}\hat{\vec{\bm{F}}}+\hat{\bm{f}}{}^{\dag}\hat{\bm{f}}\right]\notag\\
    &=\int_{V\times V}d\bm{r}d\bm{r}'\int_{\mathbb{R}^d}d\bm{k}\frac{\mathrm{e}^{\mathrm{i}\bm{k}\cdot(\bm{r}-\bm{r}')}}{(2\pi)^d}\notag\\
    &\phantom{\int_{V\times V}}\times\left[\vec{\bm{F}}{}^{\top}(\bm{r})\dvec{\mathsf{M}}{}^{\frac{1}{2}}(\bm{r})\dvec{\mathsf{M}}{}^{\frac{1}{2}}(\bm{r}')\vec{\bm{F}}(\bm{r}')\right.\notag\\
    &\left.\phantom{\dvec{\mathsf{M}}{}^{\frac{1}{2}}(\bm{r}')\vec{\bm{F}}(\bm{r}')}+\bm{f}^{\top}(\bm{r})\mathsf{m}^{\frac{1}{2}}(\bm{r})\mathsf{m}^{\frac{1}{2}}(\bm{r}')\bm{f}(\bm{r}')\right]\notag\\
    &=\int_{V\times V}d\bm{r}d\bm{r}'\delta(\bm{r}-\bm{r}')\left[\vec{\bm{F}}{}^{\top}(\bm{r})\dvec{\mathsf{M}}{}^{\frac{1}{2}}(\bm{r})\dvec{\mathsf{M}}{}^{\frac{1}{2}}(\bm{r}')\vec{\bm{F}}(\bm{r}')\right.\notag\\
    &\left.\phantom{\dvec{\mathsf{M}}{}^{\frac{1}{2}}(\bm{r}')\vec{\bm{F}}(\bm{r}')\dvec{\mathsf{M}}{}^{\frac{1}{2}}(\bm{r}')\vec{\bm{F}}}+\bm{f}^{\top}(\bm{r})\mathsf{m}^{\frac{1}{2}}(\bm{r})\mathsf{m}^{\frac{1}{2}}(\bm{r}')\bm{f}(\bm{r}')\right]\notag\\
    &=\int_{V}d\bm{r}\left[\vec{\bm{F}}{}^{\top}\dvec{\mathsf{M}}\vec{\bm{F}}+\bm{f}^{\top}\mathsf{m}\bm{f}\right]=\sigma,
    \label{deriv:RDS_wavenumberdecomp}
\end{align}
where we omit the argument $t$. We also obtain Eq.~\eqref{RDS_wavenumberdecomp_PBC} by replacing $[1/(2\pi)^d]\int_{\mathbb{R}^d}d\bm{k}$ with $[1/|V|]\sum_{\bm{k}}$ and using $\sum_{\bm{k}}\mathrm{e}^{\mathrm{i}\bm{k}\cdot(\bm{r}-\bm{r}')}/|V|=\delta(\bm{r}-\bm{r}')$ in the case of the systems with periodic boundaries.

\section{The details of the Wasserstein geometry}\label{ap:Wasserstein_geo}
In this section, we provide more details on the Wasserstein distances for RDSs. First, we justify the reformulation of the $1$-Wasserstein distance in Eq.~\eqref{another_Def_L1} in Sec.~\ref{ap:reduction_L1}. We also generalize the Kantorovich--Rubinstein duality~\eqref{FP_Kantorovich--Rubinstein} in Sec.~\ref{ap:KR_dual}. Second, we discuss the properties of the optimizer of the $2$-Wasserstein distance in Sec.~\ref{ap:optimizer_L2}. We introduce two reformulations of the $2$-Wasserstein distance in Sec.~\ref{ap:reformulate_L2}. We also confirm that the Wasserstein distances satisfy the axioms of distance in Sec.~\ref{ap:aximos_distance}. We derive the inequality between the Wasserstein distances~\eqref{compare_L1_L2} and the TSL based on the $1$-Wasserstein distance~\eqref{TSL_L1} in Sec.~\ref{ap:deriv_compare_L1_L2} and Sec.~\ref{ap:deriv_L1_TSL}. \add{We provide details of the minimum dissipation determined by the $1$-Wasserstein distance in Sec.~\ref{ap:minimum dissipation 1-Wasserstein}.}
Finally, we see the property of the $1$-Wasserstein distance for the Fisher--KPP equation in Sec~\ref{ap:L1_FisherKPP}. 

In the following, we often use $\vec{c}(0)$ and $\vec{c}(\tau)$ instead of $\vec{c}^{A}$ and $\vec{c}^{B}$ since we need to solve the optimization problem for time series $\vec{c}=\{\vec{c}(t)\}_{t\in[0,\tau]}$ such that $\vec{c}^{A}_{\mathcal{X}}=\vec{c}_{\mathcal{X}}(0)$ and $\vec{c}^{B}_{\mathcal{X}}=\vec{c}_{\mathcal{X}}(\tau)$ hold from the constraints.

\subsection{Reduction of computational complexity of the $1$-Wasserstein distance}\label{ap:reduction_L1}
Here, we derive the reduced form of the $1$-Wasserstein distance~\eqref{another_Def_L1} from the original definition~\eqref{Def_L1}. In the following, $\mathcal{J}^{\diamond}=\qty(\vec{\bm{J}}{}^{\diamond},\bm{j}^{\diamond})$ denotes an optimizer of Eq.~\eqref{Def_L1}. Letting $\mathcal{U}^{\diamond}=\qty(\vec{\bm{U}}{}^{\diamond},\bm{u}^{\diamond})$ be the optimizer of the right-hand side in Eq.~\eqref{another_Def_L1}, the following inequality holds,
\begin{align*}\label{first_inequality_for_another_Def_L1}
    \inf_{\U}|\U|_{\RD}&=|\U^{\diamond}|_{\RD}\\
    &=\int_{0}^{\tau}dt\int_{V}d\bm{r}\qty[\sum_{\alpha\in\mathcal{S}}\left\|\frac{\bm{U}^{\diamond}_{(\alpha)}}{\tau}\right\|+\sum_{\rho\in\mathcal{R}}\left|\frac{u^{\diamond}_{\rho}}{\tau}\right|]\\
    &=\int_{0}^{\tau}dt\left|\frac{\U^{\diamond}}{\tau}\right|_{\RD}\\
    &\geq\inf_{\J'}\int_{0}^{\tau}dt\left|\J'\right|_{\RD}\\&={W}_{1,\mathcal{X}}(\vec{c}(0),\vec{c}(\tau)),
\stepcounter{equation}\tag{\theequation}
\end{align*}
where we use that $\U^{\diamond}/\tau$ satisfies the conditions imposed on the minimization problem~\eqref{Def_L1} because we can rewrite the condition in Eq.~\eqref{another_condition_L1} to $\partial_t\vec{c}_{\mathcal{X}}=\qty[\Div\qty(\U^{\diamond}/\tau)]_{\mathcal{X}}$. On the other hand, we can obtain the inequality in the opposite direction by using $\J^{\diamond}$ as
\begin{align*}\label{second_inequality_for_another_Def_L1}
&{W}_{1,\mathcal{X}}(\vec{c}(0),\vec{c}(\tau))=\int_{0}^{\tau}dt\left|\J^{\diamond}\right|_{\RD}\\
&=\int_{0}^{\tau}dt\int_{V}d\bm{r}\qty[\sum_{\alpha\in\mathcal{S}}\left\|\bm{J}^{\diamond}_{(\alpha)}\right\|+\sum_{\rho\in\mathcal{R}}\left|j^{\diamond}_{\rho}\right|]
\\
&\geq\int_{V}d\bm{r}\qty[\sum_{\alpha\in\mathcal{S}}\left\|\int_{0}^{\tau}dt\,\bm{J}^{\diamond}_{(\alpha)}\right\|+\sum_{\rho\in\mathcal{R}}\left|\int_{0}^{\tau}dt \,j^{\diamond}_{\rho}\right|]\\
&=\left|\int_{0}^{\tau}dt\,\J^{\diamond}\right|_{\RD}\geq\inf_{\U}\left|\U\right|_{\RD},
\stepcounter{equation}\tag{\theequation}
\end{align*}
where we define time-integrated current as
\begin{align*}
    \int_{0}^{\tau}dt\,\J^{\diamond}\coloneqq\qty(\int_{0}^{\tau}dt\,\vec{\bm{J}}{}^{\diamond},\int_{0}^{\tau}dt\,\bm{j}^{\diamond}),
\end{align*}\label{time_integrated_current}
with 
\begin{align*}
\begin{cases}
    \qty(\displaystyle\int_{0}^{\tau}dt\,\vec{\bm{J}}{}^{\diamond})_{(\alpha),i}\coloneqq\displaystyle\int_{0}^{\tau}dt\, J_{(\alpha)i}^{\diamond}\\
    \qty(\displaystyle\int_{0}^{\tau}dt\, \bm{j}^{\diamond})_{\rho}\coloneqq\displaystyle\int_{0}^{\tau}dt \,j_{\rho}
\end{cases}.
\end{align*}
We also use that time integrated current~\eqref{time_integrated_current} satisfies the condition in Eq.~\eqref{another_condition_L1} to obtain Eq.~\eqref{another_Def_L1}. Unifying the two inequalities, Eq.~\eqref{first_inequality_for_another_Def_L1} and Eq.~\eqref{second_inequality_for_another_Def_L1}, leads to the new form of ${W}_{1,\mathcal{X}}$~\eqref{another_Def_L1}. Such a rewrite means that the optimizer of Eq.~\eqref{Def_L1} is not unique because we can take
\begin{align}
    \bm{J}_{(\alpha)}^{\diamond}(t)=\Theta_{\alpha}(t)\bm{U}_{(\alpha)}^{\diamond},\;j_{\rho}^{\diamond}(t)=\vartheta_{\rho}(t)u_{\rho}^{\diamond}
\end{align}
as the optimal current of Eq.~\eqref{Def_L1} with the arbitrary nonnegative-valued functions $\Theta_{\alpha}(t)$ and $\vartheta_{\rho}(t)$ which satisfy
\begin{align}
    \int_{0}^{\tau}dt\,\Theta_{\alpha}(t)=1,\;\int_{0}^{\tau}dt\,\vartheta_{\rho}(t)=1.
\end{align}

\add{
The optimization problem in Eq.~\eqref{another_Def_L1} provides a geometric interpretation of the $1$-Wasserstein distance. To confirm it, we rewrite the reduced form of the $1$-Wasserstein distance~\eqref{another_Def_L1} with the original time series of current $\J$, which satisfies $\vec{c}_{\mathcal{X}}(\tau)-\vec{c}_{\mathcal{X}}(0)=(\Div\int_{0}^{\tau}dt\,\J)_{\mathcal{X}}$, as follows:
\begin{align}
    {W}_{1,\mathcal{X}}(\vec{c}(0),\vec{c}(\tau))&=\inf_{\U}|\U|_{\RD}\notag\\
    &=\inf_{\U}\left|\int_{0}^{\tau}dt\,\J-\left(\int_{0}^{\tau}dt\,\J-\U\right)\right|_{\RD}\notag\\
    &=\inf_{\U^{\mathrm{cyc}}}\left|\int_{0}^{\tau}dt\,\J-\U^{\mathrm{cyc}}\right|_{\RD},
    \label{1-W as projection}
\end{align}
where infimum at the third line is taken among all $\U^{\mathrm{cyc}}$ satisfying the condition $\left(\Div\U^{\mathrm{cyc}}\right)_{\mathcal{X}}=\vecrm{0}_{\mathcal{X}}$, and $\U$ at the first and second lines satisfies Eq.~(\ref{another_condition_L1}).
Here, we use the identity $|\U|_{\RD}=|\int_{0}^{\tau}dt\,\J-(\int_{0}^{\tau}dt\,\J-\U)|_{\RD}$ in the first transform, and let $\U^{\mathrm{cyc}}$ denote $\int_{0}^{\tau}dt\,\J-\U$. Then, the condition on $\U$, namely, $\vec{c}_{\mathcal{X}}(\tau)-\vec{c}_{\mathcal{X}}(0)=(\Div\U)_{\mathcal{X}}$, yields $(\Div\U^{\mathrm{cyc}})_{\mathcal{X}}=(\Div[\int_{0}^{\tau}dt\,\J-\U])_{\mathcal{X}}=\vecrm{0}_{\mathcal{X}}$. This implies that the optimization in the last line in Eq.~\eqref{1-W as projection} is performed on all currents that are cyclic in the sense explained in Sec.~\ref{sec:ex&hk}, i.e., the candidate $\U^{\mathrm{cyc}}$ does not affect the time evolution of the internal species. Obviously, the optimizer of Eq.~\eqref{1-W as projection} is the projection of the original time-integrated current $\int_{0}^{\tau}dt\,\J$ onto the space of cyclic (time-integrated) currents with respect to the norm $|\cdot|_{\RD}$. It indicates that the $1$-Wasserstein distance is the distance between the original time-integrated current and the space of cyclic (time-integrated) currents measured with $|\cdot|_{\RD}$.
}

In the following, we consider the characteristics of the optimizer $\U^{\diamond}$. We can rewrite the reduced form of the $1$-Wasserstein distance~\eqref{another_Def_L1} with Lagrange multiplier $\vec{\phi}$, whose external part is the zero vector, as 
\begin{align}\label{L1_Lagrange_maltiplier}
    {W}_{1,\mathcal{X}}(\vec{c}(0),\vec{c}(\tau))=\inf_{\U}\sup_{\vec{\phi}|\vec{\phi}_{\mathcal{Y}}=\vecrm{0}_{\mathcal{Y}}}\mathcal{I}_{1,\mathcal{X}}\qty[\U,\vec{\phi}],
\end{align}
where the functional $\mathcal{I}_{1,\mathcal{X}}\qty[\U,\vec{\phi}]$ is defined as
\begin{align}\label{L1_functional}
    \mathcal{I}_{1,\mathcal{X}}\qty[\U,\vec{\phi}]\coloneqq\left|\U\right|_{\RD}+\cip{\vec{\phi}}{\vec{c}(\tau)-\vec{c}(0)-\Div\U}.
\end{align}
Here, we consider the supremum over the Lagrange multiplier $\vec{\phi}$ under the condition $\vec{\phi}_{\mathcal{Y}}=\vecrm{0}_{\mathcal{Y}}$  because the term $\cip{\vec{\phi}}{\vec{c}(\tau)-\vec{c}(0)-\Div\U}$ only gives a contribution for the constraint on internal species~\eqref{another_condition_L1} when $\vec{\phi}_{\mathcal{Y}}=\vecrm{0}_{\mathcal{Y}}$.
The boundary condition imposed on $\U$ lets us transform Eq.~\eqref{L1_functional} by partial integration to
\begin{align}\label{L1_functional_PI}
    \mathcal{I}_{1,\mathcal{X}}\qty[\U,\vec{\phi}]=\left|\U\right|_{\RD}+\cip{\vec{\phi}}{\vec{c}(\tau)-\vec{c}(0)}-\fip{\Grad\vec{\phi}}{\U}.
\end{align}
Then, its functional derivative and $\vec{\phi}_{\mathcal{Y}}=\vecrm{0}_{\mathcal{Y}}$ lead to
\begin{align} \label{condition_1W_optimizer_Y}
\bm{U}^{\diamond}_{(\alpha)}=\bm{0},\;u^{\diamond}_{\rho}=0 
\end{align}
for all $\alpha\in\mathcal{Y}$ and $\rho\in\mathcal{R}\setminus\mathcal{R}_{\mathcal{X}}$, and
\begin{align}
\bm{U}^{\diamond}_{(\alpha)}&=\|\bm{U}^{\diamond}_{(\alpha)}\|\bm{\nabla}_{\bm{r}}\phi^{\diamond}_{\alpha}, \notag\\
u^{\diamond}_{\rho}&=\left|u^{\diamond}_{\rho}\right|\qty(\nabla_{\mathsf{s}}\vec{\phi}{}^{\diamond})_{\rho}, \label{condition_1W_optimizer_X}
\end{align}
for all $\alpha\in\mathcal{X}$ and $\rho\in\mathcal{R}_{\mathcal{X}}$. Thus,  for all $\alpha\in\mathcal{X}$, the optimal potential $\vec{\phi}{}^{\diamond}$ satisfies $\left\|\bm{\nabla}_{\bm{r}}\phi^{\diamond}_{\alpha}\right\|=1$ unless $\bm{U}_{(\alpha)}^{\diamond}=\bm{0}$. Similarly, for all $\rho\in\mathcal{R}_{\mathcal{X}}$, $|(\nabla_{\mathsf{s}}\vec{\phi}{}^{\diamond})_{\rho}|=1$ holds unless $u_{\rho}^{\diamond}=0$. Note that these conditions indicate that the gradient of a potential determines the direction of the optimal current. In addition, these conditions lead to a new expression,
\begin{align}\label{L1_optimal_form}
    &{W}_{1,\mathcal{X}}(\vec{c}(0),\vec{c}(\tau))= \left|\U^{\diamond}\right|_{\RD}\notag \\    &=\int_{V}d\bm{r}\qty[\sum_{\alpha\in\mathcal{X}}\left\|\bm{U}_{(\alpha)}^{\diamond}\right\|+\sum_{\rho\in\mathcal{R}_{\mathcal{X}}}\left|u^{\diamond}_{\rho}\right|]\notag\\    &=\int_{V}d\bm{r}\qty[\sum_{\alpha\in\mathcal{X}}\bm{\nabla}_{\bm{r}}\phi_{\alpha}^{\diamond}\cdot\bm{U}_{(\alpha)}^{\diamond}+\sum_{\rho\in\mathcal{R}_{\mathcal{X}}}\qty(\nabla_{\mathsf{s}}\vec{\phi}{}^{\diamond})_{\rho}u^{\diamond}_{\rho}].
\end{align}
This expression shows that the value of $\bm{\nabla}_{\bm{r}}\phi_{\alpha}(\bm{r})$ does not affect ${W}_{1,\mathcal{X}}(\vec{c}(0),\vec{c}(\tau))$ if $\bm{U}^{\diamond}_{(\alpha)}(\bm{r};t)=\bm{0}$ at the position $\bm{r}$. The value $[\nabla_{\mathsf{s}}\vec{\phi}{}^{\diamond}(\bm{r})]_{\rho}$ also does not matter where $u^{\diamond}_{\rho}(\bm{r};t)=0$. 

 \subsection{Kantorovich--Rubinstein duality of the $1$-Wasserstein distance}\label{ap:KR_dual} 
\add{We here verify the Kantorovich--Rubinstein duality,
\begin{align}
    \cip{\vec{\phi}{}^{\bullet}}{\vec{c}(\tau)-\vec{c}(0)}=\left|\U^{\diamond}\right|_{\RD},
    \label{equivalenceKR}
\end{align}
where $\mathcal{U}^{\diamond}=\qty(\vec{\bm{U}}{}^{\diamond},\bm{u}^{\diamond})$ and $\vec{\phi}{}^{\bullet}$ denote the optimizer of Eq.~\eqref{another_Def_L1} and Eq.~\eqref{RDS_Kantorovich--Rubinstein}, respectively. To obtain Eq.~\eqref{equivalenceKR}, we show the inequalities
\begin{align}
    \cip{\vec{\phi}{}^{\bullet}}{\vec{c}(\tau)-\vec{c}(0)}\leq\left|\U^{\diamond}\right|_{\RD},
    \label{KR_ineq_1}
\end{align}
and
\begin{align}
    \cip{\vec{\phi}{}^{\bullet}}{\vec{c}(\tau)-\vec{c}(0)}\geq\left|\U^{\diamond}\right|_{\RD}.
    \label{KR_ineq_2}
\end{align}
}

\add{The first inequality is easily obtained as,
\begin{align}
    &\cip{\vec{\phi}{}^{\bullet}}{\vec{c}(\tau)-\vec{c}(0)}=\cip{\vec{\phi}{}^{\bullet}}{\Div\mathcal{U}^{\diamond}}=\fip{\Grad\vec{\phi}{}^{\bullet}}{\mathcal{U}^{\diamond}}\notag\\
    &=\int_{V}d\bm{r}\qty[\sum_{\alpha\in\mathcal{X}}\bm{\nabla}_{\bm{r}}\phi_{\alpha}^{\bullet}\cdot\bm{U}_{(\alpha)}^{\diamond}+\sum_{\rho\in\mathcal{R}_{\mathcal{X}}}\qty(\nabla_{\mathsf{s}}\vec{\phi}{}^{\bullet})_{\rho}u^{\diamond}_{\rho}]\notag\\
    &\leq\int_{V}d\bm{r}\qty[\sum_{\alpha\in\mathcal{X}}\left\|\bm{U}_{(\alpha)}^{\diamond}\right\|+\sum_{\rho\in\mathcal{R}_{\mathcal{X}}}\left|u^{\diamond}_{\rho}\right|]=\left|\U^{\diamond}\right|_{\RD},
\end{align}
where we use the generalized $1$-Lipschitz continuity~\eqref{RDS_1_Lip}, $\|\bm{\nabla}_{\bm{r}}\phi^{\bullet}_{\alpha}\|\leq1$ for $\alpha\in\mathcal{X}$ and $\left|(\nabla_{\mathsf{s}}\vec{\phi}^{\bullet})_{\rho}\right|\leq1$ for $\rho\in\mathcal{R}_{\mathcal{X}}$, and the conditions on $\U^{\diamond}$ in Eq.~\eqref{condition_1W_optimizer_Y}.
}

\add{To verify the second inequality, we consider the optimization problem
\begin{align}
    \sup_{\vec{\psi}}\cip{\vec{\psi}}{\vec{c}(\tau)-\vec{c}(0)},
    \label{severeKR}
\end{align}
with the following conditions: $\vec{\psi}\in\mathrm{Lip}^1_{\mathcal{X}}$ and the diffusion part of $\Grad\vec{\psi}$ satisfies the same boundary conditions as the one imposed on $\vec{\bm{U}}{}^{\diamond}$.
Using another representation of the conditions for the generalized $1$-Lipschitz continuity~\eqref{RDS_1_Lip},
\begin{align}
    \begin{cases}
        \forall\alpha\in\mathcal{X},\;\left\|\bm{\nabla}_{\bm{r}}\psi_{\alpha}\right\|^2-1\leq0\\
        \forall\rho\in\mathcal{R}_{\mathcal{X}},\;\left|\left(\nabla_{\mathsf{s}}\vec{\psi}\right)_{\rho}\right|^2-1\leq0
    \end{cases},
\end{align}
we can rewrite the optimization problem~\eqref{severeKR} with the Lagrange multipliers $\{U^{\rm L}_{\alpha}\geq0\}_{\alpha\in\mathcal{X}}$ and $\{u^{\rm L}_{\rho}\geq0\}_{\rho\in\mathcal{R}_{\mathcal{X}}}$ as
\begin{align}
    \sup_{\vec{\psi}}\inf_{\substack{\{U^{\rm L}_{\alpha}\geq0\}_{\alpha\in\mathcal{X}},\\\{u^{\rm L}_{\rho}\geq0\}_{\rho\in\mathcal{R}_{\mathcal{X}}}}}\mathcal{I}_{\mathrm{KR}}\left[\vec{\psi},\{U^{\rm L}_{\alpha}\}_{\alpha\in\mathcal{X}},\{u^{\rm L}_{\rho}\}_{\rho\in\mathcal{R}_{\mathcal{X}}}\right],
    \label{LM_severeKR}
\end{align}
with the following conditions: $\vec{\psi}_{Y}=\vecrm{0}_{Y}$ and the diffusion part of $\Grad\vec{\psi}$ satisfies the same boundary conditions as the one imposed on $\vec{\bm{U}}{}^{\diamond}$. Here, the functional $\mathcal{I}_{\mathrm{KR}}$ is defined as
\begin{align}
    \mathcal{I}_{\mathrm{KR}}\coloneqq&\cip{\vec{\psi}}{\vec{c}(\tau)-\vec{c}(0)}\notag\\&-\frac{1}{2}\int_{V}d\bm{r}\sum_{\alpha\in\mathcal{X}}U^{\rm L}_{\alpha}\left(\left\|\bm{\nabla}_{\bm{r}}\psi_{\alpha}\right\|^2-1\right)\notag\\&-\frac{1}{2}\int_{V}d\bm{r}\sum_{\rho\in\mathcal{R}_{\mathcal{X}}}u^{\rm L}_{\rho}\left(\left|\left(\nabla_{\mathsf{s}}\vec{\psi}\right)_{\rho}\right|^2-1\right).
    \label{functional_severe_KR}
\end{align}
Computing the functional derivatives of Eq.~\eqref{functional_severe_KR}, we can verify that the optimizer of Eq.~\eqref{LM_severeKR}, $(\vec{\psi}{}^{\bullet},\{U_{\alpha}^{{\rm L}\bullet}\}_{\alpha\in\mathcal{X}},\{u^{{\rm L}\bullet}_{\rho}\}_{\rho\in\mathcal{R}_\mathcal{X}})$, satisfies
\begin{align}
&c_{\alpha}(\tau)-c_{\alpha}(0) \nonumber\\
&=-\bm{\nabla}_{\bm{r}}\cdot\left(U_{\alpha}^{{\rm L}\bullet}\bm{\nabla}_{\bm{r}}\psi^{\bullet}_{\alpha}\right)+\sum_{\rho\in\mathcal{R}_{\mathcal{X}}}S_{\alpha\rho}u_{\rho}^{{\rm L}\bullet}\left(\nabla_{s}\vec{\psi}{}^{\bullet}\right)_{\rho}
    ,\label{EL_severeKR}
\end{align}
for all $\alpha\in\mathcal{X}$ as the Euler--Lagrange equation and 
\begin{align}
    \begin{cases}
        \forall\alpha\in\mathcal{X},\;U_{\alpha}^{{\rm L}\bullet}\left(\left\|\bm{\nabla}_{\bm{r}}\psi^{\bullet}_{\alpha}\right\|^2-1\right)=0\\
        \forall\rho\in\mathcal{R}_{\mathcal{X}},\;u^{{\rm L}\bullet}_{\rho}\left(\left|\left(\nabla_{\mathsf{s}}\vec{\psi}{}^{\bullet}\right)_{\rho}\right|^2-1\right)=0
    \end{cases},\label{complementary slackness}
\end{align}
as the complementary slackness conditions. These conditions lead to
\begin{align}
    &\cip{\vec{\psi}{}^{\bullet}}{\vec{c}(\tau)-\vec{c}(0)}\notag\\&=\frac{1}{2}\int_{V}d\bm{r}\sum_{\alpha\in\mathcal{X}}U^{{\rm L}\bullet}_{\alpha}(\|\bm{\nabla}_{\bm{r}}\psi^{\bullet}_{\alpha}\|^2+1)\notag\\&\phantom{=}\;+\frac{1}{2}\int_{V}d\bm{r}\sum_{\rho\in\mathcal{R}_{\mathcal{X}}}u^{{\rm L}\bullet}_{\rho}\left(\left|\left(\nabla_{\mathsf{s}}\vec{\psi}{}^{\bullet}\right)_{\rho}\right|^2+1\right)\notag\\
    &=\int_{V}d\bm{r}\left[\sum_{\alpha\in\mathcal{X}}U^{{\rm L}\bullet}_{\alpha}+\sum_{\rho\in\mathcal{R}_{\mathcal{X}}}u_{\rho}^{{\rm L}\bullet}\right],
    \label{severKR_optimizer}
\end{align}
where we substitute the Euler--Lagrange equation~\eqref{EL_severeKR} into Eq.~\eqref{LM_severeKR} in the first transform and use the complementary slackness conditions~\eqref{complementary slackness} in the second transform. The nonnegativity of the Lagrange multipliers and the complementary slackness conditions~\eqref{complementary slackness} allow us to rewrite the right-hand side of Eq.~\eqref{severKR_optimizer} as $\left|\U^{\bullet}\right|_{\RD}$, with a new time-integrated current $\U^{\bullet}=\qty(\vec{\bm{U}}{}^{\bullet},\bm{u}^{\bullet})$ defined as
\begin{align}
    \bm{U}^{\bullet}_{(\alpha)}&\coloneqq
    \begin{cases}
        U^{{\rm L}\bullet}_{\alpha}\bm{\nabla}_{\bm{r}}\psi^{\bullet}_{\alpha}&(\alpha\in\mathcal{X})\\
        \bm{0}&(\alpha\in\mathcal{Y})
    \end{cases},\notag\\
    \qty(\bm{u}^{\bullet})_{\rho}&\coloneqq
    \begin{cases}
        u^{{\rm L}\bullet}_{\rho}\left(\nabla_{\mathsf{s}}\vec{\psi}{}^{\bullet}\right)_{\rho}&(\rho\in\mathcal{R}_{\mathcal{X}})\\
        0&(\rho\in\mathcal{R}\setminus\mathcal{R}_{\mathcal{X}})
    \end{cases}.
    \label{def Ubullet}
\end{align}
Thus, we obtain
\begin{align}
    \cip{\vec{\psi}{}^{\bullet}}{\vec{c}(\tau)-\vec{c}(0)}=|\U^{\bullet}|_{\RD}\geq|\U^{\diamond}|_{\RD},
    \label{pre_ineq2}
\end{align}
where we use the fact that $\U^{\bullet}$ becomes a candidate of the optimization problem in Eq.~\eqref{another_Def_L1}. This is because the conditions on the diffusion part of $\Grad\vec{\psi}{}^{\bullet}$ let $\U^{\bullet}$ satisfy the boundary conditions imposed in the optimization problem in Eq.~\eqref{another_Def_L1}, and the Euler--Lagrange equation~\eqref{EL_severeKR} leads to $\vec{c}_{\mathcal{X}}(\tau)-\vec{c}_{\mathcal{X}}(0)=\left(\Div\U^{\bullet}\right)_{\mathcal{X}}$. 
Finally, we can verify the second inequality~\eqref{KR_ineq_2} by combining this inequality~\eqref{pre_ineq2} and the consequence of adding the conditions in the optimization 
in Eq.~\eqref{severeKR},
\begin{align}
    \cip{\vec{\phi}{}^{\bullet}}{\vec{c}(\tau)-\vec{c}(0)}\geq\cip{\vec{\psi}{}^{\bullet}}{\vec{c}(\tau)-\vec{c}(0)}.
\end{align}
We remark that the equality~\eqref{equivalenceKR} implies that $\U^{\bullet}$ is also an optimizer for the $1$-Wasserstein distance~\eqref{another_Def_L1}.}

\subsection{The optimizer of the $2$-Wasserstein distance}\label{ap:optimizer_L2}
We can rewrite the optimization problem in Eq.~\eqref{Def_L2} using Lagrange multiplier $\vec{\phi}$ as
\begin{align}\label{rewrite_L2}
    {W}_{2,\mathcal{X}}&(\vec{c}(0),\vec{c}(\tau)|\vec{b}_{\mathcal{Y}})^2=\inf_{\vec{c},\mathcal{F}}\sup_{\vec{\phi}|\vec{\phi}_{\mathcal{Y}}=\vecrm{0}_{\mathcal{Y}}}\tau\mathcal{I}_{2,\mathcal{X}}\qty[\vec{c},\mathcal{F},\vec{\phi}],
\end{align}
where $\mathcal{I}_{2,\mathcal{X}}$ is the functional defined as
\begin{align*}\label{functional_L2}
&\mathcal{I}_{2,\mathcal{X}}\qty[\vec{c},\mathcal{F},\vec{\phi}]\notag\\&\phantom{\mathcal{I}_{2,\mathcal{X}}}\coloneqq\int_{0}^{\tau}dt\qty[\fip{\mathcal{F}}{\mathcal{F}}_{\mathcal{M}_{\vec{c}}}+2\cip{\vec{\phi}}{\partial_{t}\vec{c}-\Div\mathcal{M}_{\vec{c}}\mathcal{F}}].
\stepcounter{equation}\tag{\theequation}
\end{align*}
Here, we consider the supremum over the Lagrange multiplier $\vec{\phi}$ under the condition $\vec{\phi}_{\mathcal{Y}}=\vecrm{0}_{\mathcal{Y}}$  because the term $\cip{\vec{\phi}}{\partial_{t}\vec{c}-\Div\mathcal{M}_{\vec{c}}\mathcal{F}}$ only gives a contribution for the constraint on internal species Eq.~\eqref{constraints_L2} when $\vec{\phi}_{\mathcal{Y}}=\vecrm{0}_{\mathcal{Y}}$.
Because we fix $\vec{c}(0)$ and $\vec{c}(\tau)$, impose boundary conditions on $\mathsf{M}\vec{\bm{F}}$ for internal species, and let $\vec{\phi}$ satisfy $\phi_{\alpha}=0$ for all $\alpha\in\mathcal{Y}$, partial integration yields
\begin{align*}\label{partial_integral_L2}
    &\mathcal{I}_{2,\mathcal{X}}\qty[\vec{c},\mathcal{F},\vec{\phi}] \notag \\
    &=2\cip{\vec{\phi}(\tau)}{\vec{c}(\tau)}-2\cip{\vec{\phi}(0)}{\vec{c}(0)}\\
    &\quad+\int_{0}^{\tau}dt\qty[\fip{\mathcal{F}-2\Grad\vec{\phi}}{\mathcal{F}}_{\mathcal{M}_{\vec{c}}}-2\cip{\partial_{t}\vec{\phi}}{\vec{c}}].
\stepcounter{equation}\tag{\theequation}
\end{align*}
In the following, we write the optimizer of the right-hand side of Eq.~\eqref{rewrite_L2} as $\qty(\vec{c}^{\star},\mathcal{F}^{\star},\vec{\phi}{}^{\star})$. As in case in Appendix~\ref{ap:optimal_condition}, the functional derivative of Eq.~\eqref{partial_integral_L2} leads to the conditions to be satisfied by $\qty(\vec{c}^{\star},\mathcal{F}^{\star},\vec{\phi}{}^{\star})$, 
\begin{align}
    \sum_{\beta\in\mathcal{S}}\mathsf{M}^{\star}_{(\alpha\beta)}\qty(\bm{F}^{\star}_{(\beta)}-\bm{\nabla}_{\bm{r}}\phi^{\star}_{\beta})=0
\end{align}
for all $\alpha\in\mathcal{S}$, and
\begin{align}
    m^{\star}_{\rho}\qty{f^{\star}_{\rho}-\qty(\nabla_{\mathsf{s}}\vec{\phi}{}^{\star})_{\rho}}=0
\end{align}
for all $\rho\in\mathcal{R}$, where $\mathcal{F}^{\star}=(\vec{\bm{F}}{}^{\star},\bm{f}^{\star})$, and $\mathsf{M}_{(\alpha\beta)}^{\star}=[\dvec{\mathsf{M}}{}^{\star}]_{(\alpha\beta)}$ and $m^{\star}_{\rho}=[\mathsf{m}^{\star}]_{\rho\rho}$ are given by $
\mathcal{M}_{\vec{c}^{\star}} =
    \dvec{\mathsf{M}}{}^{\star} \oplus \mathsf{m}^{\star}$.
$\dvec{\mathsf{M}}{}^{\star}$ and $m^{\star}$ indicate the mobility tensor and the edgewise Onsager coefficient matrix for the optimal concentration distribution $\vec{c}^{\star}$. These results and positive-definiteness of $\dvec{\mathsf{M}}{}^{\star}$ and $\mathsf{m}^{\star}$ make the optimal force be the gradient of potential,
\begin{align}\label{potential_condition_L2}
    \mathcal{F}^{\star}=\Grad\vec{\phi}{}^{\star}.
\end{align}
This condition means that in order to minimize EP, we should drive the system by the conservative thermodynamic force corresponding to the potential $\vec{\phi}{}^{\star}$.

\subsection{Reformulations of the $2$-Wasserstein distance}\label{ap:reformulate_L2}
Here, we introduce two reformulations of the $2$-Wasserstein distance. The condition for the optimal force in Eq.~\eqref{potential_condition_L2} let us rewrite the $2$-Wasserstein distance as the form in Eq.~\eqref{another_Def_L2},
\begin{align}\label{potential_form_L2}
    &{W}_{2,\mathcal{X}}(\vec{c}(0),\vec{c}(\tau)|\vec{b}_{\mathcal{Y}})^2\notag\\
    &=\inf_{\vec{c},\vec{\phi}|\vec{\phi}_{\mathcal{Y}}=\vecrm{0}_{\mathcal{Y}}}\tau\int_{0}^{\tau}dt\fip{\Grad\vec{\phi}}{\Grad\vec{\phi}}_{\mathcal{M}_{\vec{c}}},
\end{align}
with the conditions
\begin{align}\label{condition_potential_form_L2}
    \partial_{t}\vec{c}_{\mathcal{X}}=\qty(\Div\mathcal{M}_{\vec{c}}\Grad\vec{\phi})_{\mathcal{X}},\;\vec{c}_{\mathcal{Y}}(t)=\vec{b}_{\mathcal{Y}}.
\end{align}

We can also rewrite the $2$-Wasserstein distance as
\begin{align}\label{constant_speed_L2}
    &{W}_{2,\mathcal{X}}(\vec{c}(0),\vec{c}(\tau)|\vec{b}_{\mathcal{Y}})\notag\\&=\inf_{\vec{c},\vec{\phi}|\vec{\phi}_{\mathcal{Y}}=\vecrm{0}_{\mathcal{Y}}}\int_{0}^{\tau}dt\sqrt{\fip{\Grad\vec{\phi}}{\Grad\vec{\phi}}_{\mathcal{M}_{\vec{c}}}},
\end{align}
with the same conditions as Eq.~\eqref{condition_potential_form_L2}. 

To prove the equivalence between Eqs.~\eqref{potential_form_L2} and \eqref{constant_speed_L2}, we consider the optimizer of  Eq.~\eqref{constant_speed_L2}. Let $(\vec{c}^{\sharp},\vec{\phi}{}^{\sharp})$ denote the optimizer of Eq.~\eqref{constant_speed_L2}. For the derivation of Eq.~\eqref{constant_speed_L2}, it is sufficient to confirm
\begin{align}\label{W2_geq_cspeed}
    {W}_{2,\mathcal{X}}(\vec{c}(0),\vec{c}(\tau)|\vec{b}_{\mathcal{Y}})^2\geq\qty(\int_{0}^{\tau}dt\sqrt{\fip{\Grad\vec{\phi}{}^{\sharp}}{\Grad\vec{\phi}{}^{\sharp}}_{\mathcal{M}_{\vec{c}^{\sharp}}}})^2,
\end{align}
and
\begin{align}\label{W2_leq_cspeed}
    {W}_{2,\mathcal{X}}(\vec{c}(0),\vec{c}(\tau)|\vec{b}_{\mathcal{Y}})^2\leq\qty(\int_{0}^{\tau}dt\sqrt{\fip{\Grad\vec{\phi}{}^{\sharp}}{\Grad\vec{\phi}{}^{\sharp}}_{\mathcal{M}_{\vec{c}^{\sharp}}}})^2.
\end{align}

We can easily show the first inequality~\eqref{W2_geq_cspeed}
\begin{align}\label{CS_for_L2}
    &{W}_{2,\mathcal{X}}(\vec{c}(0),\vec{c}(\tau)|\vec{b}_{\mathcal{Y}})^2=\tau\int_{0}^{\tau}dt\fip{\Grad\vec{\phi}{}^{\star}}{\Grad\vec{\phi}{}^{\star}}_{\mathcal{M}_{\vec{c}^{\star}}}\notag\\
    &=\qty(\int_{0}^{\tau}dt)\qty(\int_{0}^{\tau}dt\fip{\Grad\vec{\phi}{}^{\star}}{\Grad\vec{\phi}{}^{\star}}_{\mathcal{M}_{\vec{c}^{\star}}})\notag\\
    &\geq\qty(\int_{0}^{\tau}dt\sqrt{\fip{\Grad\vec{\phi}{}^{\star}}{\Grad\vec{\phi}{}^{\star}}_{\mathcal{M}_{\vec{c}^{\star}}}})^2\notag\\
    &\geq\qty(\int_{0}^{\tau}dt\sqrt{\fip{\Grad\vec{\phi}{}^{\sharp}}{\Grad\vec{\phi}{}^{\sharp}}_{\mathcal{M}_{\vec{c}^{\sharp}}}})^2,
\end{align}
where we used the Cauchy--Schwarz inequality and the fact that $(\vec{c}^{\sharp},\vec{\phi}{}^{\sharp})$ denote the optimizer of Eq.~\eqref{constant_speed_L2}.

To derive the second inequality~\eqref{W2_leq_cspeed}, we use arc-length reparametrization by referring to the literatures~\cite{ambrosio2005gradient, dolbeault2009new}. Introducing a function $s_{\epsilon}(t)$ for $t \in [0, \tau]$ with sufficiently small $\epsilon>0$ as
\begin{align}
    s_{\epsilon}(t)\coloneqq\int_{0}^{t}dt'\sqrt{\epsilon+\fip{\Grad\vec{\phi}{}^{\sharp}(t')}{\Grad\vec{\phi}{}^{\sharp}(t')}_{\mathcal{M}_{\vec{c}^{\sharp}(t')}}},
\end{align}
we can define the inverse function $t_{\epsilon}\coloneqq s^{-1}_{\epsilon}$ because $d_t s_{\epsilon}(t) > 0$ holds so that $s_{\epsilon}(t)$ is an increasing function of $t$. The inverse function $t_{\epsilon}(s)$ satisfies
\begin{align} \label{jacobian_reparametrize}
    \left. d_s t_{\epsilon} (s) \right.|_{s=s_{\epsilon}(t)}=\qty(d_t s_{\epsilon})^{-1}=\frac{1}{\sqrt{\epsilon+\fip{\Grad\vec{\phi}{}^{\sharp}}{\Grad\vec{\phi}{}^{\sharp}}_{\mathcal{M}_{\vec{c}^{\sharp}}}}}.
\end{align}
We define the reparametrized concentration distribution $\vec{\omega}(s)$ as $\vec{\omega}(s)\coloneqq\vec{c}^{\sharp}(t_{\epsilon}(s))$, and a potential $\vec{\zeta}(s)$ as $\vec{\zeta}(s)\coloneqq(d_st_{\epsilon}(s))\vec{\phi}{}^{\sharp}(t_{\epsilon}(s))$. These quantities
satisfy
\begin{align}\label{reparametrized_c}
    \vec{\omega}(0)=\vec{c}^{\sharp}(0)=\vec{c}(0),\;\vec{\omega}(s_{\epsilon}(\tau))=\vec{c}^{\sharp}(\tau)=\vec{c}(\tau),
\end{align}
\begin{align}\label{reparametrized_c_ext}
    \vec{\omega}_{\mathcal{Y}}(s)=\vec{c}^{\sharp}_{\mathcal{Y}}(t_{\epsilon}(s))=\vec{b}_{\mathcal{Y}},
\end{align}
and
\begin{align}\label{reparametrized_pot}
    \partial_s \vec{\omega}_{\mathcal{X}}(s)&=(d_s t_{\epsilon}(s)) \left. \partial_t \vec{c}^{\sharp}_{\mathcal{X}}( t) \right|_{t =t_{\epsilon}(s)}\notag\\
    &=(d_s t_{\epsilon}(s)) \qty(\Div\mathcal{M}_{\vec{c}^{\sharp}(t_{\epsilon}(s))}\Grad\vec{\phi}{}^{\sharp}(t_{\epsilon}(s)))_{\mathcal{X}}\notag\\
    &=\qty(\Div\mathcal{M}_{\vec{\omega}(s)}\Grad\vec{\zeta}(s))_{\mathcal{X}}.
\end{align}
These conditions in Eq.~\eqref{reparametrized_c}, Eq.~\eqref{reparametrized_c_ext}, and Eq.~\eqref{reparametrized_pot} are the same as the conditions imposed on the optimization problem in Eq.~\eqref{potential_form_L2} if we replace the time duration $\tau$ with $s_{\epsilon}(\tau)$. From the definition in Eq.~\eqref{potential_form_L2} and Eq.~\eqref{jacobian_reparametrize}, we thus obtain
\begin{align}\label{repar_ineq}
    &{W}_{2,\mathcal{X}}(\vec{c}(0),\vec{c}(\tau)|\vec{b}_\mathcal{Y})^2\leq s_{\epsilon}(\tau)\int_{0}^{s_{\epsilon}(\tau)}ds\fip{\Grad\vec{\zeta}}{\Grad\vec{\zeta}}_{\mathcal{M}_{\vec{\omega}}}\notag\\
    &=s_{\epsilon}(\tau)\int_{0}^{\tau}dt [d_t s_{\epsilon}(t) ]\qty[ \left. d_s t_{\epsilon} (s) \right.|_{s=s_{\epsilon}(t)}]^2\fip{\Grad\vec{\phi}{}^{\sharp}}{\Grad\vec{\phi}{}^{\sharp}}_{\mathcal{M}_{\vec{c}^{\sharp}}}\notag\\
    &=s_{\epsilon}(\tau)\int_{0}^{\tau}dt\frac{\fip{\Grad\vec{\phi}{}^{\sharp}}{\Grad\vec{\phi}{}^{\sharp}}_{\mathcal{M}_{\vec{c}^{\sharp}}}}{\sqrt{\epsilon+\fip{\Grad\vec{\phi}{}^{\sharp}}{\Grad\vec{\phi}{}^{\sharp}}_{\mathcal{M}_{\vec{c}^{\sharp}}}}}\notag\\
    &\leq\qty(\int_{0}^{\tau}dt\sqrt{\epsilon+\fip{\Grad\vec{\phi}{}^{\sharp}}{\Grad\vec{\phi}{}^{\sharp}}_{\mathcal{M}_{\vec{c}^{\sharp}}}})^2.
\end{align}
Taking the limit $\epsilon\to0$ in Eq.~\eqref{repar_ineq} leads to the inequality we need to derive~\eqref{W2_leq_cspeed}. 

The optimizer of Eq.~\eqref{potential_form_L2} $(\vec{c}^{\star},\vec{\phi}{}^{\star})$ is one of the optimizers of Eq.~\eqref{constant_speed_L2} because we can easily derive
\begin{align}\label{opt_is_const_speed}
    {W}_{2,\mathcal{X}}(\vec{c}(0),\vec{c}(\tau)|\vec{b}_{\mathcal{Y}})=\int_{0}^{\tau}dt\sqrt{\fip{\Grad\vec{\phi}{}^{\star}}{\Grad\vec{\phi}{}^{\star}}_{\mathcal{M}_{\vec{c}^{\star}}}},
\end{align}
by repeating the same argument above using 
\begin{align}
    s^{\star}_{\epsilon}(t)\coloneqq\int_{0}^{t}dt'\sqrt{\epsilon+\fip{\Grad\vec{\phi}{}^{\star}(t')}{\Grad\vec{\phi}{}^{\star}(t')}_{\mathcal{M}_{\vec{c}^{\star}(t')}}},
\end{align}
instead of $s_{\epsilon}(t)$. The form of the $2$-Wasserstein distance in Eq.~\eqref{opt_is_const_speed} means that the optimizer $(\vec{c}^{\star},\vec{\phi}{}^{\star})$ satisfies 
the equality condition of the Cauchy--Schwarz inequality in Eq.~\eqref{CS_for_L2} so that the condition,
\begin{align}
    \frac{{W}_{2,\mathcal{X}}(\vec{c}(0),\vec{c}(\tau)|\vec{b}_{\mathcal{Y}})}{\tau}=\sqrt{\fip{\Grad\vec{\phi}{}^{\star}}{\Grad\vec{\phi}{}^{\star}}_{\mathcal{M}_{\vec{c}^{\star}}}},
\end{align}
holds for all $t\in[0,\tau]$. This condition means that the geodesic in Fig.~\ref{fig:RDS_TSL_scheme} has the constant speed.

\subsection{Axioms of Distance}\label{ap:aximos_distance}
Here, we confirm that the Wasserstein distances satisfy the axioms of distance: nondegenerateness, symmetry, and the triangle inequality. 

First, we prove the nondegenerateness of the Wasserstein distances. The nondegenerateness of the $1$-Wasserstein distance is ${W}_{1,\mathcal{X}}(\vec{c}(0),\vec{c}(\tau))=0\Leftrightarrow\vec{c}_{\mathcal{X}}(0)=\vec{c}_{\mathcal{X}}(\tau)$. Letting ${W}_{1,\mathcal{X}}(\vec{c}(0),\vec{c}(\tau))=0$ hold, all the elements of the optimal current are zero. Then, the constraint $\partial_t\vec{c}_{\mathcal{X}}=\qty(\Div\mathcal{J}')_{\mathcal{X}}$ leads to $\vec{c}_{\mathcal{X}}(0)=\vec{c}_{\mathcal{X}}(\tau)$. Conversely, assuming $\vec{c}_{\mathcal{X}}(0)=\vec{c}_{\mathcal{X}}(\tau)$, the current with zero elements satisfies the constraints imposed on the optimization problem for the $1$-Wasserstein distance so that ${W}_{1,\mathcal{X}}(\vec{c}(0),\vec{c}(\tau))=0$ holds. We can also prove the nondegenerateness of the $2$-Wasserstein distance by replacing the current with the force and arguing similarly.

Second, we prove the symmetry of the Wasserstein distance. The symmetry of the $1$ -Wasserstein distance is ${W}_{1,\mathcal{X}}(\vec{c}^{A},\vec{c}^{B})={W}_{1,\mathcal{X}}(\vec{c}^{B},\vec{c}^{A})$. Letting $\vec{c}^{\diamond}$ and $\J^{\diamond}$ denote the optimizer for ${W}_{1,\mathcal{X}}(\vec{c}^{A},\vec{c}^{B})$, the time reversal quantities $\vec{c}'(t)\coloneqq\vec{c}^{\diamond}(\tau-t)$ and $\mathcal{J}'(t)\coloneqq-\mathcal{J}^{\diamond}(\tau-t)$ satisfy the constraints imposed on the optimization problem for ${W}_{1,\mathcal{X}}(\vec{c}^{B},\vec{c}^{A})$. Then, we obtain an inequality,
\begin{align}
    {W}_{1,\mathcal{X}}(\vec{c}^{A},\vec{c}^{B})&=\int_{0}^{\tau}dt\left|\mathcal{J}^{\diamond}(t)\right|_{\rm RD}\notag\\
    &=\int_{0}^{\tau}dt\left|-\mathcal{J}'(\tau-t)\right|_{\rm RD}\notag\\
    &=\int_{0}^{\tau}dt\left|\mathcal{J}'(t)\right|_{\rm RD}\notag\\
    &\geq{W}_{1,\mathcal{X}}(\vec{c}^{B},\vec{c}^{A}).
\end{align}
Similarly, we can obtain the inequality in the opposite direction, ${W}_{1,\mathcal{X}}(\vec{c}^{A},\vec{c}^{B})\leq{W}_{1,\mathcal{X}}(\vec{c}^{B},\vec{c}^{A})$. We can also prove the symmetry of the $2$-Wasserstein distance in the same way. 

Third, we prove the triangle inequality of the $1$-Wasserstein distance, ${W}_{1,\mathcal{X}}(\vec{c}^{A},\vec{c}^{B})+{W}_{1,\mathcal{X}}(\vec{c}^{B},\vec{c}^{C})\geq {W}_{1,\mathcal{X}}(\vec{c}^{A},\vec{c}^{C})$. Letting the optimizer for ${W}_{1,\mathcal{X}}(\vec{c}^{A},\vec{c}^{B})$ and ${W}_{1,\mathcal{X}}(\vec{c}^{B},\vec{c}^{C})$ be $\qty(\vec{c}^{\diamond,1},\mathcal{J}^{\diamond,1})$ and $\qty(\vec{c}^{\diamond,2},\mathcal{J}^{\diamond,2})$, respectively, a new concentration distributions $\vec{c}'$ and a new current $\mathcal{J}'$ defined as 
\begin{align}
\vec{c}'(t)\coloneqq
\begin{cases}
\vec{c}^{\diamond,1}(2t) & \left(0\leq t<\dfrac{\tau}{2}\right) \\
\vec{c}^{\diamond,2}(2t-\tau) & \left(\dfrac{\tau}{2}\leq t\leq\tau\right)
\end{cases}
,
\end{align}
\begin{align}
\mathcal{J}'(t)\coloneqq
\begin{cases}
2\mathcal{J}^{\diamond,1}(2t) & \left(0\leq t<\dfrac{\tau}{2}\right) \\
2\mathcal{J}^{\diamond,2}(2t-\tau) & \left(\dfrac{\tau}{2}\leq t\leq\tau\right)
\end{cases}
,
\end{align}
satisfy the constraints imposed on the optimization problem for ${W}_{1,\mathcal{X}}(\vec{c}^{A},\vec{c}^{C})$: $\vec{c}'(0)=\vec{c}^{A}$, $\vec{c}'(\tau)=\vec{c}^{C}$, and $\partial_t\vec{c}'_{\mathcal{X}}=\qty(\Div\mathcal{J}')_{\mathcal{X}}$. Thus, we obtain the triangle inequality as
\begin{align}
    &{W}_{1,\mathcal{X}}(\vec{c}^{A},\vec{c}^{C})\leq\int_{0}^{\tau}dt\left|\mathcal{J}'(t)\right|_{\rm RD}\notag\\
    &=\int_{0}^{\frac{\tau}{2}}dt\,2\left|\mathcal{J}^{\diamond,1}(2t)\right|_{\rm RD}+\int_{\frac{\tau}{2}}^{\tau}dt\,2\left|\mathcal{J}^{\diamond,2}(2t-\tau)\right|_{\rm RD}\notag\\
    &=\int_{0}^{\tau}dt\left|\mathcal{J}^{\diamond,1}(t)\right|_{\rm RD}+\int_{0}^{\tau}dt\left|\mathcal{J}^{\diamond,2}(t)\right|_{\rm RD}\notag\\
    &={W}_{1,\mathcal{X}}(\vec{c}^{A},\vec{c}^{B})+{W}_{1,\mathcal{X}}(\vec{c}^{B},\vec{c}^{C}).
\end{align}

Finally, we prove the triangle inequality of the $2$-Wasserstein distance, ${W}_{2,\mathcal{X}}(\vec{c}^{A},\vec{c}^{B}|\vec{b}_{\mathcal{Y}})+{W}_{2,\mathcal{X}}(\vec{c}^{B},\vec{c}^{C}|\vec{b}_{\mathcal{Y}})\geq {W}_{2,\mathcal{X}}(\vec{c}^{A},\vec{c}^{C}|\vec{b}_{\mathcal{Y}})$. We need to use the reformulation of the $2$-Wasserstein distance in Eq.~\eqref{constant_speed_L2} to derive the triangle inequality~\cite{dolbeault2009new}. Letting the optimizer of the optimization problem in Eq.~\eqref{constant_speed_L2} for ${W}_{2,\mathcal{X}}(\vec{c}^{A},\vec{c}^{B}|\vec{b}_{\mathcal{Y}})$ and ${W}_{2,\mathcal{X}}(\vec{c}^{B},\vec{c}^{C}|\vec{b}_{\mathcal{Y}})$ be $\qty(\vec{c}^{\sharp,1},\vec{\phi}{}^{\sharp,1})$ and $\qty(\vec{c}^{\sharp,2},\vec{\phi}{}^{\sharp,2})$, respectively, a new concentration distributions $\vec{c}'$ and a new potential $\vec{\phi}'$ defined as 
\begin{align}
\vec{c}'(t)\coloneqq
\begin{cases}
\vec{c}^{\sharp,1}(2t) & \left(0\leq t<\dfrac{\tau}{2}\right) \\
\vec{c}^{\sharp,2}(2t-\tau) & \left(\dfrac{\tau}{2}\leq t\leq\tau\right)
\end{cases}
,
\end{align}
\begin{align}
\vec{\phi}'(t)\coloneqq
\begin{cases}
2\vec{\phi}{}^{\sharp,1}(2t) & \left(0\leq t<\dfrac{\tau}{2}\right) \\
2\vec{\phi}{}^{\sharp,2}(2t-\tau) & \left(\dfrac{\tau}{2}\leq t\leq\tau\right)
\end{cases}
,
\end{align}
satisfy the constraints imposed on the optimization problem for ${W}_{2,\mathcal{X}}(\vec{c}^{A},\vec{c}^{C}|\vec{b}_{\mathcal{Y}})$: $\vec{c}'_{\mathcal{X}}(0)=\vec{c}_{\mathcal{X}}^{A}$, $\vec{c}'_{\mathcal{X}}(\tau)=\vec{c}_{\mathcal{X}}^{C}$, $\vec{c}'_{\mathcal{Y}}(t)=\vec{b}_{\mathcal{Y}}$, and $\partial_t\vec{c}'_{\mathcal{X}}=\qty(\Div\mathcal{M}_{\vec{c}'}\Grad\vec{\phi}')_{\mathcal{X}}$. Thus, we obtain the triangle inequality as
\begin{align}\label{tri_ineq_L2}
    &{W}_{2,\mathcal{X}}(\vec{c}^{A},\vec{c}^{C}|\vec{b}_{\mathcal{Y}})\leq\int_{0}^{\tau}dt\sqrt{\fip{\Grad\vec{\phi}'}{\Grad\vec{\phi}'}_{\mathcal{M}_{\vec{c}'}}}\notag\\
    &=\int_{0}^{\frac{\tau}{2}}dt\,2\sqrt{\fip{\Grad\vec{\phi}{}^{\sharp,1}(2t)}{\Grad\vec{\phi}{}^{\sharp,1}(2t)}_{\mathcal{M}_{\vec{c}^{\sharp, 1}(2t)}}}\notag\\
    &\phantom{=}+\int_{\frac{\tau}{2}}^{\tau}dt\,2\sqrt{\fip{\Grad\vec{\phi}{}^{\sharp,2}(2t-\tau)}{\Grad\vec{\phi}{}^{\sharp,2}(2t-\tau)}_{\mathcal{M}_{\vec{c}^{\sharp, 2}(2t-\tau)}}}\notag\\
    &=\int_{0}^{\tau}dt\sqrt{\fip{\Grad\vec{\phi}{}^{\sharp,1}(t)}{\Grad\vec{\phi}{}^{\sharp,1}(t)}_{\mathcal{M}_{\vec{c}^{\sharp, 1}(t)}}}\notag\\
    &\phantom{=}+\int_{0}^{\tau}dt\sqrt{\fip{\Grad\vec{\phi}{}^{\sharp,2}(t)}{\Grad\vec{\phi}{}^{\sharp,2}(t)}_{\mathcal{M}_{\vec{c}^{\sharp, 2}(t)}}}\notag\\
    &={W}_{2,\mathcal{X}}(\vec{c}^{A},\vec{c}^{B}|\vec{b}_{\mathcal{Y}})+{W}_{2,\mathcal{X}}(\vec{c}^{B},\vec{c}^{C}|\vec{b}_{\mathcal{Y}}).
\end{align}

\add{\subsection{Derivation of the inequality between the Wasserstein distances}\label{ap:deriv_compare_L1_L2}}

\add{Here, we derive the inequality between the $1$- and $2$-Wasserstein distances~\eqref{compare_L1_L2}. In the following, we let $\left(\vec{c}{}^{\star},\F^{\star}\right)$ denote the optimizer of the minimization problem for the $2$-Wasserstein distance~\eqref{Def_L2}. The corresponding current $\J^{\star} = \left(\vec{\bm{J}}{}^{\star},\bm{j}^{\star}\right)$ is defined as $\J^{\star}=\M_{\vec{c}{}^{\star}}\F^{\star}$. We use a new current $\J^{\times}=\left(\vec{\bm{J}}{}^{\times},\bm{j}^{\times}\right)$ and a new force $\F^{\mathrm{uni}}=\left(\vec{\bm{F}}{}^{\mathrm{uni}},\bm{f}^{\mathrm{uni}}\right)$, defined as
\begin{align}
    \bm{J}^{\times}_{(\alpha)}\coloneqq
    \begin{cases}
        \bm{J}^{\star}_{(\alpha)}&(\alpha\in\mathcal{X})\notag\\
        \bm{0}&(\alpha\in\mathcal{Y})
    \end{cases},\;
    j^{\times}_{\rho}\coloneqq
    \begin{cases}
        j^{\star}_{\rho}&(\rho\in\mathcal{R}_{\mathcal{X}})\notag\\
        0&(\rho\in\mathcal{R}\setminus\mathcal{R}_{\mathcal{X}})
    \end{cases},
\end{align}
and
\begin{align}
    \bm{F}^{\mathrm{uni}}_{(\alpha)}\coloneqq
    \begin{cases}
        \dfrac{\bm{J}^{\star}_{(\alpha)}}{\|\bm{J}^{\star}_{(\alpha)}\|}&(\alpha\in\mathcal{X},\|\bm{J}^{\star}_{(\alpha)}\|\neq0)\notag\\
        \bm{0}&(\mathrm{otherwise})
    \end{cases},
\end{align}
\begin{align}
    f^{\mathrm{uni}}_{\rho}\coloneqq
    \begin{cases}
        \dfrac{j^{\star}_{\rho}}{|j^{\star}_{\rho}|}&(\rho\in\mathcal{R}_{\mathcal{X}},|j^{\star}_{\rho}|\neq0)\notag\\
        0&(\mathrm{otherwise})
    \end{cases}.
\end{align}
The current $\J^{\times}$ satisfies
\begin{align}
    \int_{0}^{\tau}dt\left|\J^{\times}\right|_{\RD}\geq W_{1,\mathcal{X}}\left(\vec{c}(0),\vec{c}(\tau)\right),
    \label{ineq1}
\end{align}
since the conditions imposed on $\left(\vec{c}{}^{\star},\F^{\star}\right)$ let $\J^{\times}$ be a candidate of the optimization problem in the definition of the $1$-Wasserstein distance~\eqref{Def_L1}. The force $\F^{\mathrm{uni}}$ also satisfies
\begin{align}
    \fip{\F^{\mathrm{uni}}}{\F^{\mathrm{uni}}}_{\M_{\vec{c}{}^{\star}}}\leq\mnorm{\M_{\vec{c}{}^{\star}}}.
    \label{ineq2}
\end{align}
We obtain this using the following two facts: the diffusion part $\vec{\bm{F}}{}^{\mathrm{uni}}(\bm{r};t)$ is a candidate of the maximization problem in the definition of $\diffmax(\bm{r};t)$~\eqref{max_mobility}, and the reaction part satisfies $|f^{\mathrm{uni}}_{\rho}(\bm{r};t)|^2\leq1$ for all $\rho\in\mathcal{R}_{\mathcal{X}}$ and $f^{\mathrm{uni}}_{\rho}(\bm{r};t)=0$ for all $\rho\in\mathcal{R}\setminus\mathcal{R}_{\mathcal{X}}$. Using the notation $\M_{\vec{c}^{\star}}=\dvec{\mathsf{M}}{}^{\star} \oplus \mathsf{m}^{\star}$, we can derive Eq.~\eqref{ineq2} from these facts as $\fip{\F^{\mathrm{uni}}}{\F^{\mathrm{uni}}}_{\M_{\vec{c}{}^{\star}}}=\int_{V}d\bm{r}[\vec{\bm{F}}{}^{\mathrm{uni}\top}\dvec{\mathsf{M}}{}^{\star}\vec{\bm{F}}{}^{\mathrm{uni}}+\sum_{\rho\in\mathcal{R}}m_{\rho}^{\star}(f^{\mathrm{uni}}_{\rho})^2]\leq\int_{V}d\bm{r}[\diffmax+\sum_{\rho\in\mathcal{R}_{\mathcal{X}}}m^{\star}_{\rho}]=\mnorm{\M_{\vec{c}{}^{\star}}}$. The force $\F^{\mathrm{uni}}$ also relates to $\left|\J^{\times}\right|_{\RD}$ as
\begin{align}
    \left|\J^{\times}\right|_{\RD}=\fip{\F^{\mathrm{uni}}}{\F^{\star}}_{\M_{\vec{c}{}^{\star}}},
    \label{1-norm_innerproduct}
\end{align}
which is verified as
\begin{align}
    \left|\J^{\times}\right|_{\RD}&=\int_{V}d\bm{r}\left[\sum_{\alpha\in\mathcal{S}}\|\bm{J}^{\times}_{(\alpha)}\|+\sum_{\rho\in\mathcal{R}}\left|j^{\times}_{\rho}\right|\right]\notag\\
    &=\int_{V}d\bm{r}\left[\sum_{\alpha\in\mathcal{X}}\|\bm{J}^{\star}_{(\alpha)}\|+\sum_{\rho\in\mathcal{R}_{\mathcal{X}}}\left|j^{\star}_{\rho}\right|\right]\notag\\&=\int_{V}d\bm{r}\left[\sum_{\alpha\in\mathcal{X}}\bm{F}^{\mathrm{uni}}_{(\alpha)}\cdot\bm{J}^{\star}_{(\alpha)}+\sum_{\rho\in\mathcal{R}_{\mathcal{X}}}f^{\mathrm{uni}}_{\rho}j^{\star}_{\rho}\right]\notag\\
    &=\fip{\F^{\mathrm{uni}}}{\J^{\star}}=\fip{\F^{\mathrm{uni}}}{\F^{\star}}_{\M_{\vec{c}^{\star}}}.
\end{align}
}

\add{To obtain the desired inequality~\eqref{compare_L1_L2}, we use the Cauchy--Schwarz inequality,
\begin{align}
    \fip{\F^{\mathrm{uni}}}{\F^{\star}}_{\M_{\vec{c}^{\star}}}&\leq\sqrt{\fip{\F^{\mathrm{uni}}}{\F^{\mathrm{uni}}}_{\M_{\vec{c}^{\star}}}\fip{\F^{\star}}{\F^{\star}}_{\M_{\vec{c}^{\star}}}}\notag\\&\leq\sqrt{\mnorm{\M_{\vec{c}^{\star}}}\fip{\F^{\star}}{\F^{\star}}_{\M_{\vec{c}^{\star}}}}.
    \label{ineq3}
\end{align}
Here, we also use the property of $\F^{\mathrm{uni}}$ in Eq.~\eqref{ineq2}.
Integrating both sides of this inequality~\eqref{ineq3}, we obtain
\begin{align}
    \int_{0}^{\tau}dt\sqrt{\mnorm{\M_{\vec{c}^{\star}}}\fip{\F^{\star}}{\F^{\star}}_{\M_{\vec{c}^{\star}}}}&\geq\int_{0}^{\tau}dt\fip{\F^{\mathrm{uni}}}{\F^{\star}}_{\M_{\vec{c}^{\star}}}\notag\\&=\int_{0}^{\tau}dt\left|\J^{\times}\right|_{\RD}\notag\\&\geq W_{1,\mathcal{X}}\left(\vec{c}(0),\vec{c}(\tau)\right),
    \label{ineq4}
\end{align}
where we use Eq.~\eqref{1-norm_innerproduct} in the first transform and Eq.~\eqref{ineq1} in the second transform.
The Cauchy--Schwarz inequality also provides the following inequality,
\begin{align}
&\left(\int_{0}^{\tau}dt\sqrt{\mnorm{\M_{\vec{c}^{\star}}}\fip{\F^{\star}}{\F^{\star}}_{\M_{\vec{c}^{\star}}}}\right)^2\notag\\
    &\leq\left(\int_{0}^{\tau}dt\,\mnorm{\M_{\vec{c}^{\star}}}\right)\left(\int_{0}^{\tau}dt\fip{\F^{\star}}{\F^{\star}}_{\M_{\vec{c}^{\star}}}\right)\notag\\
    &=\left(\frac{1}{\tau}\int_{0}^{\tau}dt\,\mnorm{\M_{\vec{c}^{\star}}}\right)\left(\tau\int_{0}^{\tau}dt\fip{\F^{\star}}{\F^{\star}}_{\M_{\vec{c}^{\star}}}\right)\notag\\
    &=\langle\mnorm{\M_{\vec{c}^{\star}}}\rangle_{\tau}{W}_{2,\mathcal{X}}\qty(\vec{c}(0),\vec{c}(\tau)\middle|\vec{c}_{\mathcal{Y}}(t))^2.
    \label{ineq5}
\end{align}
Here, $\vec{c}_{\mathcal{Y}}(t)$ does not depend on time. 
Combining Eq.~\eqref{ineq4} and Eq.~\eqref{ineq5} leads to the desired inequality~\eqref{compare_L1_L2}.
}

\add{\subsection{Derivation of thermodynamic speed limit based on the $1$-Wasserstein distance}
\label{ap:deriv_L1_TSL}}

\add{Here, we derive the TSLs in Eq.~\eqref{exEPR_L1_line_element} and Eq.~\eqref{TSL_L1} from the inequality between the Wasserstein distances in Eq.~\eqref{compare_L1_L2}. Substituting $\vec{c}^{A}=\vec{c}(t)$, $\vec{c}^{B}=\vec{c}(t+\varDelta t)$, and $\vec{b}_{\mathcal{Y}}=\vec{c}_{\mathcal{Y}}(t)$ with $\varDelta t\ll 1$ into Eq.~\eqref{compare_L1_L2}, we obtain
\begin{align}
    \frac{{W}_{1,\mathcal{X}}\qty(\vec{c}(t),\vec{c}(t+\varDelta t))^2}{\frac{1}{\varDelta t}\int_{t}^{t+\varDelta t}ds\,\mnorm{\M}}\leq{W}_{2,\mathcal{X}}\qty(\vec{c}(t),\vec{c}(t+\varDelta t)\middle|\vec{c}_{\mathcal{Y}}(t))^2.
\end{align}
Expanding the denominator of the left-hand side with respect to $\varDelta t$, we can rewrite this inequality as
\begin{align}
    &\frac{{W}_{1,\mathcal{X}}\qty(\vec{c}(t),\vec{c}(t+\varDelta t))^2}{\mnorm{\M}}+o(\varDelta t^2)\notag\\&\leq{W}_{2,\mathcal{X}}\qty(\vec{c}(t),\vec{c}(t+\varDelta t)\middle|\vec{c}_{\mathcal{Y}}(t))^2.
\end{align}
Then, dividing both sides by $\varDelta t^2$ and taking the limit $\varDelta t\to 0$ yields Eq.~\eqref{exEPR_L1_line_element}. Rewriting Eq.~\eqref{exEPR_L1_line_element} as $v_{1}\leq\sqrt{\mnorm{\M}\sigma^{\mathrm{ex}}}$ and using the Cauchy--Schwarz inequality, we obtain a part of the TSLs in Eq.~\eqref{TSL_L1} as
\begin{align}
l_{1,\tau}^2&=\qty[\int_{0}^{\tau}dt\,v_1(t)]^2\notag\\
&\leq\qty[\int_{0}^{\tau}dt\sqrt{\mnorm{\M}\sigma^{\mathrm{ex}}}]^2\notag\\
&\leq\qty(\int_{0}^{\tau}dt\,\mnorm{\M})\qty(\int_{0}^{\tau}dt\,\sigma^{\mathrm{ex}})\notag\\
&=\left\langle\mnorm{\M}\right\rangle_{\tau}\tau\Sigma^{\mathrm{ex}}_{\tau}.
\end{align}
We also obtain $l_{1,\tau}\geq{W}_{1,\mathcal{X}}(\vec{c}(0),\vec{c}(\tau))$ as a direct consequence of the triangle inequality for the $1$-Wasserstein distance. Unifying these results, we reach the TSLs in Eq.~\eqref{TSL_L1}.}

\add{Although it is not obvious which is tighter, $l_{2,\tau}^2$ or $l_{1,\tau}^2/\left\langle\mnorm{\M}\right\rangle_{\tau}$ in the TSLs, we can obtain the following inequality directly from Eq.~\eqref{exEPR_L1_line_element},
\begin{align}
    l_{2,\tau}^2\geq \dfrac{l_{1,\tau}^2}{\max_{t\in[0,\tau]}\mnorm{\M}}.
\end{align}
}

\add{\subsection{Details of the minimum dissipation formula with the $1$-Wasserstein distance~\eqref{minimumEP_1Wasserstein}}\label{ap:minimum dissipation 1-Wasserstein} }

\add{\subsubsection{Derivation of the minimum dissipation formula with the $1$-Wasserstein distance~\eqref{minimumEP_1Wasserstein}}\label{ap:minimum dissipation 1-Wasserstein derivation}}

\add{Here, we prove the minimum dissipation formula with the $1$-Wasserstein distance~\eqref{minimumEP_1Wasserstein}.}

\add{First, we verify that the right-hand side in Eq.~\eqref{minimumEP_1Wasserstein} provides a lower bound of the EP under the conditions in Eqs.~\eqref{constraint:time_evolution} and~\eqref{constraint:intensity_bound} as
\begin{align}
    \Sigma_{\tau}\left[\M',\F'\right] \geq\frac{W_{1,\mathcal{X}}\left(\vec{c}(0),\vec{c}(\tau)\right)^2}{\tau\langle\mnorm{\M'}\rangle_{\tau}}\geq\frac{W_{1,\mathcal{X}}\left(\vec{c}(0),\vec{c}(\tau)\right)^2}{\tau\mmax}.
\end{align}
Here, the first inequality is obtained in the same way as in the case of Eq.~\eqref{compare_L1_L2}. It is enough to use $(\M',\F')$ instead of $(\M_{\vec{c}^{\star}},\F^{\star})$ in Appendix~\ref{ap:deriv_compare_L1_L2}. The second inequality is due to the inequality (\ref{constraint:intensity_bound}).}

\add{Second, we construct an optimizer $\left(\M^{\diamond},\F^{\diamond}\right)$, which achieves
\begin{align}
    \Sigma_{\tau}[\M^{\diamond},\F^{\diamond}]=\frac{W_{1,\mathcal{X}}\left(\vec{c}(0),\vec{c}(\tau)\right)^2}{\tau\mmax}.
    \label{equality_optimalprotocol}
\end{align}
Let $\mathcal{U}^{\diamond}=\qty(\vec{\bm{U}}{}^{\diamond},\bm{u}^{\diamond})$ be the optimizer of the right-hand side in Eq.~\eqref{another_Def_L1}, which satisfies $W_{1,\mathcal{X}}\left(\vec{c}(0),\vec{c}(\tau)\right)=\left|\U^{\diamond}\right|_{\RD}$ and $\vec{c}_{\mathcal{X}}(\tau)-\vec{c}_{\mathcal{X}}(0)=\left(\Div\U^{\diamond}\right)_{\mathcal{X}}$.  Using $\U^{\diamond}$, we can obtain the optimal Onsager operator $\M^{\diamond} = \dvec{\mathsf{M}}{}^{\diamond}\oplus \mathsf{m}^{\diamond}$ as
\begin{align}
    \mathsf{M}^{\diamond}_{(\alpha\beta)}\coloneqq\frac{\mmax\|\bm{U}^{\diamond}_{(\alpha)}\|}{|\U^{\diamond}|_{\RD}}\delta_{\alpha\beta}\mathsf{I},\;\;m^{\diamond}_{\rho}\coloneqq\frac{\mmax|u^{\diamond}_{\rho}|}{|\U^{\diamond}|_{\RD}},
\end{align}
and the optimal force $\F^{\diamond} =(\vec{\bm{F}}^{\diamond}, \bm{f}^{\diamond})$ as
\begin{align}
    \bm{F}^{\diamond}_{(\alpha)}&\coloneqq\begin{cases}
        \dfrac{|\U^{\diamond}|_{\RD}}{\tau\mmax}\dfrac{\bm{U}^{\diamond}_{(\alpha)}}{\|\bm{U}^{\diamond}_{(\alpha)}\|}&(\bm{U}^{\diamond}_{(\alpha)}\neq\bm{0})\\
        \bm{0}&(\bm{U}^{\diamond}_{(\alpha)}=\bm{0})
    \end{cases},
    \\
    f^{\diamond}_{\rho}&\coloneqq\begin{cases}
        \dfrac{|\U^{\diamond}|_{\RD}}{\tau\mmax}\dfrac{u^{\diamond}_{\rho}}{|u^{\diamond}_{\rho}|}&(u^{\diamond}_{\rho}\neq0)\\
        0&(u^{\diamond}_{\rho}=0)
    \end{cases}.
\end{align}
The optimizer $\left(\M^{\diamond},\F^{\diamond}\right)$ satisfies the constraint~\eqref{constraint:time_evolution} because $\M^{\diamond}\F^{\diamond}=\U^{\diamond}/\tau$ holds. We can also verify the remaining condition~\eqref{constraint:intensity_bound} and the equality~\eqref{equality_optimalprotocol} by direct calculation.}

\add{\subsubsection{Operational interpretation of the optimizer of the minimization problem in Eq.~\eqref{minimumEP_1Wasserstein}}\label{ap:minimum dissipation 1-Wasserstein operation}}

\add{The optimization problem for $\M'$ and $\F'$ in Eq.~\eqref{minimumEP_1Wasserstein} can be reformulated in terms of of control parameters. This allows us to interpret the optimizer of the minimization problem in Eq.~\eqref{minimumEP_1Wasserstein} that is constructed in the previous section, $\left(\M^{\diamond},\F^{\diamond}\right)$, from an operational viewpoint. Here, we provide such an operational interpretation of the optimizer $\left(\M^{\diamond},\F^{\diamond}\right)$.}

\add{We consider the ideal dilute solution with the two assumptions: the mobility tensor has the simple form~\eqref{simple_mobility_tensor}, and the reactions obey the mass action law. Under these assumptions, the time evolution of the concentration distribution of the internal species $\alpha\in\mathcal{X}$ is given by
\begin{align}
    \partial_{t}c_{\alpha}=&\bm{\nabla}_{\bm{r}}\cdot\left\{D_{\alpha}c_{\alpha}\left(\bm{\nabla}_{\bm{r}}\ln c_{\alpha}-\bm{F}_{(\alpha)}^{\mathrm{mech}}\right)\right\}\notag\\&+\sum_{\rho\in\mathcal{R}_{\mathcal{X}}}S_{\alpha\rho}\left(\kappa^{+}_{\rho}\prod_{\alpha\in\mathcal{S}}c_{\alpha}^{\nu^{+}_{\alpha\rho}}-\kappa^{-}_{\rho}\prod_{\alpha\in\mathcal{S}}c_{\alpha}^{\nu^{-}_{\alpha\rho}}\right),
    \label{control RD}
\end{align}
where we let $\bm{F}_{(\alpha)}^{\mathrm{mech}}$ denote the force acting on the $\alpha$-th species.}

\add{Let us regard the diffusion coefficient $D_{\alpha}$, the force $\bm{F}_{(\alpha)}^{\mathrm{mech}}$, and the reaction rate constant $\kappa^{\pm}_{\rho}$ in Eq.~\eqref{control RD} as the controllable parameters. Then, using the concentration distribution $\vec{c}^{\diamond}$ introduced in Sec.~\ref{sec:minimumdissipation}, we can obtain the optimal diffusion coefficient and the optimal force as
\begin{align}
    D_{\alpha}^{\diamond}(t)\coloneqq\frac{\mmax\|\bm{U}^{\diamond}_{(\alpha)}\|}{|\U^{\diamond}|_{\RD}c^{\diamond}_{\alpha}(t)},
    \label{optimal diffusion coefficient}
\end{align}
and
\begin{align}
    \bm{F}_{(\alpha)}^{\mathrm{mech}\diamond}(t)\coloneqq\bm{F}_{(\alpha)}^{\diamond}+\bm{\nabla}_{\bm{r}}\ln c^{\diamond}_{\alpha}(t),
    \label{optimal diffusion force}
\end{align}
for all $\alpha\in\mathcal{S}$. We can also construct the optimal reaction rate constant as
\begin{align}
    \begin{cases}
        \kappa_{\rho}^{+\diamond}(t)\coloneqq\dfrac{1}{\prod_{\alpha\in\mathcal{S}}\left\{c^{\diamond}_{\alpha}(t)\right\}^{\nu^{+}_{\alpha\rho}}}\dfrac{\mathrm{e}^{f^{\diamond}_{\rho}}}{\mathrm{e}^{f^{\diamond}_{\rho}}-1}\dfrac{u^{\diamond}_{\rho}}{\tau}\\
        \kappa_{\rho}^{-\diamond}(t)\coloneqq\dfrac{1}{\prod_{\alpha\in\mathcal{S}}\left\{c^{\diamond}_{\alpha}(t)\right\}^{\nu^{-}_{\alpha\rho}}}\dfrac{1}{\mathrm{e}^{f^{\diamond}_{\rho}}-1}\dfrac{u^{\diamond}_{\rho}}{\tau}
    \end{cases},
    \label{optimal rate constant}
\end{align}
for all $\rho\in\mathcal{R}$. We can easily verify that these optimal parameters reproduce $(\M^{\diamond},\F^{\diamond})$ as
\begin{align}
    D_{\alpha}^{\diamond}c^{\diamond}_{\alpha}\delta_{\alpha\beta}\mathsf{I}=\frac{\mmax\|\bm{U}^{\diamond}_{(\alpha)}\|}{|\U^{\diamond}|_{\RD}}\delta_{\alpha\beta}\mathsf{I}=\mathsf{M}^{\diamond}_{(\alpha\beta)},
\end{align}
\begin{align}
    &\frac{\kappa_{\rho}^{+\diamond}\prod_{\alpha\in\mathcal{S}}c^{\diamond}_{\alpha}{}^{\nu^{+}_{\alpha\rho}}-\kappa_{\rho}^{-\diamond}\prod_{\alpha\in\mathcal{S}}c^{\diamond}_{\alpha}{}^{\nu^{-}_{\alpha\rho}}}{\ln\kappa_{\rho}^{+\diamond}\prod_{\alpha\in\mathcal{S}}c^{\diamond}_{\alpha}{}^{\nu^{+}_{\alpha\rho}}-\ln\kappa_{\rho}^{-\diamond}\prod_{\alpha\in\mathcal{S}}c^{\diamond}_{\alpha}{}^{\nu^{-}_{\alpha\rho}}}\notag\\&=\frac{\mmax|u^{\diamond}_{\rho}|}{|\U^{\diamond}|_{\RD}}=m^{\diamond}_{\rho},
\end{align}
\begin{align}
    \bm{F}_{(\alpha)}^{\mathrm{mech}\diamond}(t)-\bm{\nabla}_{\bm{r}}\ln c^{\diamond}_{\alpha}(t)=\bm{F}^{\diamond}_{(\alpha)},
\end{align}
and
\begin{align}
    \ln\frac{\kappa_{\rho}^{+\diamond}\prod_{\alpha\in\mathcal{S}}c^{\diamond}_{\alpha}{}^{\nu^{+}_{\alpha\rho}}}{\kappa_{\rho}^{-\diamond}\prod_{\alpha\in\mathcal{S}}c^{\diamond}_{\alpha}{}^{\nu^{-}_{\alpha\rho}}}=f^{\diamond}_{\rho}.
\end{align}
Note that the dependence of the optimal parameters on time is completely determined by $\vec{c}^{\diamond}(t)$ because the optimizer $\left(\M^{\diamond},\F^{\diamond}\right)$ is independent of time.
}

\add{\subsubsection{Conservative force as an optimizer of the minimization problem in Eq.~\eqref{minimumEP_1Wasserstein}}\label{ap:minimum dissipation 1-Wasserstein conservativeness}}

\add{We remark that we can take a conservative force as an optimal force of the minimization problem~\eqref{minimumEP_1Wasserstein} by using $\U^{\bullet}$ introduced in Appendix~\ref{ap:KR_dual} instead of $\U^{\diamond}$. In this case, a new optimal Onsager operator $\M^{\bullet}=\dvec{\mathsf{M}}^{\bullet}\oplus\mathsf{m}^{\bullet}$ is given by
\begin{align}
    \mathsf{M}^{\bullet}_{(\alpha\beta)}\coloneqq\frac{\mmax\|\bm{U}^{\bullet}_{(\alpha)}\|}{|\U^{\bullet}|_{\RD}}\delta_{\alpha\beta}\mathsf{I},\;\;m^{\bullet}_{\rho}\coloneqq\frac{\mmax|u^{\bullet}_{\rho}|}{|\U^{\bullet}|_{\RD}},
\end{align}
and we can take the conservative force
\begin{align}
\F^{\bullet}\coloneqq\Grad\left(\frac{|\U^{\bullet}|_{\RD}}{\tau\mmax}\vec{\psi}{}^{\bullet}\right),
\end{align}
as a new optimal force. They satisfy $\M^{\bullet}\F^{\bullet}=\U^{\bullet}/\tau$ due to the complementary slackness condition \eqref{complementary slackness} and the definition of $\U^{\bullet}$~\eqref{def Ubullet}. To derive $\M^{\bullet}\F^{\bullet}=\U^{\bullet}/\tau$, we used $\|\bm{U}^{\bullet}_{(\alpha)}\|= U^{{\rm L}\bullet}_{\alpha}$ for $\alpha \in \mathcal{X}$ and $\|u^{\bullet}_{\rho}\|= u^{{\rm L}\bullet}_{\rho}$ for $\rho \in \mathcal{R}_{\mathcal{X}}$.
Since the time-integrated current $\U^{\bullet}$ is an optimizer of the $1$-Wasserstein distance as remarked in Appendix~\ref{ap:KR_dual}, the Onsager operator $\M^{\bullet}$ and the force $\F^{\bullet}$ achieve the equality
\begin{align}
    \Sigma_{\tau}[\M^{\bullet},\F^{\bullet}]=\frac{W_{1,\mathcal{X}}\left(\vec{c}(0),\vec{c}(\tau)\right)^2}{\tau\mmax},
\end{align}
and satisfy the constraints in Eq.~\eqref{constraint:time_evolution} and Eq.~\eqref{constraint:intensity_bound}.}


\add{\subsection{Embedding time series of concentration distributions into Euclidean space by multidimensional scaling}\label{ap:embed}}

\add{We introduce the procedure of the multidimensional scaling~\cite{borg2007modern}. This method is used to visualize the similarity of a data set in a low-dimensional Euclidean space.
Let $[0,\tau]$ be the time interval of the reaction-diffusion dynamics. We used the multidimensional scaling for $n+1$ samples of the concentration $\vec{c}(t_a)$ where $t_a = (a/n) \tau$ and $a \in \{0, 1, \dots, n \}$. 
We aim to find the points $\{\bm{r}^{(0)}, \bm{r}^{(1)}, \dots, \bm{r}^{(n)} \}$ in the $d_0$-dimensional Euclidean space such that $\|\bm{r}^{(a)}-\bm{r}^{(b)}\| \simeq W_{1,\mathcal{X}}(\vec{c}(t_a),\vec{c}(t_b))$ is approximately satisfied for all $(a,b)\in \{0, 1, \dots, n \}^2$. In the numerical examples, we use $(d_0, n)=(2,40)$ for the Fisher--KPP equation and $(d_0, n)=(2,200)$ for the Brusselator model. The detailed procedure is as follows.}

\add{We define a $(n+1)\times (n+1)$ matrix $\mathsf{W}$ as
\begin{align}
    (\mathsf{W})_{ab}\coloneqq W_{1,\mathcal{X}}(\vec{c}(t_a),\vec{c}(t_b))^2,
\end{align}
for $(a,b)\in \{0, 1, \dots, n \}^2$. In general, when determining coordinates solely from the distance structure between each point, there is an ambiguity in the choice of the origin. To determine the origin, we apply the double centering to $\mathsf{W}$, i.e., we transform $\mathsf{W}$ with a matrix $(\mathsf{C})_{ab}\coloneqq\delta_{ab}-1/(n+1)$ as 
\begin{align}
    \mathsf{W}_{\mathrm{C}}\coloneqq-\frac{1}{2}\mathsf{C}\mathsf{W}\mathsf{C}.
\end{align}
We note that $\mathsf{W}_{\mathrm{C}}$ satisfies the following two properties: (i)  $\sum_{a}(\mathsf{W}_{\mathrm{C}})_{ab} =0$ and $\sum_{b}(\mathsf{W}_{\mathrm{C}})_{ab} =0$, and (ii) the square of the distance is recovered as
\begin{align}
    (\mathsf{W}_{\mathrm{C}})_{aa}+(\mathsf{W}_{\mathrm{C}})_{bb}-2(\mathsf{W}_{\mathrm{C}})_{ab}&=(\mathsf{W})_{ab}\notag\\&=W_{1,\mathcal{X}}(\vec{c}(t_a),\vec{c}(t_b))^2.
    \label{MS eq 2}
\end{align}
Here, the property (i) corresponds to the center of gravity being the origin, and the property (ii) corresponds to considering each component $(\mathsf{W}_{\mathrm{C}})_{ab}$ as an inner product between vectors from the origin to each point.
We can verify these properties using 
\begin{align}
    (\mathsf{W}_{\mathrm{C}})_{ab}=&-\frac{1}{2}\left[(\mathsf{W})_{ab}+\frac{\sum_{a',b'}(\mathsf{W})_{a'b'}}{(n+1)^2}\right]\notag\\&+\frac{\sum_{a'}(\mathsf{W})_{a'b}+\sum_{b'}(\mathsf{W})_{ab'}}{2(n+1)},
\end{align}
$(\mathsf{W})_{aa}=0$ and $(\mathsf{W})_{ab}=(\mathsf{W})_{ba}$.
}

\add{Since $\mathsf{W}_{\mathrm{C}}$ is symmetric, it is diagonalizable with an orthogonal matrix $\mathsf{Q}$ as
\begin{align}
    \mathsf{W}_{\mathrm{C}}=\mathsf{Q}\Lambda\mathsf{Q}^{\top},
    \label{MS eigendecomp}
\end{align}
with $\Lambda\coloneqq\mathrm{diag}(\lambda_0,\lambda_1,\cdots,\lambda_n)$. Here, we assume that the eigenvalues of $\mathsf{W}_{\mathrm{C}}$ are in descending order, i.e., $\lambda_0\geq\lambda_1\geq\cdots\geq\lambda_n$. We note that some eigenvalues can be negative because the $1$-Wasserstein distance is non-Euclidean. We can obtain $\bm{r}^{(a)}$ if $\lambda_{d_0-1}$ is nonnegative as
\begin{align}
    \left(\bm{r}^{(a)}\right)_{i}=(\mathsf{Q})_{ai}\sqrt{\lambda_i},
    \label{MS ra}
\end{align}
for $0\leq i\leq d_0-1$. Note that we cannot embed the time series in the $d_0$ dimensional Euclidean space if $\lambda_{d_0-1}$ is negative.}

\add{We also discuss the accuracy of the multidimensional scaling. To do so, we define the estimation error of the distance between $\vec{c}(t_a)$ and $\vec{c}(t_b)$ as 
\begin{align}
    \epsilon_{ab}\coloneqq|W_{1,\mathcal{X}}(\vec{c}(t_a),\vec{c}(t_b))^2-\|\bm{r}^{(a)}-\bm{r}^{(b)}\|^2|.
    \label{MS estimation error}
\end{align}
The definition of $\bm{r}^{(a)}$~\eqref{MS ra} yields
\begin{align}
    &\|\bm{r}^{(a)}-\bm{r}^{(b)}\|^2\notag\\
    &=\sum_{i=0}^{d_0-1}\lambda_i(\mathsf{Q})_{ai}^2+\sum_{i=0}^{d_0-1}\lambda_i(\mathsf{Q})_{bi}^2-2\sum_{i=0}^{d_0-1}\lambda_i(\mathsf{Q})_{ai}(\mathsf{Q})_{bi}\notag\\
    &=(\mathsf{W}_{\mathrm{C}})_{aa}+(\mathsf{W}_{\mathrm{C}})_{bb}-2(\mathsf{W}_{\mathrm{C}})_{ab}\notag\\
    &\phantom{=(\mathsf{W}_{\mathrm{C}})_{aa}\;}-\sum_{i=d_0}^{n}\lambda_i((\mathsf{Q})_{ai}-(\mathsf{Q})_{bi})^2\notag\\
    &=W_{1,\mathcal{X}}(\vec{c}(t_a),\vec{c}(t_b))^2-\sum_{i=d_0}^{n}\lambda_i((\mathsf{Q})_{ai}-(\mathsf{Q})_{bi})^2
    \label{MS eq 1}
\end{align}
where we use the element of Eq.~\eqref{MS eigendecomp}, i.e., $(\mathsf{W}_{\mathrm{C}})_{ab}=\sum_{i=0}^{n}\lambda_i(\mathsf{Q})_{ai}(\mathsf{Q})_{bi}$, in the second line. We also use Eq.~\eqref{MS eq 2} in the last transform.
Thus, we obtain
\begin{align}
    \epsilon_{ab}=\left|\sum_{i=d_0}^{n}\lambda_i((\mathsf{Q})_{ai}-(\mathsf{Q})_{bi})^2\right|.
\end{align}
We can derive the upper bound of $\epsilon_{ab}$ for $a\neq b$ as
\begin{align}
    \epsilon_{ab}&\leq \left( \max_{d_0\leq i\leq n}|\lambda_i| \right)\sum_{i=d_0}^{n}((\mathsf{Q})_{ai}-(\mathsf{Q})_{bi})^2\notag\\
    &\leq \left(\max_{d_0\leq i\leq n}|\lambda_i| \right) \sum_{i=0}^{n}((\mathsf{Q})_{ai}-(\mathsf{Q})_{bi})^2\notag\\
    &=2\max_{d_0\leq i\leq n}|\lambda_i|,
\end{align}
where we use the orthogonality of $\mathsf{Q}$, namely, $\sum_{i=0}^{n}(\mathsf{Q})_{ai}(\mathsf{Q})_{bi}=\delta_{ab}$, in the last transform. This upper bound implies that the eigenvalues truncated in the embedding determine the error. The error becomes small if the following two conditions are satisfied: (i) the dimension $d_0$ is sufficiently large to express the trajectory that we embed, and (ii) the absolute values of the negative eigenvalues are enough small, i.e., the effect of non-Euclideanity is enough small.
}

\subsection{The property of the $1$-Wasserstein distance for the Fisher--KPP equation}\label{ap:L1_FisherKPP}

\subsubsection{The equivalence of the lengths in the Fisher--KPP equation and other simple reaction-diffusion systems}\label{ap:FisherKPP_equivalence}

In this section, we consider an RDS in $V\subset\mathbb{R}^d$, which satisfies $\mathcal{X}=\{1\}$ and $\mathcal{R}_{\mathcal{X}}=\{1\}$. The dynamics of the concentration distribution of the internal species $Z_1$ is given by a simple RD equation,
\begin{align}
    \partial_{t}c_1(\bm{r};t)=-\bm{\nabla}_{\bm{r}}\cdot\bm{J}_{(1)}+j_1.
\end{align}
The way to interact with the outside of the system and the boundary conditions can be whatever is appropriate. The Fisher--KPP equation is an example of such a system.

Here, we prove that each monotonic increase,
\begin{align}
    \partial_t c_1(\bm{r};t)\geq 0\;\;\mathrm{for\;all}\;\bm{r}\in V\; \mathrm{and}\; t\in[0,\tau],
    \label{FKPP monotonically increasing}
\end{align}
or decrease, 
\begin{align}
    \partial_t c_1(\bm{r};t)\leq 0\;\;\mathrm{for\;all}\;\bm{r}\in V\; \mathrm{and}\; t\in[0,\tau],
     \label{FKPP monotonically decreasing}
\end{align}
of the concentration leads to the equivalence of the lengths,
\begin{align}
    &l_{1,\tau}={W}_{1,\mathcal{X}}(\vec{c}(0),\vec{c}(\tau))\notag\\&={L}_{\mathcal{X}}(\vec{c}(0),\vec{c}(\tau))={L}_{\mathcal{X}}^{\mathrm{tot}}(\vec{c}(0),\vec{c}(\tau)).
    \label{equivalence_lengths general}
\end{align}

\add{
We first verify the hierarchy of lengths,
\begin{align}
    &l_{1,\tau}\geq{W}_{1,\mathcal{X}}(\vec{c}(0),\vec{c}(\tau))\notag\\&\geq{L}_{\mathcal{X}}(\vec{c}(0),\vec{c}(\tau))\geq{L}_{\mathcal{X}}^{\mathrm{tot}}(\vec{c}(0),\vec{c}(\tau)),
    \label{hierarchy of lengths}
\end{align}
to obtain the equivalence~\eqref{equivalence_lengths general}. It is enough to show the second inequality in Eq.~\eqref{hierarchy of lengths} since we can easily derive the first and third inequalities from the triangle inequality. To obtain the desired inequality, we take a potential $\phi'_1(\bm{r})$ as $\phi'_1(\bm{r})=1$ if the concentration is monotonically increasing~\eqref{FKPP monotonically increasing} or $\phi'_1(\bm{r})=-1$ if the concentration is monotonically decreasing~\eqref{FKPP monotonically decreasing}. This potential $(\phi'_1(\bm{r}),\vec{0}{}^{\top}_{\mathcal{Y}}){}^{\top}\eqqcolon\vec{\phi}{}'$ belongs to the set $\mathrm{Lip}^1_{\mathcal{X}}$~\eqref{RDS_1_Lip} because $\|\bm{\nabla}_{\bm{r}}\phi'_1\|\leq 1$ and $|\phi'_1|\leq 1$ hold. Thus, the potential $\vec{\phi}{}'$ is a candidate for the Kantorovich--Rubinstein duality~\eqref{RDS_Kantorovich--Rubinstein}, which leads
\begin{align}
    {W}_{1,\mathcal{X}}(\vec{c}(0),\vec{c}(\tau))&=\sup_{\vec{\phi}\in\mathrm{Lip}^1_{\mathcal{X}}}\cip{\vec{\phi}}{\vec{c}(\tau)-\vec{c}(0)}\notag\\
    &\geq\cip{\vec{\phi}{}'}{\vec{c}(\tau)-\vec{c}(0)}\notag\\
    &=\int_{V}d\bm{r}\,\phi'_1(\bm{r})\{c_1(\bm{r};\tau)-c_1(\bm{r};0)\}\notag\\
    &=\int_{V}d\bm{r}\,\left|c_1(\bm{r};\tau)-c_1(\bm{r};0)\right|\notag\\
    &=L_{\mathcal{X}}(\vec{c}(0),\vec{c}(\tau))
    \label{FisherKPP_W_1>=L}
\end{align}
}

\add{
Second, we show that the maximum length in the hierarchy equals the minimum one as
\begin{align}
    l_{1,\tau}=L_{\mathcal{X}}^{\mathrm{tot}}(\vec{c}(0),\vec{c}(\tau)).
    \label{FKKP_equivalence_maxmin}
\end{align}
The minimization problem for the $1$-Wasserstein distance~\eqref{another_Def_L1} provides the variational form of $v_1(t)$,
\begin{align}
    v_1(t)=\inf_{\bm{J}'_{(1)},j'_1}\int_{V}d\bm{r}\,\qty{\|\bm{J}_{(1)}'\|+|j_1'|},
    \label{v1_definition}
\end{align}
with the following condition: the currents $(\bm{J}'_{(1)},j'_1)$ satisfies
\begin{align}
    \partial_t c\qty(\bm{r};t)=-\bm{\nabla}_{\bm{r}}\cdot\bm{J}_{(1)}'\qty(\bm{r};t)+j_1'\qty(\bm{r};t).
    \label{v1condition}
\end{align}
We can obtain the lower bound of $v_1$ as
\begin{align}
    v_1(t)\geq\int_V d\bm{r}\,\left|\partial_{t}c_1\qty(\bm{r};t)\right|,
    \label{FisherKPP_lowerbound_v1}
\end{align}
by replacing $(0,\tau)$ with $(t, t+\varDelta t)$ in Eq.~\eqref{FisherKPP_W_1>=L}, dividing this equation by $\varDelta t$, and taking the limit $\varDelta t\to0$. This lower bound is achievable by taking $\bm{J}_{(1)}'=\bm{0}$ and $j_1'=\partial_t c_1$ in Eq.~\eqref{v1_definition}. Here, we can easily verify that $\bm{J}_{(1)}'=\bm{0}$ and $j_1'=\partial_t c_1$ satisfy the condition~\eqref{v1condition}. Thus, we obtain
\begin{align}
    v_1(t)=\int_V d\bm{r}\,\left|\partial_{t}c_1\qty(\bm{r};t)\right|.
    \label{FisherKPP_equality_v1}
\end{align}
Integrating Eq.~\eqref{FisherKPP_equality_v1} leads to the desired equality~\eqref{FKKP_equivalence_maxmin} as
\begin{align}
    l_{1,\tau}&=\int_{0}^{\tau}dt\,v_1(t)\notag\\
    &=\int_{0}^{\tau}dt\int_{V}d\bm{r}\,\left|\partial_{t}c_1\qty(\bm{r};t)\right|\notag\\
    &=\left|\int_{0}^{\tau}dt\int_{V}d\bm{r}\,\partial_{t}c_1\qty(\bm{r};t)\right|\notag\\
    &=\left|\int_{V}d\bm{r}\,\left\{c_1(\bm{r};\tau)-c_1(\bm{r};0)\right\}\right|\notag\\
    &=L_{\mathcal{X}}^{\mathrm{tot}}(\vec{c}(0),\vec{c}(\tau))
\end{align}
where we used the fact that the sign of $\partial_tc_1$ is invariant to obtain the third line. 
Combining Eq.~\eqref{hierarchy of lengths} and Eq.~\eqref{FKKP_equivalence_maxmin} yields the desired result in Eq.~\eqref{equivalence_lengths general}.
}

\subsubsection{The $1$-Wasserstein distance for the traveling wave solution in the Fisher--KPP equation}\label{ap:FisherKPP_chemical_wave}
In this section, we focus on the Fisher--KPP equation in the $1$-dimensional Euclidean space $\mathbb{R}$. Letting the concentration of the external species $c_2(r;t)=c_2$ be homogeneous, the time evolution of $c_1$ is given by 
\begin{align}
    \partial_tc_1 &=D_1\partial_r^2c_1+ (\kappa_1^{+}c_2) c_1 \left(1-\frac{c_1}{c^{\rm eq}_1} \right),
\end{align}
where $c^{\rm eq}_1$ is the equilibrium concentration given by $c^{\rm eq}_1 = \kappa_1^{+}c_2/\kappa_1^{-}$. For each wave speed $v_{\mathrm{wave}}\geq 2 \sqrt{D_1 \kappa^+_1 c_2}$, the equation admits traveling wave solutions~\cite{murray2002mathematical}, whose form is given by
\begin{align}
    c_1(r;t)=c_\mathrm{wave}(r - v_{\mathrm{wave}}t) =c_\mathrm{wave}(x).
    \label{travelingwave_solution}
\end{align}
The function $c_\mathrm{wave}(x)$ is a monotonically decreasing function of $x$, and thus 
\begin{align}
    \partial_x c_\mathrm{wave}(x)\leq0,
    \label{travelingwave_gradient}
\end{align}
holds for all $x\in\mathbb{R}$. The function $c_\mathrm{wave}(x)$ also satisfies the boundary conditions
\begin{align}
    \lim_{x\to-\infty}c_\mathrm{wave}(x)=c^{\mathrm{eq}}_1,\;\lim_{x\to\infty}c_\mathrm{wave}(x)=0.
    \label{travelingwave_limit}
\end{align}

For these traveling wave solutions $c_1(r;t)=c_\mathrm{wave}(r - v_{\mathrm{wave}}t)$, we can obtain the closed form of the lengths, 
\begin{align}
    &l_{1,\tau}={W}_{1,\mathcal{X}}(\vec{c}(0),\vec{c}(\tau))\notag\\&={L}_{\mathcal{X}}(\vec{c}(0),\vec{c}(\tau))={L}_{\mathcal{X}}^{\mathrm{tot}}(\vec{c}(0),\vec{c}(\tau))\notag\\
    &=c^{\rm eq}_1 v_{\mathrm{wave}}\tau.
    \label{travelingwave_length}
\end{align}
We can verify it as follows. Combining Eq.~\eqref{travelingwave_solution} and Eq.~\eqref{travelingwave_gradient} leads to
\begin{align}
    \partial_tc_1(r;t)=-v_{\mathrm{wave}}\partial_{x}c_{\rm wave}(x)|_{x=r-v_{\mathrm{wave}}t}\geq0,
    \label{travelingwave_timederivative}
\end{align}
for all $r \in\mathbb{R}$ and $t \in [0, \tau]$, which lets the traveling wave solutions satisfy the monotonicity~\eqref{FKPP monotonically increasing}. Thus, the result of the previous section~\eqref{equivalence_lengths general} confirms the first, second, and third equalities in Eq.~\eqref{travelingwave_length}. We can also verify the remaining part,
\begin{align}
    l_{1,\tau}= c^{\rm eq}_1 v_{\mathrm{wave}}\tau,
    \label{v_1=v_wave}
\end{align}
using Eq.~\eqref{FisherKPP_equality_v1} as
\begin{align}
    l_{1,\tau}&=\int_{0}^{\tau}dt\int_{\mathbb{R}}dr\,\left|\partial_{t}c_1\qty(r;t)\right| \notag \\
    &= - v_{\rm wave} \int_{0}^{\tau}dt\int_{-\infty}^{\infty}dx\,\partial_{x}c_{\rm wave}(x)
    \notag \\
    &=c^{\rm eq}_1 v_{\mathrm{wave}}\tau,
\end{align}
where we used Eq.~\eqref{travelingwave_limit} in the space integration.

\section{The derivation of the trade-off relation in Eq.~\eqref{TSL_from_TUR}}\label{ap:derivation_trade-off}
We provide the derivation of Eq.~\eqref{TSL_from_TUR}. The triangle inequality for the time integration leads to
\begin{align}\label{tri_int}
    \left|\tilde{c}_{\alpha}(\bm{k};\tau)-\tilde{c}_{\alpha}(\bm{k};0)\right|&=\left|\int_{0}^{\tau}dt\qty[d_t\tilde{c}_{\alpha}(\bm{k};t)]\right|\notag\\&\leq\int_{0}^{\tau}dt\left|d_t\tilde{c}_{\alpha}(\bm{k};t)\right|.
\end{align}
The right-hand side of Eq.~\eqref{tri_int} has an upper bound from the TUR~\eqref{Fourier_TUR} as
\begin{align}\label{UB}
\int_{0}^{\tau}dt\left|d_t \tilde{c}_{\alpha}(\bm{k};t)\right|\leq\int_{0}^{\tau}dt\sqrt{\sigma^{\mathrm{ex}}}\sqrt{\bm{k}\cdot\mathsf{M}_{(\alpha\alpha)}^{\mathrm{tot}}(t)\bm{k}+\scd^{\mathrm{tot}}_{\alpha\alpha}(t)}.
\end{align}
The Cauchy--Schwarz inequality provides an upper bound for the right-hand side of Eq.~\eqref{UB} as
\begin{align}\label{CS_in_TUR}
&\int_{0}^{\tau}dt\sqrt{\sigma^{\mathrm{ex}}}\sqrt{\bm{k}\cdot\mathsf{M}_{(\alpha\alpha)}^{\mathrm{tot}}(t)\bm{k}+\scd^{\mathrm{tot}}_{\alpha\alpha}(t)}
\notag
\\&\leq\sqrt{\int_{0}^{\tau}dt\sigma^{\mathrm{ex}}}\sqrt{\int_{0}^{\tau}dt\qty[\bm{k}\cdot\mathsf{M}_{(\alpha\alpha)}^{\mathrm{tot}}(t)\bm{k}+\scd^{\mathrm{tot}}_{\alpha\alpha}(t)]}\notag
\\
&=\sqrt{\tau\Sigma_{\tau}^{\mathrm{ex}}\qty[\bm{k}\cdot\langle\mathsf{M}_{(\alpha\alpha)}^{\mathrm{tot}}\rangle_{\tau}\bm{k}+\langle\scd^{\mathrm{tot}}_{\alpha\alpha}\rangle_{\tau}]}.
\end{align}
We obtain Eq.~\eqref{TSL_from_TUR} by unifying Eqs.~\eqref{tri_int}, ~\eqref{UB} and ~\eqref{CS_in_TUR}.

\section{Complementary numerical demonstration of the thermodynamic uncertainty relation for $\tilde{c}_1(k;t)$ in the Brusselator}\label{ap:numex_anotherTUR}

\begin{figure*}
    \centering
    \includegraphics[width=\linewidth]{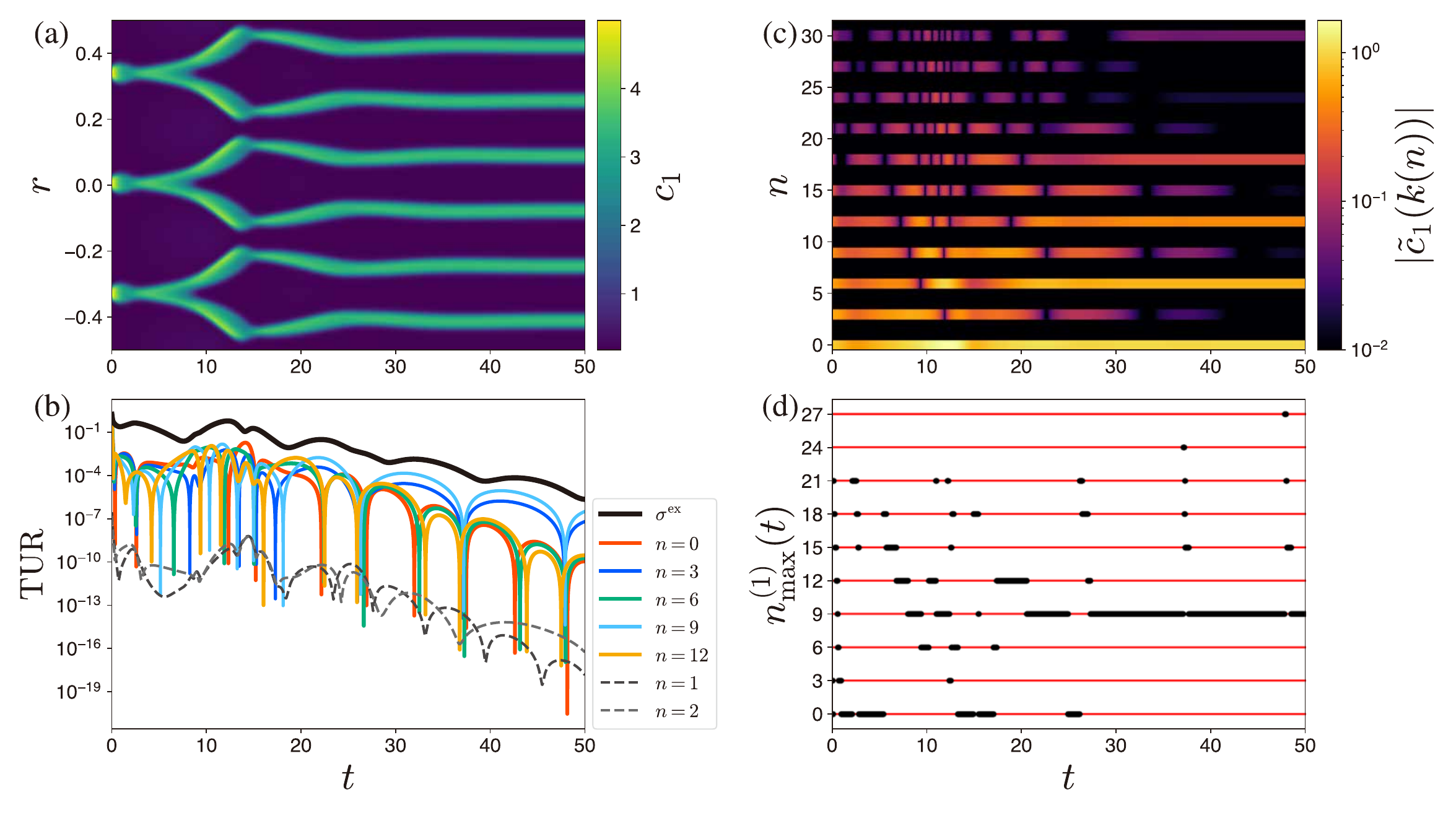}
    \caption{(a) The time series of $c_{1}$. The symmetry of the pattern changes from $3$-fold to $6$-fold. (b) The excess EPR (black line) and its lower bounds $\sigma^{\mathrm{TUR}}_{1}\qty(n;t)$ for various $n$. (c) The time series of $|\tilde{c}_{1}(k(n))|$. We omit $|\tilde{c}_{1}(k(n))|$ for $n\geq 32$. Since the symmetry of the pattern goes from $3$-fold to $6$-fold, $|\tilde{c}_1(k(n))|$ decays if $n$ is a multiple of three but not a multiple of six. (d) $n^{(1)}_{\mathrm{max}}(t)$ (black dots). Reflecting the symmetry of the pattern, $n^{(1)}_{\mathrm{max}}(t)$ is multiples of three (red lines) for all time $t$. Near the stationary pattern ($t>20$), $n^{(1)}_{\mathrm{max}}(t)$ is nine for almost all $t$.}
    \label{fig:TUR_numerics}
\end{figure*}

In Fig.~\ref{fig:TUR_numerics}, we show the TURs $\sigma^{\mathrm{TUR}}_{1}(n;t)$  for various $n$ and scatter plots of $n^{(1)}_{\mathrm{max}}(t)$, corresponding to the chemical species $Z_1$ ($\alpha=1$). Compared to the case of $\alpha=2$, the mode $n^{(1)}_{\mathrm{max}}$ tends to be larger than $n^{(2)}_{\mathrm{max}}$. This is probably because the pattern of $c_1$ has more extreme peaks than the pattern of $c_2$, resulting in a large time variation of the mode corresponding to a large $n$.

\input{biblio.bbl}

\end{document}

%% file: preamble_arxiv.tex
\usepackage{newtxtext}
\usepackage{amsmath}
\usepackage{amssymb}
\usepackage{mathtools}
\usepackage{mathrsfs}
\usepackage{scalerel}
\usepackage{physics}
\usepackage{cancel}
\usepackage{bm}
\usepackage{graphicx}
\usepackage{xcolor}
\usepackage{colortbl}
\usepackage{cases}
\usepackage[version=4]{mhchem}
\usepackage[hidelinks]{hyperref}
\hypersetup{
  colorlinks   = true, 
  urlcolor     = blue,
  linkcolor    = blue, 
  citecolor   = blue 
}

\newcommand{\add}[1]{#1}

\makeatletter
\DeclareFontFamily{OMX}{MnSymbolE}{}
\DeclareSymbolFont{MnLargeSymbols}{OMX}{MnSymbolE}{m}{n}
\SetSymbolFont{MnLargeSymbols}{bold}{OMX}{MnSymbolE}{b}{n}
\DeclareFontShape{OMX}{MnSymbolE}{m}{n}{
    <-6>  MnSymbolE5
   <6-7>  MnSymbolE6
   <7-8>  MnSymbolE7
   <8-9>  MnSymbolE8
   <9-10> MnSymbolE9
  <10-12> MnSymbolE10
  <12->   MnSymbolE12
}{}
\DeclareFontShape{OMX}{MnSymbolE}{b}{n}{
    <-6>  MnSymbolE-Bold5
   <6-7>  MnSymbolE-Bold6
   <7-8>  MnSymbolE-Bold7
   <8-9>  MnSymbolE-Bold8
   <9-10> MnSymbolE-Bold9
  <10-12> MnSymbolE-Bold10
  <12->   MnSymbolE-Bold12
}{}
\let\llangle\@undefined
\let\rrangle\@undefined
\DeclareMathDelimiter{\llangle}{\mathopen}%
                     {MnLargeSymbols}{'164}{MnLargeSymbols}{'164}
\DeclareMathDelimiter{\rrangle}{\mathclose}%
                     {MnLargeSymbols}{'171}{MnLargeSymbols}{'171}
\makeatother

\newcommand{\cip}[2]{\left\langle #1, #2 \right\rangle}
\newcommand{\fip}[2]{\left\llangle #1, #2 \right\rrangle}

\newcommand{\Grad}{\nabla}
\newcommand{\Div}{\nabla^{\dag}}
\newcommand{\F}{\mathcal{F}}
\newcommand{\J}{\mathcal{J}}
\newcommand{\U}{\mathcal{U}}
\newcommand{\M}{\mathcal{M}}

\newcommand{\RD}{\mathrm{RD}}
\newcommand{\scd}{\breve{D}}
\newcommand{\mmax}{M_{0}}
\newcommand{\mnorm}[1]{|#1|^{\mathrm{tot}}_{\mathcal{X}}}
\newcommand{\diffmax}{M^{\mathrm{max}}_{\mathcal{X}}}

\DeclareMathOperator*{\argmax}{arg\,max}
\DeclareMathOperator*{\argmin}{arg\,min}

\newcommand{\vecdisp}[1]{\kern-.15pt\overset{\kern.7pt\smash{\lower.25ex\hbox{\scalebox{.45}{$\bm{\rightarrow}$}}}\kern-.3pt}{#1}}
\newcommand{\vecsub}[1]{\overset{\smash{\lower.15ex\hbox{\kern.2pt\scalebox{.3}{$\bm{\rightarrow}$}\kern-.8pt}}}{#1}}
\renewcommand{\vec}[1]{\mathchoice{\vecdisp{#1}}{\vecdisp{#1}}{\vecsub{#1}}{\vecsub{#1}}}
\newcommand{\dvec}[1]{\overset{\smash{\lower.25ex\hbox{\scalebox{.7}[.45]{$\bm{\leftrightarrow}$}}}}{#1}}
\newcommand{\tilvec}[1]{\overset{\smash{\lower.1em\hbox{\,\protect\hstretch{.5}{\protect\vstretch{.4}{\bm{\rightsquigarrow}}}}}}{#1}}
\newcommand{\dtilvec}[1]{\overset{\smash{\lower.1em\hbox{\protect\hstretch{.5}{\protect\vstretch{.4}{\bm{\leftrightsquigarrow}}}}}}{#1}}

\newcommand{\vecrmdisp}[1]{\overset{\smash{\lower.25ex\hbox{\scalebox{.45}{$\bm{\rightarrow}$}}}}{#1}}
\newcommand{\vecrmsub}[1]{\overset{\smash{\lower.15ex\hbox{\scalebox{.3}{$\bm{\rightarrow}$}}}}{#1}}
\newcommand{\vecrm}[1]{\mathchoice{\vecrmdisp{#1}}{\vecrmdisp{#1}}{\vecrmsub{#1}}{\vecrmsub{#1}}}

%% file: biblio.bbl
%